\documentclass[12pt,english]{article}
\usepackage{mathpazo}
\usepackage[latin9]{inputenc}
\usepackage[letterpaper]{geometry}
\usepackage{babel}
\usepackage{verbatim}
\usepackage{float}
\usepackage{amsmath}
\usepackage{amssymb}
\usepackage{graphicx}
\usepackage{esint}
\usepackage[unicode=true,
 bookmarks=true,bookmarksnumbered=false,bookmarksopen=false,
 breaklinks=false,pdfborder={0 0 1},backref=false,colorlinks=false]
 {hyperref}

\makeatletter

\let\SF@@footnote\footnote
\def\footnote{\ifx\protect\@typeset@protect
    \expandafter\SF@@footnote
  \else
    \expandafter\SF@gobble@opt
  \fi
}
\expandafter\def\csname SF@gobble@opt \endcsname{\@ifnextchar[
  \SF@gobble@twobracket
  \@gobble
}
\edef\SF@gobble@opt{\noexpand\protect
  \expandafter\noexpand\csname SF@gobble@opt \endcsname}
\def\SF@gobble@twobracket[#1]#2{}
\providecommand{\tabularnewline}{\\}

\numberwithin{equation}{section}

\@ifundefined{date}{}{\date{}}

\usepackage[svgnames]{xcolor}
\usepackage{tikz}
\usetikzlibrary{decorations.markings}
\usetikzlibrary{shapes.geometric}
\usetikzlibrary{calc}
\usetikzlibrary{arrows}

\pgfdeclarelayer{edgelayer}
\pgfdeclarelayer{nodelayer}
\pgfsetlayers{edgelayer,nodelayer,main}

\tikzstyle{none}=[inner sep=0pt]

\tikzstyle{Vertex}=[circle,fill=black,draw=black]
\tikzstyle{Node}=[rectangle,fill=blue,draw=blue]
\tikzstyle{Point}=[circle,fill=red,draw=red,scale=0.33]

\tikzstyle{Edge}=[-,draw=black]
\tikzstyle{Link}=[-,dashed,draw=blue]
\tikzstyle{Segment}=[-,dotted,very thick,draw=red]
\tikzstyle{LinkArrow}=[-,draw=blue,
  postaction={decorate},decoration={markings,
  mark=at position .66 with {\arrow{latex}}}]
\tikzstyle{SegmentArrow}=[-,draw=red,
  postaction={decorate},decoration={markings,
  mark=at position .66 with {\arrow{latex}}}]
\def\centerarc [#1] (#2) (#3:#4:#5); { \draw[#1] ($(#2)+({#5*cos(#3)},{#5*sin(#3)})$) arc (#3:#4:#5); }
\newcommand{\midarrow}{\tikz \draw[-triangle 90] (0,0) -- +(.1,0);}
\newcommand{\midarrowop}{\tikz \draw[-triangle 90] (0,0) -- +(-.1,0);}

\renewcommand\[{\begin{equation}}
\renewcommand\]{\end{equation}}

\usepackage[margin=1cm]{caption}

\DeclareMathOperator{\e}{e}

\DeclareMathOperator{\ii}{i}

\DeclareMathOperator{\sign}{sign}

\makeatother

\begin{document}

\global\long\def\A{\mathbf{A}}%
\global\long\def\B{\mathbf{B}}%
\global\long\def\C{\mathbf{C}}%
\global\long\def\D{\mathbf{D}}%
\global\long\def\E{\mathbf{E}}%
\global\long\def\F{\mathbf{F}}%
\global\long\def\G{\mathbf{G}}%
\global\long\def\H{\mathbf{H}}%
\global\long\def\I{\mathbf{I}}%
\global\long\def\J{\mathbf{J}}%
\global\long\def\K{\mathbf{K}}%
\global\long\def\LL{\mathbf{L}}%
\global\long\def\M{\mathbf{M}}%
\global\long\def\N{\mathbf{N}}%
\global\long\def\OO{\mathbf{O}}%
\global\long\def\P{\mathbf{P}}%
\global\long\def\Q{\mathbf{Q}}%
\global\long\def\RR{\mathbf{R}}%
\global\long\def\SS{\mathbf{S}}%
\global\long\def\T{\mathbf{T}}%
\global\long\def\U{\mathbf{U}}%
\global\long\def\V{\mathbf{V}}%
\global\long\def\W{\mathbf{W}}%
\global\long\def\X{\mathbf{X}}%
\global\long\def\Y{\mathbf{Y}}%
\global\long\def\Z{\mathbf{Z}}%

\global\long\def\a{\mathbf{a}}%
\global\long\def\b{\mathbf{b}}%
\global\long\def\c{\mathbf{c}}%
\global\long\def\dd{\mathbf{d}}%
\global\long\def\ee{\mathbf{e}}%
\global\long\def\f{\mathbf{f}}%
\global\long\def\g{\mathbf{g}}%
\global\long\def\h{\mathbf{h}}%
\global\long\def\iii{\mathbf{i}}%
\global\long\def\j{\mathbf{j}}%
\global\long\def\k{\mathbf{k}}%
\global\long\def\l{\boldsymbol{l}}%
\global\long\def\el{\boldsymbol{\ell}}%
\global\long\def\m{\mathbf{m}}%
\global\long\def\n{\mathbf{n}}%
\global\long\def\o{\mathbf{o}}%
\global\long\def\p{\mathbf{p}}%
\global\long\def\q{\mathbf{q}}%
\global\long\def\r{\mathbf{r}}%
\global\long\def\s{\mathbf{s}}%
\global\long\def\t{\mathbf{t}}%
\global\long\def\u{\mathbf{u}}%
\global\long\def\v{\mathbf{v}}%
\global\long\def\w{\mathbf{w}}%
\global\long\def\x{\mathbf{x}}%
\global\long\def\y{\mathbf{y}}%
\global\long\def\z{\mathbf{z}}%

\global\long\def\Ga{\boldsymbol{\Gamma}}%
\global\long\def\De{\boldsymbol{\Delta}}%
\global\long\def\Th{\boldsymbol{\Theta}}%
\global\long\def\La{\boldsymbol{\Lambda}}%
\global\long\def\Xii{\boldsymbol{\Xi}}%
\global\long\def\Pii{\boldsymbol{\Pi}}%
\global\long\def\Si{\boldsymbol{\Sigma}}%
\global\long\def\Ph{\boldsymbol{\Phi}}%
\global\long\def\Ps{\boldsymbol{\Psi}}%
\global\long\def\Om{\boldsymbol{\Omega}}%

\global\long\def\al{\boldsymbol{\alpha}}%
\global\long\def\be{\boldsymbol{\beta}}%
\global\long\def\ga{\boldsymbol{\gamma}}%
\global\long\def\de{\boldsymbol{\delta}}%
\global\long\def\ep{\boldsymbol{\epsilon}}%
\global\long\def\vep{\boldsymbol{\varepsilon}}%
\global\long\def\ze{\boldsymbol{\zeta}}%
\global\long\def\et{\boldsymbol{\eta}}%
\global\long\def\th{\boldsymbol{\theta}}%
\global\long\def\io{\boldsymbol{\iota}}%
\global\long\def\ka{\boldsymbol{\kappa}}%
\global\long\def\la{\boldsymbol{\lambda}}%
\global\long\def\muu{\boldsymbol{\mu}}%
\global\long\def\nuu{\boldsymbol{\nu}}%
\global\long\def\xii{\boldsymbol{\xi}}%
\global\long\def\pii{\boldsymbol{\pi}}%
\global\long\def\rhh{\boldsymbol{\rho}}%
\global\long\def\si{\boldsymbol{\sigma}}%
\global\long\def\ta{\boldsymbol{\tau}}%
\global\long\def\ups{\boldsymbol{\upsilon}}%
\global\long\def\ph{\boldsymbol{\phi}}%
\global\long\def\vph{\boldsymbol{\varphi}}%
\global\long\def\ch{\boldsymbol{\chi}}%
\global\long\def\ps{\boldsymbol{\psi}}%
\global\long\def\om{\boldsymbol{\omega}}%

\global\long\def\AAb{\boldsymbol{\mathcal{A}}}%
\global\long\def\BBb{\boldsymbol{\mathcal{B}}}%
\global\long\def\CCb{\boldsymbol{\mathcal{C}}}%
\global\long\def\DDb{\boldsymbol{\mathcal{D}}}%
\global\long\def\EEb{\boldsymbol{\mathcal{E}}}%
\global\long\def\FFb{\boldsymbol{\mathcal{F}}}%
\global\long\def\GGb{\boldsymbol{\mathcal{G}}}%
\global\long\def\HHb{\boldsymbol{\mathcal{H}}}%
\global\long\def\IIb{\boldsymbol{\mathcal{I}}}%
\global\long\def\JJb{\boldsymbol{\mathcal{J}}}%
\global\long\def\KKb{\boldsymbol{\mathcal{K}}}%
\global\long\def\LLb{\boldsymbol{\mathcal{L}}}%
\global\long\def\MMb{\boldsymbol{\mathcal{M}}}%
\global\long\def\NNb{\boldsymbol{\mathcal{N}}}%
\global\long\def\OOb{\boldsymbol{\mathcal{O}}}%
\global\long\def\PPb{\boldsymbol{\mathcal{P}}}%
\global\long\def\QQb{\boldsymbol{\mathcal{Q}}}%
\global\long\def\RRb{\boldsymbol{\mathcal{R}}}%
\global\long\def\SSb{\boldsymbol{\mathcal{S}}}%
\global\long\def\TTb{\boldsymbol{\mathcal{T}}}%
\global\long\def\UUb{\boldsymbol{\mathcal{U}}}%
\global\long\def\VVb{\boldsymbol{\mathcal{V}}}%
\global\long\def\WWb{\boldsymbol{\mathcal{W}}}%
\global\long\def\XXb{\boldsymbol{\mathcal{X}}}%
\global\long\def\YYb{\boldsymbol{\mathcal{Y}}}%
\global\long\def\ZZb{\boldsymbol{\mathcal{Z}}}%

\global\long\def\Ab{\bar{A}}%
\global\long\def\Bb{\bar{B}}%
\global\long\def\Cb{\bar{C}}%
\global\long\def\Db{\bar{D}}%
\global\long\def\Eb{\bar{E}}%
\global\long\def\Fb{\bar{F}}%
\global\long\def\Gb{\bar{G}}%
\global\long\def\Hb{\bar{H}}%
\global\long\def\Ib{\bar{I}}%
\global\long\def\Jb{\bar{J}}%
\global\long\def\Kb{\bar{K}}%
\global\long\def\Lb{\bar{L}}%
\global\long\def\Mb{\bar{M}}%
\global\long\def\Nb{\bar{N}}%
\global\long\def\Ob{\bar{O}}%
\global\long\def\Pb{\bar{P}}%
\global\long\def\Qb{\bar{Q}}%
\global\long\def\Rb{\bar{R}}%
\global\long\def\Sb{\bar{S}}%
\global\long\def\Tb{\bar{T}}%
\global\long\def\Ub{\bar{U}}%
\global\long\def\Vb{\bar{V}}%
\global\long\def\Wb{\bar{W}}%
\global\long\def\Xb{\bar{X}}%
\global\long\def\Yb{\bar{Y}}%
\global\long\def\Zb{\bar{Z}}%

\global\long\def\ab{\bar{a}}%
\global\long\def\bb{\bar{b}}%
\global\long\def\cb{\bar{c}}%
\global\long\def\db{\bar{d}}%
\global\long\def\eb{\bar{e}}%
\global\long\def\fb{\bar{f}}%
\global\long\def\gb{\bar{g}}%
\global\long\def\hb{\bar{h}}%
\global\long\def\ib{\bar{i}}%
\global\long\def\jb{\bar{j}}%
\global\long\def\kb{\bar{k}}%
\global\long\def\lb{\bar{l}}%
\global\long\def\elb{\bar{\ell}}%
\global\long\def\mb{\bar{m}}%
\global\long\def\nb{\bar{n}}%
\global\long\def\ob{\bar{o}}%
\global\long\def\pb{\bar{p}}%
\global\long\def\qb{\bar{q}}%
\global\long\def\rb{\bar{r}}%
\global\long\def\ssb{\bar{s}}%
\global\long\def\tb{\bar{t}}%
\global\long\def\ub{\bar{u}}%
\global\long\def\vb{\bar{v}}%
\global\long\def\wb{\bar{w}}%
\global\long\def\xb{\bar{x}}%
\global\long\def\yb{\bar{y}}%
\global\long\def\zb{\bar{z}}%

\global\long\def\Gab{\bar{\Gamma}}%
\global\long\def\Deb{\bar{\Delta}}%
\global\long\def\Thb{\bar{\Theta}}%
\global\long\def\Lab{\bar{\Lambda}}%
\global\long\def\Xib{\bar{\Xi}}%
\global\long\def\Pib{\bar{\Pi}}%
\global\long\def\Sib{\bar{\Sigma}}%
\global\long\def\Phb{\bar{\Phi}}%
\global\long\def\Psb{\bar{\Psi}}%
\global\long\def\Thb{\bar{\Theta}}%

\global\long\def\alb{\bar{\alpha}}%
\global\long\def\beb{\bar{\beta}}%
\global\long\def\gab{\bar{\gamma}}%
\global\long\def\deb{\bar{\delta}}%
\global\long\def\epb{\bar{\epsilon}}%
\global\long\def\vepb{\bar{\varepsilon}}%
\global\long\def\zeb{\bar{\zeta}}%
\global\long\def\etb{\bar{\eta}}%
\global\long\def\thb{\bar{\theta}}%
\global\long\def\iob{\bar{\iota}}%
\global\long\def\kab{\bar{\kappa}}%
\global\long\def\lab{\bar{\lambda}}%
\global\long\def\mub{\bar{\mu}}%
\global\long\def\nub{\bar{\nu}}%
\global\long\def\xib{\bar{\xi}}%
\global\long\def\pib{\bar{\pi}}%
\global\long\def\rhb{\bar{\rho}}%
\global\long\def\sib{\bar{\sigma}}%
\global\long\def\tab{\bar{\tau}}%
\global\long\def\upb{\bar{\upsilon}}%
\global\long\def\phb{\bar{\phi}}%
\global\long\def\vphb{\bar{\varphi}}%
\global\long\def\chb{\bar{\chi}}%
\global\long\def\psb{\bar{\psi}}%
\global\long\def\omb{\bar{\omega}}%

\global\long\def\adt{\dot{a}}%
\global\long\def\add{\ddot{a}}%
\global\long\def\bd{\dot{b}}%
\global\long\def\bdd{\ddot{b}}%
\global\long\def\cd{\dot{c}}%
\global\long\def\cdd{\ddot{c}}%
\global\long\def\ddd{\dot{d}}%
\global\long\def\dddd{\ddot{d}}%
\global\long\def\ed{\dot{e}}%
\global\long\def\edd{\ddot{e}}%
\global\long\def\fd{\dot{f}}%
\global\long\def\fdd{\ddot{f}}%
\global\long\def\gd{\dot{g}}%
\global\long\def\gdd{\ddot{g}}%
\global\long\def\hd{\dot{h}}%
\global\long\def\hdd{\ddot{h}}%
\global\long\def\kd{\dot{k}}%
\global\long\def\kdd{\ddot{k}}%
\global\long\def\ld{\dot{l}}%
\global\long\def\ldd{\ddot{l}}%
\global\long\def\eld{\dot{\ell}}%
\global\long\def\eldd{\ddot{\ell}}%
\global\long\def\md{\dot{m}}%
\global\long\def\mdd{\ddot{m}}%
\global\long\def\nd{\dot{n}}%
\global\long\def\ndd{\ddot{n}}%
\global\long\def\od{\dot{o}}%
\global\long\def\odd{\ddot{o}}%
\global\long\def\pd{\dot{p}}%
\global\long\def\pdd{\ddot{p}}%
\global\long\def\qd{\dot{q}}%
\global\long\def\qdd{\ddot{q}}%
\global\long\def\rd{\dot{r}}%
\global\long\def\rdd{\ddot{r}}%
\global\long\def\sd{\dot{s}}%
\global\long\def\sdd{\ddot{s}}%
\global\long\def\td{\dot{t}}%
\global\long\def\tdd{\ddot{t}}%
\global\long\def\ud{\dot{u}}%
\global\long\def\udd{\ddot{u}}%
\global\long\def\vd{\dot{v}}%
\global\long\def\vdd{\ddot{v}}%
\global\long\def\wdt{\dot{w}}%
\global\long\def\wdd{\ddot{w}}%
\global\long\def\xd{\dot{x}}%
\global\long\def\xdd{\ddot{x}}%
\global\long\def\yd{\dot{y}}%
\global\long\def\ydd{\ddot{y}}%
\global\long\def\zd{\dot{z}}%
\global\long\def\zdd{\ddot{z}}%

\global\long\def\Adt{\dot{A}}%
\global\long\def\Add{\ddot{A}}%
\global\long\def\Bd{\dot{B}}%
\global\long\def\Bdd{\ddot{B}}%
\global\long\def\Cd{\dot{C}}%
\global\long\def\Cdd{\ddot{C}}%
\global\long\def\Dd{\dot{D}}%
\global\long\def\Ddd{\ddot{D}}%
\global\long\def\Ed{\dot{E}}%
\global\long\def\Edd{\ddot{E}}%
\global\long\def\Fd{\dot{F}}%
\global\long\def\Fdd{\ddot{F}}%
\global\long\def\Gd{\dot{G}}%
\global\long\def\Gdd{\ddot{G}}%
\global\long\def\Hd{\dot{H}}%
\global\long\def\Hdd{\ddot{H}}%
\global\long\def\Id{\dot{I}}%
\global\long\def\Idd{\ddot{I}}%
\global\long\def\Jd{\dot{J}}%
\global\long\def\Jdd{\ddot{J}}%
\global\long\def\Kd{\dot{K}}%
\global\long\def\Kdd{\ddot{K}}%
\global\long\def\Ld{\dot{L}}%
\global\long\def\Ldd{\ddot{L}}%
\global\long\def\Md{\dot{M}}%
\global\long\def\Mdd{\ddot{M}}%
\global\long\def\Nd{\dot{N}}%
\global\long\def\Ndd{\ddot{N}}%
\global\long\def\Od{\dot{O}}%
\global\long\def\Odd{\ddot{O}}%
\global\long\def\Pd{\dot{P}}%
\global\long\def\Pdd{\ddot{P}}%
\global\long\def\Qd{\dot{Q}}%
\global\long\def\Qdd{\ddot{Q}}%
\global\long\def\Rd{\dot{R}}%
\global\long\def\Rdd{\ddot{R}}%
\global\long\def\Sd{\dot{S}}%
\global\long\def\Sdd{\ddot{S}}%
\global\long\def\Td{\dot{T}}%
\global\long\def\Tdd{\ddot{T}}%
\global\long\def\Ud{\dot{U}}%
\global\long\def\Udd{\ddot{U}}%
\global\long\def\Vd{\dot{R}}%
\global\long\def\Vdd{\ddot{R}}%
\global\long\def\Wd{\dot{W}}%
\global\long\def\Wdd{\ddot{W}}%
\global\long\def\Xd{\dot{X}}%
\global\long\def\Xdd{\ddot{X}}%
\global\long\def\Yd{\dot{Y}}%
\global\long\def\Ydd{\ddot{Y}}%
\global\long\def\Zd{\dot{Z}}%
\global\long\def\Zdd{\ddot{Z}}%

\global\long\def\Gad{\dot{\Gamma}}%
\global\long\def\Gadd{\ddot{\Gamma}}%
\global\long\def\Ded{\dot{\Delta}}%
\global\long\def\Dedd{\ddot{\Delta}}%
\global\long\def\Thd{\dot{\Theta}}%
\global\long\def\Thdd{\ddot{\Theta}}%
\global\long\def\Lad{\dot{\Lambda}}%
\global\long\def\Ladd{\ddot{\Lambda}}%
\global\long\def\Xid{\dot{\Xi}}%
\global\long\def\Xidd{\ddot{\Xi}}%
\global\long\def\Pid{\dot{\Pi}}%
\global\long\def\Pidd{\ddot{\Pi}}%
\global\long\def\Sid{\dot{\Sigma}}%
\global\long\def\Sidd{\ddot{\Sigma}}%
\global\long\def\Phd{\dot{\Phi}}%
\global\long\def\Phdd{\ddot{\Phi}}%
\global\long\def\Psd{\dot{\Psi}}%
\global\long\def\Psdd{\ddot{\Psi}}%
\global\long\def\Thd{\dot{\Theta}}%
\global\long\def\Thdd{\ddot{\Theta}}%

\global\long\def\ald{\dot{\alpha}}%
\global\long\def\aldd{\ddot{\alpha}}%
\global\long\def\bed{\dot{\beta}}%
\global\long\def\bedd{\ddot{\beta}}%
\global\long\def\gad{\dot{\gamma}}%
\global\long\def\gadd{\ddot{\gamma}}%
\global\long\def\ded{\dot{\delta}}%
\global\long\def\dedd{\ddot{\delta}}%
\global\long\def\epd{\dot{\epsilon}}%
\global\long\def\epdd{\ddot{\epsilon}}%
\global\long\def\vepd{\dot{\varepsilon}}%
\global\long\def\vepdd{\ddot{\varepsilon}}%
\global\long\def\zed{\dot{\zeta}}%
\global\long\def\zedd{\ddot{\zeta}}%
\global\long\def\etd{\dot{\eta}}%
\global\long\def\etdd{\ddot{\eta}}%
\global\long\def\thd{\dot{\theta}}%
\global\long\def\thdd{\ddot{\theta}}%
\global\long\def\iod{\dot{\iota}}%
\global\long\def\iodd{\ddot{\iota}}%
\global\long\def\kad{\dot{\kappa}}%
\global\long\def\kadd{\ddot{\kappa}}%
\global\long\def\lad{\dot{\lambda}}%
\global\long\def\ladd{\ddot{\lambda}}%
\global\long\def\mud{\dot{\mu}}%
\global\long\def\mudd{\ddot{\mu}}%
\global\long\def\nud{\dot{\nu}}%
\global\long\def\nudd{\ddot{\nu}}%
\global\long\def\xid{\dot{\xi}}%
\global\long\def\xidd{\ddot{\xi}}%
\global\long\def\pid{\dot{\pi}}%
\global\long\def\pidd{\ddot{\pi}}%
\global\long\def\rhod{\dot{\rho}}%
\global\long\def\rhodd{\ddot{\rho}}%
\global\long\def\sid{\dot{\sigma}}%
\global\long\def\sidd{\ddot{\sigma}}%
\global\long\def\tad{\dot{\tau}}%
\global\long\def\tadd{\ddot{\tau}}%
\global\long\def\upd{\dot{\upsilon}}%
\global\long\def\updd{\ddot{\upsilon}}%
\global\long\def\phd{\dot{\phi}}%
\global\long\def\phdd{\ddot{\phi}}%
\global\long\def\vpd{\dot{\varphi}}%
\global\long\def\vpdd{\ddot{\varphi}}%
\global\long\def\chd{\dot{\chi}}%
\global\long\def\chdd{\ddot{\chi}}%
\global\long\def\psd{\dot{\psi}}%
\global\long\def\psdd{\ddot{\psi}}%
\global\long\def\omd{\dot{\omega}}%
\global\long\def\omdd{\ddot{\omega}}%

\global\long\def\BBA{\mathbb{A}}%
\global\long\def\BBB{\mathbb{B}}%
\global\long\def\BBC{\mathbb{C}}%
\global\long\def\BBD{\mathbb{D}}%
\global\long\def\BBE{\mathbb{E}}%
\global\long\def\BBF{\mathbb{F}}%
\global\long\def\BBG{\mathbb{G}}%
\global\long\def\BBH{\mathbb{H}}%
\global\long\def\BBI{\mathbb{I}}%
\global\long\def\BBJ{\mathbb{J}}%
\global\long\def\BBK{\mathbb{K}}%
\global\long\def\BBL{\mathbb{L}}%
\global\long\def\BBM{\mathbb{M}}%
\global\long\def\BBN{\mathbb{N}}%
\global\long\def\BBO{\mathbb{O}}%
\global\long\def\BBP{\mathbb{P}}%
\global\long\def\BBQ{\mathbb{Q}}%
\global\long\def\BBR{\mathbb{R}}%
\global\long\def\BBS{\mathbb{S}}%
\global\long\def\BBT{\mathbb{T}}%
\global\long\def\BBU{\mathbb{U}}%
\global\long\def\BBV{\mathbb{V}}%
\global\long\def\BBW{\mathbb{W}}%
\global\long\def\BBX{\mathbb{X}}%
\global\long\def\BBY{\mathbb{Y}}%
\global\long\def\BBZ{\mathbb{Z}}%

\global\long\def\AA{\mathcal{A}}%
\global\long\def\BB{\mathcal{B}}%
\global\long\def\CC{\mathcal{C}}%
\global\long\def\DD{\mathcal{D}}%
\global\long\def\EE{\mathcal{E}}%
\global\long\def\FF{\mathcal{F}}%
\global\long\def\GG{\mathcal{G}}%
\global\long\def\HH{\mathcal{H}}%
\global\long\def\II{\mathcal{I}}%
\global\long\def\JJ{\mathcal{J}}%
\global\long\def\KK{\mathcal{K}}%
\global\long\def\LLL{\mathcal{L}}%
\global\long\def\MM{\mathcal{M}}%
\global\long\def\NN{\mathcal{N}}%
\global\long\def\OOO{\mathcal{O}}%
\global\long\def\PP{\mathcal{P}}%
\global\long\def\QQ{\mathcal{Q}}%
\global\long\def\RRR{\mathcal{R}}%
\global\long\def\SSS{\mathcal{S}}%
\global\long\def\TT{\mathcal{T}}%
\global\long\def\UU{\mathcal{U}}%
\global\long\def\VV{\mathcal{V}}%
\global\long\def\WW{\mathcal{W}}%
\global\long\def\XX{\mathcal{X}}%
\global\long\def\YY{\mathcal{Y}}%
\global\long\def\ZZ{\mathcal{Z}}%

\global\long\def\At{\tilde{A}}%
\global\long\def\Bt{\tilde{B}}%
\global\long\def\Ct{\tilde{C}}%
\global\long\def\Dt{\tilde{D}}%
\global\long\def\Et{\tilde{E}}%
\global\long\def\Ft{\tilde{F}}%
\global\long\def\Gt{\tilde{G}}%
\global\long\def\Ht{\tilde{H}}%
\global\long\def\It{\tilde{I}}%
\global\long\def\Jt{\tilde{J}}%
\global\long\def\Kt{\tilde{K}}%
\global\long\def\Lt{\tilde{L}}%
\global\long\def\Mt{\tilde{M}}%
\global\long\def\Nt{\tilde{N}}%
\global\long\def\Ot{\tilde{O}}%
\global\long\def\Pt{\tilde{P}}%
\global\long\def\Qt{\tilde{Q}}%
\global\long\def\Rt{\tilde{R}}%
\global\long\def\St{\tilde{S}}%
\global\long\def\Tt{\tilde{T}}%
\global\long\def\Ut{\tilde{U}}%
\global\long\def\Vt{\tilde{V}}%
\global\long\def\Wt{\tilde{W}}%
\global\long\def\Xt{\tilde{X}}%
\global\long\def\Yt{\tilde{Y}}%
\global\long\def\Zt{\tilde{Z}}%

\global\long\def\at{\tilde{a}}%
\global\long\def\bt{\tilde{b}}%
\global\long\def\ct{\tilde{c}}%
\global\long\def\dt{\tilde{d}}%
\global\long\def\eet{\tilde{e}}%
\global\long\def\ft{\tilde{f}}%
\global\long\def\gt{\tilde{g}}%
\global\long\def\hht{\tilde{h}}%
\global\long\def\it{\tilde{i}}%
\global\long\def\jt{\tilde{j}}%
\global\long\def\kt{\tilde{k}}%
\global\long\def\lt{\tilde{l}}%
\global\long\def\elt{\tilde{\ell}}%
\global\long\def\mt{\tilde{m}}%
\global\long\def\nt{\tilde{n}}%
\global\long\def\ot{\tilde{o}}%
\global\long\def\pt{\tilde{p}}%
\global\long\def\qt{\tilde{q}}%
\global\long\def\rt{\tilde{r}}%
\global\long\def\st{\tilde{s}}%
\global\long\def\tt{\tilde{t}}%
\global\long\def\ut{\tilde{u}}%
\global\long\def\vt{\tilde{v}}%
\global\long\def\wt{\tilde{w}}%
\global\long\def\xt{\tilde{x}}%
\global\long\def\yt{\tilde{y}}%
\global\long\def\zt{\tilde{z}}%

\global\long\def\mfA{\mathfrak{A}}%
\global\long\def\mfB{\mathfrak{B}}%
\global\long\def\mfC{\mathfrak{C}}%
\global\long\def\mfD{\mathfrak{D}}%
\global\long\def\mfE{\mathfrak{E}}%
\global\long\def\mfF{\mathfrak{F}}%
\global\long\def\mfG{\mathfrak{G}}%
\global\long\def\mfH{\mathfrak{H}}%
\global\long\def\mfI{\mathfrak{I}}%
\global\long\def\mfJ{\mathfrak{J}}%
\global\long\def\mfK{\mathfrak{K}}%
\global\long\def\mfL{\mathfrak{L}}%
\global\long\def\mfM{\mathfrak{M}}%
\global\long\def\mfN{\mathfrak{N}}%
\global\long\def\mfO{\mathfrak{O}}%
\global\long\def\mfP{\mathfrak{P}}%
\global\long\def\mfQ{\mathfrak{Q}}%
\global\long\def\mfR{\mathfrak{R}}%
\global\long\def\mfS{\mathfrak{S}}%
\global\long\def\mfT{\mathfrak{T}}%
\global\long\def\mfU{\mathfrak{U}}%
\global\long\def\mfV{\mathfrak{V}}%
\global\long\def\mfW{\mathfrak{W}}%
\global\long\def\mfX{\mathfrak{X}}%
\global\long\def\mfY{\mathfrak{Y}}%
\global\long\def\mfZ{\mathfrak{Z}}%
\global\long\def\mfa{\mathfrak{a}}%
\global\long\def\mfb{\mathfrak{b}}%
\global\long\def\mfc{\mathfrak{c}}%
\global\long\def\mfd{\mathfrak{d}}%
\global\long\def\mfe{\mathfrak{e}}%
\global\long\def\mff{\mathfrak{f}}%
\global\long\def\mfg{\mathfrak{g}}%
\global\long\def\mfh{\mathfrak{h}}%
\global\long\def\mfi{\mathfrak{i}}%
\global\long\def\mfj{\mathfrak{j}}%
\global\long\def\mfk{\mathfrak{k}}%
\global\long\def\mfl{\mathfrak{l}}%
\global\long\def\mfm{\mathfrak{m}}%
\global\long\def\mfn{\mathfrak{n}}%
\global\long\def\mfo{\mathfrak{o}}%
\global\long\def\mfp{\mathfrak{p}}%
\global\long\def\mfq{\mathfrak{q}}%
\global\long\def\mfr{\mathfrak{r}}%
\global\long\def\mfs{\mathfrak{s}}%
\global\long\def\mft{\mathfrak{t}}%
\global\long\def\mfu{\mathfrak{u}}%
\global\long\def\mfv{\mathfrak{v}}%
\global\long\def\mfw{\mathfrak{w}}%
\global\long\def\mfx{\mathfrak{x}}%
\global\long\def\mfy{\mathfrak{y}}%
\global\long\def\mfz{\mathfrak{z}}%

\global\long\def\d{\mathrm{d}}%
\global\long\def\DDD{\mathrm{D}}%
\global\long\def\EEE{\mathrm{E}}%
\global\long\def\i{\ii}%
\global\long\def\MMM{\mathrm{M}}%
\global\long\def\OOOO{\mathrm{O}}%
\global\long\def\RRRR{\mathrm{R}}%
\global\long\def\TTT{\mathrm{T}}%
\global\long\def\UUU{\mathrm{U}}%

\global\long\def\GL{\mathrm{GL}}%
\global\long\def\ISU{\mathrm{ISU}}%
\global\long\def\ISUT{\mathrm{ISU}\left(2\right)}%
\global\long\def\SL{\mathrm{SL}}%
\global\long\def\SO{\mathrm{SO}}%
\global\long\def\SOH{\mathrm{SO}\left(3\right)}%
\global\long\def\SOT{\mathrm{SO}\left(2\right)}%
\global\long\def\Sp{\mathrm{Sp}}%
\global\long\def\SU{\mathrm{SU}}%
\global\long\def\SUT{\mathrm{SU}\left(2\right)}%
\global\long\def\UO{\mathrm{U}\left(1\right)}%
\global\long\def\gl{\mathfrak{gl}}%
\global\long\def\sl{\mathfrak{sl}}%
\global\long\def\sso{\mathfrak{so}}%
\global\long\def\soh{\mathfrak{so}\left(3\right)}%
\global\long\def\su{\mathfrak{su}}%
\global\long\def\sut{\mathfrak{su}\left(2\right)}%
\global\long\def\isut{\mathfrak{isu}\left(2\right)}%

\global\long\def\so{\Rightarrow}%
\global\long\def\os{\Leftarrow}%
\global\long\def\to{\rightarrow}%
\global\long\def\ot{\leftarrow}%
\global\long\def\soo{\Longrightarrow}%
\global\long\def\oos{\Longleftarrow}%
\global\long\def\too{\longrightarrow}%
\global\long\def\oot{\longleftarrow}%
\global\long\def\sos{\Leftrightarrow}%
\global\long\def\tot{\leftrightarrow}%
\global\long\def\soos{\Longleftrightarrow}%
\global\long\def\toot{\longleftrightarrow}%
\global\long\def\mt{\mapsto}%
\global\long\def\mtt{\longmapsto}%
\global\long\def\dn{\downarrow}%
\global\long\def\up{\uparrow}%
\global\long\def\updn{\updownarrow}%
\global\long\def\sea{\searrow}%
\global\long\def\nea{\nearrow}%
\global\long\def\nwa{\nwarrow}%
\global\long\def\swa{\swarrow}%
\global\long\def\hk{\hookrightarrow}%
\global\long\def\kh{\hookleftarrow}%
\global\long\def\soosp{\quad\Longrightarrow\quad}%
\global\long\def\oossp{\quad\Longleftarrow\quad}%
\global\long\def\soossp{\quad\Longleftrightarrow\quad}%

\global\long\def\multibrl#1{\left(#1\right.}%
\global\long\def\multibrr#1{\left.#1\right)}%
\global\long\def\multisql#1{\left[#1\right.}%
\global\long\def\multisqr#1{\left.#1\right]}%
\global\long\def\multicul#1{\left\{  #1\right.}%
\global\long\def\multicur#1{\left.#1\right\}  }%

\global\long\def\bl{\bigl|}%
\global\long\def\bll{\Bigl|}%
\global\long\def\blll{\biggl|}%
\global\long\def\bllll{\Biggl|}%

\global\long\def\ma#1#2{\left\langle #1\thinspace\middle|\thinspace#2\right\rangle }%
\global\long\def\mma#1#2#3{\left\langle #1\thinspace\middle|\thinspace#2\thinspace\middle|\thinspace#3\right\rangle }%
\global\long\def\mc#1#2{\left\{  #1\thinspace\middle|\thinspace#2\right\}  }%
\global\long\def\mmc#1#2#3{\left\{  #1\thinspace\middle|\thinspace#2\thinspace\middle|\thinspace#3\right\}  }%
\global\long\def\mr#1#2{\left(#1\thinspace\middle|\thinspace#2\right) }%
\global\long\def\mmr#1#2#3{\left(#1\thinspace\middle|\thinspace#2\thinspace\middle|\thinspace#3\right)}%

\global\long\def\pr{\parallel}%
\global\long\def\xx{\times}%
\global\long\def\dg{\lyxmathsym{\textdegree}}%
\global\long\def\sp{,\qquad}%
\global\long\def\sq{\square}%
\global\long\def\pt{\propto}%
\global\long\def\lrc{\lrcorner\thinspace}%
\global\long\def\pexp{\overrightarrow{\exp}}%
\global\long\def\dui#1#2#3{#1_{#2}{}^{#3}}%
\global\long\def\udi#1#2#3{#1^{#2}{}_{#3}}%
\global\long\def\pab{\bar{\partial}}%
\global\long\def\zr{\mathbf{0}}%
\global\long\def\on{\mathbf{1}}%
\global\long\def\na{\boldsymbol{\nabla}}%
\global\long\def\hf{\frac{1}{2}}%
\global\long\def\trd{\frac{1}{3}}%
\global\long\def\fr{\frac{1}{4}}%
\global\long\def\ei{\frac{1}{8}}%
\global\long\def\ap{\approx}%
\global\long\def\eqm{\overset{?}{=}}%
\global\long\def\fa{\forall}%
\global\long\def\ex{\exists}%
\global\long\def\xxd{\dot{\boldsymbol{x}}}%
\global\long\def\xxdd{\ddot{\boldsymbol{x}}}%
\global\long\def\ept{\tilde{\epsilon}}%

\global\long\def\AAh{\hat{\mathbf{A}}}%
\global\long\def\eeh{\hat{\mathbf{e}}}%
\global\long\def\FFh{\hat{\mathbf{F}}}%
\global\long\def\TTh{\hat{\mathbf{T}}}%
\global\long\def\Thh{\hat{\Theta}}%
\global\long\def\XXt{\tilde{\mathbf{X}}}%
\global\long\def\hr{\mathring{h}}%
\global\long\def\xxr{\mathring{\boldsymbol{x}}}%
\global\long\def\hh{\hat{h}}%
\global\long\def\xxh{\hat{\boldsymbol{x}}}%
\global\long\def\hrh{\hat{\mathring{h}}}%
\global\long\def\xxrh{\hat{\mathring{\boldsymbol{x}}}}%
\global\long\def\Sh{\hat{S}}%
\global\long\def\sxt{\frac{1}{6}}%
\global\long\def\se{\subseteq}%
\global\long\def\tHH{\widetilde{\mathcal{H}}}%
\global\long\def\bgl{\biggl|}%

\pagestyle{empty} \pagenumbering{roman}
\begin{center}
\vspace*{1cm}
\par\end{center}

\begin{center}
\textbf{\Huge{}At the Corner of Space and Time}{\Huge\par}
\par\end{center}

\begin{center}
{\Huge{}\vspace*{1cm}
}{\Huge\par}
\par\end{center}

\begin{center}
by \\
\par\end{center}

\begin{center}
\vspace*{1cm}
\par\end{center}

\begin{center}
{\Large{}Barak Shoshany }\\
\par\end{center}

\begin{center}
{\Large{}\vspace*{3cm}
}{\Large\par}
\par\end{center}

\begin{center}
A thesis \\
 presented to the University of Waterloo \\
 in fulfillment of the \\
 thesis requirement for the degree of \\
 Doctor of Philosophy \\
 in \\
 Physics \\
\par\end{center}

\begin{center}
\vspace*{2cm}
\par\end{center}

\begin{center}
Waterloo, Ontario, Canada, 2019 \\
\par\end{center}

\begin{center}
\vspace*{1cm}
\par\end{center}

\begin{center}
\copyright\ Barak Shoshany 2019 
\par\end{center}

\cleardoublepage{}

\pagestyle{plain} \setcounter{page}{2}

\begin{center}\textbf{Examining Committee Membership}\end{center}
  \noindent
The following served on the Examining Committee for this thesis. The decision of the Examining Committee is by majority vote.
  \bigskip
  
  \noindent
\begin{tabbing}
xxxxxxxxxxxxxxxxxxxxxxxxxxxxxx \=  \kill
External Examiner: \>  Karim Noui \\ 
\> Associate Professor \\
\> University of Tours \\
\end{tabbing} 
  \bigskip
  
  \noindent
\begin{tabbing}
xxxxxxxxxxxxxxxxxxxxxxxxxxxxxx \=  \kill
Supervisors: \> Laurent Freidel \\
\> Faculty \\
\> Perimeter Institute for Theoretical Physics \\
\> \\
\> Robert Myers \\
\> Faculty \\
\> Perimeter Institute for Theoretical Physics \\
\end{tabbing}
  \bigskip
  
  \noindent
  \begin{tabbing}
xxxxxxxxxxxxxxxxxxxxxxxxxxxxxx \=  \kill
Internal Members: \> Robert Mann \\
\> Professor \\
\> University of Waterloo \\
\> \\
\> John Moffat \\
\> Professor Emeritus \\
\> University of Toronto  \\
\end{tabbing}
  \bigskip
  
  \noindent
\begin{tabbing}
xxxxxxxxxxxxxxxxxxxxxxxxxxxxxx \=  \kill
Internal-External Member: \> Florian Girelli \\
\> Associate Professor \\
\> University of Waterloo \\
\end{tabbing}

\cleardoublepage{}
\begin{center}
\textbf{Author's Declaration}
\par\end{center}

This thesis consists of material all of which I authored or co-authored:
see Statement of Contributions included in the thesis. This is a true
copy of the thesis, including any required final revisions, as accepted
by my examiners.

I understand that my thesis may be made electronically available to
the public.

\cleardoublepage{}
\begin{center}
\textbf{Statement of Contributions}
\par\end{center}

This thesis is based on the papers \cite{FirstPaper}, co-authored
with Laurent Freidel and Florian Girelli, the papers \cite{Shoshany:2019ymo,Shoshany:2019kib},
of which I am the sole author, and additional unpublished material,
of which I am the sole author.

\cleardoublepage{}
\begin{center}
\textbf{Abstract}
\par\end{center}

We perform a rigorous piecewise-flat discretization of classical general
relativity in the first-order formulation, in both 2+1 and 3+1 dimensions,
carefully keeping track of curvature and torsion via holonomies. We
show that the resulting phase space is precisely that of spin networks,
the quantum states of discrete spacetime in loop quantum gravity,
with additional degrees of freedom called edge modes, which control
the gluing between cells. This work establishes, for the first time,
a rigorous proof of the equivalence between spin networks and piecewise-flat
geometries with curvature and torsion degrees of freedom. In addition,
it demonstrates that careful consideration of edge modes is crucial
both for the purpose of this proof and for future work in the field
of loop quantum gravity. It also shows that spin networks have a dual
description related to teleparallel gravity, where gravity is encoded
in torsion instead of curvature degrees of freedom. Finally, it sets
the stage for collaboration between the loop quantum gravity community
and theoretical physicists working on edge modes from other perspectives,
such as quantum electrodynamics, non-abelian gauge theories, and classical
gravity.

\cleardoublepage{}
\begin{center}
\textbf{Acknowledgments}
\par\end{center}

\textbf{Note:} Wherever a list of names is provided in this section,
it is always ordered alphabetically by last name.

First and foremost, I would like to thank my supervisor, \textbf{Laurent
Freidel}, for giving me the opportunity to study at the wonderful
Perimeter Institute and to write both my master's thesis and my PhD
thesis under his guidance, and for supporting me and believing in
me throughout the last 5 years, even when I didn't believe in myself.
His deep physical intuition has inspired me countless times. He constantly
challenged me and pushed me to expand my horizons in both mathematics
and physics, to improve my research skills as well as my presentation
and communication skills, and to be a better scientist. Needless to
say, this thesis would not have existed without him.

I am also incredibly thankful to \textbf{Florian Girelli} for collaborating
with me on this research and for his continued assistance and helpful
advice during the last 3 years. Our frequent meetings and stimulating
discussions have been an invaluable resource, and provided me not
only with the knowledge and insight necessary to succeed in writing
this thesis, but also with the motivation to overcome the many obstacles
that are an inevitable part of research.

I thank my examination committee members \textbf{Robert Mann}, \textbf{John
Moffat}, and \textbf{Robert Myers}, as well as the external examiner
\textbf{Karim Noui}, for taking time out of their very busy schedules
to read this thesis. I hope I have made it readable and interesting
enough for this to be a pleasant experience.

Special thanks go to \textbf{Wolfgang Wieland}, who patiently provided
clear and detailed explanations of complicated concepts in mathematics
and physics whenever I needed his help, and to \textbf{Marc Geiller},
\textbf{Hal Haggard}, \textbf{Seyed Faroogh Moosavian}, \textbf{Yasha
Neiman}, \textbf{Kasia Rejzner}, \textbf{Aldo Riello}, and \textbf{Vasudev
Shyam }for very helpful discussions.

I thank my students from the International Summer School for Young
Physicists (ISSYP), the Perimeter Institute Undergraduate Summer Program,
and the Perimeter Scholars International (PSI) Program over the last
5 years for consistently asking challenging questions, which forced
me to think outside the box and improve myself and my knowledge in
many different ways.

Sharing my knowledge of mathematics and physics in the numerous teaching
and outreach events, programs, and initiatives at Perimeter Institute
has served as a constant reminder of why I got into science in the
first place. For that privilege, I owe my gratitude to \textbf{Greg
Dick}, \textbf{Kelly Foyle}, \textbf{Damian Pope}, \textbf{Angela
Robinson}, \textbf{Marie Strickland}, and \textbf{Tonia Williams }from
Perimeter Institute's outreach department.

I owe great thanks to \textbf{Dan Wohns} and \textbf{Gang Xu} for
teaching me during my studies in the PSI program, for choosing me
to serve as a teaching assistant in the PSI quantum field theory and
string theory courses, for introducing me to many fun board games,
and for inviting me to conduct the Perimeter Institute Orchestra.
I also thank the members of the orchestra for agreeing to play my
composition, \emph{Music at the Planck Scale}, which was no easy task
by any means. Working on this and other music projects has made my
time at Perimeter Institute so much more fun!

I am grateful to Perimeter Institute's Academic Programs Manager,
\textbf{Debbie Guenther}, for being so amazingly helpful and supportive
in many different logistical hurdles throughout both my master's and
PhD studies. I also thank the staff at Perimeter Institute's Black
Hole Bistro, and in particular \textbf{Dan Lynch} and \textbf{Olivia
Taylor}, for providing me with delicious food on a daily basis for
5 years, and for tolerating my many unusual requests.

I would not have been where I am today without my parents, \textbf{Avner
and Judith Shoshany}, who, ever since I was born, have constantly
empowered me to follow my dreams, no matter how crazy, impossible,
or impractical they were, and always did everything they could in
order to help me achieve those dreams and more. They have my eternal
gratitude and appreciation.

Finally, I would like to thank all of my friends, and especially \textbf{Jacob
Barnett}, \textbf{Juan Cayuso}, \textbf{Daniel Guariento}, \textbf{Fiona
McCarthy}, \textbf{Yasha Neiman}, \textbf{Kasia Rejzner}, \textbf{Vasudev
Shyam}, \textbf{Matthew VonHippel}, \textbf{Dan Wohns}, and \textbf{Gang
Xu}, for much-needed distractions from the frequent frustrations of
graduate school over the last 5 years (especially in the form of playing
games\footnote{In this context I would also like to thank \textbf{Daniel Gottesman
}for creating and running his exceptional Dungeons and Dragons campaign!} or watching stand-up comedy), for providing unconditional emotional
support when needed, and/or for patiently listening to my endless
complaints.

\cleardoublepage{}

\tableofcontents{}

\cleardoublepage{}

\pagenumbering{arabic}

\section{Introduction}

\subsection{Loop Quantum Gravity}

When perturbatively quantizing gravity, one obtains a low-energy effective
theory, which breaks down at high energies. There are several different
approaches to solving this problem and obtaining a theory of quantum
gravity. String theory, for example, attempts to do so by postulating
entirely new degrees of freedom, which can then be shown to reduce
to general relativity (or some modification thereof) at the low-energy
limit. \emph{Loop quantum gravity} \cite{Rovelli:2004tv} instead
tries to quantize gravity \emph{non-perturbatively}, by quantizing
\emph{holonomies }(or \emph{Wilson loops}) instead of the metric,
in an attempt to avoid the issues arising from perturbative quantization.

The starting point of the canonical version of loop quantum gravity
\cite{thiemann_2007} is the reformulation of general relativity as
a \emph{non-abelian Yang-Mills gauge theory }on a spatial slice of
spacetime, with the \emph{gauge group }$\SUT$ related to spatial
rotations, the \emph{Yang-Mills connection }$\A$ related to the usual
connection and extrinsic curvature, and the \emph{``electric field''}
$\E$ related to the metric (or more precisely, the frame field).
Once gravity is reformulated in this way, one can utilize the existing
arsenal of techniques from Yang-Mills theory, and in particular \emph{lattice
gauge theory}, to tackle the problem of quantum gravity \cite{Henneaux:1992ig}.

This theory is quantized by considering \emph{graphs}, that is, sets
of \emph{nodes }connected by \emph{links}. One defines \emph{holonomies},
or \emph{path-ordered exponentials} of the connection, along each
link. The curvature on the spatial slice can then be probed by looking
at holonomies along loops on the graph. Without going into the technical
details, the general idea is that if we know the curvature inside
every possible loop, then this is equivalent to knowing the curvature
at every point.

The \emph{kinematical Hilbert space }of loop quantum gravity is obtained
from the set of all wave-functionals for all possible graphs, together
with an appropriate $\SUT$-invariant and diffeomorphism-invariant
inner product. The \emph{physical Hilbert space }is a subset of the
kinematical one, containing only the states invariant under all gauge
transformations -- or in other words, annihilated by all of the constraints.
Since gravity is a totally constrained system -- in the Hamiltonian
formulation, the action is just a sum of constraints -- a quantum
state annihilated by all of the constraints is analogous to a metric
which solves Einstein's equations in the classical Lagrangian formulation.

Specifically, to get from the kinematical to the physical Hilbert
space, three steps must be taken:
\begin{enumerate}
\item First, we apply the Gauss constraint to the kinematical Hilbert space.
Since the Gauss constraint generates $\SUT$ gauge transformation,
we obtain a space of $\SUT$-invariant states, called \emph{spin network
states} \cite{Ashtekar:1991kc}, which are the graphs mentioned above,
but with their links colored by irreducible representations of $\SUT$,
that is, \emph{spins }$j\in\left\{ \hf,1,\frac{3}{2},2,\ldots\right\} $.
\item Then, we apply the spatial diffeomorphism constraint. We obtain a
space of equivalence classes of spin networks under spatial diffeomorphisms,
a.k.a. \emph{knots}. These states are now abstract graphs, not localized
in space. This is analogous to how a classical geometry is an equivalence
class of metrics under diffeomorphisms.
\item Lastly, we apply the Hamiltonian constraint. This step is still not
entirely well-understood, and is one of the main open problems of
the theory.
\end{enumerate}
One of loop quantum gravity's most celebrated results is the existence
of area and volume operators. They are derived by taking the usual
integrals of area and volume forms and promoting the ``electric field''
$\E$, which is conjugate to the connection $\A$, to a functional
derivative $\delta/\delta\A$. The spin network states turn out to
be eigenstates of these operators, and they have \emph{discrete spectra
}which depend on the spins of the links. This means that loop quantum
gravity contains a \emph{quantum geometry}, which is a feature one
would expect a quantum theory of spacetime to have. It also hints
that spacetime is discrete at the Planck scale.

However, it is not clear how to rigorously define the classical geometry
related to a particular spin network state. In this thesis, we will
try to answer that question.

\subsection{Teleparallel Gravity}

The theory of general relativity famously describes gravity as a result
of the curvature of spacetime itself. Furthermore, the geometry of
spacetime is assumed to be torsionless by employing the \emph{Levi-Civita
connection}, which is torsionless by definition. While this is the
most popular formulation, there exists an alternative but mathematically
equivalent formulation called \emph{teleparallel gravity} \cite{2005physics...3046U,Maluf:2013gaa,Ferraro:2016wht},
differing from general relativity only by a boundary term. In this
formulation, one instead uses the \emph{Weitzenböck connection}, which
is flat by definition. The gravitational degrees of freedom are then
encoded in the torsion of the spacetime geometry.

In the canonical version of loop quantum gravity, as we mentioned
above, one starts by rewriting general relativity in the Hamiltonian
formulation using the Ashtekar variables. One finds a fully constrained
system, that is, the Hamiltonian is simply a sum of constraints. In
2+1 spacetime dimensions, where gravity is topological \cite{Carlip:1998uc},
there are two such constraints:
\begin{itemize}
\item The Gauss (or torsion) constraint, which imposes zero torsion everywhere,
\item The curvature (or flatness) constraint, which imposes zero curvature
everywhere.
\end{itemize}
After imposing both constraints, we will obtain the physical Hilbert
space of the theory. It does not matter which constraint is imposed
first, since the resulting physical Hilbert space will be the same.
However, on a more conceptual level, the first constraint that we
impose is used to define the kinematics of the theory, while the second
constraint will encode the dynamics. Thus, it seems natural to identify
general relativity with the quantization in which the Gauss constraint
is imposed first, and teleparallel gravity with that in which the
curvature constraint is imposed first.

Indeed, in loop quantum gravity, which is a quantization of general
relativity, the Gauss constraint is imposed first, as detailed in
the previous section. This is true in both 2+1 and 3+1 dimensions.
In 2+1D, the curvature constraint is imposed at the dynamical level
in order to obtain the Hilbert space of physical states. In 3+1D there
is no curvature constraint -- the curvature is, in general, not flat.
One instead imposes the diffeomorphism and Hamiltonian constraints
to get the physical Hilbert space, but the Gauss constraint is still
imposed first.

In \cite{clement}, an alternative choice was suggested where the
order of constraints, in 2+1D, is reversed. The curvature constraint
is imposed first by employing the \emph{group network }basis of translation-invariant
states, and the Gauss constraint is the one which encodes the dynamics.
This \emph{dual loop quantum gravity} quantization is the quantum
counterpart of teleparallel gravity, and could be used to study the
dual vacua proposed in \cite{Dittrich:2014wpa,Dittrich:2014wda}.

In this thesis, we will only deal with the classical theory. We will
show how, by discretizing the phase space of continuous gravity in
the first-order formulation (and with the Ashtekar variables in 3+1D),
one obtains a spectrum of discrete phase spaces, one of which is the
classical version of spin networks \cite{Freidel:2010aq} and the
other is a dual formulation (``dual loop gravity''), which may be
interpreted as the classical version of the group network basis, and
is intuitively related to teleparallel gravity. The latter case was
first studied in \cite{Dupuis:2017otn}, but only in 2+1 dimensions,
and only in the simple case where there are no curvature or torsion
excitations.

Another phase space of interest in the spectrum discussed above is
a mixed phase space, containing both loop gravity and its dual. In
2+1D it is intuitively related to Chern-Simons theory \cite{Horowitz:1989ng},
as we will motivate below. In this case our formalism is related to
existing results \cite{Alekseev:1993rj,Alekseev:1994pa,Alekseev:1994au,Meusburger:2003hc,Meusburger:2005mg,Meusburger:2003ta,Meusburger:2008dc}.

\subsection{Quantization, Discretization, Subdivision, and Truncation}

One of the key challenges in trying to define a theory of quantum
gravity at the quantum level is to find a regularization that does
not drastically break the fundamental symmetries of the theory. This
is a challenge in any gauge theory, but gravity is especially challenging,
for two reasons. First, one expects that the quantum theory possesses
a fundamental length scale; and second, the gauge group contains diffeomorphism
symmetry, which affects the nature of the space on which the regularization
is applied.

In gauge theories such as \emph{quantum chromodynamics} (QCD), the
only known way to satisfy these requirements, other than gauge-fixing
before regularization, is to put the theory on a \emph{lattice}, where
an effective finite-dimensional gauge symmetry survives at each scale.
One would like to devise such a scheme in the gravitational context
as well. In this thesis, we develop a step-by-step procedure to achieve
this, exploiting, among other things, the fact that first-order gravity
in 2+1 dimensions, as well as gravity in 3+1 dimensions with the Ashtekar
variables, closely resembles other gauge theories. We find not only
the spin network or holonomy-flux phase space, which is what we initially
expected, but also additional particle-like or string-like degrees
of freedom coupled to the curvature and torsion.

As explained above, in \textit{\emph{canonical loop quantum gravity}}\emph{
}(LQG), one can show that the geometric operators possess a discrete
spectrum. This is, however, only possible after one chooses the quantum
spin network states to have support on a graph. Spin network states
can be understood as describing a quantum version of \textit{discretized}
spatial geometry \cite{Rovelli:2004tv}, and the Hilbert space associated
to a graph can be related, in the classical limit, to a set of discrete
piecewise-flat geometries \cite{Dittrich:2008ar,Freidel:2010aq}.

This means that the LQG quantization scheme consists at the same time
of a \emph{quantization} and a \emph{discretization}; moreover, the
quantization of the geometric spectrum is entangled with the discretization
of the fundamental variables. It has been argued that it is essential
to disentangle these two different features \cite{Freidel:2011ue},
especially when one wants to address dynamical issues.

In \cite{Freidel:2011ue,Freidel:2013bfa}, it was suggested that one
should understand the discretization as a two-step process: a \emph{subdivision}
followed by a \emph{truncation}. In the first step one subdivides
the systems into fundamental cells, and in the second step one chooses
a truncation of degrees of freedom in each cell, which is consistent
with the symmetries of the theory. By focusing first on the classical
discretization, before any quantization takes place, several aspects
of the theory can be clarified. Let us mention some examples: 
\begin{itemize}
\item This discretization scheme allows us to study more concretely how
to recover the continuum geometry out of the classical discrete geometry
associated to the spin networks \cite{Freidel:2011ue,Freidel:2013bfa}.
In particular, since the discretization is now understood as a truncation
of the continuous degrees of freedom, it is possible to associate
a continuum geometry to the discrete data.
\item It provides a justification for the fact that, in the continuum case,
the momentum variables are equipped with a vanishing Poisson bracket,
whereas in the discrete case, the momentum variables do not commute
with each other \cite{Freidel:2011ue,Dupuis:2017otn,Dittrich:2014wpa}.
These variables need to be \emph{dressed }by the gauge connection,
as we will explain in the next section, and are now understood as
charge generators \cite{Donnelly:2016auv}.
\item As detailed above, our discretization scheme permits the discovery
and study of a dual formulation of loop gravity, both in 2+1 and 3+1
dimensions.
\end{itemize}
The separation of discretization into two distinct steps in our formalism
work as follows. First we perform a \emph{subdivision}, or decomposition
into subsystems. More precisely, we define a \emph{cellular decomposition
}on our 2D or 3D spatial manifold, where the cells can be any (convex)
polygons or polyhedra respectively. This structure has a dual structure,
which as we will see, is the spin network graph, with each cell dual
to a node, and each element on the boundary of the cell (edge in 2+1D,
side in 3+1D) dual to a link connected to that node.

Then, we perform a \emph{truncation}, or coarse-graining of the subsystems.
In this step, we assume that there is arbitrary curvature and torsion
inside each loop of the spin network. We then ``compress'' the information
about the geometry into singular codimension-2 excitations. In 2+1D
it will be stored in a 0-dimensional (point particle) excitation,
while in 3+1D it will be stored in a 1-dimensional (string) excitation.
Crucially, since the only way to probe the geometry is by looking
at the holonomies on the loops of the spin network, the observables
before and after this truncation are the same.

Another way to interpret this step is to instead assume that spacetime
is flat everywhere, with matter sources being distributive, i.e.,
given by Dirac delta functions, which then generate singular curvature
and torsion by virtue of the Einstein equation. We interpret these
distributive matter sources as point particles in 2+1D or strings
in 3+1D, and this is entirely equivalent to truncating a continuous
geometry, since holonomies cannot distinguish between continuous and
distributive geometries.

Once we performed subdivision and truncation, we can now define discrete
variables on each cell and integrate the continuous symplectic potential
in order to obtain a discrete potential, which represents the discrete
phase space. In this step, we will see that the mathematical structures
we are using conspire to cancel many terms in the potential, allowing
us to fully integrate it.

\subsection{Edge Modes}

In the subdivision process described above, some of the bulk degrees
of freedom are replaced by \emph{edge mode} degrees of freedom, which
play a key role in the construction of the full phase space and our
understanding of symmetry. This happens because dealing with subsystems
in a gauge theory requires special care with regards to boundaries,
where gauge invariance is naively broken, and thus additional degrees
of freedom must be added in order to restore it. These new degrees
of freedom transform non-trivially under new symmetry transformations
located on the edges and/or corners; we will usually ignore this distinction
and just call them ``edge modes''. The process of subdivision therefore
requires a canonical extension of the phase space, and converting
some momenta into non-commutative charge generators.

The general philosophy is presented in \cite{Donnelly:2016auv} and
exemplified in the 3+1D gravity context in \cite{Freidel:2015gpa,Freidel:2016bxd,Freidel:2018pvm}.
An intuitive reason behind this fundamental mechanism is also presented
in \cite{Rovelli:2013fga} and the general idea is, in a sense, already
present in \cite{Donnelly:2011hn}. In the 2+1D gravity context, the
edge modes have been studied in great detail in \cite{Geiller:2017xad,Geiller:2017whh}.
This phenomenon even happens when the boundary is taken to be infinity
\cite{Strominger:2017zoo}, where these new degrees of freedom are
called \emph{soft modes}.

These extra degrees of freedom, which possess their own phase space
structure and appear as \emph{dressings} of the gravitationally charged
observables, affect the commutation relations of the dressed observables.
In a precise sense, this is what happens with the fluxes in loop gravity:
the ``discretized'' fluxes are dressed by the connection degrees
of freedom, implying a different Poisson structure compared to the
continuum ones. A nice continuum derivation of this fact is given
in \cite{Cattaneo:2016zsq}.

Once this subdivision and extension of the phase space are done properly,
one has to understand the gluing of subregions as the fusion of edge
modes across the boundaries. If the boundary is trivial, this fusion
merely allows us to extend gauge-invariant observables from one region
to another. However, when several boundaries meet at a corner, there
is now the possibility to have residual degrees of freedom that come
from this fusion.

We witness exactly this phenomenon at the corners of our cellular
decomposition -- vertices in 2+1D or edges in 3+1D -- where new
degrees of freedom, in addition to the usual loop gravity ones, are
found after regluing. As we will see below, the edge modes at the
boundaries of the cells in our cellular decomposition -- edges in
2+1D or sides in 3+1D -- will cancel with the edge modes on the boundaries
of the adjacent cells. However, the modes at the corners do not have
anything to cancel with. These degrees of freedom will thus survive
the discretization process. In 2+1D, they introduce a particle-like
phase space \cite{Kirillov,Rempel:2015foa}, while in 3+1D, we interpret
them as cosmic strings \cite{PhysRevD.87.025020}.

One might expect that the geometry will be encoded in the constraints
alone, by imposing, roughly speaking, that a loop of holonomies sees
the curvature inside it and a loop of fluxes sees the torsion inside
it. As we will see, while the constraints do indeed encode the geometry,
the presence of the edge and corner modes is the reason for the inclusion
of the curvature and torsion themselves as additional phase space
variables.

After we have proven our results in 2+1D a very careful and rigorous
way, by regularizing the singularities at the vertices, we will derive
them again in a different and shorter way. In this alternative calculation,
we will show that when calculating the symplectic potential, the terms
at the corners (that is, the vertices) completely vanish if there
are no curvature or torsion excitations at the vertices. However,
if such excitations do exist, the corner terms instead turn out to
add up in exactly the right way to produce the additional phase space
variables we found before, in our more complicated calculation.

In the 3+1D case, many additional complications occur that were not
present in 2+1D, and therefore we will jump right to the alternative
analysis. A major difference, as mentioned above, is that the sources
of curvature and torsion will now be 1-dimensional strings, rather
than 0-dimensional point particles as we had in 2+1D. However, the
curvature and torsion will nonetheless still be detected by loops
of holonomies, as in the 2+1D case. Furthermore, we will discover
that, analogously, one can isolate the terms at the corners (which
are now edges), and they vanish unless the edges possess curvature
or torsion excitations. In fact, interestingly, we will obtain the
exact same discrete phase space as in the 2+1D case: a spin network
coupled to edge modes.

As we alluded in the beginning of this section, the edge modes come
equipped with new symmetries, which did not exist in the continuum
theory. We will show below that multiplying the discrete holonomies
by group elements from the right (\emph{right translations}) corresponds
to the usual gauge transformation. The new symmetries we discovered
are obtained by instead multiplying from the left (\emph{left translations}).
These symmetries leave the continuous connection invariant, and therefore
correspond to completely new degrees of freedom in the discrete variables,
which did not exist in the continuum. When the edge modes are ``frozen'',
meaning that we choose a particular value for them (which can be,
without loss of generality, the identity), the new symmetries are
broken, and the phase space reduces to the usual loop gravity phase
space (or its dual), without these additional degrees of freedom.

The conceptual shift towards an edge mode interpretation provides
a different paradigm to explore some of the key questions of loop
quantum gravity. For example, the notion of the continuum limit (in
a 3+1-dimensional theory) attached to subregions could potentially
be revisited and clarified in light of this new interpretation, and
related to the approach developed in \cite{Delcamp:2018efi,Delcamp:2016yix}.
It also strengthens, in a way, the spinor approach to LQG \cite{Girelli:2005ii,Freidel:2009ck,Livine:2011gp},
which allows one to recover the LQG formalism from spinors living
on the nodes of the graph. These spinors can be seen as a different
parametrization of the edge modes, in a similar spirit to \cite{Wieland:2018ymr,Wieland:2017ksn}.

Edge modes have recently been studied for the purpose of making proper
entropy calculations in gauge theory or, more generally, defining
local subsystems \cite{Donnelly:2011hn}. Their use could provide
some new guidance for understanding the concept of entropy in loop
quantum gravity. They are also relevant to the study of specific types
of boundary excitations in condensed matter \cite{Balachandran:1994vi},
which could generate some interesting new directions to explore in
LQG, as in \cite{Dittrich:2016typ,Delcamp:2018efi}. Very recently,
\cite{Freidel2019,Freidel2019a} showed that one can view the constraints
of 3+1D gravity in the Ashtekar formulation as a conservation of edge
charges, thus uncovering a new conceptual framework.

\subsection{Comparison to Previous Work}

\subsubsection{Combinatorial Quantization of 2+1D Gravity}

Discretization (and quantization) of 2+1D gravity was already performed
some time ago with the \emph{combinatorial quantization} of Chern-Simons
theory \cite{Alekseev:1993rj,Alekseev:1994pa,Alekseev:1994au,Meusburger:2003hc}.
While our results in 2+1D should be equivalent to this formulation,
they also add some new and important insights.

First, we work directly with the gravitational variables and the associated
geometric quantities, such as torsion and curvature. Our procedure
describes clearly how such objects should be discretized, which is
not obvious in the Chern-Simons picture; see for example \cite{Meusburger:2008bs}
where the link between the combinatorial framework and LQG was explored.

Second, we are using a different discretization procedure than the
one used in the combinatorial approach. Instead of considering the
reduced graph, we use the full graph to generate the spin network,
and assume that the equations of motion are satisfied in the cells
dual to the spin network.

Finally, we will show in detail that our discretization scheme applies,
with the necessary modifications, to the 3+1D case as well, unlike
the combinatorial approach.

\subsubsection{2+1D Gravity, Chern-Simons Theory, and Point Particles}

In \cite{Matschull1999}, it is emphasized that Chern-Simons theory
and Einstein gravity are in fact not equivalent up to a boundary term,
with the difference being that in Einstein gravity the fame field
$\ee$ is required to be invertible, unlike in Chern-Simons theory.
We will always assume in this thesis that the frame field is in fact
invertible, and often use the inverse frame field as part of our derivation,
but this will not be a problem since we are only interested in the
gravity case anyway, and in particular in the generalization to 3+1D
gravity, where the frame field is invertible as well.

The same author, in \cite{Matschull2001}, analyzed the phase space
of a toy model of discretized 2+1D gravity with point particles, which
is similar to the one we will consider here. In particular, Chapter
4 of \cite{Matschull2001} obtains results that are remarkably similar
to some of ours. The symplectic potential found in Eq. (4.48) of \cite{Matschull2001}
is reminiscent\footnote{Note that to get from the notation of \cite{Matschull2001} to ours,
one should take $z_{\lambda}\mt\x_{c}$ and $\g_{\lambda}\mt h_{c}$.} of our Eq. (\ref{eq:Theta_c}) with $\lambda=1$, which, as we will
show, is the dual (or teleparallel) loop gravity polarization, although
there it is not interpreted as such.

Another interesting treatment of point particles in 2+1D gravity was
given in \cite{Noui2004}, where the coupling to spinning particles
and the relation with the spin foam scalar product were examined,
and in \cite{Noui2004a}, where the relation between canonical loop
quantization and spin foam quantization in 3+1D was established. Finally,
in \cite{Freidel2004} it was shown, in the context of the Ponzano-Regge
model of 2+1D quantum gravity, that particles arise as curvature defects,
which is indeed what we will see in this thesis; furthermore, it was
shown this model is linked to the quantization of Chern-Simons theory.

\subsection{Outline}

This thesis is meant to be as self-contained as possible, with all
the necessary background for each chapter provided in full in the
preceding chapters. First, in Chapter \ref{subsec:Basic-Definitions-Notations},
we provide a comprehensive list of basic definitions, notations, and
conventions which will be used throughout the thesis. It is highly
recommended that the reader not skip this chapter, as some of our
notation is slightly non-standard. Then we begin the thesis itself,
which consists of four parts.

Part \ref{part:2P1-cont} presents 2+1-dimensional general relativity
in the continuum. Chapter \ref{sec:Chern-Simons-Theory-and} consists
of a self-contained derivation of Chern-Simons theory and, from it,
2+1-dimensional gravity in both the Lagrangian and Hamiltonian formulations.
It also introduces the teleparallel formulation. Chapter \ref{sec:Gauge-Transformations-and}
then discusses Euclidean gauge transformations and the role of edge
modes. Finally, in Chapter \ref{sec:Point-Particles-in} we introduce
matter degrees of freedom in the form of point particles.

In Part \ref{part:2P1-discrete}, we discretize the theory presented
in Part \ref{part:2P1-cont}. In Chapter \ref{sec:The-Discrete-Geometry},
we describe the discrete geometry we will be using. We define the
cellular decomposition and the dual spin network, including a derivation
of the spin network phase space, and then show how the continuous
geometry is truncated using holonomies. We also introduce the continuity
conditions relating the variables between different cells. Chapter
\ref{sec:Gauge-Transformations} then discusses gauge transformations
and symmetries in the discrete setting, which includes new symmetries
which did not exist in the continuous version. In Chapter \ref{sec:Discretizing-the-Symplectic-2P1}
we rigorously discretize the symplectic potential of the continuous
theory, and show how, from the continuous gravity phase space, we
obtain both the spin network and edge mode phase spaces. Chapters
\ref{sec:The-Gauss-and} and \ref{sec:The-Constraints-as} then analyzes,
in elaborate detail, the constraints obtained in the discrete theory
and the symmetries they generate. Finally, in Chapter \ref{subsec:Corner-Modes}
we repeat the derivation of Chapter \ref{sec:Discretizing-the-Symplectic-2P1}
from scratch under much simplifying assumptions.

Part \ref{part:3P1-cont} is the prelude for adapting our results
from the toy model of 2+1 dimensions to the physically relevant case
of 3+1 spacetime dimensions. In Chapter \ref{sec:Derivation-of-Ashtekar}
we provide a detailed and self-contained derivation of the Ashtekar
variables and the loop gravity Hamiltonian action, including the constraints.
Special care is taken to write everything in terms of index-free Lie-algebra-valued
differential forms, as we did in the 2+1-dimensional case, which --
in addition to being more elegant -- will greatly simplify our derivation.
Chapter \ref{sec:Cosmic-Strings-in} then introduces matter degrees
of freedom, this time in the form of cosmic strings, mirroring our
discussion of point particles in the 2+1-dimensional case.

Finally, in Part \ref{part:3P1-discrete} we discretize the 3+1-dimensional
theory. In Chapter \ref{sec:The-Discrete-Geometry-3P1} we introduce
the discrete geometry, where now the cells are 3-dimensional, and
discuss the truncation of the geometry using holonomies. In Chapter
\ref{sec:Discretizing-the-Symplectic} we perform a calculation analogous
to the one we performed in Chapters \ref{sec:Discretizing-the-Symplectic-2P1}
and later in \ref{subsec:Corner-Modes}, although it is of course
now more involved. We show that, in the 3+1-dimensional case as well,
the spin network phase space coupled to edge modes is obtained from
the continuous phase space. Moreover, we obtain a spectrum of polarizations
which includes a dual theory, as in the 2+1-dimensional case. Chapter
\ref{sec:Conclusions} summarizes the results of this thesis and presents
several avenues for potential future research.

\section{\label{subsec:Basic-Definitions-Notations}Basic Definitions, Notations,
and Conventions}

The following definitions, notations, and conventions will be used
throughout the thesis.

\subsection{\label{subsec:Lie-Group-and}Lie Group and Algebra Elements}

Let $G$ be a \emph{Lie group}, let $\mfg$ be its associated \emph{Lie
algebra}, and let $\mfg^{*}$ be the dual to that Lie algebra. The
\emph{cotangent bundle} of $G$ is the Lie group $T^{*}G\cong G\ltimes\mfg^{*}$,
where $\ltimes$ is the \emph{semidirect product}, and it has the
associated Lie algebra $\mfg\oplus\mfg^{*}$. We assume that this
group is the \emph{Euclidean} or \emph{Poincaré group}, or a generalization
thereof, and its algebra takes the form
\[
\left[\P_{i},\P_{j}\right]=0\sp\left[\J_{i},\J_{j}\right]=\dui f{ij}k\J_{k}\sp\left[\J_{i},\P_{j}\right]=\dui f{ij}k\P_{k},
\]
where:
\begin{itemize}
\item $\dui f{ij}k$ are the \emph{structure constants}, which satisfy anti-symmetry
$\dui f{ij}k=-\dui f{ji}k$ and the Jacobi identity $\dui f{[ij}l\dui f{k]l}m=0$.
\item $\J_{i}\in\mfg$ are the \emph{rotation }generators,
\item $\P_{i}\in\mfg^{*}$ are the \emph{translation }generators,
\item The indices $i,j,k$ take the values $0,1,2$ in the 2+1D case, where
they are internal spacetime indices, and $1,2,3$ in the 3+1D case,
where they are internal spatial indices.
\end{itemize}
Usually in the loop quantum gravity literature we take $G=\SUT$ such
that $\mfg^{*}=\BBR^{3}$ and
\[
\ISUT\cong\SUT\ltimes\BBR^{3}\cong T^{*}\SUT.
\]
However, here we will mostly keep $G$ abstract in order for the discussion
to be more general.

Throughout this thesis, different fonts and typefaces will distinguish
elements of different groups and algebras, or differential forms valued
in those groups and algebras, as follows:
\begin{itemize}
\item $G\ltimes\mfg^{*}$-valued forms will be written in Calligraphic font:
$\AA,\BB,\CC,...$
\item $\mfg\oplus\mfg^{*}$-valued forms will be written in bold Calligraphic
font: $\AAb,\BBb,\CCb,...$
\item $G$-valued forms will be written in regular font: $a,b,c,...$
\item $\mfg$ or $\mfg^{*}$-valued forms will be written in bold font:
$\a,\b,\c,...$
\end{itemize}

\subsection{\label{subsec:Spacetime-and-Spatial}Indices and Differential Forms}

Throughout this thesis, we will use the following conventions for
indices\footnote{The usage of lowercase Latin letters for both spatial and internal
spatial indices is somewhat confusing, but seems to be standard in
the literature, so we will use it here as well.}:
\begin{itemize}
\item In the 2+1D case:
\begin{itemize}
\item $\mu,\nu,\ldots\in\left\{ 0,1,2\right\} $ represent 2+1D spacetime
components.
\item $i,j,\ldots\in\left(0,1,2\right)$ represent 2+1D internal / Lie algebra
components.
\item $a,b,\ldots\in\left\{ 1,2\right\} $ represent 2D spatial components:
$\overbrace{0,\underbrace{1,2}_{a}}^{\mu}$.
\end{itemize}
\item In the 3+1D case:
\begin{itemize}
\item $\mu,\nu,\ldots\in\left\{ 0,1,2,3\right\} $ represent 3+1D spacetime
components.
\item $A,B,\ldots,I,J,\ldots\in\left(0,1,2,3\right)$ represent 3+1D internal
components.
\item $a,b,\ldots\in\left\{ 1,2,3\right\} $ represent 3D spatial components:
$\overbrace{0,\underbrace{1,2,3}_{a}}^{\mu}$.
\item $i,j,\ldots\in\left\{ 1,2,3\right\} $ represent 3D internal / Lie
algebra components: $\overbrace{0,\underbrace{1,2,3}_{i}}^{I}$.
\end{itemize}
\end{itemize}
We consider a 2+1D or 3+1D manifold $M$ with topology $\Sigma\xx\BBR$
where $\Sigma$ is a 2 or 3-dimensional spatial manifold and $\BBR$
represents time. Our metric signature convention is $\left(-,+,+\right)$
or $\left(-,+,+,+\right)$. In index-free notation, we denote a \emph{Lie-algebra-valued
differential form of degree $p$} (or \emph{$p$-form}) on $\Sigma$,
with one algebra index $i$ and $p$ spatial indices $a_{1},\ldots,a_{p}$,
as
\begin{equation}
\A\equiv\frac{1}{p!}A_{a_{1}\cdots a_{p}}^{i}\ta_{i}\thinspace\d x^{a_{1}}\wedge\cdots\wedge\d x^{a_{p}}\in\Omega^{p}\left(\Sigma,\mfg\right),\label{eq:index-free-notation}
\end{equation}
where $A_{a_{1}\cdots a_{p}}^{i}$ are the components and $\ta_{i}$
are the generators of the algebra $\mfg$ in which the form is valued.

Sometimes we will only care about the algebra index, and write $\A\equiv A^{i}\ta_{i}$
with the spatial indices implied, such that $A^{i}\equiv\frac{1}{p!}A_{a_{1}\cdots a_{p}}^{i}\d x^{a_{1}}\wedge\cdots\wedge\d x^{a_{p}}$
are real-valued $p$-forms. Other times we will only care about the
spacetime indices, and write $\A\equiv\frac{1}{p!}\A_{a_{1}\cdots a_{p}}\d x^{a_{1}}\wedge\cdots\wedge\d x^{a_{p}}$
with the algebra index implied, such that $\A_{a_{1}\cdots a_{p}}\equiv A_{a_{1}\cdots a_{p}}^{i}\ta_{i}$
are algebra-valued 0-forms.

\subsection{The Graded Commutator}

Given any two Lie-algebra-valued forms $\A$ and $\B$ of degrees\emph{
}$\deg\A$ and $\deg\B$ respectively, we define the \emph{graded
commutator}:
\[
\left[\A,\B\right]\equiv\A\wedge\B-\left(-1\right)^{\deg\A\deg\B}\B\wedge\A,
\]
which satisfies
\[
\left[\A,\B\right]=-\left(-1\right)^{\deg\A\deg\B}\left[\B,\A\right].
\]
If at least one of the forms has even degree, this reduces to the
usual anti-symmetric commutator; if we then interpret $\A$ and $\B$
as vectors in $\BBR^{3}$, then this is none other than the vector
cross product $\A\xx\B$. Note that $\left[\A,\B\right]$ is a Lie-algebra-valued
$\left(\deg\A+\deg\B\right)$-form.

The graded commutator satisfies the \emph{graded Leibniz rule}:
\[
\d\left[\A,\B\right]=\left[\d\A,\B\right]+\left(-1\right)^{\deg\A}\left[\A,\d\B\right].
\]
In terms of indices, with $\deg\A=p$ and $\deg\B=q$, we have
\[
\left[\A,\B\right]=\frac{1}{\left(p+q\right)!}\left[\A,\B\right]_{a_{1}\cdots a_{p}b_{1}\cdots b_{q}}^{k}\ta_{k}\d x^{a_{1}}\wedge\cdots\wedge\d x^{a_{p}}\wedge\d x^{b_{1}}\wedge\cdots\wedge\d x^{b_{q}},
\]
where
\[
\left[\A,\B\right]_{a_{1}\cdots a_{p}b_{1}\cdots b_{q}}^{k}\equiv\frac{\left(p+q\right)!}{p!q!}\udi{\epsilon}k{ij}A_{[a_{1}\cdots a_{p}}^{i}B_{b_{1}\cdots b_{q}]}^{j}.
\]
In terms of spatial indices alone, we have
\[
\left[\A,\B\right]_{a_{1}\cdots a_{p}b_{1}\cdots b_{q}}\equiv\frac{\left(p+q\right)!}{p!q!}\udi{\epsilon}k{ij}A_{[a_{1}\cdots a_{p}}^{i}B_{b_{1}\cdots b_{q}]}^{j}\ta_{k},
\]
and in terms of Lie algebra indices alone, we simply have
\[
\left[\A,\B\right]^{k}=\udi{\epsilon}k{ij}A^{i}B^{j}.
\]

\subsection{\label{subsec:The-Graded-Dot}The Graded Dot Product and the Triple
Product}

We define a \emph{dot (inner) product}, also known as the \emph{Killing
form}, on the generators of the Lie group as follows:
\begin{equation}
\J_{i}\cdot\P_{j}=\delta_{ij}\sp\J_{i}\cdot\J_{j}=\P_{i}\cdot\P_{j}=0,\label{eq:dot-product}
\end{equation}
where $\delta_{ij}$ is the Kronecker delta. Given two Lie-algebra-valued
forms $\A$ and $\B$ of degrees\emph{ }$\deg\A$ and $\deg\B$ respectively,
such that $\A\equiv A^{i}\J_{i}$ is a pure rotation and $\B\equiv B^{i}\P_{i}$
is a pure translation, we define the \emph{graded dot product}\footnote{In the general case, which will only be relevant for our discussion
of Chern-Simons theory, for $\mfg\oplus\mfg^{*}$-valued forms $\AAb\equiv\AA_{J}^{i}\J_{i}+\AA_{P}^{i}\P_{i}$
and $\BBb\equiv\BB_{J}^{i}\J_{i}+\BB_{P}^{i}\P_{i}$ we have
\[
\AAb\cdot\BBb=\delta_{ij}\left(\AA_{J}^{i}\wedge\BB_{P}^{j}+\AA_{P}^{i}\wedge\BB_{J}^{j}\right).
\]
}:
\[
\A\cdot\B\equiv\delta_{ij}A^{i}\wedge B^{j},
\]
where $\wedge$ is the usual \emph{wedge product}\footnote{Given any two differential forms $A$ and $B$, the wedge product
$A\wedge B$ is the $\left(\deg A+\deg B\right)$-form satisfying
$A\wedge B=\left(-1\right)^{\deg A\deg B}B\wedge A$ and $\d\left(A\wedge B\right)=\d A\wedge B+\left(-1\right)^{\deg A}A\wedge\d B$.}\emph{ }of differential forms. The dot product satisfies
\[
\A\cdot\B=\left(-1\right)^{\deg\A\deg\B}\B\cdot\A.
\]
Again, if at least one of the forms has even degree, this reduces
to the usual symmetric dot product. Note that $\A\cdot\B$ is a real-valued
$\left(\deg\A+\deg\B\right)$-form.

The graded dot product satisfies the graded Leibniz rule:
\[
\d\left(\A\cdot\B\right)=\d\A\cdot\B+\left(-1\right)^{\deg\A}\A\cdot\d\B.
\]
In terms of indices, with $\deg\A=p$ and $\deg\B=q$, we have
\[
\A\cdot\B=\frac{1}{\left(p+q\right)!}\left(\A\cdot\B\right)_{a_{1}\cdots a_{p}b_{1}\cdots b_{q}}\d x^{a_{1}}\wedge\cdots\wedge\d x^{a_{p}}\wedge\d x^{b_{1}}\wedge\cdots\wedge\d x^{b_{q}},
\]
where
\[
\left(\A\cdot\B\right)_{a_{1}\cdots a_{p}b_{1}\cdots b_{q}}=\frac{\left(p+q\right)!}{p!q!}\delta_{ij}A_{[a_{1}\cdots a_{p}}^{i}B_{b_{1}\cdots b_{q}]}^{j}.
\]
Since the graded dot product is a trace, and thus cyclic, it satisfies
\begin{equation}
\left(g^{-1}\A g\right)\cdot\left(g^{-1}\B g\right)=\A\cdot\B,\label{eq:cyclic}
\end{equation}
where $g$ is any group element. We will use this identity many times
throughout the thesis to simplify expressions.

Finally, by combining the dot product and the commutator, we obtain
the \emph{triple product}:
\begin{equation}
\left[\A,\B\right]\cdot\C=\A\cdot\left[\B,\C\right]=\epsilon_{ijk}A^{i}\wedge B^{j}\wedge C^{k}.\label{eq:triple-product}
\end{equation}
Note that this is a real-valued $\left(\deg\A+\deg\B+\deg\C\right)$-form.
The triple product inherits the symmetry and anti-symmetry properties
of the dot product and the commutator.

\subsection{\label{subsec:Variational-Anti-Derivations-on}Variational Anti-Derivations
on Field Space}

In addition to the familiar \emph{exterior derivative} (or \emph{differential})
$\d$ and \emph{interior product} $\iota$ on spacetime, we introduce
a \emph{variational exterior derivative} (or \emph{variational differential})
$\delta$ and a \emph{variational interior product} $I$ on field
space. These operators act analogously to $\d$ and $\iota$, and
in particular they are nilpotent, e.g. $\delta^{2}=0$, and satisfy
the graded Leibniz rule as defined above.

Degrees of differential forms are counted with respect to spacetime
and field space separately; for example, if $f$ is a 0-form then
$\d\delta f$ is a 1-form on spacetime, due to $\d$, and independently
also a 1-form on field space, due to $\delta$. The dot product defined
above also includes an implicit wedge product with respect to field-space
forms, such that e.g. $\delta\A\cdot\delta\B=-\delta\B\cdot\delta\A$
if $\A$ and $\B$ are 0-forms on field space. In this thesis, the
only place where one should watch out for the wedge product and graded
Leibniz rule on field space is when we will discuss the symplectic
form, which is a field-space 2-form; everywhere else, we will only
deal with field-space 0-forms and 1-forms.

We also define a convenient shorthand notation for the \emph{Maurer-Cartan
1-form }on field space:
\begin{equation}
\De g\equiv\delta gg^{-1},\label{eq:Maurer-Cartan}
\end{equation}
where $g$ is a $G$-valued 0-form, which satisfies
\begin{equation}
\De\left(gh\right)=\De g+g\De hg^{-1}=g\left(\De h-\De(g^{-1})\right)g^{-1},\label{eq:Delta-id-1}
\end{equation}
\begin{equation}
\De g^{-1}=-g^{-1}\De gg\sp\delta\left(\De g\right)=\hf\left[\De g,\De g\right].\label{eq:Delta-id-2}
\end{equation}
Note that $\De g$ is a $\mfg$-valued form; in fact, $\De$ can be
interpreted as a map from the Lie group $G$ to its Lie algebra $\mfg$.

\subsection{\label{subsec:Holonomies-and}$G\ltimes\protect\mfg^{*}$-valued
Holonomies and the Adjacent Subscript Rule}

A $G\ltimes\mfg^{*}$-valued holonomy from a point $a$ to a point
$b$ will be denoted as
\[
\HH_{ab}\equiv\pexp\int_{a}^{b}\AAb,
\]
where $\AAb$ is the $\mfg\oplus\mfg^{*}$-valued connection 1-form
and $\pexp$ is a \emph{path-ordered exponential}. Composition of
two holonomies works as follows:
\[
\HH_{ab}\HH_{bc}=\left(\pexp\int_{a}^{b}\AAb\right)\left(\pexp\int_{b}^{c}\AAb\right)=\pexp\int_{a}^{c}\AAb=\HH_{ac}.
\]
Therefore, in our notation, \textbf{adjacent holonomy subscripts must
always be identical}; a term such as $\HH_{ab}\HH_{cd}$ is illegal,
since one can only compose two holonomies if the second starts where
the first ends. Inversion of holonomies works as follows:
\[
\HH_{ab}^{-1}=\left(\pexp\int_{a}^{b}\AAb\right)^{-1}=\pexp\int_{b}^{a}\AAb=\HH_{ba}.
\]
For the Maurer-Cartan 1-form on field space, we \textbf{move the end
point of the holonomy to a superscript}:
\[
\De\HH_{a}^{b}\equiv\delta\HH_{ab}\HH_{ba}.
\]
On the right-hand side, the subscripts $b$ are adjacent, so the two
holonomies $\delta\HH_{ab}$ and $\HH_{ba}$ may be composed. However,
one can only compose $\De\HH_{a}^{b}$ with a holonomy that starts
at $a$, and $b$ is raised to a superscript to reflect that. For
example, $\De\HH_{a}^{b}\HH_{bc}$ is illegal, since this is actually
$\delta\HH_{ab}\HH_{ba}\HH_{bc}$ and the holonomies $\HH_{ba}$ and
$\HH_{bc}$ cannot be composed. However, $\De\HH_{a}^{b}\HH_{ac}$
is perfectly legal, and results in $\delta\HH_{ab}\HH_{ba}\HH_{ac}=\delta\HH_{ab}\HH_{bc}$.

Note that from (\ref{eq:Delta-id-1}) and (\ref{eq:Delta-id-2}) we
have
\[
\De\HH_{b}^{a}=-\HH_{ba}\De\HH_{a}^{b}\HH_{ab},
\]
\[
\De\HH_{a}^{c}=\De\left(\HH_{ab}\HH_{bc}\right)=\De\HH_{a}^{b}+\HH_{ab}\De\HH_{b}^{c}\HH_{ba}=\HH_{ab}\left(\De\HH_{b}^{c}-\De\HH_{b}^{a}\right)\HH_{ba},
\]
both of which are compatible with the adjacent subscripts rule.

\subsection{\label{subsec:The-Cartan-Decomposition}The Cartan Decomposition}

We can split a $G\ltimes\mfg^{*}$-valued (Euclidean) holonomy $\HH_{ab}$
into a \emph{rotational holonomy} $h_{ab}$, valued in $G$, and a
\emph{translational holonomy} $\x_{a}^{b}$, valued in $\mfg^{*}$.
We do this using the \emph{Cartan decomposition}
\[
\HH_{ab}\equiv\e^{\x_{a}^{b}}h_{ab}\sp\HH_{ab}\in\Omega^{0}\left(\Sigma,G\ltimes\mfg^{*}\right)\sp h_{ab}\in\Omega^{0}\left(\Sigma,G\right)\sp\x_{a}^{b}\in\Omega^{0}\left(\Sigma,\mfg^{*}\right).
\]
In the following, we will employ the useful identity
\begin{equation}
h\e^{\x}h^{-1}=\e^{h\x h^{-1}}\sp h\in\Omega^{0}\left(\Sigma,G\right)\sp\x\in\Omega^{0}\left(\Sigma,\mfg^{*}\right),\label{eq:holonomy-exp-id}
\end{equation}
which for matrix Lie algebras (such as the ones we use here) may be
proven by writing the exponential as a power series.

Taking the inverse of $\HH_{ab}$ and using (\ref{eq:holonomy-exp-id}),
we get
\[
\HH_{ab}^{-1}=\left(\e^{\x_{a}^{b}}h_{ab}\right)^{-1}=h_{ab}^{-1}\e^{-\x_{a}^{b}}=h_{ab}^{-1}\e^{-\x_{a}^{b}}\left(h_{ab}h_{ab}^{-1}\right)=\e^{-h_{ab}^{-1}\x_{a}^{b}h_{ab}}h_{ab}^{-1}.
\]
But on the other hand
\[
\HH_{ab}^{-1}=\HH_{ba}=\e^{\x_{b}^{a}}h_{ba}.
\]
Therefore, we conclude that
\begin{equation}
h_{ba}=h_{ab}^{-1}\sp\x_{b}^{a}=-h_{ab}^{-1}\x_{a}^{b}h_{ab}.\label{eq:holonomy-inv}
\end{equation}
Similarly, composing two $G\ltimes\mfg^{*}$-valued holonomies and
using (\ref{eq:holonomy-exp-id}) and (\ref{eq:holonomy-inv}), we
get
\begin{align*}
\HH_{ab}\HH_{bc} & =\left(\e^{\x_{a}^{b}}h_{ab}\right)\left(\e^{\x_{b}^{c}}h_{bc}\right)\\
 & =\e^{\x_{a}^{b}}h_{ab}\e^{\x_{b}^{c}}\left(h_{ba}h_{ab}\right)h_{bc}\\
 & =\e^{\x_{a}^{b}}\e^{h_{ab}\x_{b}^{c}h_{ba}}h_{ab}h_{bc}\\
 & =\e^{\x_{a}^{b}+h_{ab}\x_{b}^{c}h_{ba}}h_{ab}h_{bc},
\end{align*}
where we used the fact that $\mfg^{*}$ is abelian, and therefore
the exponentials may be combined linearly. On the other hand
\[
\HH_{ab}\HH_{bc}=\HH_{ac}=\e^{\x_{a}^{c}}h_{ac},
\]
so we conclude that
\begin{equation}
h_{ac}=h_{ab}h_{bc}\sp\x_{a}^{c}=\x_{a}^{b}\oplus\x_{b}^{c}\equiv\x_{a}^{b}+h_{ab}\x_{b}^{c}h_{ba}=h_{ab}\left(\x_{b}^{c}-\x_{b}^{a}\right)h_{ba},\label{eq:holonomy-comp}
\end{equation}
where in the second identity we denoted the composition of the two
translational holonomies with a $\oplus$, and used (\ref{eq:holonomy-inv})
to get the right-hand side. It is now clear why the end point of the
translational holonomy is a superscript -- again, this is for compatibility
with the adjacent subscript rule.

\clearpage{}

\part{\label{part:2P1-cont}2+1 Dimensions: The Continuous Theory}

\section{\label{sec:Chern-Simons-Theory-and}Chern-Simons Theory and 2+1D
Gravity}

\subsection{The Geometric Variables}

Consider a spacetime manifold $M$ as defined in Section \ref{subsec:Spacetime-and-Spatial}.
The geometry of spacetime is described, in the first-order formulation,
by a \emph{(co-)frame field} 1-form $e^{i}\equiv e_{\mu}^{i}\thinspace\d x^{\mu}$
and a \emph{spin connection} 1-form $\omega^{ij}\equiv\omega_{\mu}^{ij}\thinspace\d x^{\mu}$.
We can take the internal-space Hodge dual\footnote{See Footnote \ref{fn:The-Hodge-dual} for the definition of the Hodge
dual on spacetime. Here the definition is the same, except that the
star operator acts on the internal indices instead of the spacetime
indices. The trick we used here works because $\omega^{ij}$ is a
2-form on the internal space (since it is anti-symmetric), and the
Hodge dual of a 2-form in 3 dimensions is a 1-form.} of the spin connection and define a connection with only one internal
index, $A^{i}\equiv\hf\epsilon_{jk}^{i}\omega^{jk}$. We can then
identify the internal indices with algebra indices, where the connection
is valued in the rotation algebra $\mfg$ and the frame field is valued
in the translation algebra $\mfg^{*}$, and write these quantities
in convenient index-free notation:
\[
\A\equiv A_{\mu}^{i}\J_{i}\d x^{\mu}\sp\ee\equiv e_{\mu}^{i}\P_{i}\d x^{\mu}.
\]
We also collect them into the \emph{Chern-Simons connection 1-form}
$\AAb$, valued in $\mfg\oplus\mfg^{*}$:
\begin{equation}
\AAb\equiv\A+\ee\equiv A^{i}\J_{i}+e^{i}\P_{i},\label{eq:CS-split}
\end{equation}
where $\A\equiv A^{i}\J_{i}$ is the $\mfg$-valued connection 1-form
and $\ee\equiv e^{i}\P_{i}$ is the $\mfg^{*}$-valued frame field
1-form. Defining the \emph{covariant exterior derivatives }with respect
to $\AAb$ and $\A$,
\[
\d_{\AAb}\equiv\d+\left[\AAb,\cdot\right]\sp\d_{\A}\equiv\d+\left[\A,\cdot\right],
\]
we define the $\mfg\oplus\mfg^{*}$-valued \emph{curvature 2-form}
$\FFb$ as\footnote{Note that since $\AAb$ is not a tensor, the covariant derivative
acts on it with an extra $\hf$ factor which then ensures that $\FFb$
is a tensor. This also applies to $\A$ and $\F$ below.}:
\[
\FFb\equiv\d_{\AAb}\AAb=\d\AAb+\hf\left[\AAb,\AAb\right],
\]
which may be split into
\begin{equation}
\FFb\equiv\F+\T\equiv F^{i}\J_{i}+T^{i}\P_{i},\label{eq:CS-con-split}
\end{equation}
where $\F\equiv F^{i}\J_{i}$ is the $\mfg$-valued curvature 2-form
and $\T\equiv T^{i}\P_{i}$ is the $\mfg^{*}$-valued \emph{torsion
2-form}, and they are defined in terms of $\A$ and $\ee$ as
\[
\F\equiv\d_{\A}\A=\d\A+\hf\left[\A,\A\right]\sp\T\equiv\d_{\A}\ee\equiv\d\ee+\left[\A,\ee\right].
\]

\subsection{The Chern-Simons and Gravity Actions}

In our notation, the \emph{Chern-Simons action} is given by
\begin{equation}
S\left[\AAb\right]=\hf\int_{M}\AAb\cdot\left(\d\AAb+\trd\left[\AAb,\AAb\right]\right),\label{eq:CS-action}
\end{equation}
and its variation is
\[
\delta S\left[\AAb\right]=\int_{M}\left(\FFb\cdot\delta\AAb-\hf\d\left(\AAb\cdot\delta\AAb\right)\right).
\]
From this we can read the equation of motion
\begin{equation}
\FFb=0,\label{eq:eom-CS}
\end{equation}
 and, from the boundary term, the \emph{symplectic potential}
\begin{equation}
\Theta\left[\AAb\right]\equiv-\hf\int_{\Sigma}\AAb\cdot\delta\AAb,\label{eq:Theta-CS}
\end{equation}
which gives us the \emph{symplectic form}
\[
\Omega\left[\AAb\right]\equiv\delta\Theta\left[\AAb\right]=-\hf\int_{\Sigma}\delta\AAb\cdot\delta\AAb.
\]
Here, $\Sigma$ is a spatial slice as defined in Section \ref{subsec:Spacetime-and-Spatial}.
Furthermore, we can write the action\footnote{Here we use the following identities, derived from the properties
of the dot product (\ref{eq:dot-product}) and the graded commutator:
\[
\A\cdot\d\A=\ee\cdot\d\ee=\left[\ee,\ee\right]=\A\cdot\left[\A,\A\right]=\ee\cdot\left[\A,\ee\right]=0.
\]
} in terms of $\A$ and $\ee$:
\begin{equation}
S\left[\A,\ee\right]=\int_{M}\left(\ee\cdot\F-\hf\d\left(\A\cdot\ee\right)\right).\label{eq:action-dec}
\end{equation}
This is the action for 2+1D gravity, with an additional boundary term
(which is usually disregarded by assuming $M$ has no boundary). Using
the identity $\delta\F=\d_{\A}\delta\A$, we find the variation of
the action is
\[
\delta S\left[\A,\ee\right]=\int_{M}\left(\F\cdot\delta\ee+\T\cdot\delta\A-\hf\d\left(\ee\cdot\delta\A+\A\cdot\delta\ee\right)\right),
\]
and thus we see that the equations of motion are
\begin{equation}
\F=0\sp\T=0,\label{eq:eom-dec}
\end{equation}
and the (pre-)symplectic potential is
\begin{equation}
\Theta\left[\A,\ee\right]\equiv-\hf\int_{\Sigma}\left(\ee\cdot\delta\A+\A\cdot\delta\ee\right),\label{eq:Theta-dec}
\end{equation}
which corresponds to the symplectic form
\[
\Omega\left[\A,\ee\right]\equiv\delta\Theta\left[\A,\ee\right]=-\int_{\Sigma}\delta\ee\cdot\delta\A.
\]
Of course, (\ref{eq:eom-dec}) and (\ref{eq:Theta-dec}) may also
be derived from (\ref{eq:eom-CS}) and (\ref{eq:Theta-CS}).

\subsection{The Hamiltonian Formulation\protect\footnote{This derivation is based on the one in \cite{Geiller:2017xad}.}}

To go to the Hamiltonian formulation, we would like to reduce everything
to the spatial slice $\Sigma$. Since the symplectic potential is
already defined on the spatial slice, it stay!s the same. Let us write
the curvature and torsion 2-form components explicitly using the spatial
indices:
\[
\F\equiv\hf F_{\mu\nu}^{i}\J_{i}\thinspace\d x^{\mu}\wedge\d x^{\nu}\sp\T\equiv\hf T_{\mu\nu}^{i}\P_{i}\thinspace\d x^{\mu}\wedge\d x^{\nu},
\]
where
\[
F_{\mu\nu}^{i}=\partial_{[\mu}A_{\nu]}^{i}+\hf\epsilon_{jk}^{i}A_{\mu}^{j}A_{\nu}^{k}\sp T_{\mu\nu}^{i}=\partial_{[\mu}e_{\nu]}^{i}+\epsilon_{jk}^{i}A_{\mu}^{j}e_{\nu}^{k}.
\]
We also define the 3-dimensional spacetime Levi-Civita symbol $\ept^{\mu\nu\rho}$,
which is a tensor of density weight $1$ with upper indices or $-1$
with lower indices\footnote{The spacetime and spatial Levi-Civita symbols, which are tensor densities,
must be distinguished from the internal space Levi-Civita symbol $\epsilon^{ijk}$;
the internal space is flat, and thus the notion of tensor density
is irrelevant in this case.}, and is related to the 2-dimensional spatial Levi-Civita symbol by
$\ept^{ab}\equiv\ept^{0ab}$. The action becomes:
\begin{align*}
S & =\hf\int_{M}\d^{3}x\thinspace\ept^{\rho\mu\nu}e_{i\rho}F_{\mu\nu}^{i}\\
 & =\hf\int\d t\int_{\Sigma}\d^{2}x\thinspace\ept^{ab}\left(e_{i0}F_{ab}^{i}+2e_{ia}F_{b0}^{i}\right)\\
 & =\hf\int\d t\int_{\Sigma}\d^{2}x\thinspace\ept^{ab}\left(e_{i0}F_{ab}^{i}+2e_{ia}\left(\partial_{[b}A_{0]}^{i}+\hf\epsilon_{jk}^{i}A_{b}^{j}A_{0}^{k}\right)\right)\\
 & =\hf\int\d t\int_{\Sigma}\d^{2}x\thinspace\ept^{ab}\left(e_{i0}F_{ab}^{i}+e_{ia}\left(\partial_{b}A_{0}^{i}-\partial_{0}A_{b}^{i}\right)+e_{ia}\epsilon_{jk}^{i}A_{b}^{j}A_{0}^{k}\right)\\
 & =\hf\int\d t\int_{\Sigma}\d^{2}x\thinspace\ept^{ab}\left(e_{i0}F_{ab}^{i}+A_{i0}\left(\partial_{[a}e_{b]}^{i}+\epsilon_{jk}^{i}A_{a}^{j}e_{b}^{k}\right)+e_{ib}\partial_{0}A_{a}^{i}+\partial_{b}\left(e_{ia}A_{0}^{i}\right)\right)\\
 & =\hf\int\d t\left(\int_{\Sigma}\d^{2}x\thinspace\ept^{ab}\left(e_{i0}F_{ab}^{i}+A_{i0}T_{ab}^{i}+e_{ib}\partial_{0}A_{a}^{i}\right)+\partial_{b}\left(e_{ia}A_{0}^{i}\right)\right).
\end{align*}
We now pull back the connection, frame field, curvature, and torsion
to $\Sigma$, and write using index-free notation\footnote{For convenience, we use the same notation as for the 2+1-dimensional
quantities; however, since from now on we will use the 2-dimensional
quantities exclusively, this should not result in any confusion.}:
\[
\A\equiv A_{a}^{i}\J_{i}\thinspace\d x^{a}\sp\ee\equiv e_{a}^{i}\P_{i}\thinspace\d x^{a}\sp\F\equiv\hf F_{ab}^{i}\J_{i}\thinspace\d x^{a}\wedge\d x^{b}\sp\T\equiv\hf T_{ab}^{i}\P_{i}\thinspace\d x^{a}\wedge\d x^{b}.
\]
Since $\d^{2}x\thinspace\ept^{ab}=\d x^{a}\wedge\d x^{b}$, the action
becomes
\[
S=\int\d t\left(\int_{\Sigma}\left(\ee_{0}\cdot\F+\A_{0}\cdot\T+\hf\partial_{0}\A\cdot\ee\right)-\hf\d\left(\A_{0}\cdot\ee\right)\right).
\]
The third term, $\hf\partial_{0}\A\cdot\ee$, includes the only time
derivative, and from it we can read that $\A$ is the configuration
variable and $\ee$ is the conjugate momentum. Therefore, the Poisson
brackets are
\[
\left\{ A_{a}^{i}\left(x\right),e_{b}^{j}\left(y\right)\right\} =\ept_{ab}\delta^{ij}\delta\left(x,y\right).
\]
From the first two terms we can now read the constraints, implying
vanishing curvature and torsion on the spatial slice $\Sigma$:
\[
\F=0\sp\T=0.
\]
Of course, these are the same as the equations of motion (\ref{eq:eom-dec}).
Note that $\ee_{0}$ and $\A_{0}$ are \emph{Lagrange multipliers},
since they have no terms with time derivatives. We may relabel them
$\ph$ and $\th$ respectively and define the \emph{smeared Gauss
constraint }$G$ and the \emph{smeared curvature constraint }$F$:
\begin{equation}
F\left(\ph\right)\equiv\int_{\Sigma}\ph\cdot\F\sp G\left(\th\right)\equiv\int_{\Sigma}\th\cdot\T.\label{eq:smeared-const-2p1}
\end{equation}

\subsection{\label{sec:Phase-Space-Polarizations}Phase Space Polarizations and
Teleparallel Gravity}

The symplectic potential (\ref{eq:Theta-dec}) results in the symplectic
form
\[
\Omega\equiv\delta\Theta=-\int_{\Sigma}\delta\ee\cdot\delta\A.
\]
In fact, one may obtain the same symplectic form using a \textbf{family
of potentials }of the form
\begin{equation}
\Theta_{\lambda}=-\int_{\Sigma}\left(\left(1-\lambda\right)\ee\cdot\delta\A+\lambda\A\cdot\delta\ee\right),\label{eq:Theta-general}
\end{equation}
where the parameter $\lambda\in\left[0,1\right]$ determines the \emph{polarization
}of the phase space. This potential may be obtained from a family
of actions of the form
\begin{equation}
S_{\lambda}=\int_{M}\left(\ee\cdot\F-\lambda\d\left(\A\cdot\ee\right)\right),\label{eq:S_lambda}
\end{equation}
where the difference lies only in the boundary term and thus does
not affect the physics. Hence the choice of polarization does not
matter in the continuum, but it will be very important in the discrete
theory, as we will see below.

The equations of motion (or constraints, in the Hamiltonian formulation)
for any action of the form (\ref{eq:S_lambda}) are, as we have seen:
\begin{itemize}
\item The torsion (or Gauss) constraint $\T=0$,
\item The curvature constraint $\F=0$.
\end{itemize}
Now, recall that general relativity is formulated using the \emph{Levi-Civita
connection}, which is torsionless by definition. Thus, the torsion
constraint $\T=0$ can really be seen as \textbf{defining}\emph{ }the
connection $\A$ to be torsionless, and thus selecting the theory
to be general relativity. In this case, $\F=0$ is the true equation
of motion, describing the dynamics of the theory.

In the \emph{teleparallel formulation} of gravity we instead use the
\emph{Weitzenböck connection}, which is defined to be flat but not
necessarily torsionless. In this formulation, we interpret the curvature
constraint $\F=0$ as defining the connection $\A$ to be flat, while
$\T=0$ is the true equation of motion.

There are three cases of particular interest when considering the
choice of the parameter $\lambda$. The case $\lambda=0$ is the one
most suitable for 2+1D general relativity:
\begin{equation}
S_{\lambda=0}=\int_{M}\ee\cdot\F\sp\Theta_{\lambda=0}=-\int_{\Sigma}\ee\cdot\delta\A,\label{eq:STheta-0}
\end{equation}
since it indeed produces the familiar action for 2+1D gravity. The
case $\lambda=1/2$ is one most suitable for 2+1D Chern-Simons theory:
\[
S_{\lambda=\hf}=\int_{M}\left(\ee\cdot\F-\hf\d\left(\A\cdot\ee\right)\right)\sp\Theta_{\lambda=\hf}=-\hf\int_{\Sigma}\left(\ee\cdot\delta\A+\A\cdot\delta\ee\right),
\]
since it corresponds to the Chern-Simons action (\ref{eq:action-dec}).
Finally, the case $\lambda=1$ is one most suitable for 2+1D teleparallel
gravity:
\[
S_{\lambda=1}=\int_{M}\left(\ee\cdot\F-\d\left(\A\cdot\ee\right)\right)\sp\Theta_{\lambda=1}=-\int_{\Sigma}\A\cdot\delta\ee,
\]
as explained in \cite{Teleparallel}.

Further details about the different polarizations may be found in
\cite{Dupuis:2017otn}. However, the discretization procedure in that
paper did not take into account possible curvature and torsion degrees
of freedom. In this thesis, we will include these degrees of freedom
and discuss all possible polarizations of the phase space.

From now on, we will always deal with the full family of discretizations
$\lambda\in\left[0,1\right]$ in all generality, instead of choosing
a particular polarization. We will do this both in 2+1D and 3+1D.
Although the different choices of polarization are entirely equivalent
in the continuum, they become very important in the discrete theory,
and in fact, as we will see, the choices $\lambda=1$ and $\lambda=0$
lead to completely different and independent discretizations.

\section{\label{sec:Gauge-Transformations-and}Gauge Transformations and Edge
Modes}

\subsection{Euclidean Gauge Transformations}

In this section we will work with the choice $\lambda=0$, such that
the action and symplectic potential are given by (\ref{eq:STheta-0}):
\begin{equation}
S=\int_{M}\ee\cdot\F\sp\Theta=-\int_{\Sigma}\ee\cdot\delta\A,\label{eq:STheta-0-again}
\end{equation}

From the smeared constraints (\ref{eq:smeared-const-2p1}), one can
check that the curvature constraint $F$ generates translations, while
the Gauss constraint $G$ generates rotations. Together, they generate
\emph{Euclidean gauge transformations}
\begin{equation}
\A\mt g^{-1}\A g+g^{-1}\d g\sp\ee\mt g^{-1}\left(\ee+\d_{\A}\z\right)g,\label{eq:Euclid-transf-cont}
\end{equation}
where $g$ is a $G$-valued 0-form encoding rotations, and $\z$ is
a $\mfg^{*}$-valued 0-form encoding translations; pure rotations
correspond to $\z=0$ and pure translations correspond to $g=1$..
Under this transformation, the curvature and torsion transform as
\[
\F\mt g^{-1}\F g\sp\T\mt g^{-1}\left(\T+\left[\F,\z\right]\right)g,
\]
and the action (\ref{eq:STheta-0-again}) transforms as\footnote{Here we used the fact that 
\[
\left[\A,\z\right]\cdot\left[\A,\A\right]=-\z\cdot\left[\A,\left[\A,\A\right]\right]=0
\]
due to the Jacobi identity, and thus $\d_{\A}\z\cdot\F=\d\left(\z\cdot\F\right)$.}
\begin{equation}
S\mt S+\int_{\partial M}\z\cdot\F.\label{eq:action-trans}
\end{equation}
We see that the action is invariant up to a boundary term, which vanishes
on-shell due to the equation of motion $\F=0$. As for the symplectic
potential, we have\footnote{Recall our notation $\De g\equiv\delta gg^{-1}$ as defined in (\ref{eq:Maurer-Cartan}).}
\[
\delta\A\mt g^{-1}\left(\delta\A+\d_{\A}\De g\right)g,
\]
and therefore the symplectic potential in (\ref{eq:STheta-0-again})
transforms as
\[
\Theta\mt\Theta-\int_{\Sigma}\left(\ee\cdot\d_{\A}\De g+\d_{\A}\z\cdot\delta\A+\d_{\A}\z\cdot\d_{\A}\De g\right).
\]
However, we may write\footnote{Here we used the identities $\delta\F=\d_{\A}\delta\A$ and $\d_{\A}\d_{\A}\z=\left[\F,\z\right]$.}
\[
\ee\cdot\d_{\A}\De g=\T\cdot\De g-\d\left(\ee\cdot\De g\right),
\]
\[
\d_{\A}\z\cdot\delta\A=-\z\cdot\delta\F+\d\left(\z\cdot\delta\A\right),
\]
\[
\d_{\A}\z\cdot\d_{\A}\De g=\left[\F,\z\right]\cdot\De g-\d\left(\d_{\A}\z\cdot\De g\right).
\]
Using these relations, the transformed potential may be written as
\begin{equation}
\Theta\mt\Theta-\int_{\Sigma}\left(\left(\T+\left[\F,\z\right]\right)\cdot\De g-\z\cdot\delta\F\right)-\int_{\partial\Sigma}\left(\z\cdot\delta\A-\left(\ee+\d_{\A}\z\right)\cdot\De g\right).\label{eq:transformed-Theta}
\end{equation}
The first integral vanishes on-shell, after both equations of motion
$\F=\T=0$ are taken into account. The second integral is a (1-dimensional)
boundary term, which does \textbf{not} vanish on-shell. Usually we
assume that $\Sigma$ is a manifold without boundary, and therefore
this term may be neglected; however, here we will allow $\Sigma$
to have a non-empty boundary.

\subsection{\label{subsec:Edge-Modes}Edge Modes}

We may make the symplectic potential invariant under the Euclidean
transformation by defining two new fields, a $G$-valued 0-form $h$
and a $\mfg^{*}$-valued 0-form $\x$, which are defined to transform
under the gauge transformation with parameters $\left(g,\z\right)$
as follows \cite{Donnelly:2016auv,Geiller:2017xad}: 
\begin{equation}
h\mt g^{-1}h\sp\x\mt g^{-1}\left(\x-\z\right)g.\label{eq:Euclid-transf-a-b}
\end{equation}
We shall call them \emph{edge modes}. The reason for defining them
in this way is that they \textbf{cancel }the gauge transformation,
in the following sense. Using these fields, we may define the \emph{dressed
}connection and frame field, labeled\footnote{The hat, being a piece of clothing, is the natural choice to indicate
dressed variables.} by a hat:
\begin{equation}
\AAh\equiv h^{-1}\A h+h^{-1}\d h\sp\eeh\equiv h^{-1}\left(\ee+\d_{\A}\x\right)h.\label{eq:dressed-A-E}
\end{equation}
These are simply the original $\A$ and $\ee$, having\textbf{ already
undergone} a Euclidean transformation with the new fields $h$ and
$\x$ as parameters. It is easy to check that, due to the way we chose
the transformation of the fields $h$ and $\x$, the dressed connection
and frame field are invariant under any further gauge transformations,
since the transformations of $h$ and $\x$ exactly cancel out the
transformations of $\A$ and $\ee$:
\[
\AAh\mt\AAh\sp\eeh\mt\eeh.
\]
The dressed curvature and torsion are
\begin{equation}
\FFh\equiv\d_{\AAh}\AAh=\d\AAh+\hf\left[\AAh,\AAh\right]=h^{-1}\F h,\label{eq:dressed-F}
\end{equation}
\[
\TTh\equiv\d_{\AAh}\eeh=\d\eeh+\left[\AAh,\eeh\right]=h^{-1}\left(\T+\left[\F,\x\right]\right)h,
\]
and they are also invariant under Euclidean transformations,
\[
\FFh\mt\FFh\sp\TTh\mt\TTh.
\]
Replacing all of the quantities with their dressed versions, we obtain
the dressed action
\[
\hat{S}\equiv\int_{M}\eeh\cdot\FFh=\int_{M}\left(\ee+\d_{\A}\x\right)\cdot\F=S+\int_{\partial M}\x\cdot\F,
\]
note the similarity to (\ref{eq:action-trans}). By design, this action
is invariant\footnote{It was already invariant under rotations, but now it is invariant
under translations as well; this is why the edge mode $h$ does not
explicitly appear in the action.} under Euclidean transformations. Similarly, we have the dressed symplectic
potential:
\begin{align*}
\Thh & \equiv-\int_{\Sigma}\eeh\cdot\delta\AAh\\
 & =-\int_{\Sigma}\left(\ee+\d_{\A}\x\right)\cdot\left(\delta\A+\d_{\A}\De h\right)\\
 & =-\int_{\Sigma}\left(\ee\cdot\delta\A+\ee\cdot\d_{\A}\De h+\d_{\A}\x\cdot\delta\A+\d_{\A}\x\cdot\d_{\A}\De h\right),
\end{align*}
where as usual $\De h\equiv\delta hh^{-1}$, and we used the identity
\begin{equation}
\delta\AAh=h^{-1}\left(\delta\A+\d_{\A}\De h\right)h.\label{eq:delta-A-dressed}
\end{equation}
By manipulating this expression as we did for $\Theta$ in the last
chapter, we may write the dressed potential as
\begin{equation}
\Thh=\Theta-\int_{\Sigma}\left(\left(\T+\left[\F,\x\right]\right)\cdot\De h-\x\cdot\delta\F\right)-\int_{\partial\Sigma}\left(\x\cdot\delta\A-\left(\ee+\d_{\A}\x\right)\cdot\De h\right),\label{eq:dressed-Theta}
\end{equation}
note the similarity to (\ref{eq:transformed-Theta}). The first integral
vanishes on-shell, and the second integral is a boundary term.

\section{\label{sec:Point-Particles-in}Point Particles in 2+1 Dimensions}

\subsection{\label{sec:Delta-Functions-Solid-Angles}Delta Functions and Differential
Solid Angles}

Let us first prove an interesting result that we will use below. Consider
an $n$-dimensional flat Euclidean manifold. Let us define the \emph{volume
$n$-form }
\[
\epsilon\equiv\frac{1}{n!}\epsilon_{a_{1}\cdots a_{n}}\d x^{a_{1}}\wedge\cdots\wedge\d x^{a_{n}}=\d x^{1}\wedge\cdots\wedge\d x^{n}=\d^{n}x,
\]
where $\epsilon_{a_{1}\cdots a_{n}}$ is the Levi-Civita symbol. We
also define a \emph{radial coordinate }
\[
r^{2}\equiv x_{a}x^{a},
\]
such that 
\[
\d\left(r^{2}\right)=2r\thinspace\d r=2x_{a}\d x^{a}\soosp\d r=\frac{1}{r}x_{a}\d x^{a}.
\]
Taking the Hodge dual\footnote{\label{fn:The-Hodge-dual}The \emph{Hodge dual} of a $p$-form $B$
on an $n$-dimensional manifold is the $\left(n-p\right)$-form $\star B$
defined such that, for any $p$-form $A$,
\[
A\wedge\star B=\langle A,B\rangle\epsilon,
\]
where $\epsilon$ is the volume $n$-form defined above, and $\langle A,B\rangle$
is the symmetric inner product of $p$-forms, defined as
\[
\langle A,B\rangle\equiv\frac{1}{p!}A^{a_{1}\cdots a_{p}}B_{a_{1}\cdots a_{p}}.
\]
$\star$ is called the \emph{Hodge star operator}. In terms of indices,
the Hodge dual is given by
\[
\left(\star B\right)_{b_{1}\cdots b_{n-p}}=\frac{1}{p!}B_{a_{1}\cdots a_{p}}\udi{\epsilon}{a_{1}\cdots a_{p}}{b_{1}\cdots b_{n-p}},
\]
and its action on basis $p$-forms is given by
\[
\star\left(\d x^{a_{1}}\wedge\cdots\wedge\d x^{a_{p}}\right)\equiv\frac{1}{\left(n-p\right)!}\udi{\epsilon}{a_{1}\cdots a_{p}}{b_{1}\cdots b_{n-p}}\d x^{b_{1}}\wedge\cdots\wedge\d x^{b_{n-p}}.
\]
Interestingly, we have that $\star1=\epsilon$. Also, if acting with
the Hodge star on a $p$-forms twice, we get
\[
\star^{2}=\sign\left(g\right)\left(-1\right)^{p\left(n-p\right)},
\]
where $\sign\left(g\right)$ is the signature of the metric: $+1$
for Euclidean or $-1$ for Lorentzian signature.} of this 1-form, we get an $\left(n-1\right)$-form: 
\[
\star\d r=\frac{1}{\left(n-1\right)!}\frac{1}{r}x^{a}\epsilon_{aa_{1}\cdots a_{n-1}}\d x^{a_{1}}\wedge\cdots\wedge\d x^{a_{n-1}}.
\]
On the other hand, we have from the definition of the Hodge dual 
\[
\d r\wedge\star\d r=\langle\d r,\d r\rangle\epsilon=\epsilon,
\]
where 
\[
\langle\d r,\d r\rangle\equiv\left(\d r\right)_{a}\left(\d r\right)^{a}=\left(\frac{1}{r}x_{a}\right)\left(\frac{1}{r}x^{a}\right)=\frac{1}{r^{2}}x_{a}x^{a}=1.
\]
Let us calculate $\star\d r$ explicitly for $n=2$ and $n=3$. For
$n=2$ we define $x^{1}\equiv x$, $x^{2}\equiv y$, and thus obtain
the 1-form 
\[
\star\d r=\frac{1}{r}x^{a}\epsilon_{ab}\d x^{b}=\frac{1}{r}\left(x\thinspace\d y-y\thinspace\d x\right).
\]
Defining an angular coordinate $\phi$ using 
\[
\phi\equiv\arctan\left(\frac{y}{x}\right),
\]
we find by straightforward calculation that 
\[
\star\d r=r\thinspace\d\phi.
\]
Similarly, for $n=3$ with $x^{3}\equiv z$ we have the 2-form 
\[
\star\d r=\frac{1}{2r}x^{a}\epsilon_{abc}\d x^{b}\wedge\d x^{c}=\frac{1}{r}\left(x\thinspace\d y\wedge\d z+y\thinspace\d z\wedge\d x+z\d x\wedge\d y\right).
\]
Using the coordinate $\phi$ and the additional coordinate 
\[
\theta\equiv\arccos\left(\frac{z}{r}\right),
\]
we find 
\[
\star\d r=r^{2}\sin\theta\thinspace\d\theta\wedge\d\phi.
\]
Thus, we see that we indeed recover the usual volume elements for
spherical coordinates in 2 and 3 dimensions using $\d r\wedge\star\d r$.

Now, let us define, in $n$ dimensions, the $\left(n-1\right)$-form
\[
\Omega\equiv\frac{1}{r^{n-1}}\star\d r,
\]
which gives the differential solid angle of an $n$-sphere, for example:
\[
\Omega=\begin{cases}
\d\phi & n=2,\\
\sin\theta\thinspace\d\theta\wedge\d\phi & n=3.
\end{cases}
\]
Taking the exterior derivative, we get an $n$-form: 
\[
\d\Omega=\d\left(\frac{1}{r^{n-1}}\star\d r\right)=\d\left(\frac{1}{r^{n}}\right)\wedge\left(r\star\d r\right)+\frac{1}{r^{n}}\d\left(r\star\d r\right).
\]
For the first term, we have 
\[
\d\left(\frac{1}{r^{n}}\right)\wedge\left(r\star\d r\right)=\left(-\frac{n}{r^{n+1}}\d r\right)\wedge\left(r\star\d r\right)=-\frac{n}{r^{n}}\d r\wedge\star\d r=-\frac{n}{r^{n}}\epsilon.
\]
For the second term, we have 
\begin{align*}
\frac{1}{r^{n}}\d\left(r\star\d r\right) & =\frac{1}{\left(n-1\right)!}\frac{1}{r^{n}}\d\left(x^{a}\epsilon_{aa_{1}\cdots a_{n-1}}\d x^{a_{1}}\wedge\cdots\wedge\d x^{a_{n-1}}\right)\\
 & =\frac{1}{\left(n-1\right)!}\frac{1}{r^{n}}\epsilon_{aa_{1}\cdots a_{n-1}}\d x^{a}\wedge\d x^{a_{1}}\wedge\cdots\wedge\d x^{a_{n-1}}\\
 & =\frac{n}{r^{n}}\epsilon.
\end{align*}
Therefore the two terms cancel each other, and we get that 
\[
\d\Omega=0.
\]
Note that this applies everywhere except at the origin, $r=0$, since
$\Omega$ is undefined there. On the other hand, if we integrate over
the $n$-dimensional ball $B^{n}$, we find by Stokes' theorem 
\[
\int_{B^{n}}\d\Omega=\int_{S^{n-1}}\Omega=A_{n-1},
\]
where $S^{n-1}\equiv\partial B^{n}$ is the $\left(n-1\right)$-sphere
and $A_{n-1}$ is its area such that e.g. for $n=2,3$ we have: 
\[
A_{1}=2\pi\sp A_{2}=4\pi.
\]
The $n$-form $\d\Omega$ is zero everywhere except at $r=0$, yet
its integral over an $n$-dimensional volume is finite. In other words,
it behaves just like a \emph{Dirac delta function}. We thus conclude
that 
\[
\d\Omega=A_{n-1}\delta^{\left(n\right)},
\]
where $\delta^{\left(n\right)}$ is an $n$-form distribution such
that, for a 0-form $f$, 
\[
\int_{B^{n}}f\delta^{\left(n\right)}=f\left(0\right).
\]
In particular, for $n=2,3$, we find that 
\[
\d\left(\d\phi\right)=\d^{2}\phi=2\pi\delta^{\left(2\right)},
\]
\[
\d\left(\sin\theta\thinspace\d\theta\wedge\d\phi\right)=-\d^{2}\left(\cos\theta\thinspace\d\phi\right)=4\pi\delta^{\left(3\right)}.
\]

\subsection{Particles as Topological Defects}

In the following sections we will follow the formalism of \cite{Deser:1983tn,SousaGerbert1989,deSousaGerbert:1990yp,Meusburger:2005in,Hooft1988}.
Consider 2+1D polar coordinates $\left(t,r,\phi\right)$ with the
following metric:
\begin{equation}
\d s^{2}=-\left(\d t+S\thinspace\d\phi\right)^{2}+\frac{\d r^{2}}{\left(1-M\right)^{2}}+r^{2}\thinspace\d\phi^{2}.\label{eq:2p1-metric}
\end{equation}
Here, $M\in\left[0,1\right)$ is a ``mass'' and $S$ is a ``spin'',
and they are both constant real numbers. Let us now transform to the
following coordinates:
\begin{equation}
T\equiv t+S\phi\sp R\equiv\frac{r}{1-M}\sp\Phi\equiv\left(1-M\right)\phi.\label{eq:flat-coords}
\end{equation}
Then the metric becomes flat:
\[
\d s^{2}=-\d T^{2}+\d R^{2}+R^{2}\thinspace\d\Phi^{2}.
\]
However, the periodicity condition $\phi\sim\phi+2\pi$ becomes
\[
T\sim T+2\pi S\sp\Phi\sim\Phi+2\pi\left(1-M\right).
\]
The identification $\Phi\sim\Phi+2\pi\left(1-M\right)$ means that
in a plane of constant $T$, as we go around the origin, we find that
it only takes us $2\pi\left(1-M\right)$ radians to complete a full
circle, rather than $2\pi$ radians. Therefore we have obtained a
``Pac-Man''-like surface, where the angle of the ``mouth'' is
$2\pi M$, and both ends of the ``mouth'' are glued to each other.
This produces a cone with \emph{deficit angle }$2\pi M$. Note that
if $M=0$ and $S=0$, the particle is indistinguishable from flat
spacetime.

If the spin $S$ is non-zero, one end of the mouth is identified with
the other end, but at a different point in time -- there is a time
shift of $2\pi S$. This seems like it might create \emph{closed timelike
curves}, which would lead to causality violations \cite{FTL}. However,
when the spin is due to internal orbital angular momentum, the source
itself would need to be larger than the radius of any closed timelike
curves \cite{deSousaGerbert:1990yp,hooft1992causality}; thus, no
causality violations take place.

\subsection{The Frame Field and Spin Connection}

Let us now describe this geometry using a spin connection 1-form $\A\equiv A_{\mu}^{i}\J_{i}\d x^{\mu}$
and a frame field 1-form $\ee\equiv e_{\mu}^{i}\P_{i}\d x^{\mu}$
as above. For the frame field it is simplest to take
\[
e^{0}=\d T\sp e^{1}=\d R\sp e^{2}=R\thinspace\d\Phi,
\]
or in index-free notation
\[
\ee=\P_{0}\thinspace\d T+\P_{1}\thinspace\d R+R\thinspace\P_{2}\thinspace\d\Phi.
\]
To find the spin connection $\A$, we define, as above, the torsion
2-form $\T$:
\[
\T\equiv\d_{\A}\ee=\d\ee+\left[\A,\ee\right]\soosp T^{i}=\d e^{i}+\udi{\epsilon}i{jk}A^{j}\wedge e^{k}.
\]
The spin connection is the defined as the choice of $\A$ for which
$\T=0$. Explicitly, the components of the torsion are:
\[
T^{0}=A^{1}\wedge R\thinspace\d\Phi-A^{2}\wedge\d R,
\]
\[
T^{1}=A^{2}\wedge\d T-A^{0}\wedge R\thinspace\d\Phi,
\]
\[
T^{2}=R\thinspace\d^{2}\Phi+\d R\wedge\d\Phi+A^{0}\wedge\d R-A^{1}\wedge\d T.
\]
Note that we have written the 2-form $\d^{2}\Phi$ explicitly, since
$\Phi$ is not well-defined at $R=0$, and thus we are not guaranteed
to have $\d^{2}\Phi=0$. However, as we showed in Chapter \ref{sec:Delta-Functions-Solid-Angles},
\[
\d^{2}\Phi=2\pi\delta^{\left(2\right)}\left(\RR\right)\d X\wedge\d Y=2\pi\delta\left(R\right)\d R\wedge\d\Phi,
\]
where $\RR$ is the position vector with magnitude $R$, $\delta\left(R\right)\equiv R\thinspace\delta^{\left(2\right)}\left(\RR\right)$,
$X\equiv R\cos\Phi$, and $Y\equiv R\sin\Phi$. Therefore, $R\thinspace\d^{2}\Phi$
is evaluated at $R=0$, and this term vanishes.

It is easy to see that, if we want $\T$ to vanish, all of the components
of $\A$ must vanish except for $A^{0}$, which must take the value
$A^{0}=\d\Phi$ in order to cancel the $\d R\wedge\d\Phi$ term. Thus
we get that
\[
\A=\J_{0}\thinspace\d\Phi.
\]
Calculating the curvature $\F$ of the spin connection, we get
\[
\F\equiv\d_{\A}\A=\d\A+\hf\left[\A,\A\right]=\J_{0}\thinspace\d^{2}\Phi=2\pi\delta\left(R\right)\J_{0}\thinspace\d R\wedge\d\Phi,
\]
so the curvature is distributional, and vanishes everywhere except
at $R=0$.

Finally, we transform back from the coordinates $\left(T,R,\Phi\right)$
to $\left(t,r,\phi\right)$ using (\ref{eq:flat-coords}). The spin
connection becomes:
\[
\A=\left(1-M\right)\J_{0}\thinspace\d\phi,
\]
and the frame field becomes:
\[
\ee=\P_{0}\thinspace\d t+\frac{\P_{1}\thinspace\d r}{1-M}+\left(S\P_{0}+r\P_{2}\right)\d\phi.
\]
In the new coordinates, we find that the curvature and torsion are
both distributional:
\[
\F=\left(1-M\right)\J_{0}\thinspace\d^{2}\phi=2\pi\left(1-M\right)\delta\left(r\right)\J_{0}\thinspace\d r\wedge\d\phi,
\]
\[
\T=S\thinspace\P_{0}\thinspace\d^{2}\phi=2\pi S\thinspace\delta\left(r\right)\P_{0}\thinspace\d r\wedge\d\phi.
\]

Let us define $\M\equiv\left(1-M\right)\J_{0}$ and $\SS\equiv S\P_{0}$;
note that $\left[\M,\SS\right]=0$. Then we may write
\[
\A=\M\thinspace\d\phi\sp\ee=\P_{0}\thinspace\d t+\frac{1}{1-M}\P_{1}\thinspace\d r+\left(\SS+r\P_{2}\right)\d\phi,
\]
\[
\RR=2\pi\M\thinspace\delta\left(r\right)\d r\wedge\d\phi\sp\T=2\pi\SS\thinspace\delta\left(r\right)\d r\wedge\d\phi.
\]
The equations of motion (or constraints) of 2+1D gravity, $\F=\T=0$,
are satisfied everywhere except at the origin. This means that there
is a matter source (i.e. a right-hand side to the Einstein equation)
at the origin, which is of course the particle itself.

\subsection{\label{subsec:The-Dressed-Quantities}The Dressed Quantities}

The expressions for $\A$ and $\ee$ are not invariant under the gauge
transformation
\[
\A\mt h^{-1}\A h+h^{-1}\d h\sp\ee\mt h^{-1}\left(\ee+\d_{\A}\x\right)h,
\]
\[
\F\mt h^{-1}\F h\sp\T\mt h^{-1}\left(\T+\left[\F,\x\right]\right)h,
\]
where the gauge parameters are a $G$-valued 0-form $h$ and a $\mfg^{*}$-valued
0-form $\x$. When we apply these transformations, we get:
\[
\A=h^{-1}\M h\thinspace\d\phi+h^{-1}\d h,
\]
\[
\ee=h^{-1}\left(\d\x+\left(\SS+\left[\M,\x\right]\right)\d\phi\right)h,
\]
\[
\F=2\pi h^{-1}\M h\thinspace\delta\left(r\right)\d r\wedge\d\phi,
\]
\[
\T=2\pi h^{-1}\left(\SS+\left[\M,\x\right]\right)h\thinspace\delta\left(r\right)\d r\wedge\d\phi.
\]
These expressions are gauge-invariant, since any additional gauge
transformation will give the same expression with the new gauge parameters
composed with the old ones, much like the dressed variables we defined
in Section \ref{subsec:Edge-Modes}.

Below we will use a coordinate $\phi$ which is scaled by $2\pi$,
such that it has the range $\left[0,1\right)$ instead of $\left[0,2\pi\right)$.
In this case, we have that
\[
\d^{2}\Phi=\delta\left(R\right)\d R\wedge\d\Phi,
\]
and thus the expressions for the curvature and torsion are simplified:
\[
\F=h^{-1}\M h\thinspace\delta\left(r\right)\d r\wedge\d\phi,
\]
\[
\T=h^{-1}\left(\SS+\left[\M,\x\right]\right)h\thinspace\delta\left(r\right)\d r\wedge\d\phi.
\]
Finally, let us define a \emph{momentum }$\p$ and \emph{angular momentum
}$\j$:
\[
\p\equiv h^{-1}\M h,\qquad\j\equiv h^{-1}\left(\SS+\left[\M,\x\right]\right)h,
\]
which satisfy, as one would expect, the relations
\[
\p^{2}\equiv\M^{2}\sp\p\cdot\j=\M\cdot\SS.
\]
Then we have
\[
\F=\p\thinspace\delta\left(r\right)\d r\wedge\d\phi\sp\T=\j\thinspace\delta\left(r\right)\d r\wedge\d\phi.
\]
We see that the source of curvature is momentum, while the source
of torsion is angular momentum, as discussed in \cite{Meusburger:2003ta,Freidel2004,Meusburger:2005in,Meusburger:2005mg,Schroers2007,Meusburger:2008dc}.

\clearpage{}

\part{\label{part:2P1-discrete}2+1 Dimensions: The Discrete Theory}

\section{\label{sec:The-Discrete-Geometry}The Discrete Geometry}

\subsection{\label{subsec:The-Cellular-Decomposition}The Cellular Decomposition
and Its Dual}

We embed a cellular decomposition $\Delta$ and a dual cellular decomposition
$\Delta^{*}$ in our 2-dimensional spatial manifold $\Sigma$. These
structures consist of the following elements, where each element of
$\Delta$ is \textbf{uniquely dual }to an element of $\Delta^{*}$:
\begin{center}
\begin{tabular}{|c|c|c|}
\hline 
$\Delta$ &  & $\Delta^{*}$\tabularnewline
\hline 
\hline 
0-cells (\emph{vertices}) $v$ & dual to & 2-cells (\emph{faces}) $v^{*}$\tabularnewline
\hline 
1-cells (\emph{edges}) $e$ & dual to & 1-cells (\emph{links}) $e^{*}$\tabularnewline
\hline 
2-cells (\emph{cells}) $c$ & dual to & 0-cells (\emph{nodes}) $c^{*}$\tabularnewline
\hline 
\end{tabular}
\par\end{center}

\begin{figure}[!h]
\begin{centering}
\begin{tikzpicture}[scale=0.8]
	\begin{pgfonlayer}{nodelayer}
		\node [style=none] (0) at (0.25, -1.9) {$c^{*}$};
		\node [style=none] (1) at (0, -0.45) {$v$};
		\node [style=none] (2) at (1.5, 0.6) {$c^{\prime*}$};
		\node [style=none] (3) at (5.25, -3.4) {$v'$};
		\node [style=Vertex] (4) at (0, -0) {};
		\node [style=Vertex] (5) at (0, 4) {};
		\node [style=Vertex] (6) at (-5, -3) {};
		\node [style=Vertex] (7) at (5, -3) {};
		\node [style=Node] (8) at (0, -1.5) {};
		\node [style=Node] (9) at (1.25, 1) {};
		\node [style=Node] (10) at (-1.25, 1) {};
		\node [style=none] (11) at (-3.5, 3.25) {};
		\node [style=none] (12) at (3.5, 3.25) {};
		\node [style=none] (13) at (0, -4.5) {};
	\end{pgfonlayer}
	\begin{pgfonlayer}{edgelayer}
		\draw [style=Edge] (5) to (6);
		\draw [style=Edge] (6) to (7);
		\draw [style=Edge] (7) to (5);
		\draw [style=Edge] (5) to (4);
		\draw [style=Edge] (4) to (7);
		\draw [style=Edge] (4) to (6);
		\draw [style=Link] (8) to (9);
		\draw [style=Link] (9) to (10);
		\draw [style=Link] (10) to (8);
		\draw [style=Link] (10) to (11.center);
		\draw [style=Link] (9) to (12.center);
		\draw [style=Link] (8) to (13.center);
	\end{pgfonlayer}
\end{tikzpicture}
\par\end{centering}
\caption{\label{fig:Triangle}A simple piece of the cellular decomposition
$\Delta$, in black, and its dual spin network $\Gamma^{*}$, in blue.
The vertices $v$ of the 1-skeleton $\Gamma\subset\Delta$ are shown
as black circles, while the nodes $c^{*}$ of $\Gamma^{*}$ are shown
as blue squares. The edges $e\in\Gamma$ are shown as black solid
lines, while the links $e^{*}\in\Gamma^{*}$ are shown as blue dashed
lines. In particular, two nodes $c^{*}$ and $c^{\prime*}$, connected
by a link $e^{*}=\left(cc'\right)^{*}$, are labeled, as well as two
vertices $v$ and $v'$, connected by an edge $e=\left(vv'\right)=\left(cc'\right)=c\cap c'$,
which is dual to the link $e^{*}$. There is one face in the illustration,
$v^{*}$, which is the triangle enclosed by the three blue links at
the center.}
\end{figure}
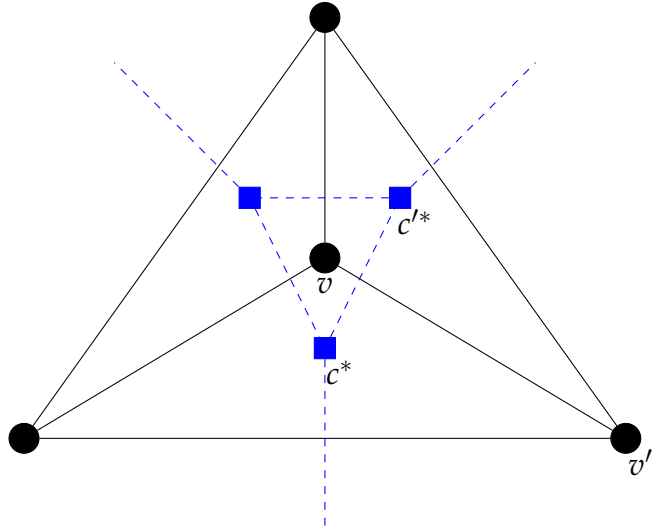

The \emph{1-skeleton graph} $\Gamma\subset\Delta$ is the set of all
vertices and edges of $\Delta$. Its dual is the \emph{spin network
graph} $\Gamma^{*}\subset\Delta^{*}$, the set of all nodes and links
of $\Delta^{*}$. Both graphs are oriented, and we write $e=\left(vv'\right)$
to indicate that the edge $e$ starts at the vertex $v$ and ends
at $v'$, and $e^{*}=\left(cc'\right)^{*}$ to indicate that the link
$e^{*}$ starts at the node $c^{*}$ and ends at $c^{\prime*}$. Furthermore,
since edges are where two cells intersect, we write $e=\left(cc'\right)\equiv\partial c\cap\partial c'$
to denote that the edge $e$ is the intersection of the boundaries
$\partial c$ and $\partial c'$ of the cells $c$ and $c'$ respectively.
If the link $e^{*}$ is dual to the edge $e$, then we have that $e=\left(cc'\right)$
and $e^{*}=\left(cc'\right)^{*}$; therefore the notation is consistent.
This construction is illustrated in Figure \ref{fig:Triangle}.

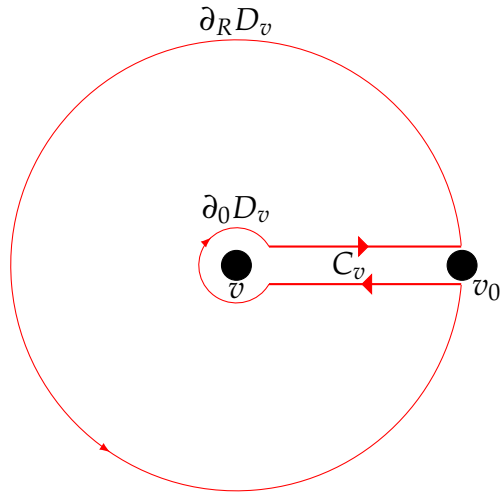
\begin{figure}[!h]
\begin{centering}
\begin{tikzpicture}
	\node [style=Vertex] at (0, 0) {};
	\node [style=none] at (0, -0.33) {$v$};
	\node [style=none] at (0, 0.75) {$\partial_{0}D_{v}$};
	\node [style=none] at (0, 3.25) {$\partial_{R}D_{v}$};
	\node [style=none] at (1.5, 0) {$C_{v}$};
	\centerarc [style=SegmentArrow] (0,0) (330:30:0.5);
	\centerarc [style=SegmentArrow] (0,0) (5:355:3);
	\draw [red,thick] ($({0.5*cos(30)},{0.5*sin(30)})$) -- node {\midarrow} ($({3*cos(5)},{0.5*sin(30)})$);
	\draw [red,thick]  ($({0.5*cos(330)},{0.5*sin(330)})$) -- node {\midarrowop} ($({3*cos(355)},{0.5*sin(330)})$);
	\node [style=Vertex] () at (3, 0) {};
	\node [style=none] () at (3.33, -0.33) {$v_{0}$};
\end{tikzpicture}
\par\end{centering}
\caption{\label{fig:Disk}The punctured disk $D_{v}$. The figure shows the
vertex $v$, cut $C_{v}$, inner boundary $\partial_{0}D_{v}$, outer
boundary $\partial_{R}D_{v}$, and reference point $v_{0}$.}
\end{figure}

For the purpose of doing calculations, it will prove useful to introduce
\emph{disks} $\Db_{v}$ around each vertex $v$. The disks have a
radius $R$, small enough that the entire disk $\Db_{v}$ is inside
the face $v^{*}$ for every $v$. We also define \emph{punctured disks}
$D_{v}$, which are obtained from the full disks $\Db_{v}$ by removing
the vertex $v$, which is at the center, and a \emph{cut }$C_{v}$,
connecting $v$ to an arbitrary point $v_{0}$ on the boundary $\partial\Db_{v}$.
Thus\footnote{Note that $v,v_{0}\in C_{v}$.}
\[
D_{v}\equiv\Db_{v}\backslash C_{v}.
\]
The punctured disks are equipped with a cylindrical coordinate system
$\left(r_{v},\phi_{v}\right)$ such that $r_{v}\in\left(0,R\right)$
and $\phi_{v}\in\left(\alpha_{v}-\hf,\alpha_{v}+\hf\right)$; note
that $\phi_{v}$ is scaled by $2\pi$, so it has a period of 1, for
notational brevity. The boundary of the punctured disk is such that
\[
\partial D_{v}=\partial_{0}D_{v}\cup C_{v}\cup\partial_{R}D_{v}\cup\Cb_{v},
\]
where $\partial_{0}D_{v}$ is the \emph{inner boundary }at $r_{v}=0$,
$C_{v}$ is the cut at $\phi_{v}=\alpha_{v}-\hf$ going from $r_{v}=0$
to $r_{v}=R$, $\partial_{R}D_{v}$ is the \emph{outer boundary }at
$r_{v}=R$, and $\Cb_{v}$ is the other side of the cut (with reverse
orientation), at $\phi_{v}=\alpha_{v}+\hf$ and going from $r_{v}=R$
to $r_{v}=0$. Note that $\partial_{R}D_{v}=\partial\Db_{v}$. The
punctured disk is illustrated in Figure \ref{fig:Disk}.

The outer boundary $\partial_{R}D_{v}$ of each disk is composed of
arcs $\left(vc_{i}\right)$ such that
\[
\partial_{R}D_{v}=\bigcup_{i=1}^{N_{v}}\left(vc_{i}\right),
\]
where $N_{v}$ is the number of cells around $v$ and the cells are
enumerated $c_{1},\ldots,c_{N_{v}}$. Similarly, the boundary $\partial c$
of the cell $c$ is composed of edges $\left(cc_{i}\right)$ and arcs
$\left(cv_{i}\right)$ such that
\[
\partial c=\bigcup_{i=1}^{N_{c}}\left(\left(cc_{i}\right)\cup\left(cv_{i}\right)\right),
\]
where $N_{c}$ is the number of cells adjacent to $c$ or, equivalently,
the number of vertices around $c$. We will use these decompositions
during the discretization process.

\subsection{\label{subsec:Classical-Spin-Networks}Classical Spin Networks}

Our goal is to show that, by discretizing the continuous phase space,
we may obtain the \emph{spin network phase space }of loop quantum
gravity. Let us therefore begin by studying this phase space.

\subsubsection{The Spin Network Phase Space}

In the previous section, we defined the spin network $\Gamma^{*}$
as a collection of links $e^{*}$ connecting nodes $c^{*}$. The kinematical
spin network phase space is isomorphic to a direct product of $T^{*}G$
cotangent bundles for each link $e^{*}\in\Gamma^{*}$:
\[
P_{\Gamma^{*}}\equiv\underset{e^{*}\in\Gamma^{*}}{\prod}T^{*}G.
\]
Since $T^{*}G\cong G\ltimes\mfg^{*}$, the phase space variables are
a group element $h_{e^{*}}\in G$ and a Lie algebra element $\X_{e^{*}}\in\mfg^{*}$
for each link $e^{*}\in\Gamma^{*}$. Under orientation reversal of
the link $e^{*}$ we have
\[
h_{e^{*-1}}=h_{e^{*}}^{-1}\sp\X_{e^{*-1}}=-h_{e^{*}}^{-1}\X_{e^{*}}h_{e^{*}}.
\]
These variables satisfy the Poisson algebra derived in the next section:
\[
\{h_{e^{*}},h_{e^{\prime*}}\}=0\sp\{X_{e^{*}}^{i},X_{e^{\prime*}}^{j}\}=\delta_{e^{*}e^{\prime*}}\udi{\epsilon}{ij}kX_{e^{*}}^{k}\sp\{X_{e^{*}}^{i},h_{e^{\prime*}}\}=\delta_{e^{*}e^{\prime*}}h_{e^{*}}\J^{i},
\]
where $e^{*}$ and $e^{\prime*}$ are two links and $\J^{i}$ are
the generators of $\mfg$.

The symplectic potential is
\begin{equation}
\Theta=\sum_{e^{*}\in\Gamma^{*}}\De h_{e^{*}}\cdot\X_{e^{*}},\label{eq:spin-network-symplectic-potential}
\end{equation}
where we used the graded dot product defined in Section \ref{subsec:The-Graded-Dot}
and the Maurer-Cartan form defined in (\ref{eq:Maurer-Cartan}). This
phase space enjoys the action of the gauge group $G^{N}$, where $N$
is the number of nodes in $\Gamma^{*}$. This action is generated
by the \emph{discrete Gauss constraint} at each node,
\[
\G_{c}\equiv\sum_{e^{*}\ni c^{*}}\X_{e^{*}},
\]
where $e^{*}\ni c^{*}$ means ``all links $e^{*}$ connected to the
node $c^{*}$''. This means that the sum of the fluxes vanishes when
summed over all the links connected to the node $c^{*}$. Later we
will see a nicer interpretation of this constraint. Given a link $e^{*}=\left(cc'\right)^{*}$,
the action of the Gauss constraint is given in terms of two group
elements $g_{c},g_{c'}\in G$, one at each node, as
\[
h_{e^{*}}\mt g_{c}h_{e^{*}}g_{c'}^{-1}\sp\X_{e^{*}}\mt g_{c}\X_{e^{*}}g_{c}^{-1}.
\]

\subsubsection{Calculation of the Poisson Brackets}

Let us calculate the Poisson brackets of the spin network phase space
$T^{*}G$, for one link. For the Maurer-Cartan form, we use the notation
\[
\th\equiv-\De h\equiv-\delta hh^{-1}.
\]
This serves two purposes: first, we can talk about the components
$\theta^{i}$ of $\th$ without the notation getting too cluttered,
and second, from (\ref{eq:Delta-id-2}), the Maurer-Cartan form thus
defined satisfies the \emph{Maurer-Cartan structure equation}
\begin{equation}
\F\left(\th\right)\equiv\delta_{\th}\th\equiv\delta\th+\hf\left[\th,\th\right]=0,\label{eq:curv-th}
\end{equation}
where $\F\left(\th\right)$ is the curvature of $\th$. Note that
$\th$ is a $\mfg$-valued 1-form on field space, and we also have
a $\mfg^{*}$-valued 0-form $\X$, the flux. We take a set of vector
fields $\q_{i}$ and $\p_{i}$ for $i\in\left\{ 1,2,3\right\} $ which
are chosen to satisfy
\[
\q_{j}\lrc\theta^{i}=\p_{j}\lrc\delta X^{i}=\delta_{J}^{i}\sp\p_{j}\lrc\theta^{i}=\q_{j}\lrc\delta X^{i}=0,
\]
where $\lrc$ is the usual interior product on differential forms\footnote{\label{fn:The-interior-product}The \emph{interior product }$V\lrc A$
of a vector $V$ with a $p$-form $A$, sometimes written $\iota_{V}A$
and sometimes called the \emph{contraction }of $V$ with $A$, is
the $\left(p-1\right)$-form with components
\[
\left(V\lrc A\right)_{a_{2}\cdots a_{p}}\equiv V^{a_{1}}A_{a_{1}\cdots a_{p}}.
\]
}. The symplectic potential on one link is taken to be
\[
\Theta=-\th\cdot\X,
\]
and its symplectic form is
\begin{align*}
\Omega\equiv\delta\Theta & =-\delta\left(\th\cdot\X\right)=-\left(\delta\th\cdot\X-\th\cdot\delta\X\right)\\
 & =-\left(-\hf\left[\th,\th\right]\cdot\X-\th\cdot\delta\X\right)\\
 & =\delta\X\cdot\th+\hf\X\cdot\left[\th,\th\right],
\end{align*}
where in the first line we used the graded Leibniz rule (on field-space
forms) and in the second line we used (\ref{eq:curv-th}). In components,
we have
\[
\Omega=\delta X_{k}\wedge\theta^{k}+\hf\epsilon_{ijk}\theta^{i}\wedge\theta^{j}X^{k}.
\]
Now, recall the definition of the \emph{Hamiltonian vector field}
of $f$: it is the vector field $\H_{f}$ satisfying
\[
\H_{f}\lrc\Omega=-\delta f.
\]
Let us contract the vector field $\q_{i}$ with $\Omega$ using $\q_{j}\lrc\theta^{i}=\delta_{J}^{i}$
and $\q_{j}\lrc\delta X^{i}=0$:
\begin{align*}
\q_{l}\lrc\Omega & =\q_{l}\lrc\left(\delta X_{k}\wedge\theta^{k}+\hf\epsilon_{ijk}\theta^{i}\wedge\theta^{j}X^{k}\right)\\
 & =-\delta X_{k}\delta_{l}^{k}+\hf\epsilon_{ijk}\delta_{l}^{i}\theta^{j}X^{k}-\hf\epsilon_{ijk}\delta_{l}^{j}\theta^{i}X^{k}\\
 & =-\delta X_{l}+\epsilon_{ljk}\theta^{j}X^{k}.
\end{align*}
Similarly, let us contract $\p_{i}$ with $\Omega$ using $\p_{j}\lrc\delta X^{i}=\delta_{J}^{i}$
and $\p_{j}\lrc\theta^{i}=0$:
\[
\p_{l}\lrc\Omega=\p_{l}\lrc\left(\delta X_{k}\wedge\theta^{k}+\hf\epsilon_{ijk}\theta^{i}\wedge\theta^{j}X^{k}\right)=\delta_{kl}\theta^{k}=\theta_{l}.
\]
Note that
\[
-\delta X_{i}=\q_{i}\lrc\Omega-\epsilon_{ijk}\theta^{j}X^{k}=\q_{i}\lrc\Omega-\epsilon_{ijk}\left(\p^{j}\lrc\Omega\right)X^{k}.
\]
Thus, we can construct the Hamiltonian vector field for $X^{i}$:
\[
\H_{X_{i}}\equiv\q_{i}-\epsilon_{ijk}\p^{j}X^{k}\sp\H_{X_{i}}\lrc\Omega=-\delta X_{i}.
\]
As for $h$, we consider explicitly the matrix components in the fundamental
representation, $h_{\ B}^{A}$. The Hamiltonian vector field for the
component $h_{\ B}^{A}$ satisfies, by definition,
\[
\H_{h_{\ B}^{A}}\lrc\Omega=-\delta h_{\ B}^{A}.
\]
If we multiply by $\left(h^{-1}\right)_{\ C}^{B}$, we get
\[
\left(\H_{h_{\ B}^{A}}\left(h^{-1}\right)_{\ C}^{B}\right)\lrc\Omega=-\delta h_{\ B}^{A}\left(h^{-1}\right)_{\ C}^{B}=\left(-\delta hh^{-1}\right)_{\ C}^{A}=\theta^{i}\left(\J_{i}\right)_{\ C}^{A}=\left(\p^{i}\left(\J_{i}\right)_{\ C}^{A}\right)\lrc\Omega.
\]
Thus we conclude that the Hamiltonian vector field for $h_{\ B}^{A}$
is
\[
\H_{h_{\ B}^{A}}=\left(h\J_{i}\right)_{\ B}^{A}\p^{i}.
\]
Now that we have found $\H_{X_{i}}$ and $\H_{h_{\ B}^{A}}$, we can
finally calculate the Poisson brackets. First, we have
\begin{align*}
\left\{ h_{\ B}^{A},h_{\ D}^{C}\right\}  & =-\Omega\left(\H_{h_{\ B}^{A}},\H_{h_{\ D}^{C}}\right)\\
 & =-\left(\delta X_{k}\wedge\theta^{k}+\hf\epsilon_{ijk}\theta^{i}\wedge\theta^{j}X^{k}\right)\left(\left(h\J_{l}\right)_{\ B}^{A}\p^{l},\left(h\J_{m}\right)_{\ D}^{C}\p^{m}\right)\\
 & =0,
\end{align*}
since $\p_{j}\lrc\theta^{i}=0$. Thus
\[
\left\{ h,h\right\} =0.
\]
Next, we have
\begin{align*}
\left\{ X^{i},X^{j}\right\}  & =-\Omega\left(\H_{X^{i}},\H_{X^{j}}\right)\\
 & =-\left(\delta X_{k}\wedge\theta^{k}+\hf\epsilon_{pqk}\theta^{p}\wedge\theta^{q}X^{k}\right)\left(\q^{i}-\epsilon_{lm}^{i}\p^{l}X^{m},\q^{j}-\epsilon_{no}^{j}\p^{n}X^{o}\right)\\
 & =\epsilon_{lm}^{i}\delta_{k}^{l}X^{m}\delta^{jk}-\delta^{ik}\epsilon_{no}^{j}\delta_{k}^{n}X^{o}-\hf\epsilon_{pqk}X^{k}\left(\delta^{ip}\delta^{jq}-\delta^{iq}\delta^{jp}\right)\\
 & =\epsilon_{k}^{ij}X^{k}-\epsilon_{k}^{ji}X^{k}-\hf\epsilon_{k}^{ij}X^{k}+\hf\epsilon_{k}^{ji}X^{k}\\
 & =\epsilon_{k}^{ij}X^{k}.
\end{align*}
Finally, we have
\begin{align*}
\left\{ X^{i},h_{\ B}^{A}\right\}  & =-\Omega\left(\H_{X^{i}},\H_{h_{\ B}^{A}}\right)\\
 & =-\left(\delta X_{k}\wedge\theta^{k}+\hf\epsilon_{ijk}\theta^{i}\wedge\theta^{j}X^{k}\right)\left(\q^{i}-\epsilon_{mn}^{i}\p^{m}X^{n},\left(h\J_{l}\right)_{\ B}^{A}\p^{l}\right)\\
 & =\delta^{ki}\left(h\J_{l}\right)_{\ B}^{A}\delta_{k}^{l}\\
 & =\left(h\J^{i}\right)_{\ B}^{A},
\end{align*}
so
\[
\left\{ X^{i},h\right\} =h\J^{i}.
\]
We conclude that the Poisson brackets are
\[
\{h,h\}=0\sp\{X^{i},X^{j}\}=\epsilon_{k}^{ij}X^{k}\sp\{X^{i},h\}=h\J^{i}.
\]
All of this was calculated on one link $e^{*}$. To get the Poisson
brackets for two phase space variables which are not necessarily on
the same link, we simply add a Kronecker delta function:
\[
\left\{ h_{e^{*}},h_{e^{\prime*}}\right\} =0\sp\left\{ X_{e^{*}}^{i},X_{e^{\prime*}}^{j}\right\} =\delta_{e^{*}e^{\prime*}}\udi{\epsilon}{ij}kX_{e^{*}}^{k}\sp\left\{ X_{e^{*}}^{i},h_{e^{\prime*}}\right\} =\delta_{e^{*}e^{\prime*}}h_{e^{*}}\J^{i}.
\]
This concludes our discussion of the spin network phase space.

\subsection{\label{subsec:Truncating-the-Geometry}Truncating the Geometry to
the Vertices}

\subsubsection{Motivation}

Before the equations of motion (i.e. the curvature and torsion constraints
$\F=\T=0$) are applied, the geometry on $\Sigma$ can have arbitrary
curvature and torsion. We would like to capture the ``essence''
of the curvature and torsion and encode them on codimension 2 defects.

For this purpose, we can imagine looking at every possible loop on
the spin network graph $\Gamma^{*}$ and taking a holonomy in $G\ltimes\mfg^{*}$
around it. This holonomy will have a part valued in $\mfg$, which
will encode the curvature, and a part valued in $\mfg^{*}$, which
will encode the torsion.

A loop of the spin network is the boundary $\partial v^{*}$ of a
face $v^{*}$. Since the face is dual to a vertex $v$, the natural
place to encode the geometry would be at the vertex. Thus, we will
place the defects at the vertices, and give them the appropriate values
in $\mfg\oplus\mfg^{*}$ obtained by the holonomies.

The disks $\Db_{v}$ defined above are in a 1-to-1 correspondence
with the faces $v^{*}$. In fact, we can imagine deforming the disks
such that they cover the faces, and their boundaries $\partial\Db_{v}$
are exactly the loops $\partial v^{*}$. Thus, we may perform calculations
on the disks instead on the faces.

This intuitive and qualitative motivation will be made precise in
the following sections.

\subsubsection{The Chern-Simons Connection on the Disks and Cells}

We define the Chern-Simons\footnote{Recall that, as explained in Section \ref{subsec:Lie-Group-and},
we use calligraphic font to denote forms valued in $G\ltimes\mfg^{*}$,
and bold calligraphic font for forms valued in its Lie algebra $\mfg\oplus\mfg^{*}$.} connection on the punctured disk $D_{v}$ as follows:
\begin{equation}
\AAb\bl_{D_{v}}\equiv\mathring{\HH}_{v}^{-1}\d\mathring{\HH}_{v}\equiv\HH_{v}^{-1}\d\HH_{v}+\HH_{v}^{-1}\MMb_{v}\HH_{v}\thinspace\d\phi_{v},\label{eq:A-CS}
\end{equation}
where:
\begin{itemize}
\item $\mathring{\HH}_{v}$ is a non-periodic $G\ltimes\mfg^{*}$-valued
0-form defined as $\mathring{\HH}_{v}\equiv\e^{\MMb_{v}\phi_{v}}\HH_{v}$,
\item $\HH_{v}$ is a periodic\footnote{By ``periodic'' we mean that, under $\phi\mt\phi+1$, the non-periodic
variable $\mathring{\HH}_{v}$ gets an additional factor of $\e^{\MMb_{v}}$
due to the term $\e^{\MMb_{v}\phi_{v}}$, while the periodic variable
$\HH_{v}$ is invariant. (Recall that we are scaling $\phi$ by $2\pi$,
so the period is $1$ and not $2\pi$.)} $G\ltimes\mfg^{*}$-valued 0-form,
\item $\MMb_{v}$ is a constant element of the \emph{Cartan subalgebra}\footnote{The Cartan subalgebra of a complex semisimple Lie algebra is a maximal
commutative subalgebra, and it is unique up to automorphisms. The
number of generators in the Cartan subalgebra is the \emph{rank }of
the algebra.} $\mfh\oplus\mfh^{*}$ of $\mfg\oplus\mfg^{*}$.
\end{itemize}
Note that this connection is related by a gauge transformation of
the form $\AAb_{0}\mt\HH_{v}^{-1}\d\HH_{v}+\HH_{v}^{-1}\AAb_{0}\HH_{v}$
to a connection $\AAb_{0}$ defined as follows:
\[
\AAb_{0}\equiv\MMb_{v}\thinspace\d\phi_{v}.
\]
The connection $\AAb_{0}$ satisfies $\left[\AAb_{0},\AAb_{0}\right]=0$,
so its curvature is $\FFb_{0}\equiv\d\AAb_{0}$. This curvature vanishes
everywhere on the \textbf{punctured }disk (which excludes the point
$v$), since $\d^{2}\phi_{v}=0$. However, at the origin of our coordinate
system, i.e. the vertex $v$, $\phi_{v}$ is not well-defined, so
we cannot guarantee that $\FFb_{0}$ vanishes at $v$ itself. Indeed,
as we have seen in chapter \ref{sec:Point-Particles-in}, the curvature
in this case will be a delta function. In addition to the rigorous
proof of Chapter \ref{sec:Delta-Functions-Solid-Angles}, we can now
demonstrate the delta-function behavior of the curvature explicitly.
If we integrate the curvature on the full disk $\Db_{v}$ using Stokes'
theorem, we get:
\[
\int_{\Db_{v}}\FFb_{0}=\oint_{\partial\Db_{v}}\AAb_{0}=\MMb_{v}\oint_{\partial\Db_{v}}\d\phi_{v}=\MMb_{v},
\]
where $\oint_{\partial\Db_{v}}\d\phi_{v}=1$ since we are using coordinates
scaled by $2\pi$, and we used the fact that $\MMb_{v}$ is constant.
We conclude that, since $\FFb_{0}$ vanishes everywhere on $D_{v}$,
and yet it integrates to a finite value at $\Db_{v}$, the curvature
$\FFb_{0}$ must take the form of a Dirac delta function centered
at $v$:
\[
\FFb_{0}=\MMb_{v}\thinspace\delta\left(v\right),
\]
where $\delta\left(v\right)$ is a distributional 2-form such that
for any 0-form $f$,
\[
\int_{\Sigma}f\thinspace\delta\left(v\right)\equiv f\left(v\right).
\]
The final step is to gauge-transform back from $\AAb_{0}$ to the
initial connection $\AAb$ defined in (\ref{eq:A-CS}). The curvature
transforms in the usual way, $\FFb_{0}\mt\HH_{v}^{-1}\FFb_{0}\HH_{v}\equiv\FFb$,
so we get
\[
\FFb\bl_{\Db_{v}}=\HH_{v}^{-1}\MMb_{v}\HH_{v}\thinspace\delta\left(v\right)\equiv\PPb_{v}\thinspace\delta\left(v\right),
\]
where we defined
\[
\PPb_{v}\equiv\HH_{v}^{-1}\MMb_{v}\HH_{v}.
\]
Note again that, while $\FFb\bl_{\Db_{v}}$ (on the \textbf{full}
disk) does \textbf{not} vanish, $\FFb\bl_{D_{v}}$ (on the \textbf{punctured}
disk) \textbf{does} vanish.

Now that we have defined $\AAb$ on the punctured disks $D_{v}$,
we may define it on the cells $c$ simply by treating the cell as
a disk without a puncture, taking $\MMb_{v}=0$ and $v\mt c$ in \ref{eq:A-CS}:
\begin{equation}
\AAb\bl_{c}\equiv\HH_{c}^{-1}\d\HH_{c},\label{eq:A-from-H}
\end{equation}
where $\HH_{c}$ is a $G\ltimes\mfg^{*}$-valued 0-form. This is a
flat connection with $\FFb=0$ everywhere inside $c$.

\subsubsection{The Connection and Frame Field on the Disks and Cells}

Now that we have defined the Chern-Simons connection 1-form $\AAb$
and found its curvature $\FFb$ on the disks, we split $\AAb$ into
a $\mfg$-valued connection 1-form $\A$ a $\mfg^{*}$-valued frame
field 1-form $\ee$ as defined in (\ref{eq:CS-split}). Similarly,
we split $\FFb$ into a $\mfg$-valued curvature 2-form $\F$ and
a $\mfg^{*}$-valued torsion 2-form $\T$ as defined in (\ref{eq:CS-con-split}).

From (\ref{eq:CS-split}) we get:
\begin{equation}
\A\bl_{D_{v}}=\mathring{h}_{v}^{-1}\d\mathring{h}_{v}\sp\ee\bl_{D_{v}}=\mathring{h}_{v}^{-1}\d\mathring{\x}_{v}\mathring{h}_{v},\label{eq:A-E-disks}
\end{equation}
where:
\begin{itemize}
\item $\mathring{h}_{v}$ is a non-periodic $G$-valued 0-form and $\mathring{\x}_{v}$
is a non-periodic $\mfg^{*}$-valued 0-form such that
\begin{equation}
\mathring{h}_{v}\equiv\e^{\M_{v}\phi_{v}}h_{v}\sp\mathring{\x}_{v}\equiv\e^{\M_{v}\phi_{v}}\left(\x_{v}+\SS_{v}\phi_{v}\right)\e^{-\M_{v}\phi_{v}},\label{eq:u-v-def}
\end{equation}
\item $h_{v}$ is a periodic $G$-valued 0-form,
\item $\x_{v}$ is a periodic $\mfg^{*}$-valued 0-form,
\item $\M_{v}$ is a constant element of the Cartan subalgebra $\mfh$ of
$\mfg$,
\item $\SS_{v}$ is a constant element of the Cartan subalgebra $\mfh^{*}$
of $\mfg^{*}$,
\item By construction $\left[\M_{v},\SS_{v}\right]=0$.
\end{itemize}
The full expressions for $\A$ and $\ee$ on $D_{v}$ in terms of
$h_{v}$ and $\x_{v}$ are as follows:
\begin{equation}
\A\bl_{D_{v}}=h_{v}^{-1}\d h_{v}+h_{v}^{-1}\M_{v}h_{v}\thinspace\d\phi_{v}\sp\ee\bl_{D_{v}}=h_{v}^{-1}\d\x_{v}h_{v}+h_{v}^{-1}\left(\SS_{v}+\left[\M_{v},\x_{v}\right]\right)h_{v}\thinspace\d\phi_{v}.\label{eq:A-E-v}
\end{equation}
Furthermore, from (\ref{eq:CS-con-split}) we get:
\[
\F\bl_{\Db_{v}}=\p_{v}\thinspace\delta\left(v\right),\qquad\T\bl_{\Db_{v}}=\j_{v}\,\delta\left(v\right),
\]
where $\p_{v},\j_{v}$ represent the \emph{momentum }and \emph{angular
momentum }respectively:
\[
\p_{v}\equiv h_{v}^{-1}\M_{v}h_{v},\qquad\j_{v}\equiv h_{v}^{-1}\left(\SS_{v}+\left[\M_{v},\x_{v}\right]\right)h_{v}.
\]
In terms of $\p_{v}$ and $\j_{v}$, we may write $\A$ and $\ee$
on the disk as follows:
\[
\A\bl_{D_{v}}=h_{v}^{-1}\d h_{v}+\p_{v}\thinspace\d\phi_{v}\sp\ee\bl_{D_{v}}=h_{v}^{-1}\d\x_{v}h_{v}+\j_{v}\thinspace\d\phi_{v}.
\]
It is clear that the first term in each definition is flat and torsionless,
while the second term (involving $\p_{v}$ and $\j_{v}$ respectively)
is the one which contributes to the curvature and torsion at $v$.
Since the punctured disk $D_{v}$ does not include $v$ itself, the
curvature and torsion vanish everywhere on it: 
\[
\F\bl_{D_{v}}=0\sp\T\bl_{D_{v}}=0.
\]
As before, while $\F$ and $\T$ do not vanish on the full disk $\Db_{v}$,
they do vanish on $D_{v}$. We call this type of geometry a \emph{piecewise
flat and torsionless geometry}\footnote{The question of whether the geometry we have defined here has a notion
of a ``continuum limit'', e.g. by shrinking the loops to points
such that the discrete defects at the vertices become continuous curvature
and torsion, is left for future work.}. Given a particular spin network $\Gamma^{*}$, and assuming that
information about the curvature and torsion may only be obtained by
taking holonomies along the loops of this spin network, the piecewise
flat and torsionless geometry carries, at least intuitively, the exact
same information as the arbitrary geometry we had before.

As for the Chern-Simons connection, the expressions for $\A$ and
$\ee$ on $c$ are obtained by taking $\M_{v}=\SS_{v}=0$ and $v\mt c$
in (\ref{eq:A-E-v}):
\begin{equation}
\A\bl_{c}=h_{c}^{-1}\d h_{c}\sp\ee\bl_{c}=h_{c}^{-1}\d\x_{c}h_{c},\label{eq:A-E-cells}
\end{equation}
where $h_{c}$ is a $G$-valued 0-form and $\x_{c}$ is a $\mfg^{*}$-valued
0-form. Of course, by construction, the curvature and torsion associated
to this connection and frame field vanish everywhere in the cell:
\[
\F\bl_{c}=0\sp\T\bl_{c}=0.
\]

\subsection{\label{subsec:The-Continuity-Conditions}The Continuity Conditions}

\subsubsection{Continuity Conditions Between Cells}

Let us consider the link $e^{*}=\left(cc'\right)^{*}$ connecting
two adjacent nodes $c^{*}$ and $c^{\prime*}$. This link is dual
to the edge $e=\left(cc'\right)=c\cap c'$, which is the boundary
between the two adjacent cells $c$ and $c'$. The connection is defined
in the union $c\cup c'$, while in each cell its restriction is encoded
in $\AAb\bl_{c}$ and $\AAb\bl_{c'}$ as defined above, in terms of
$\HH_{c}$ and $\HH_{c'}$ respectively.

\begin{figure}[!h]
\begin{centering}
\begin{tikzpicture}
	\begin{pgfonlayer}{nodelayer}
		\node [style=Vertex] (0) at (0, -2) {};
		\node [style=Vertex] (1) at (0, 2) {};
		\node [style=Vertex] (2) at (-3, -0) {};
		\node [style=Vertex] (3) at (3, -0) {};
		\node [style=Node] (4) at (-1.25, -0) {};
		\node [style=Node] (5) at (1.25, -0) {};
		\node [style=none] (6) at (-1.5, -0.3) {$c^{*}$};
		\node [style=none] (7) at (1.5, -0.3) {$c^{\prime*}$};
		\node [style=none] (8) at (0, -0.75) {};
		\node [style=none] (9) at (0, 0.25) {$(cc')$};
		\node [style=none] (10) at (-0.75, -0.87) {$\mathcal{H}_{c}(x)$};
		\node [style=none] (11) at (0.75, -0.87) {$\mathcal{H}_{c'}^{-1}(x)$};
	\end{pgfonlayer}
	\begin{pgfonlayer}{edgelayer}
		\draw [style=Edge] (1) to (2);
		\draw [style=Edge] (2) to (0);
		\draw [style=Edge] (0) to (3);
		\draw [style=Edge] (3) to (1);
		\draw [style=Edge] (1) to (0);
		\draw [style=SegmentArrow] (4) to (8.center);
		\draw [style=SegmentArrow] (8.center) to (5);
	\end{pgfonlayer}
\end{tikzpicture}
\par\end{centering}
\caption{\label{fig:ContinuityNN}To get from the node $c^{*}$ to the adjacent
node $c^{\prime*}$, we use the group element $\protect\HH_{cc'}$.
First, we choose a point $x$ somewhere on the edge $\left(cc'\right)=c\cap c'$.
Then, we take $\protect\HH_{c}\left(x\right)$ from $c^{*}$ to $x$,
following the first red arrow. Finally, we take $\protect\HH_{c'}^{-1}\left(x\right)$
from $x$ to $c^{\prime*}$, following the second red arrow. Thus
$\protect\HH_{cc'}=\protect\HH_{c}\left(x\right)\protect\HH_{c'}^{-1}\left(x\right)$.
Note that any $x\in c\cap c'$ will do, since the connection is flat
and thus all paths are equivalent.}
\end{figure}
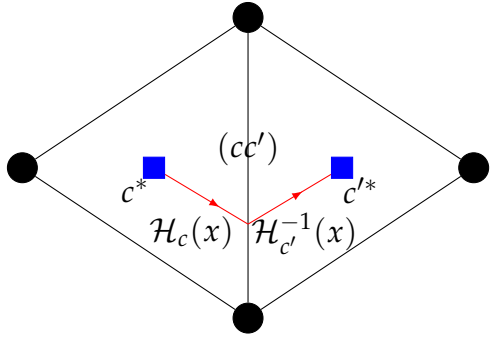

The continuity equation on the edge $\left(cc'\right)$ between the
two adjacent cells reads 
\[
\AAb\bl_{c}=\HH_{c}^{-1}\d\HH_{c}=\HH_{c'}^{-1}\d\HH_{c'}=\AAb\bl_{c'}\sp\textrm{on }\left(cc'\right)=c\cap c'.
\]
Since the connections match, this means that the group elements $\HH_{c}$
and $\HH_{c'}$ differ only by the action of a left symmetry element.
This implies that there exists a group element $\HH_{cc'}\in G\ltimes\mfg^{*}$
which is independent of $x$ and provides the change of variables
between the two parametrizations $\HH_{c}\left(x\right)$ and $\HH_{c'}\left(x\right)$
on the overlap: 
\[
\HH_{c'}\left(x\right)=\HH_{c'c}\HH_{c}\left(x\right)\sp x\in\left(cc'\right)=c\cap c'.
\]
Note that $\HH_{c'c}=\HH_{cc'}^{-1}$. Furthermore, $\HH_{cc'}$ can
be decomposed as 
\begin{equation}
\HH_{cc'}=\HH_{c}\left(x\right)\HH_{c'}^{-1}\left(x\right),\label{eq:Hcc}
\end{equation}
as illustrated in Figure \ref{fig:ContinuityNN}. 

We can decompose the continuity conditions into rotational and translational
holonomies using the rules outlined in Section \ref{subsec:The-Cartan-Decomposition}:
\begin{equation}
h_{c'}\left(x\right)=h_{c'c}h_{c}\left(x\right)\sp\x_{c'}\left(x\right)=h_{c'c}\left(\x_{c}\left(x\right)-\x_{c}^{c'}\right)h_{cc'}\sp x\in\left(cc'\right).\label{eq:cont-ccp}
\end{equation}

\subsubsection{\label{subsec:Continuity-Conditions-Disks}Continuity Conditions
Between Disks and Cells}

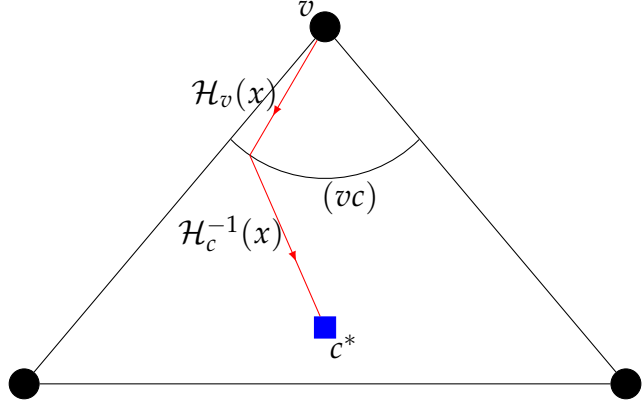
\begin{figure}[!h]
\begin{centering}
\begin{tikzpicture}
	\begin{pgfonlayer}{nodelayer}
		\node [style=Node] (0) at (0, -1) {};
		\node [style=Vertex] (1) at (0, 3) {};
		\node [style=Vertex] (2) at (-4, -1.75) {};
		\node [style=Vertex] (3) at (4, -1.75) {};
		\node [style=none] (4) at (0.25, -1.25) {$c^{*}$};
		\node [style=none] (5) at (-0.25, 3.25) {$v$};
		\node [style=none] (6) at (-1.25, 1.5) {};
		\node [style=none] (7) at (1.25, 1.5) {};
		\node [style=none] (8) at (-1, 1.29) {};
		\node [style=none] (9) at (0.33, 0.725) {$(vc)$};
		\node [style=none] (12) at (-1.25, 0.25) {$\mathcal{H}_{c}^{-1}(x)$};
		\node [style=none] (13) at (-1.2, 2.1) {$\mathcal{H}_{v}(x)$};
	\end{pgfonlayer}
	\begin{pgfonlayer}{edgelayer}
		\draw [style=Edge] (1) to (2);
		\draw [style=Edge] (2) to (3);
		\draw [style=Edge] (3) to (1);
		\draw [style=Edge, bend right=45, looseness=1.00] (6.center) to (7.center);
		\draw [style=SegmentArrow] (8.center) to (0);
		\draw [style=SegmentArrow] (1) to (8.center);
	\end{pgfonlayer}
\end{tikzpicture}
\par\end{centering}
\caption{\label{fig:ContinuityNV}To get from the vertex $v$ to the node $c^{*}$,
we use the group element $\protect\HH_{vc}$. First, we choose a point
$x$ somewhere on the arc $\left(vc\right)=\partial D_{v}\cap c$.
Then, we use $\e^{\protect\MMb_{v}\phi_{v}\left(x\right)}$ to rotate
from the cut $C_{v}$ to the angle corresponding to $x$ (rotation
not illustrated). Next, we take $\protect\HH_{v}\left(x\right)$ from
$v$ to $x$, following the first red arrow. Finally, we take $\protect\HH_{c}^{-1}\left(x\right)$
from $x$ to $c^{*}$, following the second red arrow. Thus $\protect\HH_{vc}=\e^{\protect\MMb_{v}\phi_{v}\left(x\right)}\protect\HH_{v}\left(x\right)\protect\HH_{c}^{-1}\left(x\right)$.}
\end{figure}

A similar discussion applies when one looks at the overlap $D_{v}\cap c$
between a punctured disk $D_{v}$ and a cell $c$. The boundary of
this region consists of two \emph{truncated edges} of length $R$
(the coordinate radius of the disk) touching $v$, plus an arc connecting
the two edges, which lies on the boundary of the disk $D_{v}$. In
the following we denote this arc\footnote{The arc $\left(vc\right)$ is dual to the line segment $\left(vc\right)^{*}$
connecting the vertex $v$ with the node $c^{*}$, just as the edge
$e$ is dual to the link $e^{*}$.} by $\left(vc\right)$. It is clear that the union of all such arcs
around a vertex $v$ reconstructs the outer boundary $\partial_{R}D_{v}$
of the disk, as defined in Section \ref{subsec:The-Cellular-Decomposition}:
\[
\left(vc\right)\equiv\partial_{R}D_{v}\cap c,\qquad\partial_{R}D_{v}=\bigcup_{c\ni v}\left(vc\right),
\]
where $c\ni v$ means ``all cells $c$ which have the vertex $v$
on their boundary''. In the intersection $D_{v}\cap c$ we have two
different descriptions of the connection $\AAb$. On $c$ it is described
by the $G\ltimes\mfg^{*}$-valued 0-form $\HH_{c}$, and on $D_{v}$
it is described by a $G\ltimes\mfg^{*}$-valued 0-form $\HH_{v}$.
The fact that we have a single-valued connection is expressed in the
continuity conditions 
\begin{equation}
\AAb\bl_{D_{v}}=\HH_{v}^{-1}\MMb_{v}\HH_{v}\thinspace\d\phi_{v}+\HH_{v}^{-1}\d\HH_{v}=\HH_{c}^{-1}\d\HH_{c}=\AAb\bl_{c}\sp\textrm{on }\left(vc\right)=\partial D_{v}\cap c.\label{continuity-vc}
\end{equation}
The relation between the two connections can be integrated. It means
that the elements $\HH_{v}(x)$ and $\HH_{c}(x)$ differ by the action
of the left symmetry group. In practice, this means that the integrated
continuity relation involves a (discrete) holonomy $\HH_{cv}$: 
\begin{equation}
\HH_{c}\left(x\right)=\HH_{cv}\e^{\MMb_{v}\phi_{v}\left(x\right)}\HH_{v}\left(x\right)\sp x\in\left(vc\right)\label{eq:continuity-H_cv}
\end{equation}
where $\phi_{v}\left(x\right)$ is the angle corresponding to $x$
with respect to the cut $C_{v}$. Isolating $\HH_{vc}\equiv\HH_{cv}^{-1}$,
we find 
\[
\HH_{vc}=\e^{\MMb_{v}\phi_{v}\left(x\right)}\HH_{v}\left(x\right)\HH_{c}^{-1}\left(x\right),
\]
which is illustrated in Figure \ref{fig:ContinuityNV}.

As above, we can decompose the continuity conditions into rotational
and translational holonomies using the rules outlined in Section \ref{subsec:The-Cartan-Decomposition}:
\[
h_{c}=h_{cv}\mathring{h}_{v}\sp\x_{c}=h_{cv}(\mathring{\x}_{v}-\x_{v}^{c})h_{vc}\sp\textrm{on }\left(vc\right)\sp x\in\left(vc\right).
\]

\section{\label{sec:Gauge-Transformations}Gauge Transformations and Symmetries}

\subsection{Right and Left Translations}

Let $\GG$ be a $G\ltimes\mfg^{*}$-valued 0-form. \emph{Right translations}
\begin{equation}
\HH_{c}\left(x\right)\mt\HH_{c}\left(x\right)\GG\left(x\right),\label{eq:right-tr-c}
\end{equation}
are \textit{gauge transformations} that affect the connection in the
usual way: 
\begin{equation}
\AAb\mt\GG^{-1}\AAb\GG+\GG^{-1}\d\GG.\label{eq:gauge-tr-A}
\end{equation}
We can also consider \emph{left translations} 
\begin{equation}
\HH_{c}\left(x\right)\mt\GG_{c}\HH_{c}\left(x\right),\label{eq:left-tr-c}
\end{equation}
acting on $\HH_{c}$ with a \textbf{constant }group element $\GG_{c}\in G\ltimes\mfg^{*}$.
These transformations leave the connection invariant, $\AAb\mt\AAb$.
On the one hand, they label the redundancy of our parametrization
of $\AAb$ in terms of $\HH_{c}$ as discussed above. On the other
hand, these transformations can be understood as \emph{symmetries}
of our parametrization in terms of group elements that stem from the
existence of new degrees of freedom in $\HH_{c}$ beyond the ones
in the connection $\AAb$.

This situation is similar to the situation that arises any time one
considers a gauge theory in a region with boundaries. As shown in
\cite{Donnelly:2016auv}, when we subdivide a region of space we need
to add new degrees of freedom at the boundaries of the subdivision
in order to restore gauge invariance. These degrees of freedom are
the \emph{edge modes}, which carry a non-trivial representation of
the boundary symmetry group that descends from the bulk gauge transformations.
This description is equivalent to the one in \cite{Geiller:2017xad},
which we described in Chapter \ref{sec:Gauge-Transformations-and}
in the continuous context.

Now, we can invert (\ref{eq:A-from-H}) and write $\HH_{c}$ using
a \emph{path-ordered exponential} as follows: 
\begin{equation}
\HH_{c}\left(x\right)=\HH_{c}\left(c^{*}\right)\,\pexp\int_{c^{*}}^{x}\AAb,\label{eq:H-from-A}
\end{equation}
where $\HH_{c}\left(c^{*}\right)$, the value of $\HH_{c}$ at the
node $c^{*}$, is the extra information contained in the edge mode
field $\HH_{c}$ that cannot be obtained from the connection $\AAb$.
Left translations can thus be understood as simply translating the
value of $\HH_{c}\left(c^{*}\right)$ without affecting the value
of $\AAb$.

As for the disks, we again see that gauge transformations are given
by right translations 
\begin{equation}
\HH_{v}\left(x\right)\mt\HH_{v}\left(x\right)\GG\left(x\right),\label{eq:right-tr-v}
\end{equation}
while left translations by a constant element $\GG_{v}$ in the Cartan
subgroup (which thus commutes with $\MMb_{v}$), 
\begin{equation}
\HH_{v}\left(x\right)\mt\GG_{v}\HH_{v}\left(x\right),\label{eq:left-tr-v}
\end{equation}
leave the connection invariant.

\subsection{Invariant Holonomies}

The quantity $\HH_{cc'}$, given in (\ref{eq:Hcc}) in the context
of the continuity conditions, is invariant under the right gauge transformation
(\ref{eq:right-tr-c}), since it is independent of $c$. However,
it is not invariant under the left symmetry (\ref{eq:left-tr-c})
performed at $c$ and $c'$, under which we obtain
\begin{equation}
\HH_{c'c}\mt\GG_{c'}\HH_{c'c}\GG_{c}^{-1}.\label{eq:left-tr-Hcpc}
\end{equation}
Since this symmetry leaves the connection invariant, this means that
$\HH_{cc'}$ is \emph{not} the holonomy from $c^{*}$ to $c^{\prime*}$,
along the link $(cc')^{*}$, as it is usually assumed. Instead, from
(\ref{eq:H-from-A}) and (\ref{eq:Hcc}) we have that 
\begin{equation}
\HH_{cc'}=\HH_{c}\left(c^{*}\right)\left(\pexp\int_{c^{*}}^{c^{\prime*}}\AAb\right)\HH_{c'}^{-1}\left(c'^{*}\right)\label{eq:H_ccp-pathexp}
\end{equation}
is a \emph{dressed} gauge-invariant observable. It is a gauge-invariant
version of the holonomy.

Similarly, the quantity $\HH_{vc}$ is invariant under right gauge
transformations (\ref{eq:right-tr-c}) and (\ref{eq:right-tr-v}):
\begin{equation}
\HH_{c}\left(x\right)\mt\GG_{c}\HH_{c}\left(x\right)\sp\HH_{v}\left(x\right)\mt\GG_{v}\HH_{v}\left(x\right).\label{eq:left-c-v}
\end{equation}
However, under left symmetry transformations (\ref{eq:left-tr-c})
and (\ref{eq:left-tr-v}), the connection is left invariant, and we
get 
\begin{equation}
\HH_{vc}\mt\GG_{v}\HH_{vc}\GG_{c}^{-1}.\label{eq:left-tr-Hvc}
\end{equation}

Note also that the translation in $\phi_{v}\left(x\right)$ can be
absorbed into the definition of $\HH_{v}$, so that the transformation
\begin{equation}
\phi_{v}\left(x\right)\mt\phi_{v}\left(x\right)+\beta_{v}\sp\HH_{v}\left(x\right)\mt\e^{-\MMb_{v}\beta_{v}}\HH_{v}\left(x\right),\label{transphi}
\end{equation}
is also a symmetry under which (\ref{eq:continuity-H_cv}) is invariant.
The connection $\AAb\bl_{D_{v}}$ is invariant under this symmetry.

\subsection{Dressed Holonomies and Edge Modes}

Our discussion of the transformations of the Chern-Simons connection
also applies, of course, to the connection $\A$ and frame field $\ee$.
Consider the definition $\A\bl_{c}=h_{c}^{-1}\d h_{c}$ for $\A$
in terms of $h_{c}$. Note that $\A$ is invariant under the left
action transformation $h_{c}\mt g_{c}h_{c}$ for some constant $g_{c}\in G$.
Thus, inverting the definition $\A\bl_{c}=h_{c}^{-1}\d h_{c}$ to
find $h_{c}$ in terms of $\A$, we get
\[
h_{c}\left(x\right)=h_{c}\left(c^{*}\right)\pexp\int_{c^{*}}^{x}\A,
\]
where $h_{c}\left(c^{*}\right)$ is a new degree of freedom which
does not exist in $\A$. The notation suggests that it is the holonomy
``from $c^{*}$ to itself'', but it is in general not the identity!
The notation $h_{c}\left(c^{*}\right)$ is just a placeholder for
the \emph{edge mode }which ``dresses'' the holonomy.

For the ``undressed'' holonomy -- which is simply the path-ordered
exponential from the node $c^{*}$ to some point $x$ -- we thus
have
\begin{equation}
\pexp\int_{c^{*}}^{x}\A=h_{c}^{-1}\left(c^{*}\right)h_{c}\left(x\right).\label{eq:undressed-c}
\end{equation}
Similarly, the definition $\A\bl_{D_{v}}=h_{v}^{-1}\d h_{v}+h_{v}^{-1}\M_{v}h_{v}\thinspace\d\phi_{v}$
is invariant under $h_{v}\mt g_{v}h_{v}$, but only if $g_{v}$ is
in $H$, the Cartan subgroup of $G$, since it must commute with $\M_{v}$.
Inverting the relation $\A\bl_{D_{v}}=\mathring{h}_{v}^{-1}\d\mathring{h}_{v}$,
we get
\[
\mathring{h}_{v}\left(x\right)=h_{v}\left(v\right)\pexp\int_{v}^{x}\A,
\]
where again the edge mode $h_{v}\left(v\right)$ is a new degree of
freedom. The undressed holonomy is then
\begin{equation}
\pexp\int_{v}^{x}\A=h_{v}^{-1}\left(v\right)\mathring{h}_{v}\left(x\right)=h_{v}^{-1}\left(v\right)\e^{\M_{v}\phi_{v}\left(x\right)}h_{v}\left(x\right).\label{eq:undressed-v}
\end{equation}
From (\ref{eq:undressed-c}) and (\ref{eq:undressed-v}), we may construct
general path-ordered exponentials from some point $x$ to another
point $y$ by breaking the path from $x$ to $y$ such that it passes
through an intermediate point. If that point is the node $c^{*}$,
then we get
\begin{align*}
\pexp\int_{x}^{y}\A & =\left(\pexp\int_{x}^{c^{*}}\A\right)\left(\pexp\int_{c^{*}}^{y}\A\right)\\
 & =\left(h_{c}^{-1}\left(x\right)h_{c}\left(c^{*}\right)\right)\left(h_{c}^{-1}\left(c^{*}\right)h_{c}\left(y\right)\right)\\
 & =h_{c}^{-1}\left(x\right)h_{c}\left(y\right),
\end{align*}
and if it's the vertex $v$, we similarly get
\begin{equation}
\pexp\int_{x}^{y}\A=\left(\pexp\int_{x}^{v}\A\right)\left(\pexp\int_{v}^{y}\A\right)=h_{v}^{-1}\left(x\right)\e^{\M_{v}\left(\phi_{v}\left(y\right)-\phi_{v}\left(x\right)\right)}h_{v}\left(y\right).\label{eq:exp-xy-phi}
\end{equation}
Furthermore, we may use the continuity relations (\ref{eq:continuity-ccp})
and (\ref{eq:continuity-cv}) to obtain a relation between the path-ordered
integrals and the holonomies $h_{cc'}$ and $h_{cv}$. If $y\in\left(cc'\right)$
then we can write
\begin{equation}
\pexp\int_{x}^{y}\A=h_{c}^{-1}\left(x\right)h_{cc'}h_{c'}\left(y\right),\label{eq:exp-xy-h}
\end{equation}
and if $y\in\left(cv\right)$ then we can write
\[
\pexp\int_{x}^{y}\A=h_{c}^{-1}\left(x\right)h_{cv}\mathring{h}_{v}\left(y\right)=h_{c}^{-1}\left(x\right)h_{cv}\e^{\M_{v}\phi_{v}\left(y\right)}h_{v}\left(y\right).
\]
Note that, in particular,
\begin{equation}
\pexp\int_{c^{*}}^{c^{\prime*}}\A=h_{c}^{-1}\left(c^{*}\right)h_{cc'}h_{c'}\left(c^{\prime*}\right).\label{eq:exp-ccp-h}
\end{equation}
A similar discussion applies to the translational holonomies $\x_{c}$
and $\x_{v}$, and one finds two new degrees of freedom, $\x_{c}\left(c^{*}\right)$
and $\x_{v}\left(v\right)$.

\section{\label{sec:Discretizing-the-Symplectic-2P1}Discretizing the Symplectic
Potential}

\subsection{\label{subsec:The-Choice-of}The Choice of Polarization}

Recall that there is a family of symplectic potential given by (\ref{eq:Theta-general}):
\begin{equation}
\Theta_{\lambda}=-\int_{\Sigma}\left(\left(1-\lambda\right)\ee\cdot\delta\A+\lambda\A\cdot\delta\ee\right).\label{eq:Theta-general-2}
\end{equation}
We would like to replace $\A$ and $\ee$ by their discretized expressions
given by (\ref{eq:A-E-cells}) and (\ref{eq:A-E-disks}). Before we
do this for each cell and disk individually, let us consider a toy
model where we simply take $\A=h^{-1}\d h$ and $\ee=h^{-1}\d\x h$
for some $G$-valued 0-form $h$ and $\mfg^{*}$-valued 0-form $\x$
over the entire manifold $\Sigma$. We begin by calculating the variations
of these expressions, obtaining
\[
\delta\A=\delta\left(h^{-1}\d h\right)=h^{-1}\left(\d\De h\right)h,
\]
\[
\delta\ee=\delta\left(h^{-1}\d\x h\right)=h^{-1}\left(\d\delta\x+\left[\d\x,\De h\right]\right)h,
\]
where $\De h\equiv\delta hh^{-1}$ for the Maurer-Cartan form on field
space as defined in (\ref{eq:Maurer-Cartan}). Thus, we have
\[
\Theta_{\lambda}=-\int_{\Sigma}\left(\left(1-\lambda\right)\d\x\cdot\d\De h+\lambda\d hh^{-1}\cdot\left(\d\delta\x+\left[\d\x,\De h\right]\right)\right),
\]
where we used the cyclicity of the dot product, (\ref{eq:cyclic}),
to cancel some group elements. Now, the first term is very simple;
in fact, it is clearly an exact 2-form, and thus may be easily integrated.
However, the second term is complicated, and it is unclear if it can
be integrated. Nevertheless, we know that every choice of $\lambda$
leads to the \textbf{same} symplectic form:
\[
\Omega=\delta\Theta_{\lambda}=-\int_{\Sigma}\delta\ee\cdot\delta\A=-\int_{\Sigma}\left(\d\delta\x+\left[\d\x,\De h\right]\right)\cdot\d\De h.
\]
Furthermore, we have seen from (\ref{eq:S_lambda}) that the difference
between different polarizations amounts to the addition of a boundary
term and is equivalent to an integration by parts. Thus, we employ
the following trick. First we take $\lambda=0$ in $\Theta_{\lambda}$,
so that it becomes the 2+1D gravity polarization:
\[
\Theta=-\int_{\Sigma}\ee\cdot\delta\A.
\]
Then, in the discretization process, we obtain
\[
\Theta=-\int_{\Sigma}\d\x\cdot\d\De h.
\]
The integrand is an exact 2-form, and thus may be integrated in two
equivalent ways:
\[
\d\x\cdot\d\De h=\d\left(\x\cdot\d\De h\right)=-\d\left(\d\x\cdot\De h\right).
\]
Note that the 1-forms $\x\cdot\d\De h$ and $\d\x\cdot\De h$ differ
only by a boundary term of the form $\d\left(\x\cdot\De h\right)$,
and they may be obtained from each other with integration by parts,
just as for the different polarizations. In fact, we may write:
\begin{equation}
\ee\cdot\delta\A=\d\x\cdot\d\De h=\lambda\d(\x\cdot\d\De h)-\left(1-\lambda\right)\d(\d\x\cdot\De h).\label{eq:dxdDh}
\end{equation}
We claim that, even though technically both options are equivalent
discretizations of the $\lambda=0$ polarization in (\ref{eq:Theta-general-2}),
there is in fact reason to believe that the choice of $\lambda$ in
(\ref{eq:Theta-general-2}) corresponds to the same choice of $\lambda$
in (\ref{eq:dxdDh})! We will motivate this by showing that the choice
$\lambda=0$ corresponds to the usual loop gravity polarization, which
is associated with usual general relativity, while the choice $\lambda=1$
corresponds to a dual polarization which, as we will see, is associated
with teleparallel gravity.

\subsection{Decomposing the Spatial Manifold}

As we have seen, the spatial manifold $\Sigma$ is decomposed into
cells $c$ and disks $D_{v}$. The whole manifold $\Sigma$ may be
recovered by taking the union of the cells with the \emph{closures
}of the disks (recall that the vertices $v$ are not in $D_{v}$,
they are on their boundaries):
\[
\Sigma=\left(\bigcup_{c}c\right)\cup\left(\bigcup_{v}D_{v}\cup\partial D_{v}\right).
\]
Here, we are assuming that the cells and punctured disks are disjoint;
the disks ``eat into'' the cells. We can thus split $\Theta$ into
contributions from each cell $c$ and punctured disk $D_{v}$:
\[
\Theta=\sum_{c}\Theta_{c}+\sum_{v}\Theta_{D_{v}},
\]
where
\[
\Theta_{c}=-\int_{c}\ee\cdot\delta\A\sp\Theta_{D_{v}}=-\int_{D_{v}}\ee\cdot\delta\A.
\]
Given the discretizations (\ref{eq:A-E-cells}) and (\ref{eq:A-E-disks}),
we replace $h,\x$ in (\ref{eq:dxdDh}) with $h_{c},\x_{c}$ or $\mathring{h}_{v},\mathring{\x}_{v}$
respectively, and then integrate using Stokes' theorem to obtain:
\begin{equation}
\Theta_{c}=\int_{\partial c}\left(\left(1-\lambda\right)\d\x_{c}\cdot\De h_{c}-\lambda\x_{c}\cdot\d\De h_{c}\vphantom{\bll}\right),\label{eq:Theta_c}
\end{equation}
\[
\Theta_{D_{v}}=\int_{\partial D_{v}}\left(\left(1-\lambda\right)\d\mathring{\x}_{v}\cdot\De\mathring{h}_{v}-\lambda\mathring{\x}_{v}\cdot\d\De\mathring{h}_{v}\vphantom{\bll}\right).
\]
In the next few sections, we will manipulate these expressions so
that they can be integrated once again to obtain truly discrete symplectic
potentials.

\subsection{The Vertex and Cut Contributions}

\subsubsection{Calculating the Integral}

The boundary $\partial D_{v}$ splits into three contributions: one
from the inner boundary $\partial_{0}D_{v}$ (which is the vertex
$v$), one from the cut $C_{v}$, and one from the outer boundary
$\partial_{R}D_{v}$. Thus we have
\[
\Theta_{D_{v}}=-\Theta_{\partial_{0}D_{v}}-\Theta_{C_{v}}+\Theta_{\partial_{R}D_{v}},
\]
where the minus sign comes from the fact that orientation of the outer
boundary is opposite to that of the inner boundary. Here we will discuss
the first two terms, while the contribution from the outer boundary
$\partial_{R}D_{v}$ will be calculated in Section \ref{subsec:Edge-Arc}.

Writing the terms in the integrand explicitly in terms of $\x_{v},h_{v}$
using (\ref{eq:u-v-def}), and making use of the identities
\[
\d\mathring{\x}_{v}=\e^{\M_{v}\phi_{v}}\left(\d\x_{v}+\left(\SS_{v}+\left[\M_{v},\x_{v}\right]\right)\d\phi_{v}\right)\e^{-\M_{v}\phi_{v}},
\]
\[
\De\mathring{h}_{v}=\e^{\M_{v}\phi_{v}}\left(\delta\M_{v}\phi_{v}+\De h_{v}\right)\e^{-\M_{v}\phi_{v}},
\]
\[
\d\De\mathring{h}_{v}=\e^{\M_{v}\phi_{v}}\left(\d\De h_{v}+\left(\delta\M_{v}+\left[\M_{v},\De h_{v}\right]\right)\d\phi_{v}\right)\e^{-\M_{v}\phi_{v}},
\]
we get
\[
\d\mathring{\x}_{v}\cdot\De\mathring{h}_{v}=\left(\d\x_{v}+\left(\SS_{v}+\left[\M_{v},\x_{v}\right]\right)\d\phi_{v}\right)\cdot\left(\delta\M_{v}\phi_{v}+\De h_{v}\right),
\]
\[
\mathring{\x}_{v}\cdot\d\De\mathring{h}_{v}=\left(\x_{v}+\SS_{v}\phi_{v}\right)\cdot\left(\d\De h_{v}+\left(\delta\M_{v}+\left[\M_{v},\De h_{v}\right]\right)\d\phi_{v}\right).
\]
The integral on the inner boundary $\partial_{0}D_{v}$ is easily
calculated, since $\x_{v}$ and $h_{v}$ obtain the constant values
$\x_{v}\left(v\right)$ and $h_{v}\left(v\right)$ on the inner boundary.
Hence $\d\x_{v}\left(v\right)=\d\De h_{v}\left(v\right)=0$, and these
expressions simplify to\footnote{Here we used the identity $\left[\A,\B\right]\cdot\C=\A\cdot\left[\B,\C\right]$
to get $\left[\M_{v},\x_{v}\right]\cdot\delta\M_{v}=\x_{v}\cdot\left[\delta\M_{v},\M_{v}\right]=0$
and $\SS_{v}\cdot\left[\M_{v},\De h_{v}\right]=\De h_{v}\cdot\left[\SS_{v},\M_{v}\right]=0$.}
\[
\d\mathring{\x}_{v}\cdot\De\mathring{h}_{v}\bl_{\partial_{0}D_{v}}=\left(\phi_{v}\SS_{v}\cdot\delta\M_{v}+\left(\SS_{v}+\left[\M_{v},\x_{v}\left(v\right)\right]\right)\cdot\De h_{v}\left(v\right)\right)\d\phi_{v},
\]
\[
\mathring{\x}_{v}\cdot\d\De\mathring{h}_{v}\bl_{\partial_{0}D_{v}}=\left(\phi_{v}\SS_{v}\cdot\delta\M_{v}+\x_{v}\left(v\right)\cdot\left(\delta\M_{v}+\left[\M_{v},\De h_{v}\left(v\right)\right]\right)\right)\d\phi_{v}.
\]
To evaluate the contribution from the inner boundary, we integrate
from $\phi_{v}=\alpha_{v}-1/2$ to $\phi_{v}=\alpha_{v}+1/2$. Then
since
\[
\int_{\alpha_{v}-1/2}^{\alpha_{v}+1/2}\d\phi_{v}=1\sp\int_{\alpha_{v}-1/2}^{\alpha_{v}+1/2}\phi_{v}\thinspace\d\phi_{v}=\alpha_{v},
\]
we get:
\begin{align*}
\Theta_{\partial_{0}D_{v}} & =\left(1-2\lambda\right)\alpha_{v}\SS_{v}\cdot\delta\M_{v}+\left(1-\lambda\right)\left(\SS_{v}+\left[\M_{v},\x_{v}\left(v\right)\right]\right)\cdot\De h_{v}\left(v\right)+\\
 & \qquad-\lambda\x_{v}\left(v\right)\cdot\left(\delta\M_{v}+\left[\M_{v},\De h_{v}\left(v\right)\right]\right),
\end{align*}
which may be written as
\begin{align*}
\Theta_{\partial_{0}D_{v}} & =\left(1-2\lambda\right)\alpha_{v}\SS_{v}\cdot\delta\M_{v}+\left(1-\lambda\right)\SS_{v}\cdot\De h_{v}\left(v\right)-\lambda\x_{v}\left(v\right)\cdot\delta\M_{v}+\\
 & \qquad+\left[\M_{v},\x_{v}\left(v\right)\right]\cdot\De h_{v}\left(v\right).
\end{align*}
Next, we have the cut $C_{v}$. Since $\d\phi_{v}=0$ on the cut,
we have a significant simplification:
\[
\d\mathring{\x}_{v}\cdot\De\mathring{h}_{v}\bl_{C_{v}}=\d\x_{v}\cdot\left(\delta\M_{v}\phi_{v}+\De h_{v}\right),
\]
\[
\mathring{\x}_{v}\cdot\d\De\mathring{h}_{v}\bl_{C_{v}}=\left(\x_{v}+\SS_{v}\phi_{v}\right)\cdot\d\De h_{v}.
\]
In fact, the cut has two sides: one at $\phi_{v}=\alpha_{v}-1/2$
and another at $\phi_{v}=\alpha_{v}+1/2$, with opposite orientation.
Let us label them $C_{v}^{-}$ and $C_{v}^{+}$ respectively. Any
term that does not depend explicitly on $\phi_{v}$ will vanish when
we take the difference between both sides of the cut, since they only
differ by the value of $\phi_{v}$. Thus only the terms $\d\x_{v}\cdot\delta\M_{v}\phi_{v}$
and $\SS_{v}\cdot\d\De h_{v}\phi_{v}$ survive. The relevant contribution
from each side of the cut is therefore:
\begin{align*}
\Theta_{C_{v}^{\pm}} & =\int_{r=0}^{R}\left(\left(1-\lambda\right)\d\x_{v}\cdot\delta\M_{v}\phi_{v}-\lambda\SS_{v}\cdot\d\De h_{v}\phi_{v}\right)\blll_{\phi_{v}=\alpha_{v}\pm1/2}\\
 & =\left(\alpha_{v}\pm\hf\right)\left(\left(1-\lambda\right)\delta\M_{v}\cdot\int_{r=0}^{R}\d\x_{v}-\lambda\SS_{v}\cdot\int_{r=0}^{R}\d\De h_{v}\right)\\
 & =\left(\alpha_{v}\pm\hf\right)\left(\left(1-\lambda\right)\delta\M_{v}\cdot\left(\x_{v}\left(v_{0}\right)-\x_{v}\left(v\right)\right)-\lambda\SS_{v}\cdot\left(\De h_{v}\left(v_{0}\right)-\De h_{v}\left(v\right)\right)\right),
\end{align*}
where the point at $r=0$ is the vertex $v$, and the point at $r=R$
and $\phi_{v}=\alpha_{v}\pm1/2$ is labeled $v_{0}$. Taking the difference
between both sides of the cut, we thus get the total contribution:
\begin{align*}
\Theta_{C_{v}} & =\Theta_{C_{v}^{+}}-\Theta_{C_{v}^{-}}\\
 & =\left(1-\lambda\right)\left(\x_{v}\left(v_{0}\right)-\x_{v}\left(v\right)\right)\cdot\delta\M_{v}-\lambda\SS_{v}\cdot\left(\De h_{v}\left(v_{0}\right)-\De h_{v}\left(v\right)\right).
\end{align*}
Adding up the contributions from the inner boundary and the cut, we
obtain the vertex symplectic potential $\Theta_{v}\equiv-\left(\Theta_{\partial_{0}D_{v}}+\Theta_{C_{v}}\right)$:
\begin{align}
\Theta_{v} & =-\left(1-2\lambda\right)\alpha_{v}\SS_{v}\cdot\delta\M_{v}-\SS_{v}\cdot\left(\De h_{v}\left(v\right)-\lambda\De h_{v}\left(v_{0}\right)\right)+\label{eq:Theta_v}\\
 & \qquad+\left(\x_{v}\left(v\right)-\left(1-\lambda\right)\x_{v}\left(v_{0}\right)\right)\cdot\delta\M_{v}-\left[\M_{v},\x_{v}\left(v\right)\right]\cdot\De h_{v}\left(v\right).
\end{align}

\subsection{\label{subsec:The-Particle-Potential}The ``Particle'' Potential}

\subsubsection{Simplifying the Potential}

Let $\x_{v}^{\parallel}\left(v_{0}\right)$ be the component of $\x_{v}\left(v_{0}\right)$
parallel to $\SS_{v}$:
\begin{equation}
\x_{v}\left(v_{0}\right)\equiv\x_{v}^{\parallel}\left(v_{0}\right)+\x_{v}^{\perp}\left(v_{0}\right)\sp\x_{v}^{\parallel}\left(v_{0}\right)\equiv\left(\x_{v}\left(v_{0}\right)\cdot\J_{0}\right)\P_{0},
\end{equation}
where $\J_{0}$ and $\P_{0}$ are the Cartan generator of rotations
and translations respectively, and we remind the reader that the dot
product is defined in (\ref{eq:dot-product}) as $\J_{i}\cdot\P_{j}=\delta_{ij}$
and $\J_{i}\cdot\J_{j}=\P_{i}\cdot\P_{j}=0$. Similarly, let $\De^{\parallel}h_{v}\left(v_{0}\right)$
be the component of $\De h_{v}\left(v_{0}\right)$ parallel to $\M_{v}$:
\[
\De h_{v}\left(v_{0}\right)\equiv\De^{\parallel}h_{v}\left(v_{0}\right)+\De^{\bot}h_{v}\left(v_{0}\right)\sp\De^{\parallel}h_{v}\left(v_{0}\right)\equiv\left(\De h_{v}\left(v_{0}\right)\cdot\P_{0}\right)\J_{0}.
\]
Let us now define a $\mfg$-valued 0-form $\De H_{v}$, which is a
1-form on field space (i.e. a variation\footnote{\label{fn:H_v-foot}Despite the suggestive notation, in principle
$\De H_{v}$ need not be of the form $\delta H_{v}H_{v}^{-1}$ for
some $G$-valued 0-form $H_{v}$. It can instead be of the form $\delta\h_{v}$
for some $\mfg$-valued 0-form $\h_{v}$. Its precise form is left
implicit, and we merely assume that there is a solution for either
$H_{v}$ or $\h_{v}$ in terms of $h_{v}\left(v\right)$ and $h_{v}\left(v_{0}\right)$.}):
\begin{equation}
\De H_{v}\equiv\De h_{v}\left(v\right)-\lambda\De^{\parallel}h_{v}\left(v_{0}\right),\label{eq:DeltaH_v}
\end{equation}
and a $\mfg^{*}$-valued 0-form $\X_{v}$ called the \emph{vertex
flux}:
\begin{equation}
\X_{v}\equiv\x_{v}\left(v\right)-\left(1-\lambda\right)\x_{v}^{\parallel}\left(v_{0}\right)-\left(1-2\lambda\right)\alpha_{v}\SS_{v}.\label{eq:X_v}
\end{equation}
Then since $\SS_{v}\cdot\De h_{v}\left(v_{0}\right)=\SS_{v}\cdot\De^{\parallel}h_{v}\left(v_{0}\right)$
we have
\[
\SS_{v}\cdot\left(\De h_{v}\left(v\right)-\lambda\De h_{v}\left(v_{0}\right)\right)=\SS_{v}\cdot\De H_{v},
\]
and since $\x_{v}\left(v_{0}\right)\cdot\delta\M_{v}=\x_{v}^{\parallel}\left(v_{0}\right)\cdot\delta\M_{v}$
we have
\[
\left(\x_{v}\left(v\right)-\left(1-\lambda\right)\x_{v}\left(v_{0}\right)-\left(1-2\lambda\right)\alpha_{v}\SS_{v}\right)\cdot\delta\M_{v}=\X_{v}\cdot\delta\M_{v}.
\]
Furthermore, since $\left[\M_{v},\x_{v}^{\parallel}\left(v_{0}\right)\right]=\left[\M_{v},\SS_{v}\right]=0$
and $\left[\M_{v},\X_{v}\right]\cdot\De^{\parallel}h_{v}\left(v_{0}\right)=0$
we have
\[
\left[\M_{v},\x_{v}\left(v\right)\right]\cdot\De h_{v}\left(v\right)=\left[\M_{v},\X_{v}\right]\cdot\De H_{v}.
\]
Therefore (\ref{eq:Theta_v}) becomes
\begin{equation}
\Theta_{v}=\X_{v}\cdot\delta\M_{v}-\left(\SS_{v}+\left[\M_{v},\X_{v}\right]\right)\cdot\De H_{v}.\label{eq:Theta_v-simplified}
\end{equation}
This potential resembles that of a point particle with mass $\M_{v}$
and spin $\SS_{v}$. Note that the free parameter $\lambda$ has been
absorbed into $\X_{v}$ and $\De H_{v}$, so this potential is obtained
independently of the value of $\lambda$ and thus the choice of polarization!

\subsubsection{Properties of the Potential}

Omitting the subscript $v$ for brevity, we will now study the relativistic
particle symplectic potential 
\begin{equation}
\Theta\equiv\X\cdot\delta\M-\left(\SS+\left[\M,\X\right]\right)\cdot\De H,\label{eq:relativistic-particle-potential}
\end{equation}
with the symplectic form 
\[
\Omega\equiv\delta\Theta=\delta\X\cdot\delta\M-\left(\delta\SS+\left[\delta\M,\X\right]+\left[\M,\delta\X\right]\right)\cdot\De H-\hf\left(\SS+\left[\M,\X\right]\right)\cdot\left[\De H,\De H\right].
\]
We define the \emph{momentum} $\p$ and \emph{angular momentum} $\j$
of the particle: 
\[
\p\equiv H^{-1}\M H\in\mfg^{*},\qquad\j\equiv H^{-1}\left(\SS+\left[\M,\X\right]\right)H\in\mfg,
\]
which have the variational differentials 
\[
\delta\p=H^{-1}\left(\delta\M+\left[\M,\De H\right]\right)H,
\]
\[
\delta\j=H^{-1}\left(\delta\SS+\left[\delta\M,\X\right]+\left[\M,\delta\X\right]+\left[\SS+\left[\M,\X\right],\De H\right]\right)H.
\]
We also define the ``position'' 
\[
\q\equiv H^{-1}\X H\in\mfg,
\]
in terms of which the symplectic potential may be written as 
\[
\Theta=\q\cdot\delta\p-\SS\cdot\De H.
\]

\subsubsection{Right Translations (Gauge Transformations)}

Let 
\[
\HH\equiv\left(\X,H\right)=e^{\X}H\in G\ltimes\mfg^{*}\sp\GG\equiv\left(g,\z\right)\in G\ltimes\mfg^{*}\soosp\HH\GG=\e^{\X+H\z H^{-1}}Hg.
\]
This is a right translation, with parameter $\GG$, of the group element
$\HH$, which corresponds to a gauge transformation: 
\[
H\mt Hg,\qquad\X\mt\X+H\z H^{-1},\qquad\M\mt\M\sp\SS\mt\SS.
\]
It is interesting to translate this action onto the physical variables
$(\p,\q,\j)$ which transform as 
\[
\p\to g^{-1}\p g,\qquad\q\to\z+g^{-1}\q g,\qquad\j\to g^{-1}\j g.
\]
This shows that the parameter $g$ labels a rotation of the physical
variables, while $\z$ labels a translation of the physical position
$\q$. Taking $g\equiv\e^{\g}$, we may consider transformations labeled
by $\g+\z\in\mfg\oplus\mfg^{*}$ with $\z\in\mfg^{*}$ a translation
parameter and $\g\in\mfg$ a rotation parameter, given by the infinitesimal
version of the gauge transformation\footnote{\label{fn:Lie-derivative}The transformations will be given by the
action of the Lie derivative $\LLL_{\a}\equiv I_{\a}\delta+\delta I_{\a}$
where $I_{\a}$ is the variational interior product with respect to
$\a$. In the literature the notation $\delta_{\a}$ is often used
instead, but we avoid it in order to prevent confusion with the variational
exterior derivative $\delta$.}: 
\[
\LLL_{\g,\z}h=h\g,\qquad\LLL_{\g,\z}\X=g\x g^{-1},\qquad\LLL_{\g,\z}\M=0\sp\LLL_{\g,\z}\SS=0.
\]
Let $I$ denote the interior product on field space, associated with
the variational exterior derivative $\delta$, as explained in Section
\ref{subsec:Variational-Anti-Derivations-on}. Then one finds that
this transformation is Hamiltonian: 
\[
I_{\g,\z}\Omega=-\delta H_{\g,\z}\sp H_{\g,\z}\equiv-\left(\p\cdot\z+\j\cdot\g\right).
\]
The Poisson bracket between two such Hamiltonians is given by 
\[
\left\{ H_{\g,\z},H_{\g',\z'}\right\} =\LLL_{\g,\z}H_{\g',\z'}=H_{\left(\left[\g,\z'\right]+\left[\z,\g'\right],\left[\g,\g'\right]\right)},
\]
which reproduces, as expected, the symmetry algebra $\mfg\oplus\mfg^{*}$.

\subsubsection{Left Translations (Symmetry Transformations)}

Similarly, let 
\[
\HH\equiv\left(h,\X\right)\in G\ltimes\mfg^{*}\sp\GG\equiv\left(g,\z\right)\in G\ltimes\mfg^{*}\soosp\GG\HH=\e^{\z+g\X g^{-1}}gh.
\]
This is a left translation, with parameter $\GG$, of the group element
$\HH$, which corresponds to a symmetry that leaves the connection
invariant: 
\[
h\mt gh,\qquad\X\mt\z+g\X g^{-1},\qquad\M\mt g\M g^{-1},\qquad\SS\mt g\left(\SS+\left[\z,\M\right]\right)g^{-1}.
\]
Note that it commutes with the right translation. The infinitesimal
transformation, with $g\equiv\e^{\g}$ and $\g+\z\in\mfg\oplus\mfg^{*}$
as above, is 
\begin{equation}
\LLL_{\g,\z}h=\g h,\qquad\LLL_{\g,\z}\X=\z+\left[\g,\X\right],\qquad\LLL_{\g,\z}\M=\left[\g,\M\right],\qquad\LLL_{\g,\z}\SS=\left[\g,\SS\right]+\left[\z,\M\right].\label{eq:transf-z-beta}
\end{equation}
Once again, we can prove that this transformation is Hamiltonian:
\[
I_{\g,\z}\Omega=-\delta H_{\g,\z}\sp H_{\g,\z}\equiv-\left(\M\cdot\z+\SS\cdot\g\right).
\]
This follows from the fact that 
\[
\LLL_{\g,\z}\left(\SS+\left[\M,\X\right]\right)=\left[\g,\SS+\left[\M,\X\right]\right],
\]
which implies that these transformations leave the momentum and angular
momentum invariant: $\LLL_{\g,\z}\p=0=\LLL_{\g,\z}\j.$

\subsubsection{Restriction to the Cartan Subalgebra}

In the case discussed here, where $\M\in\mfh^{*}$ and $\SS\in\mfh$
are in the Cartan subalgebra, we need to restrict the parameter of
the left translation transformation to be in $\mfh\oplus\mfh^{*}$.
A particular class of transformations of this type is when the parameter
is itself a function of $\M$ and $\SS$, which we shall denote $F\left(\M,\SS\right)$.
One finds that the infinitesimal transformation 
\begin{equation}
\delta_{F}h=\frac{\partial F}{\partial\SS}h\sp\delta_{F}\y=\frac{\partial F}{\partial\M}+\left[\frac{\partial F}{\partial\SS},\X\right]\sp\delta_{F}\M=0\sp\delta_{F}\SS=0,\label{eq:transf-F}
\end{equation}
is Hamiltonian: 
\[
I_{\delta_{F}}\Omega=-\delta H_{F}\sp H_{F}\equiv-F\left(\M,\SS\right).
\]
In particular, taking 
\[
F\left(\M,\SS\right)\equiv\frac{\xi}{2}\M^{2}+\chi\M\cdot\SS\sp\xi,\chi\in\BBR,
\]
we obtain the Hamiltonian transformation 
\begin{equation}
\delta_{F}h=\M\chi h\sp\delta_{F}\X=\M\xi+\left(\SS+\left[\M,\X\right]\right)\chi\sp\delta_{F}\M=0\sp\delta_{F}\SS=0,\label{eq:transf-F-Casimir}
\end{equation}
corresponding to (\ref{eq:transf-z-beta}) with 
\[
\z=\frac{\partial F}{\partial\M}=\M\xi+\SS\chi\sp\be=\frac{\partial F}{\partial\SS}=\M\chi.
\]
This may be integrated to 
\[
h\mt\e^{\M\chi}h,\qquad\X\mt\e^{\M\chi}\left(\M\xi+\SS\chi+\X\right)\e^{-\M\chi},\qquad\M\mt\M,\qquad\SS\mt\SS.
\]
The Hamiltonians $\M^{2}$ and $\M\cdot\SS$ represent the Casimir
invariants of the algebra $\mfg\oplus\mfg^{*}$.

\subsection{\label{subsec:Edge-Arc}The Edge and Arc Contributions}

To summarize our progress so far, we now have
\[
\Theta=\sum_{c}\Theta_{c}+\sum_{v}\Theta_{\partial_{R}D_{v}}+\sum_{v}\Theta_{v},
\]
where
\[
\Theta_{c}=\int_{\partial c}\left(\left(1-\lambda\right)\d\x_{c}\cdot\De h_{c}-\lambda\x_{c}\cdot\d\De h_{c}\right),
\]
\[
\Theta_{\partial_{R}D_{v}}=\int_{\partial_{R}D_{v}}\left(\left(1-\lambda\right)\d\mathring{\x}_{v}\cdot\De\mathring{h}_{v}-\lambda\mathring{\x}_{v}\cdot\d\De\mathring{h}_{v}\right),
\]
and $\Theta_{v}$ is given by (\ref{eq:Theta_v-simplified}).

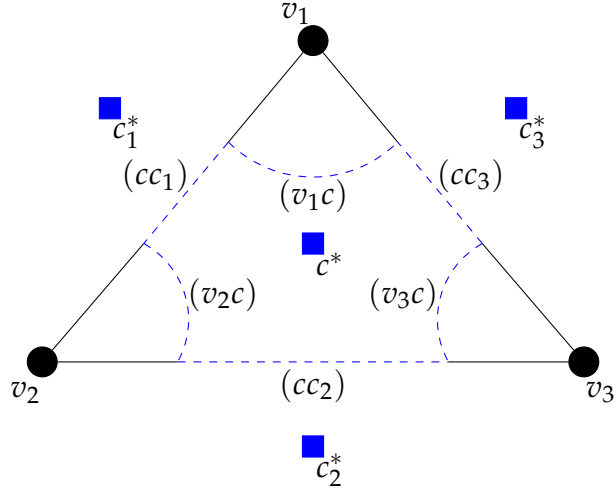
\begin{figure}[!h]
\begin{centering}
\begin{tikzpicture}[scale=0.9]
	\begin{pgfonlayer}{nodelayer}
		\node [style=Node] (0) at (0, -0) {};
		\node [style=Vertex] (1) at (0, 3) {};
		\node [style=Vertex] (2) at (-4, -1.75) {};
		\node [style=Vertex] (3) at (4, -1.75) {};
		\node [style=none] (4) at (0.25, -0.3) {$c^{*}$};
		\node [style=none] (5) at (-0.25, 3.4) {$v_1$};
		\node [style=none] (6) at (-1.25, 1.5) {};
		\node [style=none] (7) at (1.25, 1.5) {};
		\node [style=none] (8) at (2.5, -0) {};
		\node [style=none] (9) at (2, -1.75) {};
		\node [style=none] (10) at (-2.5, -0) {};
		\node [style=none] (11) at (-2, -1.75) {};
		\node [style=none] (12) at (-4.25, -2.2) {$v_2$};
		\node [style=none] (13) at (4.25, -2.2) {$v_3$};
		\node [style=none] (14) at (0, 0.7) {$(v_1c)$};
		\node [style=none] (15) at (-1.33, -0.75) {$(v_2c)$};
		\node [style=none] (16) at (1.33, -0.75) {$(v_3c)$};
		\node [style=none] (17) at (-2.33, 1) {$(cc_1)$};
		\node [style=none] (18) at (0, -2.1) {$(cc_2)$};
		\node [style=none] (19) at (2.33, 1) {$(cc_3)$};
		\node [style=Node] (20) at (3, 2) {};
		\node [style=Node] (21) at (-3, 2) {};
		\node [style=Node] (22) at (0, -3) {};
		\node [style=none] (23) at (3.25, 1.65) {$c_3^{*}$};
		\node [style=none] (24) at (-2.75, 1.65) {$c_1^{*}$};
		\node [style=none] (25) at (0.25, -3.35) {$c_2^{*}$};
	\end{pgfonlayer}
	\begin{pgfonlayer}{edgelayer}
		\draw [style=Link, bend right=45, looseness=1.00] (6.center) to (7.center);
		\draw [style=Link, bend left=45, looseness=1.00] (10.center) to (11.center);
		\draw [style=Link, bend right=45, looseness=1.00] (8.center) to (9.center);
		\draw [style=Link] (8.center) to (7.center);
		\draw [style=Link] (6.center) to (10.center);
		\draw [style=Link] (11.center) to (9.center);
		\draw [style=Edge] (1) to (6.center);
		\draw [style=Edge] (10.center) to (2);
		\draw [style=Edge] (2) to (11.center);
		\draw [style=Edge] (9.center) to (3);
		\draw [style=Edge] (3) to (8.center);
		\draw [style=Edge] (7.center) to (1);
	\end{pgfonlayer}
\end{tikzpicture}
\par\end{centering}
\caption{\label{fig:CellBoundary}The blue square in the center is the node
$c^{*}$. It is dual to the cell $c$, outlined in black. In this
simple example, we have $N=3$ vertices $v_{1},v_{2},v_{3}$ along
the boundary $\partial c$, dual to 3 disks $D_{v_{1}},D_{v_{2}},D_{v_{3}}$.
Only the wedge $D_{v_{i}}\cap c$ is shown for each disk. The cell
$c$ is adjacent to 3 cells $c_{i}$ (not shown) dual to the 3 nodes
$c_{i}^{*}$, in blue.}
\end{figure}

In order to simplify $\Theta_{\partial_{R}D_{v}}$, we recall from
Section \ref{subsec:The-Cellular-Decomposition} that the boundary
$\partial c$ of the cell $c$ is composed of edges $\left(cc_{i}\right)$
and arcs $\left(cv_{i}\right)$ such that
\[
\partial c=\bigcup_{i=1}^{N_{c}}\left(\left(cc_{i}\right)\cup\left(cv_{i}\right)\right),
\]
while the outer boundary $\partial_{R}D_{v}$ of the disk $D_{v}$
is composed of arcs $\left(vc_{i}\right)$ such that
\[
\partial_{R}D_{v}=\bigcup_{i=1}^{N_{v}}\left(vc_{i}\right),
\]
where $N_{v}$ is the number of cells around $v$. This is illustrated
in Figure \ref{fig:CellBoundary}.

Importantly, in terms of orientation, $\left(cc'\right)=\left(c'c\right)^{-1}$
and $\left(cv\right)=\left(vc\right)^{-1}$. We thus see that each
edge $\left(cc'\right)$ is integrated over exactly twice, once from
the integral over $\partial c$ and once from the integral over $\partial c'$
with opposite orientation, and similarly each arc $\left(cv\right)$
is integrated over twice, once from $\partial c$ and once from $\partial_{R}D_{v}$
with opposite orientation. Hence we may rearrange the sums and integrals
as follows:
\[
\Theta=\sum_{\left(cc'\right)}\Theta_{cc'}+\sum_{\left(vc\right)}\Theta_{vc}+\sum_{v}\Theta_{v},
\]
where
\[
\Theta_{cc'}\equiv\int_{\left(cc'\right)}\left(\left(1-\lambda\right)(\d\x_{c}\cdot\De h_{c}-\d\x_{c'}\cdot\De h_{c'})-\lambda(\x_{c}\cdot\d\De h_{c}-\x_{c'}\cdot\d\De h_{c'})\vphantom{\mathring{h}_{v}}\right),
\]
\[
\Theta_{vc}\equiv\int_{\left(vc\right)}\left(\left(1-\lambda\right)(\d\mathring{\x}_{v}\cdot\De\mathring{h}_{v}-\d\x_{c}\cdot\De h_{c})-\lambda(\mathring{\x}_{v}\cdot\d\De\mathring{h}_{v}-\x_{c}\cdot\d\De h_{c})\right).
\]
We now use the continuity conditions derived in Section \ref{subsec:The-Continuity-Conditions}:
\begin{equation}
h_{c'}=h_{c'c}h_{c}\sp\x_{c'}=h_{c'c}(\x_{c}-\x_{c}^{c'})h_{cc'}\sp\textrm{on }\left(cc'\right),\label{eq:continuity-ccp}
\end{equation}
\begin{equation}
h_{c}=h_{cv}\mathring{h}_{v}\sp\x_{c}=h_{cv}(\mathring{\x}_{v}-\x_{v}^{c})h_{vc}\sp\textrm{on }\left(vc\right),\label{eq:continuity-cv}
\end{equation}
where $h_{cc'}$, $h_{cv}$, $\x_{c}^{c'}$ and $\x_{c}^{v}$ are
all constant and satisfy, as derived in Section \ref{subsec:The-Cartan-Decomposition},
\begin{equation}
h_{cc'}=h_{c'c}^{-1}\sp h_{vc}=h_{cv}^{-1}\sp\x_{c}^{c'}=-h_{cc'}\x_{c'}^{c}h_{c'c}\sp\x_{c}^{v}=-h_{cv}\x_{v}^{c}h_{vc}.\label{eq:h-y-inverse}
\end{equation}
By plugging these relations into $\Theta_{cc'}$ and $\Theta_{vc}$
and simplifying, using the identities
\[
\De h_{c'}=h_{c'c}\left(\De h_{c}-\De h_{c}^{c'}\right)h_{cc'}\sp\De h_{c}=h_{cv}\left(\De\mathring{h}_{v}-\De h_{v}^{c}\right)h_{vc},
\]
where $\De h_{c}^{c'}\equiv\delta h_{cc'}h_{c'c}$ and $\De h_{v}^{c}\equiv\delta h_{vc}h_{cv}$,
we find:
\begin{equation}
\Theta_{cc'}=\left(1-\lambda\right)\De h_{c}^{c'}\cdot\int_{\left(cc'\right)}\d\x_{c}-\lambda\x_{c}^{c'}\cdot\int_{\left(cc'\right)}\d\De h_{c},\label{eq:Theta_ccp}
\end{equation}
\begin{equation}
\Theta_{vc}=\left(1-\lambda\right)\De h_{v}^{c}\cdot\int_{\left(vc\right)}\d\mathring{\x}_{v}-\lambda\x_{v}^{c}\cdot\int_{\left(vc\right)}\d\De\mathring{h}_{v}.\label{eq:Theta_vc}
\end{equation}

\subsection{Holonomies and Fluxes}

\begin{figure}[!h]
\begin{centering}
\begin{tikzpicture}[scale=0.9]
	\begin{pgfonlayer}{nodelayer}
		\node [style=Node] (0) at (0, -0) {};
		\node [style=Vertex] (1) at (0, 3) {};
		\node [style=Vertex] (2) at (-4, -1.75) {};
		\node [style=Vertex] (3) at (4, -1.75) {};
		\node [style=none] (4) at (0.25, -0.3) {$c^{*}$};
		\node [style=none] (5) at (-0.25, 3.4) {$v_1$};
		\node [style=Point] (6) at (-1.25, 1.5) {};
		\node [style=Point] (7) at (1.25, 1.5) {};
		\node [style=Point] (8) at (2.5, -0) {};
		\node [style=Point] (9) at (2, -1.75) {};
		\node [style=Point] (10) at (-2.5, -0) {};
		\node [style=Point] (11) at (-2, -1.75) {};
		\node [style=none] (12) at (-4.25, -2.17) {$v_2$};
		\node [style=none] (13) at (4.25, -2.17) {$v_3$};
		\node [style=Node] (14) at (3.5, 2.5) {};
		\node [style=Node] (15) at (-3.5, 2.5) {};
		\node [style=Node] (16) at (0, -3.5) {};
		\node [style=none] (17) at (3.8, 2.1) {$c_3^{*}$};
		\node [style=none] (18) at (-3.8, 2.1) {$c_1^{*}$};
		\node [style=none] (19) at (0.25, -3.9) {$c_2^{*}$};
		\node [style=none] (20) at (1.25, 1.75) {$\tau_{v_1c}=\tau_{cc_3}$};
		\node [style=none] (21) at (-1.25, 1.75) {$\sigma_{v_1c}=\sigma_{cc_1}$};
		\node [style=none] (22) at (-2.5, 0.25) {$\tau_{v_2c}=\tau_{cc_1}$};
		\node [style=none] (23) at (-2, -2) {$\sigma_{v_2c}=\sigma_{cc_2}$};
		\node [style=none] (24) at (2, -2) {$\tau_{v_3c}=\tau_{cc_2}$};
		\node [style=none] (25) at (2.5, 0.25) {$\sigma_{v_3c}=\sigma_{cc_3}$};
	\end{pgfonlayer}
	\begin{pgfonlayer}{edgelayer}
		\draw [style=LinkArrow, bend right=45, looseness=1.00] (6) to (7);
		\draw [style=LinkArrow, bend right=45, looseness=1.00] (11) to (10);
		\draw [style=LinkArrow, bend right=45, looseness=1.00] (8) to (9);
		\draw [style=LinkArrow] (8) to (7);
		\draw [style=LinkArrow] (6) to (10);
		\draw [style=LinkArrow] (11) to (9);
		\draw [style=Edge] (1) to (6);
		\draw [style=Edge] (10) to (2);
		\draw [style=Edge] (2) to (11);
		\draw [style=Edge] (9) to (3);
		\draw [style=Edge] (3) to (8);
		\draw [style=Edge] (7) to (1);
	\end{pgfonlayer}
\end{tikzpicture}
\par\end{centering}
\caption{\label{fig:IntersectionPoints}The intersection points (red circles)
of truncated edges and arcs along the oriented boundary $\partial c$
(blue arrows).}
\end{figure}
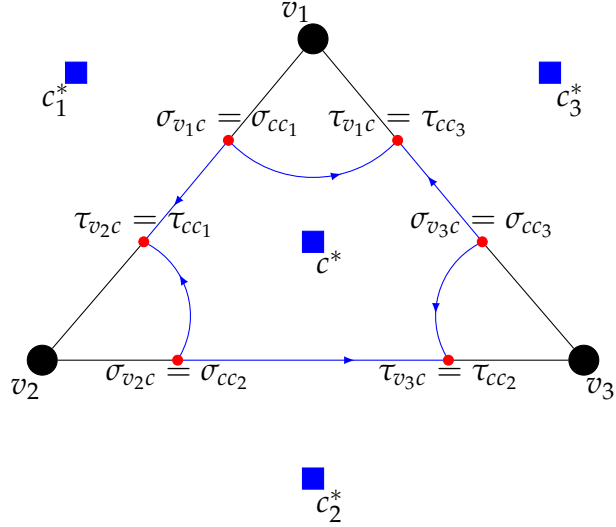

Let us label the source and target points of the edge $\left(cc'\right)$
as $\sigma_{cc'}$ and $\tau_{cc'}$ respectively, and the source
and target points of the arc $\left(vc\right)$ as $\sigma_{vc}$
and $\tau_{vc}$ respectively, where $\sigma$ stands for ``source''
and $\tau$ for ``target'':
\[
\left(cc'\right)\equiv\left(\sigma_{cc'}\tau_{cc'}\right)\sp\left(vc\right)\equiv\left(\sigma_{vc}\tau_{vc}\right).
\]
This labeling is illustrated in Figure \ref{fig:IntersectionPoints}.
We now define holonomies and fluxes on the edges and their dual links,
and on the arcs and their dual line segments.

\subsubsection{Holonomies on the Links and Segments}

The rotational holonomy $h_{cc'}$ comes from the continuity relations
(\ref{eq:continuity-ccp}). Its role is relating the variables $h_{c},\x_{c}$
on the cell $c$ to the variables $h_{c'},\x_{c'}$ on the cell $c'$.
Now, in the relation $h_{c}\left(x\right)=h_{cc'}h_{c'}\left(x\right)$,
the holonomy on the left-hand side is from the node $c^{*}$ to a
point $x$ on the edge $\left(cc'\right)$. Therefore, the holonomy
on the right-hand side should also take us from $c^{*}$ to $x$.
Since $h_{c'}\left(x\right)$ is the holonomy from $c^{\prime*}$
to $x$, we see that $h_{cc'}$ must take us from $c^{*}$ to $c^{\prime*}$.
In other words, the holonomy $h_{cc'}$ is exactly the holonomy from
$c^{*}$ to $c^{\prime*}$, along\footnote{Since the geometry is flat, the actual path taken does not matter,
only that it starts at $c^{*}$ and ends at $c^{\prime*}$. We may
therefore assume without loss of generality that the path taken by
$h_{cc'}$ is, in fact, along the link $\left(cc'\right)^{*}$.} the link $\left(cc'\right)^{*}$, which was discussed in Section
\ref{subsec:The-Cartan-Decomposition}.

Thus we define\footnote{The change from lower-case $h$ to upper-case $H$ is only symbolic
here, but it will become more meaningful when we define other holonomies
and fluxes below.} \emph{holonomies along the links $\left(cc'\right)^{*}$}:
\begin{equation}
H_{cc'}\equiv h_{cc'}\sp\De H_{c}^{c'}\equiv\delta H_{cc'}H_{c'c}.\label{eq:hol-link}
\end{equation}
Similarly, the holonomy $h_{vc}$ comes from the continuity relations
(\ref{eq:continuity-cv}), and it takes us from the vertex $v$ to
the node $c^{*}$. We define $\left(vc\right)^{*}$ to be the line
segment connecting $v$ to $c^{*}$; it is dual to the arc $\left(vc\right)$
and its inverse is $\left(cv\right)^{*}$. We then define \emph{holonomies
along the segments $\left(vc\right)^{*}$}:
\begin{equation}
H_{vc}\equiv h_{vc}\sp\De H_{v}^{c}\equiv\delta H_{vc}H_{cv}.\label{eq:hol-seg}
\end{equation}
The inverse holonomies follow immediately from the relations $h_{cc'}^{-1}=h_{c'c}$
and $h_{vc}^{-1}=h_{cv}$:
\[
H_{cc'}^{-1}=H_{c'c}\sp H_{vc}^{-1}=H_{cv}.
\]

\subsubsection{Fluxes on the Edges and Arcs}

From the integral in the first term of (\ref{eq:Theta_ccp}), we are
inspired to define \emph{fluxes along the edges }$\left(cc'\right)$:
\begin{equation}
\XXt_{c}^{c'}\equiv\int_{\left(cc'\right)}\d\x_{c}=\x_{c}\left(\tau_{cc'}\right)-\x_{c}\left(\sigma_{cc'}\right).\label{eq:flux-edge}
\end{equation}
The tilde specifies that the flux $\XXt_{c}^{c'}$ is on the edge
$\left(cc'\right)$ dual to the link $\left(cc'\right)^{*}$; the
flux $\X_{c}^{c'}$, to be defined below, is on the link, and similarly
we will define $\Ht_{cc'}$ to be the holonomy on the edge, while
$H_{cc'}$ is the holonomy on the link.

The flux $\XXt_{c}^{c'}$ is a composition of two translational holonomies.
The holonomy $-\x_{c}\left(\sigma_{cc'}\right)$ takes us from the
point $\sigma_{cc'}$ to the node $c^{*}$, and then the holonomy
$\x_{c}\left(\tau_{cc'}\right)$ takes us from $c^{*}$ to $\tau_{cc'}$.
Hence, the composition of these holonomies is a translational holonomy
from $\sigma_{cc'}$ to $\tau_{cc'}$, that is, along\footnote{Again, since the geometry is flat, the path passing through the node
$c^{*}$ is equivalent to the path going along the edge $\left(cc'\right)$.} the edge $\left(cc'\right)$, as claimed.

To find the inverse flux we use $\left(cc'\right)=\left(c'c\right)^{-1}$,
$\sigma_{cc'}=\tau_{c'c}$ and (\ref{eq:continuity-ccp}):
\[
\XXt_{c'}^{c}\equiv\int_{\left(c'c\right)}\d\x_{c'}=\x_{c'}\left(\tau_{c'c}\right)-\x_{c'}\left(\sigma_{c'c}\right)=h_{c'c}\left(\x_{c}\left(\sigma_{cc'}\right)-\x_{c}\left(\tau_{cc'}\right)\right)h_{cc'}=-H_{c'c}\XXt_{c}^{c'}H_{cc'}.
\]
Similarly, from the first integral in (\ref{eq:Theta_vc}) we are
inspired to define \emph{fluxes along the arcs} $\left(vc\right)$:
\begin{equation}
\XXt_{v}^{c}\equiv\int_{\left(vc\right)}\d\mathring{\x}_{v}=\mathring{\x}_{v}\left(\tau_{vc}\right)-\mathring{\x}_{v}\left(\sigma_{vc}\right).\label{eq:flux-arc-vc}
\end{equation}
Note that this time, the two translational holonomies are composed
at $v$. As for the inverse, we define $\XXt_{c}^{v}$ as follows
and use (\ref{eq:continuity-cv}) to find a relation with $\XXt_{v}^{c}$,
taking into account the fact that $\left(cv\right)=\left(vc\right)^{-1}$
and $\sigma_{cv}=\tau_{vc}$:
\begin{equation}
\XXt_{c}^{v}\equiv\int_{\left(cv\right)}\d\x_{c}=\x_{c}\left(\tau_{cv}\right)-\x_{c}\left(\sigma_{cv}\right)=h_{cv}\left(\mathring{\x}_{v}\left(\sigma_{vc}\right)-\mathring{\x}_{v}\left(\tau_{vc}\right)\right)h_{vc}=-H_{cv}\XXt_{v}^{c}H_{vc}.\label{eq:flux-arc-cv}
\end{equation}
In conclusion, we have the relations
\[
\XXt_{c'}^{c}=-H_{c'c}\XXt_{c}^{c'}H_{cc'}\sp\XXt_{c}^{v}=-H_{cv}\XXt_{v}^{c}H_{vc}.
\]

\subsubsection{Holonomies on the Edges and Arcs}

The holonomies and fluxes defined thus far will be used in the $\lambda=0$
polarization. In the $\lambda=1$ (dual) polarization, let us define
\emph{holonomies along the edges} $\left(cc'\right)$ and \emph{holonomies
along the arcs} $\left(vc\right)$:
\begin{equation}
\Ht_{cc'}\equiv h_{c}^{-1}\left(\sigma_{cc'}\right)h_{c}\left(\tau_{cc'}\right)\sp\De\Ht_{c}^{c'}\equiv\delta\Ht_{cc'}\Ht_{c'c},\label{eq:Ht-ccp}
\end{equation}
\begin{equation}
\Ht_{vc}\equiv\mathring{h}_{v}^{-1}\left(\sigma_{vc}\right)\mathring{h}_{v}\left(\tau_{vc}\right)\sp\De\Ht_{v}^{c}\equiv\delta\Ht_{vc}\Ht_{cv}.\label{eq:Ht-vc}
\end{equation}
As with $\XXt_{c}^{c'}$, the holonomy $\Ht_{cc'}$ starts from $\sigma_{cc'}$,
goes to $c^{*}$ via $h_{c}^{-1}\left(\sigma_{cc'}\right)$, and then
goes to $\tau_{cc'}$ via $h_{c}\left(\tau_{cc'}\right)$. Therefore
it is indeed a holonomy along the edge $\left(cc'\right)$. Similarly,
the holonomy $\Ht_{vc}$ starts from $\sigma_{vc}$, goes to $v$
via $\mathring{h}_{v}^{-1}\left(\sigma_{vc}\right)$, and then goes
to $\tau_{vc}$ via $\mathring{h}_{v}\left(\tau_{vc}\right)$. Therefore
it is indeed a holonomy along the arc $\left(vc\right)$.

The difference compared to $\XXt_{c}^{c'}$ is that in $\Ht_{cc'}$
we have rotational instead of translational holonomies, and the composition
of holonomies is (non-abelian) multiplication instead of addition.
As before, the tilde specifies that the holonomy is on the edges or
arcs and not the dual links or segments.

The variations of these holonomies are:
\begin{equation}
\De\Ht_{c}^{c'}=h_{c}^{-1}\left(\sigma_{cc'}\right)\left(\De h_{c}\left(\tau_{cc'}\right)-\De h_{c}\left(\sigma_{cc'}\right)\right)h_{c}\left(\sigma_{cc'}\right)=h_{c}^{-1}\left(\sigma_{cc'}\right)\left(\int_{\left(cc'\right)}\d\De h_{c}\right)h_{c}\left(\sigma_{cc'}\right),\label{eq:DeltaHccp}
\end{equation}
\begin{equation}
\De\Ht_{v}^{c}=\mathring{h}_{v}^{-1}\left(\sigma_{vc}\right)\left(\De\mathring{h}_{v}\left(\tau_{vc}\right)-\De\mathring{h}_{v}\left(\sigma_{vc}\right)\right)\mathring{h}_{v}\left(\sigma_{vc}\right)=\mathring{h}_{v}^{-1}\left(\sigma_{vc}\right)\left(\int_{\left(vc\right)}\d\De\mathring{h}_{v}\right)\mathring{h}_{v}\left(\sigma_{vc}\right).\label{eq:DeltaHvc}
\end{equation}
Thus, we see that they relate to the integrals in the second terms
of (\ref{eq:Theta_ccp}) and (\ref{eq:Theta_vc}).

Since $\left(cc'\right)=\left(c'c\right)^{-1}$, it is obvious that
$\Ht_{cc'}^{-1}=\Ht_{c'c}$. Furthermore, by combining (\ref{eq:Ht-vc})
with (\ref{eq:continuity-cv}) we may obtain an expression for $\Ht_{vc}$
in terms of $h_{c}$:
\[
\Ht_{vc}=h_{c}^{-1}\left(\sigma_{vc}\right)h_{c}\left(\tau_{vc}\right).
\]
If we now define
\begin{equation}
\Ht_{cv}\equiv h_{c}^{-1}\left(\sigma_{cv}\right)h_{c}\left(\tau_{cv}\right),\label{eq:Ht-vc-h_c}
\end{equation}
then using the relations $\sigma_{cv}=\tau_{vc}$ and $\tau_{cv}=\sigma_{vc}$,
which come from the fact that $\left(vc\right)=\left(cv\right)^{-1}$,
it is easy to see that $\Ht_{vc}^{-1}=\Ht_{cv}$. In conclusion, the
inverses of these holonomies satisfy the relationships
\[
\Ht_{cc'}^{-1}=\Ht_{c'c}\sp\Ht_{vc}^{-1}=\Ht_{cv}.
\]

\subsubsection{Fluxes on the Links and Segments}

Just as we defined the holonomies on the links and segments from the
variables $h_{cc'}$ and $h_{vc}$, which were used in the continuity
relations (\ref{eq:continuity-ccp}) and (\ref{eq:continuity-cv}),
we can similarly define the fluxes on the links and segments from
the variables $\x_{c}^{c'}$ and $\x_{v}^{c}$. These will, again,
be used in the dual polarization.

Let us define \emph{fluxes along the links $\left(cc'\right)^{*}$
and segments $\left(vc\right)^{*}$}:
\begin{equation}
\X_{c}^{c'}\equiv h_{c}^{-1}\left(\sigma_{cc'}\right)\x_{c}^{c'}h_{c}\left(\sigma_{cc'}\right)\sp\X_{v}^{c}\equiv\mathring{h}_{v}^{-1}\left(\sigma_{vc}\right)\x_{v}^{c}\mathring{h}_{v}\left(\sigma_{vc}\right).\label{eq:fluxes-links}
\end{equation}
The factors of $h_{c}\left(\sigma_{cc'}\right)$ and $\mathring{h}_{v}\left(\sigma_{vc}\right)$
are needed because they appear alongside the integrals in the variations
(\ref{eq:DeltaHccp}) and (\ref{eq:DeltaHvc}). Thus, if we want the
second terms in (\ref{eq:Theta_ccp}) and (\ref{eq:Theta_vc}) to
look like we want them to, we must include these extra factors in
the definition of the fluxes. The fluxes are still translational holonomies
between two cells (in the case of $\x_{c}^{c'}$) or a cell and a
disk (in the case of $\x_{v}^{c}$), but they contain an extra rotation
at the starting point.

The inverse link flux $\X_{c'}^{c}$ follows from (\ref{eq:continuity-ccp}),
(\ref{eq:h-y-inverse}) and $\sigma_{cc'}=\tau_{c'c}$, while the
inverse segment flux $\X_{c}^{v}\equiv h_{c}^{-1}\left(\sigma_{cv}\right)\x_{c}^{v}h_{c}\left(\sigma_{cv}\right)$
follows from (\ref{eq:continuity-cv}), (\ref{eq:h-y-inverse}) and
$\sigma_{cv}=\tau_{vc}$:
\[
\X_{c'}^{c}=-\Ht_{cc'}^{-1}\X_{c}^{c'}\Ht_{cc'}\sp\X_{c}^{v}=-\Ht_{vc}^{-1}\X_{v}^{c}\Ht_{vc}.
\]

\subsection{\label{subsec:The-Discretized-Symplectic}The Discretized Symplectic
Potential}

With the holonomies and fluxes defined above, we find that we can
write the symplectic potential on the edges and arcs, (\ref{eq:Theta_ccp})
and (\ref{eq:Theta_vc}), as:
\[
\Theta_{cc'}=\left(1-\lambda\right)\XXt_{c}^{c'}\cdot\De H_{c}^{c'}-\lambda\X_{c}^{c'}\cdot\De\Ht_{c}^{c'},
\]
\[
\Theta_{vc}=\left(1-\lambda\right)\XXt_{v}^{c}\cdot\De H_{v}^{c}-\lambda\X_{v}^{c}\cdot\De\Ht_{v}^{c}.
\]
The full symplectic potential becomes:
\begin{align*}
\Theta & =\sum_{\left(cc'\right)}\left(\left(1-\lambda\right)\XXt_{c}^{c'}\cdot\De H_{c}^{c'}-\lambda\X_{c}^{c'}\cdot\De\Ht_{c}^{c'}\right)+\\
 & \qquad+\sum_{\left(vc\right)}\left(\left(1-\lambda\right)\XXt_{v}^{c}\cdot\De H_{v}^{c}-\lambda\X_{v}^{c}\cdot\De\Ht_{v}^{c}\right)+\\
 & \qquad+\sum_{v}\left(\X_{v}\cdot\delta\M_{v}-\left(\SS_{v}+\left[\M_{v},\X_{v}\right]\right)\cdot\De H_{v}\right).
\end{align*}
Notice how the holonomies and fluxes are always dual to each other:
one with tilde (on the edges/arcs) and one without tilde (on the links/segments).
For the $\lambda=0$ polarization, the holonomies are on the links
$\left(cc'\right)^{*}$ and segments $\left(vc\right)^{*}$ and the
fluxes are on their dual edges $\left(cc'\right)$ and arcs $\left(vc\right)$.
This polarization corresponds to the usual loop gravity picture. For
the $\lambda=1$ (dual) polarization, we have the opposite case: the
fluxes are on the links $\left(cc'\right)^{*}$ and segments $\left(vc\right)^{*}$
and the holonomies are on their dual edges $\left(cc'\right)$ and
arcs $\left(vc\right)$. For any other choice of $\lambda$, we have
a combination of both polarizations.

The phase space corresponding to $\X\cdot\De H$ for some flux $\X$
and holonomy $H$ is called the \emph{holonomy-flux phase space},
and it is the classical phase space of the spin networks which appear
in loop quantum gravity.

\section{\label{sec:The-Gauss-and}The Gauss and Curvature Constraints}

We have seen that, in the continuum, the constraints are $\F=\T=0$.
Let us see how they translate to constraints on the discrete phase
space. There will be two types of constraints: the \emph{curvature
constraints }which corresponds to $\F=0$, and the \emph{Gauss constraints
}which correspond to $\T=0$. The constraints will be localized in
three different types of places: on the cells, on the disks, and on
the faces. After deriving all of the constraints and showing that
they are identically satisfied in our construction, we will summarize
and interpret them. The reader who is not interested in the details
of the calculation may wish to skip to Section \ref{subsec:Summary-and-Interpretation}.

\subsection{Derivation of the Constraints on the Cells}

\subsubsection{The Gauss Constraint on the Cells}

The cell Gauss constraint $\G_{c}$ will impose the torsionlessness
condition $\T\equiv\d_{\A}\ee=0$ inside the cells:
\[
0=\G_{c}\equiv\int_{c}h_{c}\left(\d_{\A}\ee\right)h_{c}^{-1}=\int_{c}\d\left(h_{c}\ee h_{c}^{-1}\right)=\int_{\partial c}h_{c}\ee h_{c}^{-1}=\int_{\partial c}\d\x_{c}.
\]
As we have seen, $\partial c$ is composed of edges $\left(cc_{i}\right)$
and arcs $\left(cv_{i}\right)$ such that
\[
\partial c=\bigcup_{i=1}^{N_{c}}\left(\left(cc_{i}\right)\cup\left(cv_{i}\right)\right).
\]
Therefore we can split the integral as follows:
\[
\G_{c}=\sum_{c'\ni c}\int_{\left(cc'\right)}\d\x_{c}+\sum_{v\ni c}\int_{\left(cv\right)}\d\x_{c},
\]
where $c'\ni c$ means ``all cells $c'$ adjacent to $c$'' and
$v\ni c$ means ``all vertices $v$ adjacent to $c$''.

Using the fluxes defined in (\ref{eq:flux-edge}) and (\ref{eq:flux-arc-cv}),
we get 
\begin{equation}
\G_{c}=\sum_{c'\ni c}\XXt_{c}^{c'}+\sum_{v\ni c}\XXt_{c}^{v}=0.\label{eq:Gauss-cell}
\end{equation}
This constraint is satisfied identically in our construction. Indeed,
from (\ref{eq:flux-edge}) and (\ref{eq:flux-arc-cv}) we have
\[
\XXt_{c}^{c'}=\x_{c}\left(\tau_{cc'}\right)-\x_{c}\left(\sigma_{cc'}\right)\sp\XXt_{c}^{v}=\x_{c}\left(\tau_{cv}\right)-\x_{c}\left(\sigma_{cv}\right).
\]
Since $\tau_{cc_{i}}=\sigma_{cv_{i}}$ and $\tau_{cv_{i}}=\sigma_{cc_{i+1}}$
(the end of an edge is the beginning of an arc and the end of an arc
is the beginning of an edge), and $\tau_{cv_{N_{c}}}=\sigma_{cc_{1}}$
(the end of the last arc is the beginning of the first edge), it is
easy to see that the sum $\sum_{c'\ni c}\XXt_{c}^{c'}+\sum_{v\ni c}\XXt_{c}^{v}$
evaluates to zero.

\subsubsection{The Curvature Constraint on the Cells}

The cell curvature constraint $F_{c}$ will impose that $\F\equiv\d\A+\hf\left[\A,\A\right]=0$
inside the cells. An equivalent condition is that the holonomy around
the cell evaluates to the identity:
\[
1=F_{c}\equiv\pexp\int_{\partial c}\A.
\]
Since $\partial c=\bigcup_{i=1}^{N_{c}}\left(\left(cc_{i}\right)\cup\left(cv_{i}\right)\right)$,
we may decompose this as a product of path-ordered exponentials over
edges and arcs:
\[
F_{c}=\prod_{i=1}^{N_{c}}\left(\pexp\int_{\left(cc_{i}\right)}\A\right)\left(\pexp\int_{\left(cv_{i}\right)}\A\right).
\]
Furthermore, since the geometry is flat, we may deform the paths so
that instead of going along the edges and arcs, it passes through
the node $c^{*}$. From (\ref{eq:undressed-c}) we have that
\[
\pexp\int_{c^{*}}^{x}\A=h_{c}^{-1}\left(c^{*}\right)h_{c}\left(x\right),
\]
so
\[
\pexp\int_{\left(cc_{i}\right)}\A=\pexp\int_{\sigma_{cc_{i}}}^{\tau_{cc_{i}}}\A=\left(\pexp\int_{\sigma_{cc_{i}}}^{c^{*}}\A\right)\left(\pexp\int_{c^{*}}^{\tau_{cc_{i}}}\A\right)=h_{c}^{-1}\left(\sigma_{cc_{i}}\right)h_{c}\left(\tau_{cc_{i}}\right)=\Ht_{cc_{i}},
\]
where we used the definition (\ref{eq:Ht-ccp}) of the holonomy on
the edge. Note that the contribution from $h_{c}\left(c^{*}\right)$
cancels. Similarly, we find
\begin{equation}
\pexp\int_{\left(cv_{i}\right)}\A=h_{c}^{-1}\left(\sigma_{cv_{i}}\right)h_{c}\left(\tau_{cv_{i}}\right)=\Ht_{cv_{i}},\label{eq:exp-cv}
\end{equation}
where we used (\ref{eq:Ht-vc-h_c}). Hence we obtain
\begin{equation}
F_{c}=\prod_{i=1}^{N_{c}}\Ht_{cc_{i}}\Ht_{cv_{i}}=1.\label{eq:curv-con-cells}
\end{equation}
This is the curvature constraint on the cells. It is easy to show
that it is satisfied identically in our construction. Indeed, using
again the relations $\tau_{cc_{i}}=\sigma_{cv_{i}}$, $\tau_{cv_{i}}=\sigma_{cc_{i+1}}$
and $\tau_{cv_{N_{c}}}=\sigma_{cc_{1}}$, we immediately see that
\[
\prod_{i=1}^{N_{c}}\Ht_{cc_{i}}\Ht_{cv_{i}}=\prod_{i=1}^{N_{c}}\left(h_{c}^{-1}\left(\sigma_{cc_{i}}\right)h_{c}\left(\tau_{cc_{i}}\right)\right)\left(h_{c}^{-1}\left(\sigma_{cv_{i}}\right)h_{c}\left(\tau_{cv_{i}}\right)\right)=1,
\]
as desired.

\subsection{Derivation of the Constraints on the Disks}

Since we have placed the curvature and torsion excitations inside
the disks, the constraints on the disks must involve these excitations
-- namely, $\M_{v}$ and $\SS_{v}$. We will now see that this is
indeed the case.

\subsubsection{The Gauss Constraint on the Disks}

The disk Gauss constraint $\G_{v}$ will impose the torsionlessness
condition $\T\equiv\d_{\A}\ee=0$ inside the \textit{punctured}\footnote{As we have seen, we only have $\T=0$ inside the \emph{punctured }disk
$D_{v}$; at the vertex $v$ itself there is torsion, but $v$ is
not part of $D_{v}$. Instead, it is on its (inner) boundary. As can
be seen from Figure \ref{fig:Disk}, the path we take here, as given
by (\ref{eq:breakdown}), does not enclose the vertex, and therefore
the interior of the path is indeed torsionless.} disks:
\[
0=\G_{v}\equiv\int_{D_{v}}\mathring{h}_{v}\left(\d_{\A}\ee\right)\mathring{h}_{v}^{-1}=\int_{D_{v}}\d\left(\mathring{h}_{v}\ee\mathring{h}_{v}^{-1}\right)=\int_{\partial D_{v}}\mathring{h}_{v}\ee\mathring{h}_{v}^{-1}=\int_{\partial D_{v}}\d\mathring{\x}_{v}.
\]
The boundary $\partial D_{v}$ is composed of the inner boundary $\partial_{0}D_{v}$,
the outer boundary $\partial_{R}D_{v}$, and the cut $C_{v}$:
\begin{equation}
\partial D_{v}=\partial_{0}D_{v}\cup\partial_{R}D_{v}\cup C_{v}.\label{eq:breakdown}
\end{equation}
Hence
\[
\G_{v}=\int_{\partial_{R}D_{v}}\d\mathring{\x}_{v}-\int_{\partial_{0}D_{v}}\d\mathring{\x}_{v}-\int_{C_{v}}\d\mathring{\x}_{v},
\]
where the minus signs represent the relative orientations of each
piece. On the inner boundary $\partial_{0}D_{v}$, we use the fact
that $\x_{v}$ takes the constant value $\x_{v}\left(v\right)$ to
obtain
\begin{align*}
\int_{\partial_{0}D_{v}}\d\mathring{\x}_{v} & =\e^{\M_{v}\phi_{v}}\left(\x_{v}\left(v\right)+\SS_{v}\phi_{v}\right)\e^{-\M_{v}\phi_{v}}\bll_{\phi_{v}=\alpha_{v}-\hf}^{\alpha_{v}+\hf}\\
 & =\SS_{v}+\e^{\M_{v}\left(\alpha_{v}-\hf\right)}\left(\e^{\M_{v}}\x_{v}\left(v\right)\e^{-\M_{v}}-\x_{v}\left(v\right)\right)\e^{-\M_{v}\left(\alpha_{v}-\hf\right)}.
\end{align*}
The outer boundary $\partial_{R}D_{v}$ splits into arcs, and we use
the definition (\ref{eq:flux-arc-vc}) of the flux:
\[
\int_{\partial_{R}D_{v}}\d\mathring{\x}_{v}=\sum_{c\in v}\int_{\left(vc\right)}\d\mathring{\x}_{v}=\sum_{c\in v}\XXt_{v}^{c}.
\]
On the cut $C_{v}$, we have contributions from both sides, one at
$\phi_{v}=\alpha_{v}-\hf$ and another at $\phi_{v}=\alpha_{v}+\hf$
with opposite orientation. Since $\d\phi_{v}=0$ on the cut, we have:
\[
\d\mathring{\x}_{v}\bl_{C_{v}}=\e^{\M_{v}\phi_{v}}\d\x_{v}\e^{-\M_{v}\phi_{v}},
\]
and thus 
\begin{align*}
\int_{C_{v}}\d\mathring{\x}_{v} & =\int_{r=0}^{R}\left(\e^{\M_{v}\phi_{v}}\d\x_{v}\e^{-\M_{v}\phi_{v}}\bll_{\phi_{v}=\alpha_{v}-\hf}^{\phi_{v}=\alpha_{v}+\hf}\right)\\
 & =\e^{\M_{v}\left(\alpha_{v}-\hf\right)}\left(\e^{\M_{v}}\left(\x_{v}\left(v_{0}\right)-\x_{v}\left(v\right)\right)\e^{-\M_{v}}-\left(\x_{v}\left(v_{0}\right)-\x_{v}\left(v\right)\right)\right)\e^{-\M_{v}\left(\alpha_{v}-\hf\right)},
\end{align*}
since $\x_{v}$ has the value $\x_{v}\left(v_{0}\right)$ at $r=R$
and $\x_{v}\left(v\right)$ at $r=0$ on the cut.

Adding up the integrals, we find that the Gauss constraint on the
disk is 
\begin{equation}
\G_{v}=\sum_{c\in v}\XXt_{v}^{c}-\SS_{v}-\e^{\M_{v}\left(\alpha_{v}-\hf\right)}\left(\e^{\M_{v}}\x_{v}\left(v_{0}\right)\e^{-\M_{v}}-\x_{v}\left(v_{0}\right)\right)\e^{-\M_{v}\left(\alpha_{v}-\hf\right)}=0.\label{eq:G_v}
\end{equation}
In fact, since this constraint is used as a generator of symmetries
(as we will see below), it automatically comes dotted with a Cartan
element $\be_{v}$, which commutes with $\e^{\M_{v}}$. Therefore,
the last term may be ignored, and the constraint simplifies to
\[
\be_{v}\cdot\G_{v}=\be_{v}\cdot\left(\sum_{c\in v}\XXt_{v}^{c}-\SS_{v}\right)=0.
\]
Thus it may also be written
\begin{equation}
\sum_{c\in v}\XXt_{v}^{c}=\SS_{v}.\label{eq:Gauss-disk}
\end{equation}
To see that this constraint is satisfied identically in our construction,
let us combine (\ref{eq:flux-arc-vc}) with (\ref{eq:u-v-def}) to
obtain
\[
\XXt_{v}^{c}=\SS_{v}\left(\tau_{vc}-\sigma_{vc}\right)+\e^{\M_{v}\tau_{vc}}\x_{v}\left(\tau_{vc}\right)\e^{-\M_{v}\tau_{vc}}-\e^{\M_{v}\sigma_{vc}}\x_{v}\left(\sigma_{vc}\right)\e^{-\M_{v}\sigma_{vc}},
\]
where we used a slight abuse of notation by using $\sigma_{vc}$ and
$\tau_{vc}$ to denote the corresponding angles, $\sigma_{vc}\equiv\phi_{v}\left(\sigma_{vc}\right)$
and $\tau_{vc}\equiv\phi_{v}\left(\tau_{vc}\right)$. Let us now sum
over the fluxes for each arc. Since $\tau_{vc_{i}}=\sigma_{vc_{i+1}}$
(each arc ends where the next one starts) and $\tau_{vc_{N_{v}}}=\sigma_{vc_{1}}+1$
(the last arc ends a full circle after the first arc began\footnote{Recall that we are using scaled angles such that a full circle corresponds
to 1 instead of $2\pi$!}), we get
\[
\sum_{i=1}^{N_{v}}\XXt_{v}^{c_{i}}=\SS_{v}+\e^{\M_{v}\sigma_{vc_{1}}}\left(\e^{\M_{v}}\x_{v}\left(\sigma_{vc_{1}}\right)\e^{-\M_{v}}-\x_{v}\left(\sigma_{vc_{1}}\right)\right)\e^{-\M_{v}\sigma_{vc_{1}}}.
\]
Choosing without loss of generality the point $v_{0}$ to be at the
beginning of the first edge, $v_{0}=\sigma_{vc_{1}}$, and recalling
that this point corresponds to the angle $\phi_{v}=\alpha_{v}-\hf$,
we indeed obtain precisely the constraint (\ref{eq:G_v}).

\subsubsection{The Curvature Constraint on the Disks}

The disk curvature constraint $F_{v}$ will impose that $\F\equiv\d\A+\hf\left[\A,\A\right]=0$
inside the punctured disks\footnote{Again, we only have $\F=0$ inside the \emph{punctured }disk $D_{v}$;
at the vertex $v$ itself, there is curvature. However, the path of
integration does not enclose the vertex, and therefore the interior
of the path is indeed flat.}. An equivalent condition is that the holonomy around the punctured
disk evaluates to the identity:
\[
1=F_{v}\equiv\pexp\int_{\partial D_{v}}\A=\pexp\left(\int_{C_{v}^{-}}\A\right)\pexp\left(\int_{\partial_{R}D_{v}}\A\right)\pexp\left(\int_{C_{v}^{+}}\A\right)\pexp\left(\int_{\partial_{0}D_{v}}\A\right).
\]
Let us describe the path of integration step by step, referring to
Figure \ref{fig:Disk}:
\begin{itemize}
\item We start at $v$, at the polar coordinates $r_{v}=0$ and $\phi_{v}=\alpha_{v}-1/2$.
\item We take the path $C_{v}^{-}$ along the cut at $\phi_{v}=\alpha_{v}-1/2$
from $r_{v}=0$ to $r_{v}=R$.
\item We go around the outer boundary $\partial_{R}D_{v}$ of the disk at
$r_{v}=R$ from $\phi_{v}=\alpha_{v}-1/2$ to $\phi_{v}=\alpha_{v}+1/2$.
\item We take the path $C_{v}^{+}$ along the cut at $\phi_{v}=\alpha_{v}+1/2$
from $r_{v}=R$ to $r_{v}=0$.
\item Finally, we go around the inner boundary $\partial_{0}D_{v}$ of the
disk at $r_{v}=0$ from $\phi_{v}=\alpha_{v}+1/2$ to $\phi_{v}=\alpha_{v}-1/2$,
back to our starting point.
\end{itemize}
Let us evaluate each term individually. On $C_{v}^{-}$ and $C_{v}^{+}$
we have\footnote{Note that the angle $\phi_{v}\left(x\right)$ in the term $\e^{\M_{v}\phi_{v}\left(x\right)}$
in (\ref{eq:undressed-v}) refers to the \emph{difference in angles}
between the starting point and the final point, therefore it vanishes
in this case since the path along the cut is purely radial.} from (\ref{eq:undressed-v}):
\[
\pexp\left(\int_{C_{v}^{-}}\A\right)=\pexp\int_{v}^{v_{0}}\A=h_{v}^{-1}\left(v\right)h_{v}\left(v_{0}\right),
\]
\[
\pexp\left(\int_{C_{v}^{+}}\A\right)=\pexp\int_{v_{0}}^{v}\A=h_{v}^{-1}\left(v_{0}\right)h_{v}\left(v\right).
\]

On the inner boundary we have, again using (\ref{eq:undressed-v}),
\[
\pexp\int_{\partial_{0}D_{v}}\A=\pexp\int_{v\left(\phi_{v}=\alpha_{v}+1/2\right)}^{v\left(\phi_{v}=\alpha_{v}-1/2\right)}\A=h_{v}^{-1}\left(v\right)\e^{-\M_{v}}h_{v}\left(v\right),
\]
since $h_{v}$ is periodic. The minus sign comes from the fact that
we are going from a larger angle to a smaller angle. Finally, on the
outer boundary we have, splitting into arcs and then using (\ref{eq:exp-cv})
and $\left(vc\right)=\left(cv\right)^{-1}$,
\[
\pexp\int_{\partial_{R}D_{v}}\A=\prod_{c\in v}\left(\pexp\int_{\left(vc\right)}\A\right)=\prod_{c\in v}\Ht_{vc}.
\]
In conclusion, the curvature constraint on the disks is
\[
F_{v}=h_{v}^{-1}\left(v\right)h_{v}\left(v_{0}\right)\left(\prod_{c\in v}\Ht_{vc}\right)h_{v}^{-1}\left(v_{0}\right)\e^{-\M_{v}}h_{v}\left(v\right)=1.
\]
In fact, we can multiply both sides by $h_{v}^{-1}\left(v_{0}\right)h_{v}\left(v\right)$
from the left and $h_{v}^{-1}\left(v\right)h_{v}\left(v_{0}\right)$
and obtain, after redefining $F_{v}$,
\[
F_{v}\equiv\left(\prod_{c\in v}\Ht_{vc}\right)h_{v}^{-1}\left(v_{0}\right)\e^{-\M_{v}}h_{v}\left(v_{0}\right)=1.
\]
This may be written more suggestively as
\[
\prod_{c\in v}\Ht_{vc}=h_{v}^{-1}\left(v_{0}\right)\e^{\M_{v}}h_{v}\left(v_{0}\right).
\]
Let us now show that this constraint is satisfied identically in our
construction. From (\ref{eq:Ht-vc}) we have
\[
\Ht_{vc}\equiv\mathring{h}_{v}^{-1}\left(\sigma_{vc}\right)\mathring{h}_{v}\left(\tau_{vc}\right),
\]
and using the definition $\mathring{h}_{v}\equiv\e^{\M_{v}\phi_{v}}h_{v}$
from (\ref{eq:u-v-def}) we get
\[
\Ht_{vc}=h_{v}^{-1}\left(\sigma_{vc}\right)\e^{\M_{v}\left(\phi_{v}\left(\tau_{vc}\right)-\phi_{v}\left(\sigma_{vc}\right)\right)}h_{v}\left(\tau_{vc}\right).
\]
Now, consider the product
\[
\prod_{c\in v}\Ht_{vc}=\prod_{i=1}^{N_{v}}h_{v}^{-1}\left(\sigma_{vc_{i}}\right)\e^{\M_{v}\left(\phi_{v}\left(\tau_{vc_{i}}\right)-\phi_{v}\left(\sigma_{vc_{i}}\right)\right)}h_{v}\left(\tau_{vc_{i}}\right).
\]
This is a telescoping product; the term $h_{v}\left(\tau_{vc_{i}}\right)$
always cancels the term $h_{v}^{-1}\left(\sigma_{vc_{i+1}}\right)$
in the next factor in the product. After the cancellations take place,
we are left only with $h_{v}^{-1}\left(\sigma_{vc_{1}}\right)$, the
product of exponents 
\[
\prod_{i=1}^{N_{v}}\e^{\M_{v}\left(\phi_{v}\left(\tau_{vc_{i}}\right)-\phi_{v}\left(\sigma_{vc_{i}}\right)\right)}=\e^{\M_{v}},
\]
where we used the fact that the angles sum to 1, and $h_{v}\left(\tau_{vc_{N_{v}}}\right)=h_{v}\left(\sigma_{vc_{1}}\right)$.
In conclusion:
\[
\prod_{c\in v}\Ht_{vc}=h_{v}^{-1}\left(\sigma_{vc_{1}}\right)\e^{\M_{v}}h_{v}\left(\sigma_{vc_{1}}\right).
\]
If we then choose, without loss of generality, the point $v_{0}$
(which defines the cut $C_{v}$) to be at $\sigma_{vc_{1}}$ (where
$c_{1}$ is an arbitrarily chosen cell), we get
\[
\prod_{c\in v}\Ht_{vc}=h_{v}^{-1}\left(v_{0}\right)\e^{\M_{v}}h_{v}\left(v_{0}\right),
\]
and we see that the constraint is indeed identically satisfied.

\subsection{Derivation of the Constraints on the Faces}

We have seen that the Gauss constraints, as we have defined them,
involve the fluxes on the edges and arcs. Since these fluxes are not
part of the phase space for $\lambda=1$, these constraints cannot
be imposed in that case. Similarly, the curvature constraints involve
the holonomies on the edges and arcs and therefore will not work for
the case $\lambda=0$. This is a result of formulating both constraints
on the cells and disks, which then requires us to use the holonomies
and fluxes on the edges and arcs which are on their boundaries.

Alternatively, instead of demanding that the torsion and curvature
vanish on the cells and disks, we may demand that they vanish on the
faces $v^{*}$ created by the spin network links. Since the (closures
of the) faces cover the entire spatial manifold $\Sigma$, this is
entirely equivalent.

This alternative form is obtained by deforming (or expanding) the
disks such that they coincide with the faces. The inner boundary $\partial_{0}D_{v}\to\partial_{0}v^{*}$
is still the vertex $v$. The outer boundary $\partial_{R}D_{v}\to\partial_{R}v^{*}$
now consists of links $\left(c_{i}c_{i+1}\right)^{*}$, where $i\in\left\{ 1,\ldots,N_{v}\right\} $
and $c_{N_{v}+1}\equiv c_{1}$. The point $v_{0}$ on the outer boundary
can now be identified, without loss of generality, with the node $c_{1}^{*}$.
Thus, the cut $C_{v}\to C_{v^{*}}$ now extends from $v$ to $c_{1}^{*}$.

Since the spatial manifold $\Sigma$ is now composed solely of the
union of the closures of the faces, and not cells and disks, we only
need one type of Gauss constraint and one type of curvature constraint.
Let us derive them now.

\subsubsection{The Gauss Constraint on the Faces}

The face Gauss constraint $\G_{v^{*}}$ will impose the torsionlessness
condition $\T\equiv\d_{\A}\ee=0$ inside the faces:
\[
0=\G_{v^{*}}\equiv\int_{v^{*}}\mathring{h}_{v}\left(\d_{\A}\ee\right)\mathring{h}_{v}^{-1}=\int_{v^{*}}\d\left(\mathring{h}_{v}\ee\mathring{h}_{v}^{-1}\right)=\int_{\partial v^{*}}\mathring{h}_{v}\ee\mathring{h}_{v}^{-1}=\int_{\partial v^{*}}\d\mathring{\x}_{v}.
\]
The boundary $\partial v^{*}$ is composed of the inner boundary $\partial_{0}v^{*}$,
the outer boundary $\partial_{R}v^{*}$, and the cut $C_{v^{*}}$:
\begin{equation}
\G_{v^{*}}=\int_{\partial_{R}v^{*}}\d\mathring{\x}_{v}-\int_{\partial_{0}v^{*}}\d\mathring{\x}_{v}-\int_{C_{v^{*}}}\d\mathring{\x}_{v},\label{eq:G_f_v-decomp}
\end{equation}
where the minus signs represent the relative orientations of each
piece. On the inner boundary $\partial_{0}v^{*}$, we use the fact
that $\x_{v}$ takes the constant value $\x_{v}\left(v\right)$ to
obtain as for $\partial_{0}D_{v}$ above:
\[
\int_{\partial_{0}v^{*}}\d\mathring{\x}_{v}=\SS_{v}+\e^{\M_{v}\left(\alpha_{v}-\hf\right)}\left(\e^{\M_{v}}\x_{v}\left(v\right)\e^{-\M_{v}}-\x_{v}\left(v\right)\right)\e^{-\M_{v}\left(\alpha_{v}-\hf\right)}.
\]
On the cut $C_{v}$, we have as before
\[
\int_{C_{v}}\d\mathring{\x}_{v}=\e^{\M_{v}\left(\alpha_{v}-\hf\right)}\left(\e^{\M_{v}}\left(\x_{v}\left(v_{0}\right)-\x_{v}\left(v\right)\right)\e^{-\M_{v}}-\left(\x_{v}\left(v_{0}\right)-\x_{v}\left(v\right)\right)\right)\e^{-\M_{v}\left(\alpha_{v}-\hf\right)}.
\]
The outer boundary $\partial_{R}v^{*}$ splits into links:
\begin{equation}
\int_{\partial_{R}v^{*}}\d\mathring{\x}_{v}=\sum_{i=1}^{N_{v}}\int_{c_{i}^{*}}^{c_{i+1}^{*}}\d\mathring{\x}_{v}=\sum_{i=1}^{N_{v}}\left(\mathring{\x}_{v}\left(c_{i+1}^{*}\right)-\mathring{\x}_{v}\left(c_{i}^{*}\right)\right).\label{eq:outer-1}
\end{equation}
Now, (\ref{eq:continuity-cv}) can be inverted\footnote{Note that (\ref{eq:continuity-cv}) is only valid on the arc $\left(vc\right)$,
which is the boundary between $c$ and $D_{v}$. However, since we
have expanded the disks, the arcs now coincide with the links, with
every arc $\left(vc\right)$ intersecting the two links connected
to the node $c^{*}$. Thus the equation is still valid at $c^{*}$
itself.} to get
\[
\mathring{\x}_{v}=h_{vc}\x_{c}h_{cv}+\x_{v}^{c}.
\]
Plugging into (\ref{eq:outer-1}), we get
\[
\int_{\partial_{R}v^{*}}\d\mathring{\x}_{v}=\sum_{i=1}^{N_{v}}\left(h_{vc_{i+1}}\x_{c_{i+1}}\left(c_{i+1}^{*}\right)h_{c_{i+1}v}-h_{vc_{i}}\x_{c_{i}}\left(c_{i}^{*}\right)h_{c_{i}v}+\x_{v}^{c_{i+1}}-\x_{v}^{c_{i}}\right).
\]
In fact, we can get rid of the first two terms, since the sum is telescoping:
each term of the from $h_{vc_{i}}\x_{c_{i}}\left(c_{i}^{*}\right)h_{c_{i}v}$
for $i=j$ is canceled\footnote{Of course, $\x_{v}^{c_{i+1}}$ and $\x_{v}^{c_{i}}$ also cancel each
other, but we choose to leave them.} by a term of the form $h_{vc_{i+1}}\x_{c_{i+1}}\left(c_{i+1}^{*}\right)h_{c_{i+1}v}$
for $i=j-1$. Thus we get
\begin{equation}
\int_{\partial_{R}v^{*}}\d\mathring{\x}_{v}=\sum_{i=1}^{N_{v}}\left(\x_{v}^{c_{i+1}}-\x_{v}^{c_{i}}\right).\label{eq:outer-2}
\end{equation}

Next, we note that from (\ref{eq:continuity-ccp}) we have
\[
h_{cc'}=h_{c}h_{c'}^{-1}\sp\x_{c}^{c'}=\x_{c}-h_{cc'}\x_{c'}h_{c'c},
\]
and if we plug in (\ref{eq:continuity-cv}) for $h_{c}$, $h_{c'}$,
$\x_{c}$ and $\x_{c'}$ we get 
\begin{equation}
h_{cc'}=h_{cv}h_{vc'},\label{eq:hccp-decomp}
\end{equation}
\begin{equation}
\x_{c}^{c'}=h_{cv}(\mathring{\x}_{v}-\x_{v}^{c})h_{vc}-h_{cc'}h_{c'v}(\mathring{\x}_{v}-\x_{v}^{c'})h_{vc'}h_{c'c}.\label{eq:xccp-decomp}
\end{equation}
From (\ref{eq:hccp-decomp}) we see that $h_{cc'}h_{c'v}=h_{cv}$.
Plugging this into (\ref{eq:xccp-decomp}), we get the simplified
expression
\begin{equation}
\x_{c}^{c'}=h_{cv}\left(\x_{v}^{c'}-\x_{v}^{c}\right)h_{vc}.\label{eq:xccp-decomp-simp}
\end{equation}
Therefore, we may rewrite (\ref{eq:outer-2}) as:
\begin{equation}
\int_{\partial_{R}v^{*}}\d\mathring{\x}_{v}=\sum_{i=1}^{N_{v}}h_{vc_{i}}\x_{c_{i}}^{c_{i+1}}h_{c_{i}v}.\label{eq:outer-3}
\end{equation}
Finally, we recall from (\ref{eq:fluxes-links}) the definition of
the fluxes on the links:
\[
\X_{c}^{c'}\equiv h_{c}^{-1}\left(\sigma_{cc'}\right)\x_{c}^{c'}h_{c}\left(\sigma_{cc'}\right)=h_{c}^{-1}\left(v_{0}\right)\x_{c}^{c'}h_{c}\left(v_{0}\right).
\]
In the second equality we use the fact that, since we have deformed
the disks, the source point $\sigma_{cc'}$ of the edge $\left(cc'\right)$
lies on the spin network itself, and we can further deform the edge
such that $\sigma_{cc'}=v_{0}$. Plugging into (\ref{eq:outer-3}),
we obtain
\[
\int_{\partial_{R}v^{*}}\d\mathring{\x}_{v}=\sum_{i=1}^{N_{v}}h_{vc_{i}}h_{c_{i}}\left(v_{0}\right)\X_{c_{i}}^{c_{i+1}}h_{c_{i}}^{-1}\left(v_{0}\right)h_{c_{i}v}.
\]
Finally, from (\ref{eq:continuity-cv}) we have $h_{vc}h_{c}=\mathring{h}_{v}$,
and we get
\begin{equation}
\int_{\partial_{R}v^{*}}\d\mathring{\x}_{v}=\mathring{h}_{v}\left(v_{0}\right)\left(\sum_{i=1}^{N_{v}}\X_{c_{i}}^{c_{i+1}}\right)\mathring{h}_{v}^{-1}\left(v_{0}\right).\label{eq:outer-4}
\end{equation}
Adding up the integrals in (\ref{eq:G_f_v-decomp}), we obtain the
Gauss constraint on the faces:
\begin{align}
\G_{v^{*}} & =\mathring{h}_{v}\left(v_{0}\right)\left(\sum_{i=1}^{N_{v}}\X_{c_{i}}^{c_{i+1}}\right)\mathring{h}_{v}^{-1}\left(v_{0}\right)-\SS_{v}+\label{eq:Gauss-faces}\\
 & \qquad-\e^{\M_{v}\left(\alpha_{v}-\hf\right)}\left(\e^{\M_{v}}\x_{v}\left(v_{0}\right)\e^{-\M_{v}}-\x_{v}\left(v_{0}\right)\right)\e^{-\M_{v}\left(\alpha_{v}-\hf\right)}=0.\nonumber 
\end{align}
Just like the Gauss constraint on the disks, this can be simplified
by noting that the constraint comes dotted with an element $\be_{v^{*}}$
of the Cartan subalgebra, which commutes with $\M_{v}$:
\[
\be_{v^{*}}\cdot\G_{v^{*}}=\be_{v^{*}}\cdot\left(h_{v}\left(v_{0}\right)\left(\sum_{i=1}^{N_{v}}\X_{c_{i}}^{c_{i+1}}\right)h_{v}^{-1}\left(v_{0}\right)-\SS_{v}\right)=0,
\]
where we used the fact that $\mathring{h}_{v}=\e^{\M_{v}\phi_{v}}h_{v}$
and the $\e^{\M_{v}\phi_{v}}$ part commutes with $\be_{v^{*}}$.
Thus, Gauss constraint on the faces may be rewritten in a simplified
way:
\[
\G_{v^{*}}\equiv\sum_{i=1}^{N_{v}}\X_{c_{i}}^{c_{i+1}}-h_{v}^{-1}\left(v_{0}\right)\SS_{v}h_{v}\left(v_{0}\right)=0.
\]
Let us now show that this constraint is satisfied identically. We
have from the definition of $\mathring{\x}_{v}$:
\begin{align*}
\int_{\partial_{R}v^{*}}\d\mathring{\x}_{v} & =\sum_{i=1}^{N_{v}}\int_{c_{i}^{*}}^{c_{i+1}^{*}}\d\mathring{\x}_{v}=\sum_{i=1}^{N_{v}}\left(\mathring{\x}_{v}\left(c_{i+1}^{*}\right)-\mathring{\x}_{v}\left(c_{i}^{*}\right)\right)\\
 & =\sum_{i=1}^{N_{v}}\multibrl{\e^{\M_{v}\phi_{v}\left(c_{i+1}^{*}\right)}\x_{v}\left(c_{i+1}^{*}\right)\e^{-\M_{v}\phi_{v}\left(c_{i+1}^{*}\right)}-\e^{\M_{v}\phi_{v}\left(c_{i}^{*}\right)}\x_{v}\left(c_{i}^{*}\right)\e^{-\M_{v}\phi_{v}\left(c_{i}^{*}\right)}+}\\
 & \qquad\multibrr{+\SS_{v}\left(\phi_{v}\left(c_{i+1}^{*}\right)-\phi_{v}\left(c_{i}^{*}\right)\right)\vphantom{\e^{\M_{v}\phi_{v}\left(c_{i+1}^{*}\right)}}}.
\end{align*}
The sum is telescoping, and every term cancels the previous one. However,
in the term with $i=N_{v}$, we have
\[
\phi_{v}\left(c_{N_{v}+1}^{*}\right)=\phi_{v}\left(c_{1}^{*}\right)+1,
\]
since $\phi_{v}$, unlike $\x_{v}$, is not periodic. Therefore, the
first and last terms do not cancel each other. If we furthermore choose
$v_{0}\equiv c_{1}^{*}$, we get
\[
\int_{\partial_{R}v^{*}}\d\mathring{\x}_{v}=\SS_{v}+\e^{\M_{v}\phi_{v}\left(v_{0}\right)}\left(\e^{\M_{v}}\x_{v}\left(v_{0}\right)\e^{-\M_{v}}-\x_{v}\left(v_{0}\right)\right)\e^{-\M_{v}\phi_{v}\left(v_{0}\right)}.
\]
Then, using (\ref{eq:outer-4}) we immediately obtain (\ref{eq:Gauss-faces}),
as desired.

\subsubsection{The Curvature Constraint on the Faces}

The face curvature constraint $F_{v^{*}}$ will impose that $\F\equiv\d\A+\hf\left[\A,\A\right]=0$
inside the faces. As before, an equivalent condition is that the holonomy
around the face evaluates to the identity:
\[
1=F_{v^{*}}\equiv\pexp\int_{\partial v^{*}}\A=\pexp\left(\int_{C_{v}^{-}}\A\right)\pexp\left(\int_{\partial_{R}v^{*}}\A\right)\pexp\left(\int_{C_{v}^{+}}\A\right)\pexp\left(\int_{\partial_{0}v^{*}}\A\right).
\]
On $C_{v}^{-}$ and $C_{v}^{+}$ we have as before
\[
\pexp\left(\int_{C_{v}^{-}}\A\right)=\pexp\int_{v}^{v_{0}}\A=h_{v}^{-1}\left(v\right)h_{v}\left(v_{0}\right),
\]
\[
\pexp\left(\int_{C_{v}^{+}}\A\right)=\pexp\int_{v_{0}}^{v}\A=h_{v}^{-1}\left(v_{0}\right)h_{v}\left(v\right).
\]

On the inner boundary we have
\[
\pexp\int_{\partial_{0}v^{*}}\A=\pexp\int_{v\left(\phi_{v}=\alpha_{v}+1/2\right)}^{v\left(\phi_{v}=\alpha_{v}-1/2\right)}\A=h_{v}^{-1}\left(v\right)\e^{-\M_{v}}h_{v}\left(v\right).
\]
Finally, we decompose the outer boundary (which is now a loop on the
spin network) into links:
\[
\pexp\int_{\partial_{R}v^{*}}\A=\prod_{i=1}^{N_{v}}\left(\pexp\int_{c_{i}^{*}}^{c_{i+1}^{*}}\A\right).
\]
From (\ref{eq:exp-ccp-h}) we know that
\[
\pexp\int_{c^{*}}^{c^{\prime*}}\A=h_{c}^{-1}\left(c^{*}\right)h_{cc'}h_{c'}\left(c^{\prime*}\right),
\]
and therefore
\[
\pexp\int_{\partial_{R}v^{*}}\A=\prod_{i=1}^{N_{v}}h_{c_{i}}^{-1}\left(c_{i}^{*}\right)h_{c_{i}c_{i+1}}h_{c_{i+1}}\left(c_{i+1}^{*}\right)=h_{c_{1}}^{-1}\left(v_{0}\right)\left(\prod_{i=1}^{N_{v}}h_{c_{i}c_{i+1}}\right)h_{c_{1}}\left(v_{0}\right),
\]
where we used the choice $v_{0}\equiv c_{1}^{*}$ and the fact that
the product is telescoping, that is, each term $h_{c_{i+1}}\left(c_{i+1}^{*}\right)$
cancels the term $h_{c_{i+1}}^{-1}\left(c_{i+1}^{*}\right)$ which
follows it, except the first and last terms, which have nothing to
cancel with.

Joining the integrals, we get
\[
h_{v}^{-1}\left(v\right)h_{v}\left(v_{0}\right)h_{c_{1}}^{-1}\left(v_{0}\right)\left(\prod_{i=1}^{N_{v}}h_{c_{i}c_{i+1}}\right)h_{c_{1}}\left(v_{0}\right)h_{v}^{-1}\left(v_{0}\right)\e^{-\M_{v}}h_{v}\left(v\right)=1.
\]
From (\ref{eq:continuity-cv}) we find that
\[
h_{c_{1}}\left(v_{0}\right)h_{v}^{-1}\left(v_{0}\right)=h_{c_{1}v},
\]
and thus
\[
h_{v}^{-1}\left(v\right)h_{vc_{1}}\left(\prod_{i=1}^{N_{v}}h_{c_{i}c_{i+1}}\right)h_{c_{1}v}\e^{-\M_{v}}h_{v}\left(v\right)=1.
\]
For the last step, since we have the identity on the right-hand side,
we may cycle the group elements and rewrite the constraint as follows:
\[
F_{v^{*}}\equiv\left(\prod_{i=1}^{N_{v}}h_{c_{i}c_{i+1}}\right)h_{c_{1}v}\e^{-\M_{v}}h_{vc_{1}}=1.
\]
Switching to the notation of (\ref{eq:hol-link}) and (\ref{eq:hol-seg}),
we rewrite this as
\[
F_{v^{*}}\equiv\left(\prod_{i=1}^{N_{v}}H_{c_{i}c_{i+1}}\right)H_{c_{1}v}\e^{-\M_{v}}H_{vc_{1}}=1.
\]
An even nicer form of this constraint is
\begin{equation}
\prod_{i=1}^{N_{v}}H_{c_{i}c_{i+1}}=H_{c_{1}v}\e^{\M_{v}}H_{vc_{1}}.\label{eq:curv-con-faces}
\end{equation}
In other words, the loop of holonomies on the left-hand side would
be the identity if there is no curvature, that is, $\M_{v}=0$.

\begin{figure}[!h]
\begin{centering}
\begin{tikzpicture}[scale=1.5]
	\node [style=Vertex] (vbottom) at (0, -2) {};
	\node [style=Vertex] (vtop) at (0, 2) {};
	\node [style=Vertex] (vleft) at (-3, -0) {};
	\node [style=Vertex] (vright) at (3, -0) {};
	\node [style=Node] (cleft) at (-1.25, -0) {};
	\node [style=Node] (cright) at (1.25, -0) {};
	\node [style=none] at (0, 2.25) {$v$};
	\node [style=none] at (-1.4, -0.25) {$c^{*}$};
	\node [style=none] at (1.4, -0.25) {$c^{\prime*}$};
	\node [style=none] at (0, -0.2) {$h_{cc'}$};
	\node [style=none] at (-1, 0.9) {$h_{cv}$};
	\node [style=none] at (1, 0.9) {$h_{vc'}$};
	\node [style=none, scale=0.7] at (0, 1.6) {$\mathrm{e}^{\mathbf{M}_v\phi_{v}^{cc'}}$};
	\draw [style=Edge] (vtop) to (vleft);
	\draw [style=Edge] (vleft) to (vbottom);
	\draw [style=Edge] (vbottom) to (vright);
	\draw [style=Edge] (vright) to (vtop);
	\draw [style=Edge] (vtop) to (vbottom);
	\draw [style=SegmentArrow] (cleft) to (cright);
	\centerarc [style=SegmentArrow] (0,2) (230:310:0.25);
	\draw [style=SegmentArrow] (cleft) to ($(0,2)+({0.25*cos(230)},{0.25*sin(230)})$);
	\draw [style=SegmentArrow] ($(0,2)+({0.25*cos(310)},{0.25*sin(310)})$) to (cright);
\end{tikzpicture}
\par\end{centering}
\caption{\label{fig:Curvature}The holonomy from $c^{*}$ to $c^{\prime*}$
going either directly or through the vertex $v$.}
\end{figure}
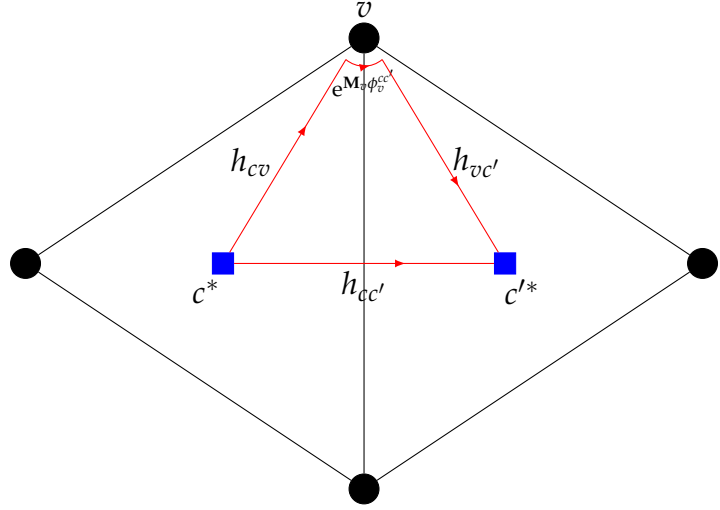

To show that this constraint is satisfied identically, we use (\ref{eq:exp-xy-phi})
with $x=c^{*}$ and $y=c^{\prime*}$:
\[
\pexp\int_{c^{*}}^{c^{\prime*}}\A=h_{v}^{-1}\left(c^{*}\right)\e^{\M_{v}\left(\phi_{v}\left(c^{\prime*}\right)-\phi_{v}\left(c^{*}\right)\right)}h_{v}\left(c^{\prime*}\right).
\]
Comparing with (\ref{eq:exp-ccp-h}), we see that
\[
h_{c}^{-1}\left(c^{*}\right)h_{cc'}h_{c'}\left(c^{\prime*}\right)=h_{v}^{-1}\left(c^{*}\right)\e^{\M_{v}\left(\phi_{v}\left(c^{\prime*}\right)-\phi_{v}\left(c^{*}\right)\right)}h_{v}\left(c^{\prime*}\right),
\]
and therefore
\[
h_{cc'}=h_{cv}\e^{\M_{v}\left(\phi_{v}\left(c^{\prime*}\right)-\phi_{v}\left(c^{*}\right)\right)}h_{vc'}.
\]
This is illustrated in Figure \ref{fig:Curvature}. We now use this
to rewrite the left-hand side of (\ref{eq:curv-con-faces}) as follows:
\[
\prod_{i=1}^{N_{v}}h_{c_{i}c_{i+1}}=\prod_{i=1}^{N_{v}}h_{c_{i}v}\e^{\M_{v}\left(\phi_{v}\left(c_{i+1}^{*}\right)-\phi_{v}\left(c_{i}^{*}\right)\right)}h_{vc_{i+1}}.
\]
Again, we have a telescoping product, and after canceling terms we
are left with
\[
\prod_{i=1}^{N_{v}}h_{c_{i}c_{i+1}}=h_{c_{1}v}\e^{\M_{v}}h_{vc_{1}},
\]
which is exactly (\ref{eq:curv-con-faces}) after using (\ref{eq:hol-link})
and (\ref{eq:hol-seg}).

\subsection{\label{subsec:Summary-and-Interpretation}Summary and Interpretation}

In conclusion, we have obtained\footnote{One might wonder about the appearance of $h_{v}\left(v_{0}\right)$
in (\ref{eq:summary_G_f_v}) and (\ref{eq:summary_F_v}), since the
true phase space variable is $H_{v}$, defined implicitly in (\ref{eq:DeltaH_v})
as a function of $h_{v}\left(v\right)$ and $h_{v}\left(v_{0}\right)$.
It is possible that there is an expression for these two constraints
in terms of $H_{v}$ instead of $h_{v}\left(v_{0}\right)$, but since
we only have an \textbf{implicit }definition for $H_{v}$ in terms
of its variation $\De H_{v}$, it is unclear how to obtain it. For
now, we simply assume that both $H_{v}$ and $h_{v}\left(v_{0}\right)$
are phase space variables. See also footnote \ref{fn:H_v-foot}.} Gauss constraints $\G_{c},\G_{v},\G_{v^{*}}$ and curvature constraints
$F_{c},F_{v},F_{v^{*}}$ for each cell $c$, disk $D_{v}$ and face
$v^{*}$:
\begin{equation}
\G_{c}\equiv\sum_{i=1}^{N_{c}}\left(\XXt_{c}^{c_{i}}+\XXt_{c}^{v_{i}}\right)=0,\label{eq:summary_G_c}
\end{equation}
\begin{equation}
\G_{v}\equiv\sum_{i=1}^{N_{v}}\XXt_{v}^{c_{i}}-\SS_{v}=0,\label{eq:summary_G_v}
\end{equation}
\begin{equation}
\G_{v^{*}}\equiv\sum_{i=1}^{N_{v}}\X_{c_{i}}^{c_{i+1}}-h_{v}^{-1}\left(v_{0}\right)\SS_{v}h_{v}\left(v_{0}\right)=0,\label{eq:summary_G_f_v}
\end{equation}
\begin{equation}
F_{c}\equiv\prod_{i=1}^{N_{c}}\Ht_{cc_{i}}\Ht_{cv_{i}}=1,\label{eq:summary_F_c}
\end{equation}
\begin{equation}
F_{v}\equiv\left(\prod_{i=1}^{N_{v}}\Ht_{vc_{i}}\right)h_{v}^{-1}\left(v_{0}\right)\e^{-\M_{v}}h_{v}\left(v_{0}\right)=1,\label{eq:summary_F_v}
\end{equation}
\begin{equation}
F_{v^{*}}\equiv\left(\prod_{i=1}^{N_{v}}H_{c_{i}c_{i+1}}\right)H_{c_{1}v}\e^{-\M_{v}}H_{vc_{1}}=1.\label{eq:summary_F_f_v}
\end{equation}
The Gauss constraint on the cell $c$ can also be written as
\begin{equation}
\sum_{c'\ni c}\XXt_{c}^{c'}=-\sum_{v\ni c}\XXt_{c}^{v}.\label{eq:G_c-split}
\end{equation}
It tells us that the sum of fluxes along the edges and arcs surrounding
$c$ is zero, as expected given that the interior of $c$ is flat.
Alternatively, we may say that the sum of fluxes along the edges is
prevented from summing to zero by the presence of the fluxes on the
arcs.

The Gauss constraint on the punctured disk $D_{v}$ can also be written
as
\begin{equation}
\sum_{c\in v}\XXt_{v}^{c}=\SS_{v}.\label{eq:G_v-split}
\end{equation}
It tells us that the sum of fluxes on the arcs of the disk is prevented
from summing to zero due to the torsion at the vertex $v$, as encoded
in the parameter $\SS_{v}$. Note that if $\SS_{v}=0$, that is, there
is no torsion at $v$, then the constraint becomes simply $\sum_{c\in v}\XXt_{v}^{c}=0.$

Importantly, notice that the sum $\sum_{v\ni c}\XXt_{c}^{v}$ on the
right-hand side of (\ref{eq:G_c-split}) is over all the fluxes on
the arcs surrounding a particular cell $c$, while the sum $\sum_{c\in v}\XXt_{v}^{c}$
on the left-hand side of (\ref{eq:G_v-split}) is over all the fluxes
on the arcs surrounding a particular disk $D_{v}$. While the sums
look alike at first sight, they are completely different and one cannot
be exchanged for the other.

The Gauss constraint on the face $v^{*}$ can also be written as
\[
\sum_{i=1}^{N_{v}}\X_{c_{i}}^{c_{i+1}}=h_{v}^{-1}\left(v_{0}\right)\SS_{v}h_{v}\left(v_{0}\right).
\]
It tells us that the sum of fluxes on the link forming the boundary
of the face is prevented from summing to zero due to the torsion at
the vertex $v$, as encoded in the parameter $\SS_{v}$.

The curvature constraint on the cell $c$ is
\[
\prod_{i=1}^{N_{c}}\Ht_{cc_{i}}\Ht_{cv_{i}}=1.
\]
It is analogous to the cell Gauss constraint, and imposes that the
product of holonomies along the boundary of the cell is the identity.

The curvature constraint on the punctured disk $D_{v}$ can also be
written as
\[
\prod_{c\in v}\Ht_{vc}=h_{v}^{-1}\left(v_{0}\right)\e^{\M_{v}}h_{v}\left(v_{0}\right).
\]
On the left-hand side, we have a loop of holonomies around the vertex
$v$. If $\M_{v}=0$, that is, there is no curvature at $v$, then
the right-hand side becomes the identity, as we would expect. Otherwise,
it is a quantity which depends on the curvature. The curvature constraint
on the disks is thus analogous to the Gauss constraint on the disks,
with torsion replaced by curvature.

Finally, the curvature constraint on the face $v^{*}$ can also be
written as
\[
\prod_{i=1}^{N_{v}}H_{c_{i}c_{i+1}}=H_{c_{1}v}\e^{\M_{v}}H_{vc_{1}}.
\]
It has the same meaning as the one on the disks, except that the loop
of holonomies around the vertex $v$ is now composed of links instead
of arcs.

\section{\label{sec:The-Constraints-as}The Constraints as Generators of Symmetries}

Now that we have obtained the Gauss and curvature constraints on the
cells, disks, and faces, we would like to derive the symmetries that
they generate. Recall again\footnote{We first encountered Hamiltonian vector fields in Section \ref{subsec:Classical-Spin-Networks}.}
that, given a symplectic form $\Omega$, the \emph{Hamiltonian vector
field} of $f$ is the vector field $\H_{f}$ satisfying
\[
I_{\H_{f}}\Omega=-\delta f.
\]
This can be interpreted as the variational interior product $I_{\H_{f}}$
producing a transformation on the phase space represented by $\Omega$,
which is \emph{generated }by the constraint $f$. We will now show
that the Gauss constraint generates rotations, and the curvature constraint
generates translations. In other words, we will find transformations,
given by the Lie derivative $\LLL_{\a}\equiv I_{\a}\delta+\delta I_{\a}$
for some Hamiltonian vector field corresponding to a transformation
with some parameter $\a$ (using a slight abuse of notation), that
are of the schematic form
\[
I_{\be}\Omega\propto-\be\cdot\delta\G\sp I_{\z}\Omega\propto-\z\cdot\De F_{c},
\]
where $\be$ is a rotation parameter and $\z$ is a translation parameter.
As before, the reader who is not interested in the calculation itself
may skip directly to the results in Section \ref{subsec:Conclusions-Sym}.

\subsection{The Discrete Symplectic Form}

The discrete symplectic potential we have found is
\begin{align*}
\Theta & =\sum_{\left(cc'\right)}\left(\left(1-\lambda\right)\XXt_{c}^{c'}\cdot\De H_{c}^{c'}-\lambda\X_{c}^{c'}\cdot\De\Ht_{c}^{c'}\right)+\\
 & \qquad+\sum_{\left(vc\right)}\left(\left(1-\lambda\right)\XXt_{v}^{c}\cdot\De H_{v}^{c}-\lambda\X_{v}^{c}\cdot\De\Ht_{v}^{c}\right)+\\
 & \qquad+\sum_{v}\left(\X_{v}\cdot\delta\M_{v}-\left(\SS_{v}+\left[\M_{v},\X_{v}\right]\right)\cdot\De H_{v}\right).
\end{align*}
In the second line, we can use (\ref{eq:flux-arc-cv}), that is, $\XXt_{c}^{v}=-H_{cv}\XXt_{v}^{c}H_{vc}$,
to write
\[
\XXt_{v}^{c}\cdot\De H_{v}^{c}=\left(-H_{vc}\XXt_{c}^{v}H_{cv}\right)\cdot\left(\delta H_{vc}H_{cv}\right)=\XXt_{c}^{v}\cdot\De H_{c}^{v}.
\]
Thus, the labels $c$ and $v$ may be freely exchanged. Using the
identity $\delta\De H=\hf\left[\De H,\De H\right]$, we find that
the corresponding symplectic form $\Omega\equiv\delta\Theta$ is
\begin{align*}
\Omega & =\sum_{\left(cc'\right)}\left(1-\lambda\right)\left(\delta\XXt_{c}^{c'}\cdot\De H_{c}^{c'}+\hf\XXt_{c}^{c'}\cdot\left[\De H_{c}^{c'},\De H_{c}^{c'}\right]\right)+\\
 & \qquad-\sum_{\left(cc'\right)}\lambda\left(\delta\X_{c}^{c'}\cdot\De\Ht_{c}^{c'}+\hf\X_{c}^{c'}\cdot\left[\De\Ht_{c}^{c'},\De\Ht_{c}^{c'}\right]\right)+\\
 & \qquad+\sum_{\left(vc\right)}\left(1-\lambda\right)\left(\delta\XXt_{v}^{c}\cdot\De H_{v}^{c}+\hf\XXt_{v}^{c}\cdot\left[\De H_{v}^{c},\De H_{v}^{c}\right]\right)+\\
 & \qquad-\sum_{\left(vc\right)}\lambda\left(\delta\X_{v}^{c}\cdot\De\Ht_{v}^{c}+\hf\X_{v}^{c}\cdot\left[\De\Ht_{v}^{c},\De\Ht_{v}^{c}\right]\right)+\\
 & \qquad+\sum_{v}\left(\delta\X_{v}\cdot\delta\M_{v}-\left(\delta\SS_{v}+\left[\delta\M_{v},\X_{v}\right]+\left[\M_{v},\delta\X_{v}\right]\right)\cdot\De H_{v}\right)+\\
 & \qquad-\sum_{v}\hf\left(\SS_{v}+\left[\M_{v},\X_{v}\right]\right)\cdot\left[\De H_{v},\De H_{v}\right].
\end{align*}
We now look for transformations with parameters $g_{c}\equiv\e^{\be_{c}},g_{v}\equiv\e^{\be_{v}},$
$\z_{c}$ and $\z_{v}$ such that:
\[
I_{\be_{c}}\Omega\propto-\be_{c}\cdot\delta\G_{c}\sp I_{\be_{v}}\Omega\propto-\be_{v}\cdot\delta\G_{v},
\]
\[
I_{\z_{c}}\Omega\propto-\z_{c}\cdot\De F_{c}\sp I_{\z_{v}}\Omega\propto-\z_{v}\cdot\De F_{v}.
\]
We will see that the proportionality coefficients will be $\lambda$-dependent.

\subsection{The Gauss Constraints as Generators of Rotations}

\subsubsection{The Gauss Constraint on the Cells}

Let us consider the rotation transformation with parameter $\be_{c}$
defined by\footnote{As we explained in Footnote \ref{fn:Lie-derivative}, in the literature
the notation $\delta_{\a}$ is often used for the transformation with
respect to the parameter $\a$, but we avoid it in order to prevent
confusion with the variational exterior derivative $\delta$. The
transformation is indeed given by the action of the Lie derivative
$\LLL_{\a}\equiv I_{\a}\delta+\delta I_{\a}$, as indicated here.}
\[
\LLL_{\be_{c}}H_{cc'}=\be_{c}H_{cc'}\sp\LLL_{\be_{c}}H_{cv}=\be_{c}H_{cv},
\]
\[
\LLL_{\be_{c}}\XXt_{c}^{c'}=[\be_{c},\XXt_{c}^{c'}]\sp\LLL_{\be_{c}}\XXt_{c}^{v}=[\be_{c},\XXt_{c}^{v}],
\]
such that any other variables are unaffected; this includes variables
unrelated to the particular $c$ of choice, as well as the dual variables
$\Ht_{cc'}$, $\Ht_{cv}$, $\X_{c}^{c'}$, and $\X_{c}^{v}$.

Applying it to $\Omega$ and using the identity $I_{\be_{c}}\De H_{c}^{c'}=I_{\be_{c}}\De H_{c}^{v}=\be_{c}$,
we get:
\begin{align*}
I_{\be_{c}}\Omega & =\sum_{c'\ni c}\left(1-\lambda\right)\left([\be_{c},\XXt_{c}^{c'}]\cdot\De H_{c}^{c'}-\delta\XXt_{c}^{c'}\cdot\be_{c}+\XXt_{c}^{c'}\cdot[\be_{c},\De H_{c}^{c'}]\right)+\\
 & \qquad+\sum_{v\ni c}\left(1-\lambda\right)\left([\be_{c},\XXt_{c}^{v}]\cdot\De H_{c}^{v}-\delta\XXt_{c}^{v}\cdot\be_{c}+\XXt_{c}^{v}\cdot[\be_{c},\De H_{c}^{v}]\right).
\end{align*}
However, the first and last triple products in each line cancel each
other, and we are left with:
\begin{align*}
I_{\be_{c}}\Omega & =-\left(1-\lambda\right)\be_{c}\cdot\left(\sum_{c'\ni c}\delta\XXt_{c}^{c'}+\sum_{v\ni c}\delta\XXt_{c}^{v}\right)=-\left(1-\lambda\right)\be_{c}\cdot\delta\G_{c}.
\end{align*}
Hence this transformation is generated by the cell Gauss constraint
$\G_{c}$, given by (\ref{eq:summary_G_c}), as long as $\lambda\ne1$.

\subsubsection{The Gauss Constraint on the Disks}

Next we consider the rotation transformation with parameter $\be_{v}$
defined by
\[
\LLL_{\be_{v}}H_{vc}=\be_{v}H_{vc}\sp\LLL_{\be_{v}}\XXt_{v}^{c}=[\be_{v},\XXt_{v}^{c}],
\]
\[
\LLL_{\be_{v}}H_{v}=\left(1-\lambda\right)\be_{v}H_{v}\sp\LLL_{\be_{v}}\X_{v}=\left(1-\lambda\right)[\be_{v},\X_{v}],
\]
such that any other variables are unaffected; this includes variables
unrelated to the particular $v$ of choice, as well as the dual variables
$\Ht_{vc}$ and $\X_{v}^{c}$. Importantly, we choose the 0-form $\be_{v}$
to be valued in the Cartan subalgebra, so it commutes with $\M_{v}$
and $\SS_{v}$.

Applying the transformation to $\Omega$ and using the identities
$I_{\be_{v}}\De H_{v}^{c}=\be_{v}$ and $I_{\be_{v}}\De H_{v}=\left(1-\lambda\right)\be_{v}$,
we get:
\begin{align*}
I_{\be_{v}}\Omega & =\left(1-\lambda\right)\sum_{c\in v}\left([\be_{v},\XXt_{v}^{c}]\cdot\De H_{v}^{c}-\delta\XXt_{v}^{c}\cdot\be_{v}+\XXt_{v}^{c}\cdot\left[\be_{v},\De H_{v}^{c}\right]\right)+\\
 & \qquad+\left(1-\lambda\right)\left([\be_{v},\X_{v}]\cdot\delta\M_{v}-\left[\M_{v},[\be_{v},\X_{v}]\right]\cdot\De H_{v}\right)+\\
 & \qquad+\left(1-\lambda\right)\left(\left(\delta\SS_{v}+\left[\delta\M_{v},\X_{v}\right]+\left[\M_{v},\delta\X_{v}\right]\right)\cdot\be_{v}-\left(\SS_{v}+\left[\M_{v},\X_{v}\right]\right)\cdot\left[\be_{v},\De H_{v}\right]\right).
\end{align*}
Isolating $\be_{v}$ and using the fact that it commutes with $\M_{v}$
and $\SS_{v}$, we see that most terms cancel\footnote{In this calculation, we make use of the Jacobi identity:
\[
\left[\be_{v},\left[\M_{v},\X_{v}\right]\right]+\left[\M_{v},\left[\X_{v},\be_{v}\right]\right]=-\left[\X_{v},\left[\be_{v},\M_{v}\right]\right]=0.
\]
}, and we get:
\[
I_{\be_{v}}\Omega=-\left(1-\lambda\right)\be_{v}\cdot\left(\sum_{c\in v}\delta\XXt_{v}^{c}-\delta\SS_{v}\right)=-\left(1-\lambda\right)\be_{v}\cdot\G_{v}.
\]
Hence this transformation is generated by the disk Gauss constraint
$\G_{v}$, given by (\ref{eq:summary_G_v}), as long as $\lambda\ne1$.

\subsubsection{The Gauss Constraint on the Faces}

Lastly, we consider the rotation transformation with parameter $\be_{v^{*}}$
defined by
\[
\LLL_{\be_{v^{*}}}\Ht_{cc'}=-\be_{v^{*}}\Ht_{cc'}\sp\LLL_{\be_{v^{*}}}\X_{c}^{c'}=-[\be_{v^{*}},\X_{c}^{c'}],
\]
\[
\LLL_{\be_{v^{*}}}H_{v}=\lambda\bar{\be}_{v^{*}}H_{v}\sp\LLL_{\be_{v^{*}}}\X_{v}=\lambda[\bar{\be}_{v^{*}},\X_{v}],
\]
such that any other variables are unaffected, including variables
unrelated to the particular $v$ of choice as well as the dual variables
$H_{cc'}$ and $\XXt_{c}^{c'}$, and such that
\[
\be_{v^{*}}\equiv h_{v}^{-1}\left(v_{0}\right)\bar{\be}_{v^{*}}h_{v}\left(v_{0}\right),
\]
where $\bar{\be}_{v^{*}}$ is valued in the Cartan subalgebra. Applying
the transformation to $\Omega$, we get after a calculation analogous
to the one we did for the disks,
\begin{align*}
I_{\be_{v^{*}}}\Omega & =-\lambda\left(\be_{v^{*}}\cdot\sum_{c'\in c}\delta\X_{c}^{c'}-\bar{\be}_{v^{*}}\cdot\delta\SS_{v}\right)\\
 & =-\lambda\be_{v^{*}}\cdot\left(\sum_{c'\in c}\delta\X_{c}^{c'}-h_{v}^{-1}\left(v_{0}\right)\delta\SS_{v}h_{v}\left(v_{0}\right)\right).
\end{align*}
The variation of the Gauss constraint (\ref{eq:summary_G_f_v}) is
\[
\delta\G_{v^{*}}=\sum_{i=1}^{N_{v}}\delta\X_{c_{i}}^{c_{i+1}}-h_{v}^{-1}\left(v_{0}\right)\left(\delta\SS_{v}+\left[\SS_{v},\De h_{v}\left(v_{0}\right)\right]\right)h_{v}\left(v_{0}\right),
\]
but since $\bar{\be}_{v^{*}}$ is in the Cartan we have $\bar{\be}_{v^{*}}\cdot\left[\SS_{v},\De h_{v}\left(v_{0}\right)\right]=0$,
so this simplifies to
\[
\be_{v^{*}}\cdot\delta\G_{v^{*}}=\be_{v^{*}}\cdot\left(\sum_{i=1}^{N_{v}}\delta\X_{c_{i}}^{c_{i+1}}-h_{v}^{-1}\left(v_{0}\right)\delta\SS_{v}h_{v}\left(v_{0}\right)\right).
\]
Thus, in conclusion,
\[
I_{\be_{v^{*}}}\Omega=-\lambda\be_{v^{*}}\cdot\delta\G_{v^{*}},
\]
and this transformation is generated by the face Gauss constraint
$\G_{v}$, given by (\ref{eq:summary_G_f_v}), as long as $\lambda\ne0$.

\subsection{The Curvature Constraints as Generators of Translations}

\subsubsection{The Curvature Constraint on the Cells}

For the curvature constraint on the cells, we would like to find a
translation transformation with parameter $\z_{c}$ such that
\[
I_{\z_{c}}\Omega=-\z_{c}\cdot\De F_{c}.
\]
First, we should calculate $\De F_{c}$. Recall that
\[
F_{c}\equiv\prod_{i=1}^{N_{c}}\Ht_{cc_{i}}\Ht_{cv_{i}}=1.
\]
To simplify the calculation, let us define $K_{i}\equiv\Ht_{cc_{i}}\Ht_{cv_{i}}$
such that we may write
\[
F_{c}=\prod_{i=1}^{N}K_{i}=K_{1}\cdots K_{N},
\]
where we omit the subscript $c$ on $N_{c}$ for brevity. Then
\begin{align*}
\delta F_{c} & =\delta K_{1}K_{2}\cdots K_{N}+K_{1}\delta K_{2}K_{3}\cdots K_{N}+\cdots+\\
 & \qquad+K_{1}\cdots K_{N-2}\delta K_{N-1}K_{N}+K_{1}\cdots K_{N-1}\delta K_{N}\\
 & =\De K_{1}K_{1}K_{2}\cdots K_{N}+K_{1}\De K_{2}K_{2}K_{3}\cdots K_{N}+\cdots+\\
 & \qquad+K_{1}\cdots K_{N-2}\De K_{N-1}K_{N-1}K_{N}+K_{1}\cdots K_{N-1}\De K_{N}K_{N},
\end{align*}
where $\De K_{i}\equiv\delta K_{i}K_{i}^{-1}$. Hence 
\begin{align*}
\De F_{c} & \equiv\delta F_{c}F_{c}^{-1}\\
 & =\De K_{1}+K_{1}\De K_{2}K_{1}^{-1}+\cdots+\\
 & \qquad+\left(K_{1}\cdots K_{N-2}\right)\De K_{N-1}\left(K_{1}\cdots K_{N-2}\right)^{-1}+\\
 & \qquad+\left(K_{1}\cdots K_{N-1}\right)\De K_{N}\left(K_{1}\cdots K_{N-1}\right)^{-1}\\
 & \equiv\sum_{i=1}^{N}\left(K_{1}\cdots K_{i-1}\right)\De K_{i}\left(K_{1}\cdots K_{i-1}\right)^{-1},
\end{align*}
where $K_{1}\cdots K_{i-1}\equiv1$ for $i=1$. For conciseness, we
may define $\chi_{i}$ such that $\chi_{1}\equiv1$ and, for $i>1$,
\[
\chi_{i}\equiv K_{1}\cdots K_{i-1}=\Ht_{cc_{1}}\Ht_{cv_{1}}\cdots\Ht_{cc_{i-1}}\Ht_{cv_{i-1}},
\]
and write
\[
\De F_{c}=\sum_{i=1}^{N}\chi_{i}\De K_{i}\chi_{i}^{-1}.
\]

Plugging in $K_{i}\equiv\Ht_{cc_{i}}\Ht_{cv_{i}}$ back, and using
the identity
\[
\De K_{i}=\De\Ht_{c}^{c_{i}}+\Ht_{cc_{i}}\De\Ht_{c}^{v_{i}}\Ht_{c_{i}c}
\]
we get
\begin{equation}
\De F_{c}=\sum_{i=1}^{N}\chi_{i}\left(\De\Ht_{c}^{c_{i}}+\Ht_{cc_{i}}\De\Ht_{c}^{v_{i}}\Ht_{c_{i}c}\right)\chi_{i}^{-1}.\label{eq:DeltaF_c}
\end{equation}
Now, if we transform only the dual fluxes $\X_{c}^{c'}$ and $\X_{c}^{v}$
(for a particular $c$), then we get
\[
I_{\z_{c}}\Omega=-\lambda\sum_{i=1}^{N_{c}}\left(\LLL_{\z_{c}}\X_{c}^{c_{i}}\cdot\De\Ht_{c}^{c_{i}}+\LLL_{\z_{c}}\X_{c}^{v_{i}}\cdot\De\Ht_{c}^{v_{i}}\right).
\]
Comparing with (\ref{eq:DeltaF_c}), we see that if we take
\[
\LLL_{\z_{c}}\X_{c}^{c_{i}}=\chi_{i}^{-1}\z_{c}\chi_{i}\sp\LLL_{\z_{c}}\X_{c}^{v_{i}}=\Ht_{c_{i}c}\chi_{i}^{-1}\z_{c}\chi_{i}\Ht_{cc_{i}},
\]
we will obtain
\[
I_{\z_{c}}\Omega=-\lambda\z_{c}\cdot\De F_{c},
\]
as required. Hence this transformation is generated by the cell curvature
constraint $F_{c}$, given by (\ref{eq:summary_F_c}), as long as
$\lambda\ne0$.

\subsubsection{The Curvature Constraint on the Disks}

As in the cell case, we would like to find a translation transformation
with parameter $\z_{v}$ such that
\[
I_{\z_{v}}\Omega=-\z_{v}\cdot\De F_{v},
\]
where 
\[
F_{v}\equiv\left(\prod_{i=1}^{N_{v}}\Ht_{vc_{i}}\right)h_{v}^{-1}\left(v_{0}\right)\e^{-\M_{v}}h_{v}\left(v_{0}\right)=1.
\]
First, we should calculate $\De F_{v}$. Let us define, omitting the
subscript $v$ on $N_{v}$ for brevity,
\[
K_{i}\equiv\Ht_{vc_{i}}\sp i\in\left\{ 1,\ldots,N\right\} ,
\]
\[
K_{N+1}\equiv h_{v}^{-1}\left(v_{0}\right)\e^{-\M_{v}}h_{v}\left(v_{0}\right),
\]
and
\[
\chi_{1}\equiv1\sp\chi_{i}\equiv K_{1}\cdots K_{i-1}.
\]
Then we may calculate similarly to the previous section
\[
F_{v}=\prod_{i=1}^{N+1}K_{i}\soosp\De F_{v}=\sum_{i=1}^{N+1}\chi_{i}\De K_{i}\chi_{i}^{-1}.
\]
Note that for $i=N+1$ we have
\[
\chi_{N+1}\equiv K_{1}\cdots K_{N}=F_{v}K_{N+1}^{-1}=F_{v}h_{v}^{-1}\left(v_{0}\right)\e^{\M_{v}}h_{v}\left(v_{0}\right),
\]
and since we are imposing $F_{v}=1$, we get simply
\[
\chi_{N+1}=h_{v}^{-1}\left(v_{0}\right)\e^{\M_{v}}h_{v}\left(v_{0}\right).
\]
Furthermore, using the fact that
\[
\De K_{N+1}=h_{v}^{-1}\left(v_{0}\right)\left(\e^{-\M_{v}}\De h_{v}\left(v_{0}\right)\e^{\M_{v}}-\De h_{v}\left(v_{0}\right)-\delta\M_{v}\right)h_{v}\left(v_{0}\right),
\]
we see that
\[
\chi_{N+1}\De K_{N+1}\chi_{N+1}^{-1}=h_{v}^{-1}\left(v_{0}\right)\left(\De h_{v}\left(v_{0}\right)-\e^{\M_{v}}\De h_{v}\left(v_{0}\right)\e^{-\M_{v}}-\delta\M_{v}\right)h_{v}\left(v_{0}\right).
\]
Therefore, we finally obtain the result
\[
\De F_{v}=\sum_{i=1}^{N_{v}}\chi_{i}\De\Ht_{v}^{c_{i}}\chi_{i}^{-1}+h_{v}^{-1}\left(v_{0}\right)\left(\De h_{v}\left(v_{0}\right)-\e^{\M_{v}}\De h_{v}\left(v_{0}\right)\e^{-\M_{v}}-\delta\M_{v}\right)h_{v}\left(v_{0}\right).
\]
Now, let us take
\[
\z_{v}\equiv h_{v}^{-1}\left(v_{0}\right)\bar{\z}_{v}h_{v}\left(v_{0}\right),
\]
where $\bar{\z}_{v}$ is a 0-form valued in the Cartan subalgebra,
and calculate $\z_{v}\cdot\De F_{v}$. We find that, since $\left[\bar{\z}_{v},\M_{v}\right]=0$,
the terms $\De h_{v}\left(v_{0}\right)-\e^{\M_{v}}\De h_{v}\left(v_{0}\right)\e^{-\M_{v}}$
cancel out and we are left with
\begin{equation}
\z_{v}\cdot\De F_{v}=\z_{v}\cdot\left(\sum_{i=1}^{N_{v}}\chi_{i}\De\Ht_{v}^{c_{i}}\chi_{i}^{-1}-h_{v}^{-1}\left(v_{0}\right)\delta\M_{v}h_{v}\left(v_{0}\right)\right).\label{eq:DeltaF_v}
\end{equation}
We may now derive the appropriate transformation. If we transform
only the segment flux $\X_{v}^{c}$ and the vertex flux $\X_{v}$
(for a particular $v$), then we get
\[
I_{\z_{v}}\Omega=-\lambda\sum_{i=1}^{N_{v}}\LLL_{\z_{v}}\X_{v}^{c_{i}}\cdot\De\Ht_{v}^{c_{i}}+\LLL_{\z_{v}}\X_{v}\cdot\left(\delta\M_{v}+\left[\M_{v},\De H_{v}\right]\right).
\]
Comparing with (\ref{eq:DeltaF_v}), we see that if we take
\[
\LLL_{\z_{v}}\X_{v}^{c_{i}}=\chi_{i}^{-1}\z_{v}\chi_{i}\sp\LLL_{\z_{v}}\X_{v}=\lambda\bar{\z}_{v},
\]
we will obtain, since $\bar{\z}_{v}\cdot\left[\M_{v},\De H_{v}\right]=0$,
\[
I_{\z_{v}}\Omega=-\lambda\z_{v}\cdot\left(\sum_{i=1}^{N_{v}}\chi_{i}\De\Ht_{v}^{c_{i}}\chi_{i}^{-1}-h_{v}^{-1}\left(v_{0}\right)\delta\M_{v}h_{v}\left(v_{0}\right)\right)=-\lambda\z_{v}\cdot\De F_{v},
\]
as required. Hence this transformation is generated by the disk curvature
constraint $F_{v}$, given by (\ref{eq:summary_F_v}), as long as
$\lambda\ne0$.

\subsubsection{The Curvature Constraint on the Faces}

We would now like to find a translation transformation with parameter
$\z_{v^{*}}$ such that
\[
I_{\z_{v^{*}}}\Omega=-\z_{v^{*}}\cdot\De F_{v^{*}},
\]
where 
\[
F_{v^{*}}\equiv\left(\prod_{i=1}^{N_{v}}H_{c_{i}c_{i+1}}\right)H_{c_{1}v}\e^{-\M_{v}}H_{vc_{1}}=1.
\]
As before, to calculate $\De F_{v^{*}}$ we define, omitting the subscript
$v$ on $N_{v}$ for brevity,
\[
K_{i}\equiv H_{c_{i}c_{i+1}}\sp i\in\left\{ 1,\ldots,N\right\} ,
\]
\[
K_{N+1}\equiv H_{c_{1}v}\e^{-\M_{v}}H_{vc_{1}},
\]
\[
\chi_{1}\equiv1\sp\chi_{i}\equiv K_{1}\cdots K_{i-1}.
\]
Then a similar calculation to the previous chapter gives
\[
\De F_{v^{*}}=\sum_{i=1}^{N_{v}}\chi_{i}\De H_{c_{i}}^{c_{i+1}}\chi_{i}^{-1}+H_{c_{1}v}\left(\De H_{v}^{c_{1}}-\e^{\M_{v}}\De H_{v}^{c_{1}}\e^{\M_{v}}-\delta\M_{v}\right)H_{vc_{1}},
\]
and if we take
\[
\z_{v^{*}}\equiv H_{c_{1}v}\bar{\z}_{v^{*}}H_{vc_{1}},
\]
where $\bar{\z}_{v^{*}}$ is a 0-form valued in the Cartan subalgebra,
we get
\begin{equation}
\z_{v^{*}}\cdot\De F_{v^{*}}=\z_{v^{*}}\cdot\left(\sum_{i=1}^{N_{v}}\chi_{i}\De H_{c_{i}}^{c_{i+1}}\chi_{i}^{-1}-H_{c_{1}v}\delta\M_{v}H_{vc_{1}}\right).\label{eq:DeltaF_f_v}
\end{equation}
We may now derive the appropriate transformation. If we transform
only the edge flux $\XXt_{c}^{c'}$ and the vertex flux $\X_{v}$
(for a particular $v$), then we get
\[
I_{\z_{v^{*}}}\Omega=\left(1-\lambda\right)\sum_{i=1}^{N_{v}}\LLL_{\z_{v^{*}}}\XXt_{c_{i}}^{c_{i+1}}\cdot\De H_{c}^{c'}+\LLL_{\z_{v^{*}}}\X_{v}\cdot\left(\delta\M_{v}+\left[\M_{v},\De H_{v}\right]\right).
\]
Comparing with (\ref{eq:DeltaF_f_v}), we see that if we take
\[
\LLL_{\z_{v^{*}}}\XXt_{c_{i}}^{c_{i+1}}=-\chi_{i}^{-1}\z_{v^{*}}\chi_{i}\sp\LLL_{\z_{v}}\X_{v}=\left(1-\lambda\right)H_{vc_{1}}\z_{v^{*}}H_{c_{1}v},
\]
we will obtain
\[
I_{\z_{v^{*}}}\Omega=-\left(1-\lambda\right)\z_{v^{*}}\cdot\left(\sum_{i=1}^{N_{v}}\chi_{i}\De H_{c_{i}}^{c_{i+1}}\chi_{i}^{-1}-H_{c_{1}v}\delta\M_{v}H_{vc_{1}}\right)=-\left(1-\lambda\right)\z_{v^{*}}\cdot\De F_{v^{*}},
\]
as required. Hence this transformation is generated by the face curvature
constraint $F_{v}$, given by (\ref{eq:summary_F_v}), as long as
$\lambda\ne0$.

\subsection{\label{subsec:Conclusions-Sym}Conclusions}

We have found that the Gauss constraints $\G_{c},\G_{v},\G_{v^{*}}$
and curvature constraints $F_{c}$, $F_{v}$, $F_{v^{*}}$ for each
cell $c$, disk $D_{v}$ and face $v^{*}$, given by (\ref{eq:summary_G_c}),
(\ref{eq:summary_G_v}), (\ref{eq:summary_G_f_v}), (\ref{eq:summary_F_c}),
(\ref{eq:summary_F_v}) and (\ref{eq:summary_F_f_v}), generate transformations
with rotation parameters $\be_{c},\be_{v},\be_{v^{*}}$ and translations
parameters $\z_{c},\z_{v},\z_{v^{*}}$ as follows:
\[
I_{\be_{c}}\Omega=-\left(1-\lambda\right)\be_{c}\cdot\delta\G_{c}\sp I_{\be_{v}}\Omega=-\left(1-\lambda\right)\be_{v}\cdot\delta\G_{v}\sp I_{\be_{v^{*}}}\Omega=-\lambda\be_{v^{*}}\cdot\delta\G_{v^{*}},
\]
\[
I_{\z_{c}}\Omega=-\lambda\z_{c}\cdot\De F_{c}\sp I_{\z_{v}}\Omega=-\lambda\z_{v}\cdot\De F_{v}\sp I_{\z_{v^{*}}}\Omega=-\left(1-\lambda\right)\z_{v^{*}}\cdot\De F_{v^{*}}.
\]
The Gauss constraint on the cell $c$ generates rotations of the holonomies
on the links $\left(cc'\right)^{*}$ and segments $\left(cv\right)^{*}$
connected to the node $c^{*}$ and the fluxes on the edges $\left(cc'\right)$
and arcs $\left(cv\right)$ surrounding $c$:
\[
\LLL_{\be_{c}}H_{cc'}=\be_{c}H_{cc'}\sp\LLL_{\be_{c}}H_{cv}=\be_{c}H_{cv},
\]
\[
\LLL_{\be_{c}}\XXt_{c}^{c'}=[\be_{c},\XXt_{c}^{c'}]\sp\LLL_{\be_{c}}\XXt_{c}^{v}=[\be_{c},\XXt_{c}^{v}],
\]
where $\be_{c}$ is a $\mfg^{*}$-valued 0-form.

The Gauss constraint on the disk $D_{v}$ generates rotations of the
holonomies on the segments $\left(vc\right)^{*}$ connected to the
vertex $v$ and the fluxes on the arcs $\left(vc\right)$ surrounding
$D_{v}$, as well as the holonomy and flux on the vertex $v$ itself:
\[
\LLL_{\be_{v}}H_{vc}=\be_{v}H_{vc}\sp\LLL_{\be_{v}}\XXt_{v}^{c}=[\be_{v},\XXt_{v}^{c}],
\]
\[
\LLL_{\be_{v}}H_{v}=\left(1-\lambda\right)\be_{v}H_{v}\sp\LLL_{\be_{v}}\X_{v}=\left(1-\lambda\right)[\be_{v},\X_{v}],
\]
where $\be_{v}$ is a 0-form valued in the Cartan subalgebra $\mfh^{*}$
of $\mfg^{*}$.

The Gauss constraint on the face $v^{*}$ generates rotations of the
fluxes on the links $\left(cc'\right)^{*}$ surrounding $v^{*}$ and
the holonomies on their dual edges $\left(cc'\right)$, as well as
the holonomy and flux on the vertex $v$ itself:
\[
\LLL_{\be_{v^{*}}}\Ht_{cc'}=-\be_{v^{*}}\Ht_{cc'}\sp\LLL_{\be_{v^{*}}}\X_{c}^{c'}=-[\be_{v^{*}},\X_{c}^{c'}],
\]
\[
\LLL_{\be_{v^{*}}}H_{v}=\lambda\bar{\be}_{v^{*}}H_{v}\sp\LLL_{\be_{v^{*}}}\X_{v}=\lambda[\bar{\be}_{v^{*}},\X_{v}],
\]
where $\bar{\be}_{v^{*}}$ is a 0-form valued in the Cartan subalgebra
$\mfh^{*}$ of $\mfg^{*}$ and $\be_{v^{*}}\equiv h_{v}^{-1}\left(v_{0}\right)\bar{\be}_{v^{*}}h_{v}\left(v_{0}\right)$.

The curvature constraint on the cell $c$ generates translations\footnote{Note that the curvature constraints do not transform any holonomies,
since the holonomies are unaffected by translations.} of the fluxes on the links $\left(cc'\right)^{*}$ and segments $\left(cv\right)^{*}$
connected to the node $c^{*}$:
\[
\LLL_{\z_{c}}\X_{c}^{c_{i}}=\chi_{i}^{-1}\z_{c}\chi_{i}\sp\LLL_{\z_{c}}\X_{c}^{v_{i}}=\Ht_{c_{i}c}\chi_{i}^{-1}\z_{c}\chi_{i}\Ht_{cc_{i}},
\]
where
\[
\chi_{1}\equiv1\sp\chi_{i}=\Ht_{cc_{1}}\Ht_{cv_{1}}\cdots\Ht_{cc_{i-1}}\Ht_{cv_{i-1}},
\]
and $\z_{c}$ is a $\mfg$-valued 0-form.

The curvature constraint on the disk $D_{v}$ generates translations
of the fluxes on the segments $\left(vc\right)^{*}$ connected to
the vertex $v$, as well as the flux on the vertex $v$ itself:
\[
\LLL_{\z_{v}}\X_{v}^{c_{i}}=\chi_{i}^{-1}\z_{v}\chi_{i}\sp\LLL_{\z_{v}}\X_{v}=\lambda\bar{\z}_{v},
\]
where
\[
\chi_{1}\equiv1\sp\chi_{i}\equiv\Ht_{vc_{i}}\cdots\Ht_{vc_{i-1}},
\]
$\bar{\z}_{v}$ is a 0-form valued in the Cartan subalgebra $\mfh$
of $\mfg$, and $\z_{v}\equiv h_{v}^{-1}\left(v_{0}\right)\bar{\z}_{v}h_{v}\left(v_{0}\right)$.

The curvature constraint on the face $v^{*}$ generates translations
of the fluxes on the edges $\left(cc'\right)$ dual to the links surrounding
the face $v^{*}$, as well as the flux on the vertex $v$ itself:

\[
\LLL_{\z_{v^{*}}}\XXt_{c_{i}}^{c_{i+1}}=-\chi_{i}^{-1}\z_{v^{*}}\chi_{i}\sp\LLL_{\z_{v}}\X_{v}=\left(1-\lambda\right)H_{vc_{1}}\z_{v^{*}}H_{c_{1}v},
\]
where
\[
\chi_{1}\equiv1\sp\chi_{i}\equiv H_{c_{1}c_{2}}\cdots H_{c_{i-1}c_{i}},
\]
and $\z_{v^{*}}$ is a 0-form valued in the Cartan subalgebra $\mfh$
of $\mfg$.

Importantly, in the case $\lambda=0$, the usual loop gravity polarization,
the curvature constraints on the cells and disks do not generate any
transformations since $I_{\z_{c}}\Omega=I_{\z_{v}}\Omega=0$. Similarly,
for the case $\lambda=1$, the dual polarization, the Gauss constraints
on the cells and disks do not generate any transformations since $I_{\be_{c}}\Omega=I_{\be_{v}}\Omega=0$.
Of course, the reason for this is that, as we noted earlier, these
constraints are formulated in the first place in terms of holonomies
and fluxes which only exist in a particular polarization. Thus for
$\lambda=0$ we must instead use the curvature constraint on the faces,
and for $\lambda=1$ we must instead use the Gauss constraint on the
faces.

In the hybrid polarization with $\lambda=1/2$, all of the discrete
variables exist: there are holonomies and fluxes on both the links/edges
and the arcs/segments. Therefore, in this polarization all 6 types
of constraints may be consistently formulated using the available
variables, and all of them generate transformations.

Note that, if $\lambda=0$ or $\lambda=1$, those transformations
for which $I_{\a}\Omega=0$ for some parameter $\a$ are \emph{gauge
symmetries}; they are in the \emph{kernel }of the symplectic form,
and should therefore be divided out. The transformations for which
$I_{\a}\Omega\ne0$ are also symmetries, but not gauge symmetries.

\section{\label{subsec:Corner-Modes}Shrinking the Disks: Focusing on the
Corners}

\subsection{Introduction}

In the Chapter \ref{sec:Discretizing-the-Symplectic-2P1} we rigorously
calculated the discrete symplectic potential by using disks to regularize
the singularities at the punctures. In this chapter, we will take
a different approach. Instead of disks, we will just have points;
this may be understood as the limit where the radius of the disks
goes to zero, $R\to0$, in the derivation above. In other words, we
shrink the disks to points.

Not only will this calculation be shorter -- since we will only have
the edge terms and nothing more -- and much more elegant, we will
also be able to precisely illustrate the role of \emph{corner modes
}in our derivation. The premise here is that the edge modes actually
\textbf{cancel }between the cells; this is why we don't get any contributions
from the edges themselves. They cancel simply because the contribution
from one cell exactly cancels the contribution from the other cell.

In a flat and torsionless geometry, the \textbf{corners }of the cells
-- that is, the vertices -- will also cancel; the contributions
from each of the cells surrounding a vertex will add up to zero simply
because this is a sum of holonomies going in a loop around the vertex,
and if there is no curvature or torsion, the sum of holonomies is
zero, as we would expect. However, if we introduce curvature and/or
torsion, the sum of holonomies now probes these curvature and torsion,
and the contributions will no longer add up to zero. This illustrates
the importance of corner modes in our derivation.

The methodology of this chapter is as follows. First, in Section \ref{subsec:The-Symplectic-Potential}
we show that the dressed symplectic potential (\ref{eq:dressed-Theta}),
which we derived back in Section \ref{subsec:Edge-Modes}, reduces
on-shell to a pure boundary contribution. In fact, only the \textbf{dressing
itself }survives, since the undressed symplectic potential vanishes
on-shell. This is a cleaner and shorter way of deriving the starting
point for the discretized potential.

Then, in Section \ref{subsec:From-Cells-to} we will decompose this
boundary contribution into contributions from each edge. Here we will
notice that our earlier formalism forces us to choose, for each edge
$\left(cc'\right)$, a particular cell $c$; to alleviate this asymmetry
and facilitate the derivation in the next sections, we will symmetrize
the symplectic potential by taking equal contributions from both $c$
and $c'$ for each edge.

In Section \ref{subsec:From-Edges-to} we will manipulate this potential
and bring it to a form that will then allow us to clearly \textbf{separate
the edge contributions from the corner contributions}. Each edge will
provide us with two distinct corner contributions, one from each of
the vertices on the boundary of that edge. We will then show how all
the contributions from the edges connected to a particular vertex
add up to a total corner contributions at that vertex, starting with
a simple example of three edges in Section \ref{subsec:Isolating-the-Vertex}
and then generalizing to $N$ edges in Section \ref{subsec:Generalizing-to-}. 

Finally, in Section \ref{subsec:Analysis-of-the} we will analyze
our results, where we will see that the edge terms make up the spin
network phase space on each edge (and its dual link), while the corner
terms make up the point particle phase space on each vertex.

\subsection{\label{subsec:The-Symplectic-Potential}The Symplectic Potential
On-Shell}

Consider the dressed symplectic potential (\ref{eq:dressed-Theta}):
\[
\Thh=\int_{\Sigma}\ee\cdot\delta\A-\int_{\Sigma}\left(\left(\T+\left[\F,\x\right]\right)\cdot\De h-\x\cdot\delta\F\right)-\int_{\partial\Sigma}\left(\x\cdot\delta\A-\left(\ee+\d_{\A}\x\right)\cdot\De h\right),
\]

One possible solution to the equations of motion $\F=\T=0$ is the
trivial solution $\A=\ee=0$. Of course, if that is a solution, then
any solutions related to it by the gauge transformations (\ref{eq:Euclid-transf-cont})
are also solutions. Therefore, on-shell we can take $\A=\ee=0$, and
we get:
\[
\Thh=\int_{\partial\Sigma}\d\x\cdot\De h.
\]
Now, we decompose $\Sigma$ into cells as described in Section \ref{subsec:The-Cellular-Decomposition}.
Then
\[
\Thh=\sum_{c}\int_{\partial c}\d\x_{c}\cdot\De h_{c},
\]
where $h_{c}$ and $\x_{c}$ are the edge modes on the cell $c$.
Note also that
\[
\d\x_{c}\cdot\d\De h_{c}=\d\left(\x_{c}\cdot\d\De h_{c}\right)=-\d\left(\d\x_{c}\cdot\De h_{c}\right),
\]
so we can write
\[
\d\x_{c}\cdot\d\De h_{c}=\d\left(\lambda\x_{c}\cdot\d\De h_{c}-\left(1-\lambda\right)\d\x_{c}\cdot\De h_{c}\right),
\]
with $\lambda\in\left[0,1\right]$. As explained in Section \ref{subsec:The-Choice-of},
$\lambda=0$ corresponds to the usual polarization while $\lambda=1$
corresponds to the dual or teleparallel polarization. Then the dressed
potential becomes (ignoring the overall minus sign)
\[
\Thh=\sum_{c}\int_{\partial c}\left(\lambda\x_{c}\cdot\d\De h_{c}-\left(1-\lambda\right)\d\x_{c}\cdot\De h_{c}\right).
\]
This potential is, of course, exactly the same potential we discussed
in the previous chapters, except that $h_{c}$ and $\x_{c}$ are now
interpreted as edge modes instead of holonomies.

\subsection{\label{subsec:From-Cells-to}From Cells to Edges}

As above, the boundary of each cell, $\partial c$, may be decomposed
into individual edges:
\[
\partial c=\bigcup_{i=1}^{N_{c}}\left(cc_{i}\right),
\]
where $c_{i}$ with $i\in\left\{ 1,\ldots,N_{c}\right\} $ are all
the cells adjacent to $c$, which therefore share edges $\left(cc_{i}\right)$
with it. Thus, we may split the integrals on the boundaries into integrals
on the edges, taking into account that there are two contributions
to each edge $\left(cc'\right)$, one from the cell $c$ and one from
the cells $c'$:
\begin{equation}
\Thh=\sum_{c}\int_{\partial c}I_{c}=\sum_{\left(cc'\right)}\int_{\left(cc'\right)}\left(I_{c'}-I_{c}\right),\label{eq:sum-c-1}
\end{equation}
where $I_{c}$ are the integrands:
\[
I_{c}\equiv\lambda\x_{c}\cdot\d\De h_{c}-\left(1-\lambda\right)\d\x_{c}\cdot\De h_{c}.
\]
As before, $h_{c'}$ and $\x_{c'}$ are related to $h_{c}$ and $\x_{c}$
via the continuity conditions (\ref{eq:cont-ccp}), with constant
parameters $h_{cc'}$ and $\x_{c}^{c'}$:
\[
h_{c'}=h_{c'c}h_{c}\sp\x_{c'}=\x_{c'}^{c}\oplus\x_{c}=\x_{c'}^{c}+h_{c'c}\x_{c}h_{cc'}=h_{c'c}\left(\x_{c}-\x_{c}^{c'}\right)h_{cc'},
\]
where
\[
h_{c'c}=h_{cc'}^{-1}\sp\x_{c'}^{c}=-h_{c'c}\x_{c}^{c'}h_{cc'}.
\]
From (\ref{eq:Delta-id-1}), we have
\[
\De h_{c'}=\De\left(h_{c'c}h_{c}\right)=h_{c'c}\left(\De h_{c}-\De h_{c}^{c'}\right)h_{cc'}.
\]
Furthermore, since $h_{cc'}$ and $\x_{c}^{c'}$ are constant, we
have
\[
\d\De h_{c'}=h_{c'c}\d\De h_{c}h_{cc'}\sp\d\x_{c'}=\d\left(h_{c'c}(\x_{c}-\x_{c}^{c'})h_{cc'}\right)=h_{c'c}\d\x_{c}h_{cc'}.
\]
Plugging into the integrand, we get
\[
I_{c'}=\lambda\left(\x_{c}-\x_{c}^{c'}\right)\cdot\d\De h_{c}-\left(1-\lambda\right)\d\x_{c}\cdot\left(\De h_{c}-\De h_{c}^{c'}\right),
\]
and we see that we can cancel some terms:
\[
I_{c'}-I_{c}=-\lambda\x_{c}^{c'}\cdot\d\De h_{c}+\left(1-\lambda\right)\d\x_{c}\cdot\De h_{c}^{c'}.
\]
So far, this is the same expression we dealt with in previous chapters.
Note, however, that now all the expressions are based at $c$, and
none are based at $c'$. We can equivalently write an expression where
they are based at $c'$, simply by exchanging $c$ and $c'$ in the
above expression (multiplying by an overall minus sign so that we
still have $I_{c'}-I_{c}$ on the right-hand side):
\[
I_{c'}-I_{c}=\lambda\x_{c'}^{c}\cdot\d\De h_{c'}-\left(1-\lambda\right)\d\x_{c'}\cdot\De h_{c'}^{c}.
\]
By adding both versions, we obtain a \textbf{symmetric} term:
\[
I_{c'}-I_{c}=\hf\left(\left(1-\lambda\right)\left(\d\x_{c}\cdot\De h_{c}^{c'}-\d\x_{c'}\cdot\De h_{c'}^{c}\right)-\lambda\left(\x_{c}^{c'}\cdot\d\De h_{c}-\x_{c'}^{c}\cdot\d\De h_{c'}\right)\right).
\]
Integrating, we get
\begin{align*}
\int_{\left(cc'\right)}\left(I_{c'}-I_{c}\right) & =\hf\left(1-\lambda\right)\left(\left(\int_{\left(cc'\right)}\d\x_{c}\right)\cdot\De h_{c}^{c'}-\left(\int_{\left(cc'\right)}\d\x_{c'}\right)\cdot\De h_{c'}^{c}\right)+\\
 & \qquad-\hf\lambda\left(\x_{c}^{c'}\cdot\left(\int_{\left(cc'\right)}\d\De h_{c}\right)-\x_{c'}^{c}\cdot\left(\int_{\left(cc'\right)}\d\De h_{c'}\right)\right)\\
 & =\hf\left(1-\lambda\right)\left(\left(\x_{c}\left(v'\right)-\x_{c}\left(v\right)\right)\cdot\De h_{c}^{c'}-\left(\x_{c'}\left(v'\right)-\x_{c'}\left(v\right)\right)\cdot\De h_{c'}^{c}\right)+\\
 & \qquad-\hf\lambda\left(\x_{c}^{c'}\cdot\left(\De h_{c}\left(v'\right)-\De h_{c}\left(v\right)\right)-\x_{c'}^{c}\cdot\left(\De h_{c'}\left(v'\right)-\De h_{c'}\left(v\right)\right)\right).
\end{align*}

\subsection{\label{subsec:From-Edges-to}From Edges to Vertices}

For each edge $\left(cc'\right)$ we have, isolating the terms based
at $c$,
\[
\Theta_{cc'}^{\left(c\right)}\equiv\left(1-\lambda\right)\left(\x_{c}\left(v'\right)-\x_{c}\left(v\right)\right)\cdot\De h_{c}^{c'}-\lambda\x_{c}^{c'}\cdot\left(\De h_{c}\left(v'\right)-\De h_{c}\left(v\right)\right).
\]
Using the continuity conditions to write $\De h_{c}\left(v'\right),\x_{c}\left(v'\right)$
in terms of $\De h_{c'}\left(v'\right),\x_{c'}\left(v'\right)$,
\[
\De h_{c}\left(v'\right)=h_{cc'}\left(\De h_{c'}\left(v'\right)-\De h_{c'}^{c}\right)h_{c'c}\sp\x_{c}\left(v'\right)=h_{cc'}\left(\x_{c'}\left(v'\right)-\x_{c'}^{c}\right)h_{c'c},
\]
we get
\begin{align*}
\Theta_{cc'}^{\left(c\right)} & =\left(1-\lambda\right)\left(h_{cc'}\left(\x_{c'}\left(v'\right)-\x_{c'}^{c}\right)h_{c'c}-\x_{c}\left(v\right)\right)\cdot\De h_{c}^{c'}+\\
 & \qquad-\lambda\x_{c}^{c'}\cdot\left(h_{cc'}\left(\De h_{c'}\left(v'\right)-\De h_{c'}^{c}\right)h_{c'c}-\De h_{c}\left(v\right)\right)\\
 & =\left(1-\lambda\right)\left(\left(\x_{c'}\left(v'\right)-\x_{c'}^{c}\right)\cdot h_{c'c}\De h_{c}^{c'}h_{cc'}-\x_{c}\left(v\right)\cdot\De h_{c}^{c'}\right)+\\
 & \qquad-\lambda\left(h_{c'c}\x_{c}^{c'}h_{cc'}\cdot\left(\De h_{c'}\left(v'\right)-\De h_{c'}^{c}\right)-\x_{c}^{c'}\cdot\De h_{c}\left(v\right)\right)\\
 & =\left(1-\lambda\right)\left(-\left(\x_{c'}\left(v'\right)-\x_{c'}^{c}\right)\cdot\De h_{c'}^{c}-\x_{c}\left(v\right)\cdot\De h_{c}^{c'}\right)+\\
 & \qquad-\lambda\left(-\x_{c'}^{c}\cdot\left(\De h_{c'}\left(v'\right)-\De h_{c'}^{c}\right)-\x_{c}^{c'}\cdot\De h_{c}\left(v\right)\right)\\
 & =\left(1-2\lambda\right)\x_{c}^{c'}\cdot\De h_{c}^{c'}-\left(1-\lambda\right)\left(\x_{c'}\left(v'\right)\cdot\De h_{c'}^{c}+\x_{c}\left(v\right)\cdot\De h_{c}^{c'}\right)+\\
 & \qquad+\lambda\left(\x_{c'}^{c}\cdot\De h_{c'}\left(v'\right)+\x_{c}^{c'}\cdot\De h_{c}\left(v\right)\right)\\
 & =\frac{1-2\lambda}{2}\x_{c}^{c'}\cdot\De h_{c}^{c'}+\lambda\x_{c}^{c'}\cdot\De h_{c}\left(v\right)-\left(1-\lambda\right)\x_{c}\left(v\right)\cdot\De h_{c}^{c'}+\\
 & \qquad+\frac{1-2\lambda}{2}\x_{c'}^{c}\cdot\De h_{c'}^{c}+\lambda\x_{c'}^{c}\cdot\De h_{c'}\left(v'\right)-\left(1-\lambda\right)\x_{c'}\left(v'\right)\cdot\De h_{c'}^{c},
\end{align*}
where we used the identities (see again Sections \ref{subsec:Holonomies-and}
and \ref{subsec:The-Cartan-Decomposition}) 
\[
h_{c'c}\De h_{c}^{c'}h_{cc'}=-\De h_{c'}^{c}\sp h_{c'c}\x_{c}^{c'}h_{cc'}=-\x_{c'}^{c},
\]
from which we find 
\[
\x_{c'}^{c}\cdot\De h_{c'}^{c}=\x_{c}^{c'}\cdot\De h_{c}^{c'},
\]
so this expression is invariant under the exchange $c\tot c'$. Now
we can symmetrize this by restoring the terms based at $c'$:
\begin{align*}
\Theta_{cc'} & \equiv\Theta_{cc'}^{\left(c\right)}-\Theta_{cc'}^{\left(c'\right)}\\
 & =\frac{1-2\lambda}{2}\x_{c}^{c'}\cdot\De h_{c}^{c'}+\left(1-\lambda\right)\left(\x_{c}\left(v'\right)-\x_{c}\left(v\right)\right)\cdot\De h_{c}^{c'}+\lambda\x_{c}^{c'}\cdot\left(\De h_{c}\left(v\right)-\De h_{c}\left(v'\right)\right)+\\
 & \qquad+\frac{1-2\lambda}{2}\x_{c'}^{c}\cdot\De h_{c'}^{c}+\left(1-\lambda\right)\left(\x_{c'}\left(v\right)-\x_{c'}\left(v'\right)\right)\cdot\De h_{c'}^{c}+\lambda\x_{c'}^{c}\cdot\left(\De h_{c'}\left(v'\right)-\De h_{c'}\left(v\right)\right).
\end{align*}

\subsection{\label{subsec:Isolating-the-Vertex}Isolating the Vertex Terms}

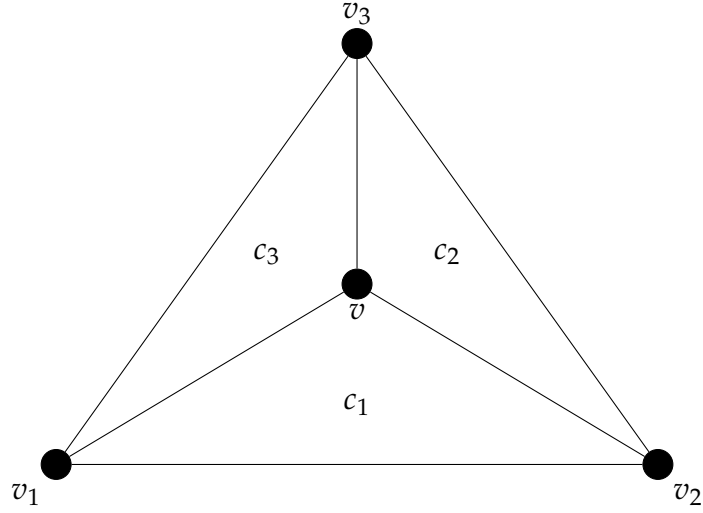
\begin{figure}
\begin{centering}
\begin{tikzpicture}[scale=0.8]
	\begin{pgfonlayer}{nodelayer}
		\node [style=none] (1) at (0, -0.45) {$v$};
		\node [style=Vertex] (4) at (0, 0) {};
		\node [style=Vertex] (5) at (0, 4) {};
		\node [style=Vertex] (6) at (-5, -3) {};
		\node [style=Vertex] (7) at (5, -3) {};
		\node [style=none] (8) at (-1.5, 0.5) {$c_3$};
		\node [style=none] (9) at (0, -2) {$c_1$};
		\node [style=none] (10) at (1.5, 0.5) {$c_2$};
		\node [style=none] (11) at (5.5, -3.5) {$v_2$};
		\node [style=none] (12) at (0, 4.5) {$v_3$};
		\node [style=none] (13) at (-5.5, -3.5) {$v_1$};
	\end{pgfonlayer}
	\begin{pgfonlayer}{edgelayer}
		\draw [style=Edge] (5) to (6);
		\draw [style=Edge] (6) to (7);
		\draw [style=Edge] (7) to (5);
		\draw [style=Edge] (5) to (4);
		\draw [style=Edge] (4) to (7);
		\draw [style=Edge] (4) to (6);
	\end{pgfonlayer}
\end{tikzpicture}
\par\end{centering}
\caption{\label{fig:ThreeCells}Three cells $c_{1},c_{2},c_{3}$ surrounding
a single vertex $v$.}
\end{figure}

Consider three cells $c_{i}$, $i\in\left\{ 1,2,3\right\} $, as illustrated
in Figure \ref{fig:ThreeCells}. As we calculated above, the potentials
on the three edges $\left(c_{1}c_{2}\right)$, $\left(c_{2}c_{3}\right)$,
and $\left(c_{3}c_{1}\right)$, after cancellations, are
\[
\Theta_{c_{1}c_{2}}=\left(1-\lambda\right)\left(\x_{c_{1}}\left(v_{2}\right)-\x_{c_{1}}\left(v\right)\right)\cdot\De h_{c_{1}}^{c_{2}}-\lambda\x_{c_{1}}^{c_{2}}\cdot\left(\De h_{c_{1}}\left(v_{2}\right)-\De h_{c_{1}}\left(v\right)\right),
\]
\[
\Theta_{c_{2}c_{3}}=\left(1-\lambda\right)\left(\x_{c_{2}}\left(v_{3}\right)-\x_{c_{2}}\left(v\right)\right)\cdot\De h_{c_{2}}^{c_{3}}-\lambda\x_{c_{2}}^{c_{3}}\cdot\left(\De h_{c_{2}}\left(v_{3}\right)-\De h_{c_{2}}\left(v\right)\right),
\]
\[
\Theta_{c_{3}c_{1}}=\left(1-\lambda\right)\left(\x_{c_{3}}\left(v_{1}\right)-\x_{c_{3}}\left(v\right)\right)\cdot\De h_{c_{3}}^{c_{1}}-\lambda\x_{c_{3}}^{c_{1}}\cdot\left(\De h_{c_{3}}\left(v_{1}\right)-\De h_{c_{3}}\left(v\right)\right).
\]
Symmetrizing these terms as above, we get
\begin{align*}
\Theta_{c_{1}c_{2}} & =\left(\frac{1-2\lambda}{2}\x_{c_{1}}^{c_{2}}\cdot\De h_{c_{1}}^{c_{2}}+\lambda\x_{c_{1}}^{c_{2}}\cdot\De h_{c_{1}}\left(v\right)-\left(1-\lambda\right)\x_{c_{1}}\left(v\right)\cdot\De h_{c_{1}}^{c_{2}}\right)+\\
 & \qquad+\left(\frac{1-2\lambda}{2}\x_{c_{2}}^{c_{1}}\cdot\De h_{c_{2}}^{c_{1}}+\lambda\x_{c_{2}}^{c_{1}}\cdot\De h_{c_{2}}\left(v_{2}\right)-\left(1-\lambda\right)\x_{c_{2}}\left(v_{2}\right)\cdot\De h_{c_{2}}^{c_{1}}\right),
\end{align*}
\begin{align*}
\Theta_{c_{2}c_{3}} & =\left(\frac{1-2\lambda}{2}\x_{c_{2}}^{c_{3}}\cdot\De h_{c_{2}}^{c_{3}}+\lambda\x_{c_{2}}^{c_{3}}\cdot\De h_{c_{2}}\left(v\right)-\left(1-\lambda\right)\x_{c_{2}}\left(v\right)\cdot\De h_{c_{2}}^{c_{3}}\right)+\\
 & \qquad+\left(\frac{1-2\lambda}{2}\x_{c_{3}}^{c_{2}}\cdot\De h_{c_{3}}^{c_{2}}+\lambda\x_{c_{3}}^{c_{2}}\cdot\De h_{c_{3}}\left(v_{3}\right)-\left(1-\lambda\right)\x_{c_{3}}\left(v_{3}\right)\cdot\De h_{c_{3}}^{c_{2}}\right),
\end{align*}
\begin{align*}
\Theta_{c_{3}c_{1}} & =\left(\frac{1-2\lambda}{2}\x_{c_{3}}^{c_{1}}\cdot\De h_{c_{3}}^{c_{1}}+\lambda\x_{c_{3}}^{c_{1}}\cdot\De h_{c_{3}}\left(v\right)-\left(1-\lambda\right)\x_{c_{3}}\left(v\right)\cdot\De h_{c_{3}}^{c_{1}}\right)+\\
 & \qquad+\left(\frac{1-2\lambda}{2}\x_{c_{1}}^{c_{3}}\cdot\De h_{c_{1}}^{c_{3}}+\lambda\x_{c_{1}}^{c_{3}}\cdot\De h_{c_{1}}\left(v_{1}\right)-\left(1-\lambda\right)\x_{c_{1}}\left(v_{1}\right)\cdot\De h_{c_{1}}^{c_{3}}\right).
\end{align*}
From this symmetric form, we can clearly see that each edge term $\Theta_{c_{i}c_{i+1}}$
has a contribution from its source vertex $v$ and target vertex $v_{i+1}$,
and both contributions are symmetric. Therefore, we conclude that
the second line in each expression belongs to the contribution of
the potential on the edge to the potential on the vertices $v_{1},v_{2},v_{3}$.
We choose to focus only on the central vertex $v$, so we can look
only at the contributions to $v$ \textbf{without loss of generality}.

Let us collect all of the terms involving the vertex $v$:
\begin{align*}
\Theta_{v} & =\frac{1-2\lambda}{2}\x_{c_{1}}^{c_{2}}\cdot\De h_{c_{1}}^{c_{2}}+\lambda\x_{c_{1}}^{c_{2}}\cdot\De h_{c_{1}}\left(v\right)-\left(1-\lambda\right)\x_{c_{1}}\left(v\right)\cdot\De h_{c_{1}}^{c_{2}}+\\
 & \qquad+\frac{1-2\lambda}{2}\x_{c_{2}}^{c_{3}}\cdot\De h_{c_{2}}^{c_{3}}+\lambda\x_{c_{2}}^{c_{3}}\cdot\De h_{c_{2}}\left(v\right)-\left(1-\lambda\right)\x_{c_{2}}\left(v\right)\cdot\De h_{c_{2}}^{c_{3}}+\\
 & \qquad+\frac{1-2\lambda}{2}\x_{c_{3}}^{c_{1}}\cdot\De h_{c_{3}}^{c_{1}}+\lambda\x_{c_{3}}^{c_{1}}\cdot\De h_{c_{3}}\left(v\right)-\left(1-\lambda\right)\x_{c_{3}}\left(v\right)\cdot\De h_{c_{3}}^{c_{1}}.
\end{align*}

\subsection{\label{subsec:Generalizing-to-}Generalizing to $N$ Cells}

In general, for $N$ cells $c_{1},\ldots,c_{N}$ around a vertex $v$,
and setting $c_{N+1}\equiv c_{1}$, we find
\[
\Theta_{v}=\sum_{i=1}^{N}\left(\frac{1-2\lambda}{2}\x_{c_{i}}^{c_{i+1}}\cdot\De h_{c_{i}}^{c_{i+1}}+\lambda\x_{c_{i}}^{c_{i+1}}\cdot\De h_{c_{i}}\left(v\right)-\left(1-\lambda\right)\x_{c_{i}}\left(v\right)\cdot\De h_{c_{i}}^{c_{i+1}}\right).
\]
The first term, $\x_{c_{i}}^{c_{i+1}}\cdot\De h_{c_{i}}^{c_{i+1}}$,
is simply the holonomy-flux term on the link between the two cells.

In order to manipulate this expression, we bring all of the expression
to the vertex, using the continuity conditions
\[
h_{c_{i}}=h_{c_{i}v}h_{v}\sp\De h_{c_{i}}=\De h_{c_{i}}^{v}+h_{c_{i}v}\De h_{v}h_{vc_{i}}\sp\x_{c_{i}}=\x_{c_{i}}^{v}+h_{c_{i}v}\x_{v}h_{vc_{i}}.
\]
Note that these conditions are only valid on the arcs, but since we
shrunk the disks, they are now valid at the vertex. We get:
\begin{align*}
\Theta_{v} & =\sum_{i=1}^{N}\frac{1-2\lambda}{2}\x_{c_{i}}^{c_{i+1}}\cdot\De h_{c_{i}}^{c_{i+1}}+\\
 & \qquad+\lambda\sum_{i=1}^{N}\x_{c_{i}}^{c_{i+1}}\cdot\left(\De h_{c_{i}}^{v}+h_{c_{i}v}\De h_{v}\left(v\right)h_{vc_{i}}\right)-\left(1-\lambda\right)\sum_{i=1}^{N}\left(\x_{c_{i}}^{v}+h_{c_{i}v}\x_{v}\left(v\right)h_{vc_{i}}\right)\cdot\De h_{c_{i}}^{c_{i+1}}.
\end{align*}
We can take the terms which do not depend on $i$ out of the sum and
rewrite this as
\begin{align*}
\Theta_{v} & =\sum_{i=1}^{N}\left(\frac{1-2\lambda}{2}\x_{c_{i}}^{c_{i+1}}\cdot\De h_{c_{i}}^{c_{i+1}}+\lambda\x_{c_{i}}^{c_{i+1}}\cdot\De h_{c_{i}}^{v}-\left(1-\lambda\right)\x_{c_{i}}^{v}\cdot\De h_{c_{i}}^{c_{i+1}}\right)+\\
 & \qquad+\lambda\De h_{v}\left(v\right)\cdot\sum_{i=1}^{N}h_{vc_{i}}\x_{c_{i}}^{c_{i+1}}h_{c_{i}v}-\left(1-\lambda\right)\x_{v}\left(v\right)\cdot\sum_{i=1}^{N}h_{vc_{i}}\De h_{c_{i}}^{c_{i+1}}h_{c_{i}v}.
\end{align*}
Now, let $N=2$, such that $c_{3}=c_{1}$, then we calculate
\begin{align*}
\De\left(h_{vc_{1}}h_{c_{1}c_{2}}h_{c_{2}c_{3}}h_{c_{3}v}\right) & =\delta\left(h_{vc_{1}}h_{c_{1}c_{2}}h_{c_{2}c_{3}}h_{c_{3}v}\right)\left(h_{vc_{3}}h_{c_{3}c_{2}}h_{c_{2}c_{1}}h_{c_{1}v}\right)\\
 & =\De h_{v}^{c_{1}}+h_{vc_{1}}\De h_{c_{1}}^{c_{2}}h_{c_{1}v}+h_{vc_{1}}h_{c_{1}c_{2}}\De h_{c_{2}}^{c_{3}}h_{c_{2}c_{1}}h_{c_{1}v}+\\
 & \qquad+h_{vc_{1}}h_{c_{1}c_{2}}h_{c_{2}c_{3}}\De h_{c_{3}}^{v}h_{c_{3}c_{2}}h_{c_{2}c_{1}}h_{c_{1}v}\\
 & =\De h_{v}^{c_{1}}+h_{vc_{1}}\De h_{c_{1}}^{c_{2}}h_{c_{1}v}+h_{vc_{2}}\De h_{c_{2}}^{c_{3}}h_{c_{2}v}+h_{vc_{3}}\De h_{c_{3}}^{v}h_{c_{3}v}\\
 & =\De h_{v}^{c_{1}}+\sum_{i=1}^{N}h_{vc_{i}}\De h_{c_{i}}^{c_{i+1}}h_{c_{i}v}+h_{vc_{1}}\De h_{c_{1}}^{v}h_{c_{1}v}\\
 & =\sum_{i=1}^{N}h_{vc_{i}}\De h_{c_{i}}^{c_{i+1}}h_{c_{i}v},
\end{align*}
where in the last line we used the fact that $h_{vc_{1}}\De h_{c_{1}}^{v}h_{c_{1}v}=-\De h_{v}^{c_{1}}$.
Noticing a pattern, we conclude that in the general case of $N$ cells
we should have
\[
\De\left(h_{vc_{1}}\left(\prod_{i=1}^{N}h_{c_{i}c_{i+1}}\right)h_{c_{N+1}v}\right)=\sum_{i=1}^{N}h_{vc_{i}}\De h_{c_{i}}^{c_{i+1}}h_{c_{i}v}.
\]
Similarly, for the translational holonomies we have the addition rule
(\ref{eq:holonomy-comp}),
\[
\x_{a}^{b}\oplus\x_{b}^{c}\equiv\x_{a}^{b}+h_{ab}\x_{b}^{c}h_{ba},
\]
so we can calculate that
\begin{align*}
\x_{v}^{c_{1}}\oplus\x_{c_{1}}^{c_{2}}\oplus\x_{c_{2}}^{c_{3}}\oplus\x_{c_{3}}^{v} & =\x_{v}^{c_{1}}+h_{vc_{1}}\left(\x_{c_{1}}^{c_{2}}\oplus\x_{c_{2}}^{c_{3}}\oplus\x_{c_{3}}^{v}\right)h_{c_{1}v}\\
 & =\x_{v}^{c_{1}}+h_{vc_{1}}\left(\x_{c_{1}}^{c_{2}}+h_{c_{1}c_{2}}\left(\x_{c_{2}}^{c_{3}}\oplus\x_{c_{3}}^{v}\right)h_{c_{2}c_{1}}\right)h_{c_{1}v}\\
 & =\x_{v}^{c_{1}}+h_{vc_{1}}\left(\x_{c_{1}}^{c_{2}}+h_{c_{1}c_{2}}\left(\x_{c_{2}}^{c_{3}}+h_{c_{2}c_{3}}\x_{c_{3}}^{v}h_{c_{3}c_{2}}\right)h_{c_{2}c_{1}}\right)h_{c_{1}v}\\
 & =\x_{v}^{c_{1}}+h_{vc_{1}}\x_{c_{1}}^{c_{2}}h_{c_{1}v}+h_{vc_{2}}\left(\x_{c_{2}}^{c_{3}}+h_{c_{2}c_{3}}\x_{c_{3}}^{v}h_{c_{3}c_{2}}\right)h_{c_{2}v}\\
 & =\x_{v}^{c_{1}}+h_{vc_{1}}\x_{c_{1}}^{c_{2}}h_{c_{1}v}+h_{vc_{2}}\x_{c_{2}}^{c_{3}}h_{c_{2}v}+h_{vc_{3}}\x_{c_{3}}^{v}h_{c_{3}v}\\
 & =\x_{v}^{c_{1}}+\sum_{i=1}^{N}h_{vc_{i}}\x_{c_{i}}^{c_{i+1}}h_{c_{i}v}+h_{vc_{1}}\x_{c_{1}}^{v}h_{c_{1}v}\\
 & =\sum_{i=1}^{N}h_{vc_{i}}\x_{c_{i}}^{c_{i+1}}h_{c_{i}v},
\end{align*}
where in the last line we used the fact that $h_{vc_{1}}\x_{c_{1}}^{v}h_{c_{1}v}=-\x_{v}^{c_{1}}$.
We conclude that in the general case we have
\[
\x_{v}^{c_{1}}\oplus\left(\bigoplus_{i=1}^{N}\x_{c_{i}}^{c_{i+1}}\right)\oplus\x_{c_{N+1}}^{v}=\sum_{i=1}^{N}h_{vc_{i}}\x_{c_{i}}^{c_{i+1}}h_{c_{i}v}.
\]
As we have seen above, the rotational and translational holonomies
detect curvature and torsion as follows:
\begin{equation}
h_{vc_{1}}\left(\prod_{i=1}^{N}h_{c_{i}c_{i+1}}\right)h_{c_{N+1}v}=\e^{\M_{v}},\label{eq:detect-M}
\end{equation}
\begin{equation}
\x_{v}^{c_{1}}\oplus\left(\bigoplus_{i=1}^{N}\x_{c_{i}}^{c_{i+1}}\right)\oplus\x_{c_{N+1}}^{v}=\SS_{v}+\left[\M_{v},\left(1-\lambda\right)\x_{v}\left(v\right)\right].\label{eq:detect-S}
\end{equation}
Since $\De\left(\e^{\M_{v}}\right)=\delta\M_{v}$, we conclude that
\begin{equation}
\sum_{i=1}^{N}h_{vc_{i}}\De h_{c_{i}}^{c_{i+1}}h_{c_{i}v}=\delta\M_{v}\sp\sum_{i=1}^{N}h_{vc_{i}}\x_{c_{i}}^{c_{i+1}}h_{c_{i}v}=\SS_{v}+\left[\M_{v},\x_{v}\left(v\right)\right].\label{eq:detect-MS}
\end{equation}
Then the vertex potential becomes
\begin{align}
\Theta_{v} & =\sum_{i=1}^{N}\left(\frac{1-2\lambda}{2}\x_{c_{i}}^{c_{i+1}}\cdot\De h_{c_{i}}^{c_{i+1}}+\lambda\x_{c_{i}}^{c_{i+1}}\cdot\De h_{c_{i}}^{v}-\left(1-\lambda\right)\x_{c_{i}}^{v}\cdot\De h_{c_{i}}^{c_{i+1}}\right)+\nonumber \\
 & \qquad+\lambda\De h_{v}\left(v\right)\cdot\left(\SS_{v}+\left[\M_{v},\x_{v}\left(v\right)\right]\right)-\left(1-\lambda\right)\x_{v}\left(v\right)\cdot\delta\M_{v}.\label{eq:Theta-v-corner}
\end{align}

\subsection{\label{subsec:Analysis-of-the}Analysis of the Vertex Potential}

\subsubsection{First Line}

The first line is simply the usual edge potential we found above,
written in another form. Recall that in Section \ref{subsec:The-Discretized-Symplectic}
we found
\[
\Theta_{cc'}=\left(1-\lambda\right)\XXt_{c}^{c'}\cdot\De H_{c}^{c'}-\lambda\X_{c}^{c'}\cdot\De\Ht_{c}^{c'}.
\]
Here we have instead
\[
\Theta_{c_{i}c_{i+1}}^{\left(v\right)}=\frac{1-2\lambda}{2}\x_{c_{i}}^{c_{i+1}}\cdot\De h_{c_{i}}^{c_{i+1}}+\lambda\x_{c_{i}}^{c_{i+1}}\cdot\De h_{c_{i}}^{v}-\left(1-\lambda\right)\x_{c_{i}}^{v}\cdot\De h_{c_{i}}^{c_{i+1}},
\]
which relates to the link $\left(c_{i}c_{i+1}\right)^{*}$ but contains
terms which involve a particular vertex $v$. Since each edge $\left(cc'\right)$
is related to exactly two vertices, let's call them $v$ and $v'$,
the total contribution to each edge $\left(cc'\right)$ will be (with
a minus sign due to opposite orientation)
\[
\Theta_{cc'}=\Theta_{cc'}^{\left(v\right)}-\Theta_{cc'}^{\left(v'\right)}=\lambda\x_{c}^{c'}\cdot\left(\De h_{c}^{v}-\De h_{c}^{v'}\right)-\left(1-\lambda\right)\left(\x_{c}^{v}-\x_{c}^{v'}\right)\cdot\De h_{c}^{c'}.
\]
We now note that we may define holonomies and fluxes on the edges
as follows:
\begin{equation}
h_{vv'}=h_{vc}h_{cv'}\soosp\De h_{v}^{v'}=h_{vc}\left(\De h_{c}^{v'}-\De h_{c}^{v}\right)h_{cv},\label{eq:vvp-split-h}
\end{equation}
\begin{equation}
\x_{v}^{v'}=\x_{v}^{c}\oplus\x_{c}^{v'}=h_{vc}\left(\x_{c}^{v'}-\x_{c}^{v}\right)h_{cv}.\label{eq:vvp-split-x}
\end{equation}
Then the edge potential becomes
\[
\Theta_{cc'}=\left(1-\lambda\right)h_{cv}\x_{v}^{v'}h_{vc}\cdot\De h_{c}^{c'}-\lambda\x_{c}^{c'}\cdot h_{cv}\De h_{v}^{v'}h_{vc}.
\]
Remarkably, we have achieved a more precise definition of the spin
network and dual spin network phase space -- using the explicit holonomies
on the edges, $h_{vv'}$ and $\x_{v}^{v'}$, instead of the more complicated
expressions as we had before! We can now write
\[
\Theta_{cc'}=\left(1-\lambda\right)\XXt_{c}^{c'}\cdot\De H_{c}^{c'}-\lambda\X_{c}^{c'}\cdot\De\Ht_{c}^{c'},
\]
where we define
\[
\XXt_{c}^{c'}\equiv h_{cv}\x_{v}^{v'}h_{vc}\sp\De\Ht_{c}^{c'}\equiv h_{cv}\De h_{v}^{v'}h_{vc}.
\]

\subsubsection{Second Line}

The second line of the vertex potential is similar to the particle
potential that we obtained with the puncture picture, except in this
case it appears simply due to the gluing between the cells, without
any mention of disks or delta functions. We see that the contributions
to the symplectic potential at $v$ would cancel if there was no curvature
or torsion there, but in the general case they do not cancel, and
additional degrees of freedom are created.

In order to obtain exactly the same term, we simply absorb the $\lambda$-dependent
terms into new variables, in a similar\footnote{Note that here we do not have to worry about the terms at $v_{0}$,
since the disks are already shrunk.} way to the redefinitions we did in Section \ref{subsec:The-Particle-Potential}
\[
\X_{v}\equiv\left(1-\lambda\right)\x_{v}\left(v\right)\sp H_{v}\equiv\lambda h_{v}\left(v\right),
\]
and get
\[
\Theta_{v}=\X_{v}\cdot\delta\M_{v}-\left(\SS_{v}+\left[\M_{v},\X_{v}\right]\right)\cdot\De H_{v}.
\]

\subsubsection{Conclusion}

In conclusion, the full symplectic potential we have obtained is
\begin{equation}
\Theta=\sum_{\left(cc'\right)}\left(\left(1-\lambda\right)\XXt_{c}^{c'}\cdot\De H_{c}^{c'}-\lambda\X_{c}^{c'}\cdot\De\Ht_{c}^{c'}\right)+\sum_{v}\left(\X_{v}\cdot\delta\M_{v}-\left(\SS_{v}+\left[\M_{v},\X_{v}\right]\right)\cdot\De H_{v}\right).\label{eq:Theta-corner}
\end{equation}
The first term consists of pure edge contributions, and it describes
a spin network phase space and a dual spin network phase space on
each pair of edge and its dual link. The second term consists of pure
corner contributions, and it describes a point particle phase space
on each vertex.

The separation between edge and corner contributions clearly plays
a crucial role in understanding the discrete phase space and its relation
to the continuous phase space, as well as the relation between spin
networks and piecewise-flat-and-torsionless geometries. We see that
spin networks only ``know'' about the edge contributions, but properly
taking the curvature and torsion into account -- together with the
redundancies of the discretization and the symmetries associated with
them, as we have discussed above -- requires one to acknowledge the
corner contributions as well. 

In the rest of this thesis, we will use the methodology of this chapter
to derive similar results in the more complicated setting of 3+1D
gravity.

\clearpage{}

\part{\label{part:3P1-cont}3+1 Dimensions: The Continuous Theory}

In the previous chapters of this thesis, we discretized continuous
gravity in 2+1 dimensions and obtained the spin network phase space.
Of course, in reality, our universe has 3+1 dimensions. Therefore,
in order for our discussion to apply to the real works, we would like
to generalize it to 3+1 dimensions.

In 2+1 dimensions, gravity is topological. The equations of motion,
$\F=\T=0$, simply require that there is no curvature or torsion anywhere.
Gravity itself has no dynamical degrees of freedom, and there are
no gravitational waves or gravitons. The only degrees of freedom we
took into account in our discussion were the point particle degrees
of freedom, which appeared as codimension-2 point singularities in
an otherwise completely featureless spacetime. Dividing the spatial
slice into cells yielded additional artificial degrees of freedom
which governed the transformations between cells.

In 3+1 dimensions, gravity is \textbf{not }topological. In order to
apply our calculation to this case, we must make some simplifying
assumptions:
\begin{itemize}
\item In 2+1 dimensions, the geometry inside each 2-dimensional cell was
flat and torsionless. In 3+1 dimensions, we will \textbf{impose by
hand }that the geometry inside each 3-dimensional cell should be flat
and torsionless.
\item In 2+1 dimensions, the matter sources were given as codimension-2
singularities, which were particles on the 0-dimensional vertices
of the 2-dimensional cells. In 3+1 dimensions, the matter sources
will also be given as codimension-2 singularities, which will now
be \emph{strings }on the 1-dimensional edges of the 3-dimensional
cells.
\item In 2+1 dimensions, we made use of the first-order formalism, with
the connection $\A$ and frame field $\ee$ given as Lie-algebra-valued
1-forms. Among other things, this allowed us to perform calculations
elegantly and compactly using index-free notation. In 3+1 dimensions,
the closest thing to this is the formulation of gravity using \emph{Ashtekar
variables}. Therefore, we will work in this formulation.
\end{itemize}
In Chapter \ref{sec:Derivation-of-Ashtekar} we will derive the Ashtekar
variables, which we will use for the remainder of this thesis. Then,
in Chapter \ref{sec:Cosmic-Strings-in}, we will discuss cosmic strings
in 3+1 dimensions, mirroring the discussion of point particles in
2+1 dimensions in Chapter \ref{sec:Point-Particles-in}. Finally,
in Chapter \ref{sec:Discretizing-the-Symplectic} we will discretize
the symplectic potential. The discretization will be more involved
than that of Chapter \ref{sec:Discretizing-the-Symplectic-2P1}, but
the final result will be remarkably similar.

\section{\label{sec:Derivation-of-Ashtekar}Derivation of the Ashtekar Variables}

Let us now begin the lengthy but important task of deriving the Ashtekar
variables. We will start by describing the first-order formulation
of 3+1D gravity, introducing the spin connection and frame field in
Section \ref{subsec:The-Spin-Connection}, the Holst action in Section
\ref{subsec:The-Holst-Action}, and the Hamiltonian formulation in
Section \ref{subsec:The-Hamiltonian-Formulation}.

In Section \ref{subsec:The-Ashtekar-Variables} we will define the
Ashtekar variables themselves, along with useful identities. We will
then proceed, in Section \ref{subsec:The-Action-in}, to rewrite the
Hamiltonian action of first-order gravity using these variables, and
define the Gauss, vector, and scalar constraints. We will derive the
symplectic potential in in Section \ref{subsec:The-Symplectic-Potential-Ashtekar},
and describe the smeared constraints and the symmetries they generate
in Section \ref{subsec:The-Constraints-as}. Finally, Section \ref{subsec:Summary-Ashtekar}
will summarize the results; impatient readers may wish to skip directly
to that section.

\subsection{\label{subsec:The-Spin-Connection}The Spin Connection and Frame
Field}

Let $M=\Sigma\xx\BBR$ be a 3+1-dimensional spacetime manifold, where
$\Sigma$ is a 3-dimensional spatial slice and $\BBR$ represents
time. Please see Section \ref{subsec:Spacetime-and-Spatial} for details
and conventions.

We define a spacetime $\mathfrak{so}\left(3,1\right)$ \emph{spin
connection} 1-form $\omega_{\mu}^{IJ}$ and a \emph{frame field }1-form
$e_{\mu}^{I}$. Here we will use \emph{partially index-free notation},
where only the internal-space indices of the forms are written explicitly:
\[
e^{I}\equiv e_{\mu}^{I}\thinspace\d x^{\mu}\sp\omega^{IJ}\equiv\omega_{\mu}^{IJ}\thinspace\d x^{\mu}.
\]
The frame field is related to the familiar metric by:
\[
g=\eta_{IJ}e^{I}\otimes e^{J}\soosp g_{\mu\nu}=\eta_{IJ}e_{\mu}^{I}e_{\nu}^{J},
\]
where $\eta_{IJ}$ is the Minkowski metric acting on the internal
space indices. Thus, the internal space is flat, and the curvature
is entirely encoded in the fields $e^{I}$; we will see below that
$\omega^{IJ}$ is completely determined by $e^{I}$. We also have
an \emph{inverse frame field}\footnote{Usually the vector $e_{I}^{\mu}$ is called the frame field and the
1-form $e_{\mu}^{I}$ is called the coframe field, but we will ignore
that subtlety here.} $e_{I}^{\mu}$, a vector, which satisfies:
\[
e_{I}^{\mu}e_{\nu}^{I}=\delta_{\nu}^{\mu}\sp e_{I}^{\mu}e_{\mu}^{J}=\delta_{I}^{J}\sp g_{\mu\nu}e_{I}^{\mu}e_{J}^{\nu}=\eta_{IJ}.
\]
We can view $e_{I}^{\mu}$ as a set of four 4-vectors, $e_{1}$, $e_{2}$,
$e_{3}$, and $e_{4}$, which form an \emph{orthonormal basis }(in
Lorentzian signature) with respect to the usual inner product:
\[
\langle x,y\rangle\equiv g_{\mu\nu}x^{\mu}y^{\mu}\soosp\langle e_{I},e_{J}\rangle=\eta_{IJ}.
\]
The familiar \emph{Levi-Civita connection} $\Gamma_{\mu\nu}^{\lambda}$
is related to the spin connection and frame field by
\[
\Gamma_{\mu\nu}^{\lambda}=\omega_{\mu J}^{I}e_{I}^{\lambda}e_{\nu}^{J}+e_{I}^{\lambda}\partial_{\mu}e_{\nu}^{I},
\]
such that there is a \emph{covariant derivative} $\nabla_{\mu}$,
which acts on both spacetime and internal indices, and is \emph{compatible}
with (i.e. annihilates) the frame field:
\begin{equation}
\nabla_{\mu}e_{\nu}^{I}\equiv\partial_{\mu}e_{\nu}^{I}-\Gamma_{\mu\nu}^{\lambda}e_{\lambda}^{I}+\omega_{\mu J}^{I}e_{\nu}^{J}=0.\label{eq:e-comp}
\end{equation}
Now, if we act with the covariant derivative on the internal-space
Minkowski metric $\eta_{IJ}$, we find:
\[
\nabla_{\mu}\eta^{IJ}=\partial_{\mu}\eta^{IJ}+\omega_{\mu K}^{I}\eta^{KJ}+\omega_{\mu K}^{J}\eta^{IK}.
\]
Of course, $\eta^{IJ}$ is constant in spacetime, so $\partial_{\mu}\eta^{IJ}=0$.
If we furthermore demand that the spin connection is metric-compatible
with respect to the internal-space metric, that is $\nabla_{\mu}\eta^{IJ}=0$,
then we get
\[
0=\omega_{\mu K}^{I}\eta^{KJ}+\omega_{\mu K}^{J}\eta^{IK}=\omega_{\mu}^{IJ}+\omega_{\mu}^{JI}=2\omega_{\mu}^{(IJ)}.
\]
We thus conclude that the spin connection must be anti-symmetric in
its internal indices:
\[
\omega_{\mu}^{(IJ)}=0\soosp\omega_{\mu}^{IJ}=\omega_{\mu}^{[IJ]}.
\]
Let us also define the \emph{covariant differential }$\d_{\omega}$
as follows:
\[
\d_{\omega}\phi\equiv\d\phi\sp\d_{\omega}X^{I}\equiv\d X^{I}+\udi{\omega}IJ\wedge X^{J},
\]
where $\phi$ is a scalar in the internal space and $X^{I}$ is a
vector in the internal space. With this we may define the \emph{torsion
2-form}:
\[
T^{I}\equiv\d_{\omega}e^{I}=\d e^{I}+\udi{\omega}IJ\wedge e^{J},
\]
and the \emph{curvature 2-form}:
\[
\udi FIJ\equiv\d_{\omega}\udi{\omega}IJ=\d\udi{\omega}IJ+\udi{\omega}IK\wedge\udi{\omega}KJ.
\]
Note that $\d_{\omega}$, unlike $\d$, is \textbf{not} nilpotent.
Instead, it satisfies the \emph{first Bianchi identity}
\begin{align}
\d_{\omega}^{2}X^{I} & =\d\left(\d X^{I}+\udi{\omega}IK\wedge X^{K}\right)+\udi{\omega}IJ\wedge\left(\d X^{J}+\udi{\omega}JK\wedge X^{K}\right)\nonumber \\
 & =\left(\d\udi{\omega}IK\wedge X^{K}-\udi{\omega}IK\wedge\d X^{K}\right)+\left(\udi{\omega}IJ\wedge\d X^{J}+\udi{\omega}IJ\wedge\udi{\omega}JK\wedge X^{K}\right)\nonumber \\
 & =\d\udi{\omega}IK\wedge X^{K}+\udi{\omega}IJ\wedge\udi{\omega}JK\wedge X^{K}\nonumber \\
 & =\left(\d\udi{\omega}IK+\udi{\omega}IJ\wedge\udi{\omega}JK\right)\wedge X^{K}\nonumber \\
 & =\udi FIK\wedge X^{K}.\label{eq:Bianchi}
\end{align}

\subsection{\label{subsec:The-Holst-Action}The Holst Action}

\subsubsection{The Action and its Variation}

The action of 3+1D gravity (with zero cosmological constant) is given
by the \emph{Holst action}:\footnote{Usually there is also a factor of $1/\kappa$ in front of the action,
where $\kappa\equiv8\pi G$ and $G$ is \emph{Newton's constant}.
However, here we take $\kappa\equiv1$ for brevity.}
\[
S\equiv\fr\int_{M}\left(\star+\frac{1}{\gamma}\right)e_{I}\wedge e_{J}\wedge F^{IJ},
\]
where $\star$ is the internal-space \emph{Hodge dual} such that
\[
\star\left(e_{I}\wedge e_{J}\right)\equiv\hf\epsilon_{IJKL}e^{K}\wedge e^{L},
\]
$\gamma\in\BBR\backslash\left\{ 0\right\} $ is called the \emph{Barbero-Immirzi
parameter}, and
\[
\udi FIJ\equiv\d_{\omega}\udi{\omega}IJ=\d\udi{\omega}IJ+\udi{\omega}IK\wedge\udi{\omega}KJ.
\]
is the \emph{curvature 2-form }defined above. Let us derive the equation
of motion and symplectic potential from the Holst action. Taking the
variation, we get
\[
\delta S=\fr\int_{M}\left(2\left(\star+\frac{1}{\gamma}\right)\delta e_{I}\wedge e_{J}\wedge F^{IJ}+\left(\star+\frac{1}{\gamma}\right)e_{I}\wedge e_{J}\wedge\delta F^{IJ}\right).
\]
In the second term, we use the identity $\delta F^{IJ}=\d_{\omega}\left(\delta\omega^{IJ}\right)$
and integrate by parts to get
\begin{align*}
\left(\star+\frac{1}{\gamma}\right)e_{I}\wedge e_{J}\wedge\delta F^{IJ} & =\left(\star+\frac{1}{\gamma}\right)e_{I}\wedge e_{J}\wedge\d_{\omega}\left(\delta\omega^{IJ}\right)\\
 & =\d_{\omega}\left(\left(\star+\frac{1}{\gamma}\right)e_{I}\wedge e_{J}\wedge\delta\omega^{IJ}\right)-2\left(\star+\frac{1}{\gamma}\right)\d_{\omega}e_{I}\wedge e_{J}\wedge\delta\omega^{IJ}.
\end{align*}
Thus the variation becomes
\[
\delta S=\hf\int_{M}\left(\left(\star+\frac{1}{\gamma}\right)\delta e_{I}\wedge e_{J}\wedge F^{IJ}-\left(\star+\frac{1}{\gamma}\right)\d_{\omega}e_{I}\wedge e_{J}\wedge\delta\omega^{IJ}\right)+\Theta,
\]
where the \emph{symplectic potential} $\Theta$ is the boundary term:
\begin{equation}
\Theta\equiv\fr\int_{\Sigma}\left(\star+\frac{1}{\gamma}\right)e_{I}\wedge e_{J}\wedge\delta\omega^{IJ}.\label{eq:symplectic-Holst}
\end{equation}

\subsubsection{The $\delta\omega$ Variation and the Definition of the Spin Connection}

From the variation with respect to $\delta\omega$ we see that the
\emph{torsion 2-form} must vanish:
\begin{equation}
T^{I}\equiv\d_{\omega}e^{I}=\d e^{I}+\omega_{\ J}^{I}\wedge e^{J}=0.\label{eq:torsion-zero}
\end{equation}
In fact, we can take this equation of motion as a \textbf{definition
}of $\omega$. In other words, the only independent variable in our
theory is going to be the frame field $e^{I}$, and the spin connection
$\omega^{IJ}$ is going to be completely determined by $e^{I}$. Once
$\omega$ is defined in this way, it automatically satisfies this
equation of motion (or equivalently, there is no variation with respect
to $\delta\omega$ in the first place since $\omega$ is not an independent
variable). The formulation where $e$ and $\omega$ are independent
is called \emph{first-order}, and when $\omega$ depends on $e$ it
is called \emph{second-order}.

Note that in the usual metric formulation of general relativity, the
Levi-Civita connection $\Gamma_{\alpha\beta}^{\mu}$ is also taken
to be torsionless, but in the teleparallel formulation we instead
use a connection (the Weitzenböck connection) which is flat but has
torsion; see Chapter \ref{sec:Phase-Space-Polarizations} for more
details.

Let us look at the anti-symmetric part of the compatibility condition
(\ref{eq:e-comp}):
\[
\nabla_{[\mu}e_{\nu]I}=\partial_{[\mu}e_{\nu]I}+\omega_{[\mu|IL|}e_{\nu]}^{L}=0.
\]
Note that the term $\Gamma_{\mu\nu}^{\lambda}e_{\lambda}^{I}$ vanishes
automatically from this equation since $\Gamma_{\left[\mu\nu\right]}^{\lambda}=0$
from requiring that the Levi-Civita connection is torsion-free. Also,
the anti-symmetrizer in $\omega_{[\mu|IL|}e_{\nu]}^{L}$ acts on the
spacetime indices only (i.e. $\mu$ and $\nu$ are not inside the
anti-symmetrizer). Contracting with $e_{J}^{\mu}e_{K}^{\nu}$, we
get
\[
e_{J}^{\mu}e_{K}^{\nu}\left(\partial_{[\mu}e_{\nu]I}+\omega_{[\mu|IL|}e_{\nu]}^{L}\right)=0.
\]
We now permute the indices $I,J,K$ in this equation:
\[
e_{I}^{\mu}e_{J}^{\nu}\left(\partial_{[\mu}e_{\nu]K}+\omega_{[\mu|KL|}e_{\nu]}^{L}\right)=0,
\]
\[
e_{K}^{\mu}e_{I}^{\nu}\left(\partial_{[\mu}e_{\nu]J}+\omega_{[\mu|JL|}e_{\nu]}^{L}\right)=0.
\]
Taking the sum of the last two equations minus the first one, we get:
\begin{align*}
0 & =e_{I}^{\mu}e_{J}^{\nu}\partial_{[\mu}e_{\nu]K}+e_{K}^{\mu}e_{I}^{\nu}\partial_{[\mu}e_{\nu]J}-e_{J}^{\mu}e_{K}^{\nu}\partial_{[\mu}e_{\nu]I}+e_{I}^{\mu}e_{J}^{\nu}\omega_{[\mu|KL|}e_{\nu]}^{L}+e_{K}^{\mu}e_{I}^{\nu}\omega_{[\mu|JL|}e_{\nu]}^{L}-e_{J}^{\mu}e_{K}^{\nu}\omega_{[\mu|IL|}e_{\nu]}^{L}\\
 & =e_{I}^{\mu}e_{J}^{\nu}\partial_{[\mu}e_{\nu]K}+e_{K}^{\mu}e_{I}^{\nu}\partial_{[\mu}e_{\nu]J}-e_{J}^{\mu}e_{K}^{\nu}\partial_{[\mu}e_{\nu]I}+\\
 & \qquad+\hf\left(\omega_{\mu KJ}e_{I}^{\mu}-\omega_{\nu KI}e_{J}^{\nu}\right)+\hf\left(\omega_{\mu JI}e_{K}^{\mu}-\omega_{\nu JK}e_{I}^{\nu}\right)-\hf\left(\omega_{\mu IK}e_{J}^{\mu}-\omega_{\nu IJ}e_{K}^{\nu}\right)\\
 & =e_{I}^{\mu}e_{J}^{\nu}\partial_{[\mu}e_{\nu]K}+e_{K}^{\mu}e_{I}^{\nu}\partial_{[\mu}e_{\nu]J}-e_{J}^{\mu}e_{K}^{\nu}\partial_{[\mu}e_{\nu]I}+\omega_{\mu\left(IJ\right)}e_{K}^{\mu}-\omega_{\mu\left(KI\right)}e_{J}^{\mu}-\omega_{\mu\left[JK\right]}e_{I}^{\mu}.
\end{align*}
Since $\omega_{\mu\left(IJ\right)}=0$, the two symmetric terms cancel,
and we get
\[
\omega_{\mu JK}e_{I}^{\mu}=e_{I}^{\mu}e_{J}^{\nu}\partial_{[\mu}e_{\nu]K}+e_{K}^{\mu}e_{I}^{\nu}\partial_{[\mu}e_{\nu]J}-e_{J}^{\mu}e_{K}^{\nu}\partial_{[\mu}e_{\nu]I}.
\]
Finally, we multiply by $e_{\lambda}^{I}$ to get 
\begin{align*}
\omega_{\lambda JK} & =e_{\lambda}^{I}\left(e_{I}^{\mu}e_{J}^{\nu}\partial_{[\mu}e_{\nu]K}+e_{K}^{\mu}e_{I}^{\nu}\partial_{[\mu}e_{\nu]J}-e_{J}^{\mu}e_{K}^{\nu}\partial_{[\mu}e_{\nu]I}\right)\\
 & =e_{J}^{\nu}\partial_{[\lambda}e_{\nu]K}+e_{K}^{\mu}\partial_{[\mu}e_{\lambda]J}-e_{\lambda}^{I}e_{J}^{\mu}e_{K}^{\nu}\partial_{[\mu}e_{\nu]I}.
\end{align*}
Rearranging and relabeling the indices, we obtain the slightly more
elegant form:
\begin{align*}
\omega_{\mu}^{IJ} & =e^{\lambda I}\partial_{[\mu}e_{\lambda]}^{J}+e^{\lambda J}\partial_{[\lambda}e_{\mu]}^{I}-e_{\mu K}e^{\lambda I}e^{\sigma J}\partial_{[\lambda}e_{\sigma]}^{K}\\
 & =e^{\lambda I}\partial_{[\mu}e_{\lambda]}^{J}-e^{\lambda J}\partial_{[\mu}e_{\lambda]}^{I}-e_{\mu K}e^{\lambda I}e^{\sigma J}\partial_{[\lambda}e_{\sigma]}^{K}\\
 & =2e^{\lambda[I}\partial_{[\mu}e_{\lambda]}^{J]}-e_{\mu K}e^{\lambda I}e^{\sigma J}\partial_{[\lambda}e_{\sigma]}^{K},
\end{align*}
where the first term contains an anti-symmetrizer in both the spacetime
and internal space indices. Thus, $\omega$ is completely determined
by $e$, just as $\Gamma$ is completely determined by $g$ in the
usual metric formulation.

\subsubsection{The $\delta e$ Variation and the Einstein Equation}

From the variation with respect to $\delta e$ we get
\[
e_{J}\wedge\left(\star+\frac{1}{\gamma}\right)F^{IJ}=0.
\]
Note that, from the Bianchi identity (\ref{eq:Bianchi}), we have
$e_{J}\wedge F^{IJ}=\d_{\omega}^{2}e_{J}=0$ by the torsion condition
(\ref{eq:torsion-zero}). In other words, the $\gamma$-dependent
term vanishes on-shell, i.e., when the torsion vanishes. We are therefore
left with 
\[
e_{J}\wedge\star F^{IJ}=0,
\]
which is the \emph{Einstein equation} $R_{\mu\nu}-\hf g_{\mu\nu}R=0$
in first-order form. Note that this equation is independent of $\gamma$;
therefore, the $\gamma$-dependent term in the action does not affect
the physics, at least not at the level of the classical equation of
motion.

Let us prove that this is indeed the Einstein equation. We have
\begin{align*}
0=e_{J}\wedge\star F^{IJ} & =\eta_{JK}e^{K}\wedge\star F^{IJ}\\
 & =\hf\eta_{JK}\epsilon_{LM}^{IJ}e^{K}\wedge F^{LM}\\
 & =\hf\epsilon_{KLM}^{I}e^{K}\wedge F^{LM}\\
 & =\hf\epsilon_{KLM}^{I}e_{\rho}^{K}F_{\mu\nu}^{LM}\d x^{\rho}\wedge\d x^{\mu}\wedge\d x^{\nu}.
\end{align*}
Taking the spacetime Hodge dual of this 3-form, we get
\[
0=\star\left(e_{J}\wedge\star F^{IJ}\right)=\frac{1}{3!\cdot2}\epsilon_{\alpha}^{\rho\mu\nu}\epsilon_{KLM}^{I}e_{\rho}^{K}F_{\mu\nu}^{LM}\d x^{\alpha}.
\]
Of course, we can throw away the numerical factor of $1/3!\cdot2$,
and look at the components of the 1-form:
\[
\epsilon_{\alpha}^{\rho\mu\nu}\epsilon_{KLM}^{I}e_{\rho}^{K}F_{\mu\nu}^{LM}=0.
\]
The relation between the \emph{Riemann tensor}\footnote{The Riemann tensor satisfies the symmetry $R_{\mu\nu\alpha\beta}=R_{\alpha\beta\mu\nu}$,
so we can write it as $R_{\mu\nu}^{\alpha\beta}$ with the convention
that, if the indices are lowered, each pair could be either the first
or second pair of indices, as long as they are adjacent. In other
words, $g_{\alpha\gamma}g_{\beta\delta}R_{\mu\nu}^{\gamma\delta}=R_{\mu\nu\alpha\beta}$
or equivalently $g_{\alpha\gamma}g_{\beta\delta}R_{\mu\nu}^{\gamma\delta}=R_{\alpha\beta\mu\nu}$.} on spacetime and the curvature 2-form is:
\[
F_{\mu\nu}^{LM}=e_{\gamma}^{L}e_{\delta}^{M}R_{\mu\nu}^{\gamma\delta}.
\]
Plugging in, we get
\[
\epsilon_{\alpha}^{\rho\mu\nu}\epsilon_{KLM}^{I}e_{\rho}^{K}e_{\gamma}^{L}e_{\delta}^{M}R_{\mu\nu}^{\gamma\delta}=0.
\]
Multiplying by $e_{I}^{\beta}$, and using the relation
\[
\epsilon_{KLM}^{I}e_{I}^{\beta}e_{\rho}^{K}e_{\gamma}^{L}e_{\delta}^{M}=\epsilon_{\rho\gamma\delta}^{\beta},
\]
we get, after raising $\alpha$ and lowering $\beta$,
\[
\epsilon^{\rho\mu\nu\alpha}\epsilon_{\rho\gamma\delta\beta}R_{\mu\nu}^{\gamma\delta}=0.
\]
Finally, we use the identity
\[
\epsilon^{\rho\mu\nu\alpha}\epsilon_{\rho\gamma\delta\beta}=-2\left(\delta_{\gamma}^{[\mu}\delta_{\delta}^{\nu]}\delta_{\beta}^{\alpha}+\delta_{\gamma}^{[\alpha}\delta_{\delta}^{\mu]}\delta_{\beta}^{\nu}+\delta_{\gamma}^{[\nu}\delta_{\delta}^{\alpha]}\delta_{\beta}^{\mu}\right),
\]
where the minus sign comes from the Lorentzian signature of the metric,
to get:
\[
R_{\beta}^{\alpha}-\hf\delta_{\beta}^{\alpha}R=0,
\]
where we defined the \emph{Ricci tensor} and \emph{Ricci scalar}:
\[
R_{\beta}^{\alpha}\equiv R_{\mu\beta}^{\mu\alpha}\sp R\equiv R_{\mu}^{\mu}.
\]
Lowering $\alpha$, we see that we have indeed obtained the Einstein
equation,
\[
R_{\alpha\beta}-\hf g_{\alpha\beta}R=0,
\]
as desired.

\subsection{\label{subsec:The-Hamiltonian-Formulation}The Hamiltonian Formulation}

\subsubsection{\label{subsec:The-3+1-Split}The 3+1 Split and the Time Gauge}

To go to the Hamiltonian formulation, we split our spacetime manifold
$M$ into space $\Sigma$ and time $\BBR$. We remind the reader that,
as detailed in Section \ref{subsec:Spacetime-and-Spatial}, the spacetime
and spatial indices on both real space and the internal space are
related as follows:
\[
\overbrace{0,\underbrace{1,2,3}_{a}}^{\mu}\sp\overbrace{0,\underbrace{1,2,3}_{i}}^{I}.
\]
Let us decompose the 1-form $e^{I}\equiv e_{\mu}^{I}\d x^{\mu}$:
\[
e^{0}\equiv e_{\mu}^{0}\thinspace\d x^{\mu}=e_{0}^{0}\thinspace\d x^{0}+e_{a}^{0}\thinspace\d x^{a}\sp e^{i}\equiv e_{\mu}^{i}\thinspace\d x^{\mu}=e_{0}^{i}\thinspace\d x^{0}+e_{a}^{i}\thinspace\d x^{a}.
\]
Here we merely changed notation from 3+1D spacetime indices $I,\mu$
to 3D spatial indices $i,a$. However, now we are going to impose
a partial gauge fixing, the \emph{time gauge}, given by
\begin{equation}
e_{a}^{0}=0.\label{eq:time-gauge}
\end{equation}
We also define
\begin{equation}
e_{0}^{0}\equiv N\sp e_{0}^{i}\equiv N^{i},\label{eq:lapse-shift}
\end{equation}
where $N$ is called the \emph{lapse} and $N^{i}$ is called the \emph{shift},
as in the ADM formalism. In other words, we have:
\[
e^{0}=N\thinspace\d x^{0}\sp e^{i}=N^{i}\thinspace\d x^{0}+e_{a}^{i}\thinspace\d x^{a},
\]
or in matrix form,
\[
e_{\mu}^{I}=\left(\begin{array}{cccc}
N &  & N^{i} & \thinspace\\
\\
0 &  & e_{a}^{i}\\
\\
\end{array}\right).
\]
As we will soon see, $N$ and $N^{i}$ are non-dynamical \emph{Lagrange
multipliers}, so we are left with $e_{a}^{i}$ as the only dynamical
degrees of freedom of the frame field -- although they will be further
reduced by the internal gauge symmetry.

\subsubsection{The Hamiltonian}

In order to derive the Hamiltonian, we are going to have to sacrifice
the elegant index-free differential form language (for now) and write
everything in terms of indices. This will allow us to perform the
3+1 split in those indices. Writing the differential forms explicitly
in coordinate basis, that is, $e^{I}\equiv e_{\mu}^{I}\d x^{\mu}$
and so on, we get:
\begin{align*}
e^{I}\wedge e^{J}\wedge F^{KL} & =\left(e_{\mu}^{I}\d x^{\mu}\right)\wedge\left(e_{\nu}^{J}\d x^{\nu}\right)\wedge\left(F_{\rho\sigma}^{KL}\d x^{\rho}\wedge\d x^{\sigma}\right)\\
 & =e_{\mu}^{I}e_{\nu}^{J}F_{\rho\sigma}^{KL}\d x^{\mu}\wedge\d x^{\nu}\wedge\d x^{\rho}\wedge\d x^{\sigma}.
\end{align*}
Note that $\d x^{\mu}\wedge\d x^{\nu}\wedge\d x^{\rho}\wedge\d x^{\sigma}$
is a wedge produce of 1-forms, and is therefore completely anti-symmetric
in the indices $\mu\nu\rho\sigma$, just like the Levi-Civita symbol\footnote{\label{fn:Density}The tilde on the Levi-Civita symbol signifies that
it is not a tensor but a \emph{tensor density}. The symbol is defined
as
\[
\ept_{\mu\nu\rho\sigma}\equiv\begin{cases}
+1 & \textrm{if }\left(\mu\nu\rho\sigma\right)\textrm{ is an even permutation of }\left(0123\right),\\
-1 & \textrm{if }\left(\mu\nu\rho\sigma\right)\textrm{ is an odd permutation of }\left(0123\right),\\
0 & \textrm{if any two indices are the same}.
\end{cases}
\]
By definition this quantity has the same values in every coordinate
system, and thus it cannot be a tensor. Let us define a \emph{tensor
density }$\Tt$ as a quantity related to a proper tensor $T$ by
\[
\Tt=\left|g\right|^{-w/2}T,
\]
where $g$ is the determinant of the metric and $w$ is called the
\emph{density weight}. It can be shown that
\[
\ept_{\mu\nu\rho\sigma}\equiv g^{-1/2}\epsilon_{\mu\nu\rho\sigma},
\]
and therefore the Levi-Civita symbol is a tensor density of weight
$+1$.} $\ept^{\mu\nu\rho\sigma}$. Thus we can write:
\[
\d x^{\mu}\wedge\d x^{\nu}\wedge\d x^{\rho}\wedge\d x^{\sigma}=-\ept^{\mu\nu\rho\sigma}\d x^{0}\wedge\d x^{1}\wedge\d x^{2}\wedge\d x^{3},
\]
where the minus sign comes from the fact that $\sign\left(g\right)=-1$,
and we defined $\ept^{\mu\nu\rho\sigma}\equiv\sign\left(g\right)\ept_{\mu\nu\rho\sigma}$.
To see that this relation is satisfied, simply plug in values for
$\mu,\nu,\rho,\sigma$ and compare both sides. For example, for $\left(\mu\nu\rho\sigma\right)=\left(0123\right)$
we have:
\[
-\ept^{0123}=\ept_{0123}=+1,
\]
and both sides are satisfied. We thus have
\begin{align*}
e^{I}\wedge e^{J}\wedge F^{KL} & =-\ept^{\mu\nu\rho\sigma}e_{\mu}^{I}e_{\nu}^{J}F_{\rho\sigma}^{KL}\d x^{0}\wedge\d x^{1}\wedge\d x^{2}\wedge\d x^{3}\\
 & =-\ept^{\mu\nu\rho\sigma}e_{\mu}^{I}e_{\nu}^{J}F_{\rho\sigma}^{KL}\d t\wedge\d^{3}x.
\end{align*}
Plugging this into the Holst action, we get:\footnote{We chose to write down the internal space Minkowski metric $\eta_{IJ}$
explicitly so that internal space indices $I,J,\ldots$ on differential
forms can always be upstairs and spacetime indices $\mu,\nu,\ldots$
can always be downstairs. This will also remind us that terms with
$I,J=0$ in the summation should get a minus sign, since $\eta_{00}=-1$.}
\begin{align*}
S & =\fr\int_{M}\left(\star+\frac{1}{\gamma}\right)e_{I}\wedge e_{J}\wedge F^{IJ}\\
 & =\fr\int_{M}\left(\hf\epsilon_{IJKL}e^{I}\wedge e^{J}\wedge F^{KL}+\frac{1}{\gamma}\eta_{IK}\eta_{JL}e^{I}\wedge e^{J}\wedge F^{KL}\right)\\
 & =-\fr\int\d t\int\d^{3}x\thinspace\ept^{\mu\nu\rho\sigma}\left(\hf\epsilon_{IJKL}e_{\mu}^{I}e_{\nu}^{J}F_{\rho\sigma}^{KL}+\frac{1}{\gamma}\eta_{IK}\eta_{JL}e_{\mu}^{I}e_{\nu}^{J}F_{\rho\sigma}^{KL}\right)\\
 & =-\fr\int\d t\int\d^{3}x\thinspace\ept^{0abc}\multibrl{2\cdot\hf\epsilon_{IJKL}e_{0}^{I}e_{a}^{J}F_{bc}^{KL}+2\cdot\hf\epsilon_{IJKL}e_{a}^{I}e_{b}^{J}F_{0c}^{KL}+}\\
 & \qquad\qquad\multibrr{+2\cdot\frac{1}{\gamma}\eta_{IK}\eta_{JL}e_{0}^{I}e_{a}^{J}F_{bc}^{KL}+2\cdot\frac{1}{\gamma}\eta_{IK}\eta_{JL}e_{a}^{I}e_{b}^{J}F_{0c}^{KL}},
\end{align*}
where the factors of 2 come from, for example,
\begin{align*}
\ept^{0abc}\hf\epsilon_{IJKL}e_{0}^{I}e_{a}^{J}F_{bc}^{KL}+\ept^{a0bc}\hf\epsilon_{IJKL}e_{a}^{I}e_{0}^{J}F_{bc}^{KL} & =\ept^{0abc}\hf\epsilon_{IJKL}e_{0}^{I}e_{a}^{J}F_{bc}^{KL}+\ept^{a0bc}\hf\epsilon_{IJKL}e_{0}^{J}e_{a}^{I}F_{bc}^{KL}\\
 & =\ept^{0abc}\hf\epsilon_{IJKL}e_{0}^{I}e_{a}^{J}F_{bc}^{KL}+\ept^{a0bc}\hf\epsilon_{JIKL}e_{0}^{I}e_{a}^{J}F_{bc}^{KL}\\
 & =\ept^{0abc}\hf\epsilon_{IJKL}e_{0}^{I}e_{a}^{J}F_{bc}^{KL}+\ept^{0abc}\hf\epsilon_{IJKL}e_{0}^{I}e_{a}^{J}F_{bc}^{KL}\\
 & =2\cdot\ept^{0abc}\hf\epsilon_{IJKL}e_{0}^{I}e_{a}^{J}F_{bc}^{KL}.
\end{align*}
Next, we define the 3-dimensional Levi-Civita symbol as $\ept^{abc}\equiv\ept^{0abc}$.
Then we get:
\begin{align*}
S & =-\hf\int\d t\int\d^{3}x\thinspace\ept^{abc}\multibrl{\hf\epsilon_{IJKL}e_{0}^{I}e_{a}^{J}F_{bc}^{KL}+\hf\epsilon_{IJKL}e_{a}^{I}e_{b}^{J}F_{0c}^{KL}+}\\
 & \qquad\qquad\multibrr{+\frac{1}{\gamma}\eta_{IK}\eta_{JL}e_{0}^{I}e_{a}^{J}F_{bc}^{KL}+\frac{1}{\gamma}\eta_{IK}\eta_{JL}e_{a}^{I}e_{b}^{J}F_{0c}^{KL}}.
\end{align*}
We do the same in the internal indices, defining\footnote{Here, the Levi-Civita symbol $\epsilon^{IJKL}$ is actually a tensor,
not a tensor density, since we are in a flat space -- so we omit
the tilde.} $\epsilon^{ijk}\equiv\epsilon^{0ijk}$. For the first two terms,
we simply take $\left(IJKL\right)=\left(0ijk\right),\left(i0jk\right),\left(ij0k\right),\left(ijk0\right)$
in the sum, which we can do due to the Levi-Civita symbol $\epsilon_{IJKL}$:
\[
\hf\ept^{abc}\epsilon_{IJKL}e_{0}^{I}e_{a}^{J}F_{bc}^{KL}=\hf\ept^{abc}\epsilon_{ijk}\left(\left(e_{0}^{0}e_{a}^{i}-e_{0}^{i}e_{a}^{0}\right)F_{bc}^{jk}+2e_{0}^{i}e_{a}^{j}F_{bc}^{0k}\right),
\]
\[
\hf\ept^{abc}\epsilon_{IJKL}e_{a}^{I}e_{b}^{J}F_{0c}^{KL}=\ept^{abc}\epsilon_{ijk}\left(e_{a}^{0}e_{b}^{i}F_{0c}^{jk}+e_{a}^{i}e_{b}^{j}F_{0c}^{0k}\right).
\]
For the next two terms, we use the fact that
\[
\eta_{IJ}=\begin{cases}
-1 & I=J=0,\\
+1 & I=J\ne0,\\
0 & \textrm{otherwise},
\end{cases}
\]
to split into the following four distinct cases:
\[
\eta_{IK}\eta_{JL}=\begin{cases}
\eta_{00}\eta_{00}=+1,\\
\eta_{00}\eta_{ij}=-\delta_{ij},\\
\eta_{ij}\eta_{00}=-\delta_{ij},\\
\eta_{ik}\eta_{jl}=+\delta_{ik}\delta_{jl}.
\end{cases}
\]
Thus we get (using the fact that $F^{00}=0$ since it's anti-symmetric):
\begin{align*}
\frac{1}{\gamma}\ept^{abc}\eta_{IK}\eta_{JL}e_{0}^{I}e_{a}^{J}F_{bc}^{KL} & =\frac{1}{\gamma}\ept^{abc}\left(e_{0}^{0}e_{a}^{0}F_{bc}^{00}-\delta_{ij}e_{0}^{0}e_{a}^{i}F_{bc}^{0j}-\delta_{ij}e_{0}^{i}e_{a}^{0}F_{bc}^{j0}+\delta_{ik}\delta_{jl}e_{0}^{i}e_{a}^{j}F_{bc}^{kl}\right)\\
 & =\frac{1}{\gamma}\ept^{abc}\left(\delta_{ij}\left(e_{0}^{i}e_{a}^{0}-e_{0}^{0}e_{a}^{i}\right)F_{bc}^{0j}+\delta_{ik}\delta_{jl}e_{0}^{i}e_{a}^{j}F_{bc}^{kl}\right),
\end{align*}
\begin{align*}
\frac{1}{\gamma}\ept^{abc}\eta_{IK}\eta_{JL}e_{a}^{I}e_{b}^{J}F_{0c}^{KL} & =\frac{1}{\gamma}\ept^{abc}\left(e_{a}^{0}e_{b}^{0}F_{0c}^{00}-\delta_{ij}e_{a}^{0}e_{b}^{i}F_{0c}^{0j}-\delta_{ij}e_{a}^{i}e_{b}^{0}F_{0c}^{j0}+\delta_{ik}\delta_{jl}e_{a}^{i}e_{b}^{j}F_{0c}^{kl}\right)\\
 & =\frac{1}{\gamma}\ept^{abc}\left(2\delta_{ij}e_{a}^{i}e_{b}^{0}F_{0c}^{0j}+\delta_{ik}\delta_{jl}e_{a}^{i}e_{b}^{j}F_{0c}^{kl}\right).
\end{align*}
Now, as indicated above, we impose the \emph{time gauge }(\ref{eq:time-gauge})
and define the \emph{lapse }and \emph{shift }(\ref{eq:lapse-shift}):
\[
e_{a}^{0}=0\sp e_{0}^{0}\equiv N\sp e_{0}^{i}\equiv N^{i}\equiv N^{d}e_{d}^{i},
\]
where we have converted the shift into a spatial vector $N^{d}$ instead
of an internal space vector. Plugging in, we get
\[
\hf\ept^{abc}\epsilon_{IJKL}e_{0}^{I}e_{a}^{J}F_{bc}^{KL}=\hf\ept^{abc}\epsilon_{ijk}\left(Ne_{a}^{i}F_{bc}^{jk}+2N^{d}e_{d}^{i}e_{a}^{j}F_{bc}^{0k}\right),
\]
\[
\hf\ept^{abc}\epsilon_{IJKL}e_{a}^{I}e_{b}^{J}F_{0c}^{KL}=\ept^{abc}\epsilon_{ijk}e_{a}^{i}e_{b}^{j}F_{0c}^{0k},
\]
\[
\frac{1}{\gamma}\ept^{abc}\eta_{IK}\eta_{JL}e_{0}^{I}e_{a}^{J}F_{bc}^{KL}=\frac{1}{\gamma}\ept^{abc}\left(\delta_{ik}\delta_{jl}N^{d}e_{d}^{i}e_{a}^{j}F_{bc}^{kl}-\delta_{ij}Ne_{a}^{i}F_{bc}^{0j}\right),
\]
\[
\frac{1}{\gamma}\ept^{abc}\eta_{IK}\eta_{JL}e_{a}^{I}e_{b}^{J}F_{0c}^{KL}=\frac{1}{\gamma}\ept^{abc}\delta_{ik}\delta_{jl}e_{a}^{i}e_{b}^{j}F_{0c}^{kl}.
\]
The action thus becomes, after taking out a factor of $1/\gamma$
and isolating terms proportional to $N$ and $N^{d}$:
\begin{align*}
S & =-\frac{1}{2\gamma}\int\d t\int\d^{3}x\thinspace\ept^{abc}\multisql{\left(\delta_{ik}\delta_{jl}e_{a}^{i}e_{b}^{j}F_{0c}^{kl}+\gamma\epsilon_{ijk}e_{a}^{i}e_{b}^{j}F_{0c}^{0k}\right)+}\\
 & \qquad+N^{d}\left(\delta_{ik}\delta_{jl}e_{d}^{i}e_{a}^{j}F_{bc}^{kl}+\gamma\epsilon_{ijk}e_{d}^{i}e_{a}^{j}F_{bc}^{0k}\right)+\\
 & \qquad\multisqr{-N\left(\delta_{ik}e_{a}^{i}F_{bc}^{0k}-\hf\gamma\epsilon_{ikl}e_{a}^{i}F_{bc}^{kl}\right)}.
\end{align*}

\subsection{\label{subsec:The-Ashtekar-Variables}The Ashtekar Variables}

\subsubsection{The Densitized Triad and Related Identities}

Let us define the \emph{densitized triad}, which is a rank $\left(1,0\right)$
tensor of density weight\footnote{See Footnote \ref{fn:Density} for the definition of a tensor density.
The densitized triad has weight $-1$ since $\det\left(e\right)=\sqrt{\det\left(g\right)}$
has weight $-1$.} $-1$:
\[
\Et_{i}^{a}\equiv\det\left(e\right)e_{i}^{a}.
\]
The inverse triad $e_{i}^{a}$ is related to the inverse metric $g^{ab}$
via
\[
g^{ab}=e_{i}^{a}e_{j}^{b}\delta^{ij}.
\]
Multiplying by $\det\left(g\right)=\det\left(e\right)^{2}$ we get
\[
\det\left(g\right)g^{ab}=\Et_{i}^{a}\Et_{j}^{b}\delta^{ij}.
\]
We now prove some identities. First, consider the determinant identity
for a 3-dimensional matrix,
\[
\epsilon_{ijk}e_{a}^{i}e_{b}^{j}e_{c}^{k}=\det\left(e\right)\ept_{abc}.
\]
Multiplying by $e_{l}^{a}$ and using $e_{a}^{i}e_{l}^{a}=\delta_{l}^{i}$,
we get
\[
\epsilon_{ljk}e_{b}^{j}e_{c}^{k}=\epsilon_{ijk}e_{a}^{i}e_{b}^{j}e_{c}^{k}e_{l}^{a}=\det\left(e\right)e_{l}^{a}\ept_{abc}=\Et_{l}^{a}\ept_{abc}.
\]
Next, multiplying by $\ept^{bcd}$ and using the identity
\[
\ept_{abc}\ept^{bcd}=2\delta_{a}^{d},
\]
we get
\[
\ept^{bcd}\epsilon_{ljk}e_{b}^{j}e_{c}^{k}=\Et_{l}^{a}\ept_{abc}\ept^{bcd}=2\Et_{l}^{d}.
\]
Renaming indices, we obtain the identity
\[
\Et_{i}^{a}=\hf\ept^{abc}\epsilon_{ijk}e_{b}^{j}e_{c}^{k}.
\]
Similarly, one may prove the identity\textbf{
\[
e_{a}^{i}=\frac{\epsilon^{ijk}\ept_{abc}\Et_{j}^{b}\Et_{k}^{c}}{2\det\left(e\right)}.
\]
}Since
\[
\det\left(\Et\right)=\det\left(\det\left(e\right)e_{i}^{a}\right)=\left(\det\left(e\right)\right)^{2},
\]
we obtain an expression for the triad 1-form solely in terms of the
densitized triad:
\[
e_{a}^{i}=\frac{\epsilon^{ijk}\ept_{abc}\Et_{j}^{b}\Et_{k}^{c}}{2\sqrt{\det\left(\Et\right)}}.
\]
Contracting with $\ept^{ade}$, we get 
\begin{align*}
\ept^{ade}e_{a}^{i} & =\frac{\epsilon^{imn}\left(\ept^{ade}\ept_{abc}\right)\Et_{m}^{b}\Et_{n}^{c}}{2\sqrt{\det\left(\Et\right)}}\\
 & =\frac{\epsilon^{imn}\left(\delta_{b}^{d}\delta_{c}^{e}-\delta_{c}^{d}\delta_{b}^{e}\right)\Et_{m}^{b}\Et_{n}^{c}}{2\sqrt{\det\left(\Et\right)}}\\
 & =\frac{\epsilon^{imn}\Et_{m}^{d}\Et_{n}^{e}}{\sqrt{\det\left(\Et\right)}},
\end{align*}
from which we find that
\[
\ept^{abc}e_{a}^{i}=\frac{\epsilon^{ijk}\Et_{j}^{b}\Et_{k}^{c}}{\sqrt{\det\left(E\right)}}.
\]
In conclusion, we have the following definitions and identities:
\[
\Et_{i}^{a}\equiv\det\left(e\right)e_{i}^{a}=\hf\ept^{abc}\epsilon_{ijk}e_{b}^{j}e_{c}^{k},
\]
\[
\epsilon^{ijm}\Et_{m}^{c}=\ept^{abc}e_{a}^{i}e_{b}^{j}\sp\ept^{abc}e_{a}^{j}=e_{p}^{b}e_{q}^{c}\epsilon^{jpq}\det\left(e\right),
\]
\begin{equation}
e_{a}^{i}=\frac{\epsilon^{ijk}\ept_{abc}\Et_{j}^{b}\Et_{k}^{c}}{2\sqrt{\det\left(E\right)}}\sp\ept^{abc}e_{a}^{i}=\frac{\epsilon^{ijk}\Et_{j}^{b}\Et_{k}^{c}}{\sqrt{\det\left(E\right)}}.\label{eq:epsilon-e-id}
\end{equation}

\subsubsection{\label{subsec:The-Ashtekar-Barbero-Connection}The Ashtekar-Barbero
Connection}

Since we have performed a 3+1 split of the spin connection $\omega_{\mu}^{IJ}$,
we can use its individual components to define a new connection on
the spatial slice.

First, we use the fact that the spatial part of the spin connection,
$\omega_{a}^{ij}$, is anti-symmetric in the internal indices, and
thus it behaves as a 2-form on the internal space. This means that
we can take its Hodge dual, and obtain a \emph{dual spin connection}
$\Gamma_{a}^{i}$:
\[
\Gamma_{a}^{i}\equiv-\hf\epsilon_{jk}^{i}\omega_{a}^{jk}\soossp\omega_{a}^{jk}=-\epsilon_{i}^{jk}\Gamma_{a}^{i}.
\]
The minus sign here is meant to make the Gauss law, which we will
derive shortly, have the same relative sign as the Gauss law from
2+1D gravity and Yang-Mills theory; note that, in some other sources,
$\Gamma_{a}^{i}$ is defined without this minus sign.

Importantly, instead of two internal indices, $\Gamma_{a}^{i}$ only
has one. We can do this only in 3 dimensions, since the Hodge dual
takes a $k$-form into a $\left(3-k\right)$-form. We are lucky that
we do, in fact, live in a 3+1-dimensional spacetime, otherwise this
simplification would not have been possible!

Next, we define the \emph{extrinsic curvature} $K_{a}^{i}$:
\[
K_{a}^{i}\equiv\omega_{a}^{i0}=-\omega_{a}^{0i}.
\]
Again, this definition differs by a minus sign from some other sources.
Note that we will extend both definitions to $a=0$, for brevity only;
$\Gamma_{0}$ and $K_{0}$ will not be dynamical variables, as we
shall see.

Using the dual spin connection and the extrinsic curvature, we may
now define the \emph{Ashtekar-Barbero connection} $A_{a}^{i}$:
\[
A_{a}^{i}\equiv\Gamma_{a}^{i}+\gamma K_{a}^{i}.
\]
The original spin connection $\omega_{\mu}^{IJ}$ was 1-form on spacetime
which had two internal indices, and was valued in the Lie algebra
of the Lorentz group, also known as $\mathfrak{so}\left(3,1\right)$.
In short, it was an $\mathfrak{so}\left(3,1\right)$-valued 1-form
on spacetime\footnote{The generators of the Lorentz algebra are $L^{IJ}$ with $I,J\in\left\{ 0,1,2,3\right\} $,
and they are anti-symmetric in $I$ and $J$. They are related to
rotations $J^{I}$ and boosts $K^{I}$ by $J^{I}=\hf\udi{\epsilon}I{JK}L^{JK}$
and $K^{I}=L^{0I}$.}. The three quantities we have defined, $\Gamma_{a}^{i}$, $K_{a}^{i}$,
and $A_{a}^{i}$, resulted from reducing both spacetime and the internal
space from 3+1 dimensions to 3 dimensions. Thus, they are 1-forms
on 3-dimensional space, not spacetime, and the internal space is now
invariant under $\mathfrak{so}\left(3\right)$ only.

Since the Lie algebras $\mathfrak{so}\left(3\right)$ and $\sut$
are isomorphic, and since in Yang-Mills theory we use $\sut$, we
might as well use $\sut$ as the symmetry of our internal space instead
of $\mathfrak{so}\left(3\right)$. Thus, the quantities $\Gamma_{a}^{i}$,
$K_{a}^{i}$ and $A_{a}^{i}$ are all $\sut$-valued 1-forms on 3-dimensional
space. We can also, however, work more generally with some unspecified
(compact) Lie algebra $\mfg$, similar to what we did in the previous
chapters. We will again use index-free notation, as defined in (\ref{eq:index-free-notation}).
In particular, we will write for the connection, frame field, dual
connection and extrinsic curvature:
\[
\A\equiv A_{a}^{i}\ta_{i}\thinspace\d x^{a}\sp\ee\equiv e_{a}^{i}\ta_{i}\thinspace\d x^{a}\sp\Ga\equiv\Gamma_{a}^{i}\ta_{i}\thinspace\d x^{a}\sp\K\equiv K_{a}^{i}\ta_{i}\thinspace\d x^{a},
\]
where $\ta_{i}$ are the generators of $\mfg$.

\subsubsection{The Dual Spin Connection in Terms of the Frame Field}

Recall that in the Lagrangian formulation we had the torsion equation
of motion
\[
T^{I}\equiv\d_{\omega}e^{I}=\d e^{I}+\omega_{\ J}^{I}\wedge e^{J}=0.
\]
Explicitly, the components of the 2-form $T^{I}$ are:
\[
\hf T_{\mu\nu}^{I}=\partial_{[\mu}e_{\nu]}^{I}+\eta_{JK}\omega_{[\mu}^{IJ}e_{\nu]}^{K}.
\]
Taking the spatial components after a 3+1 split in both spacetime
and the internal space, we get
\[
\hf T_{ab}^{i}=\partial_{[a}e_{b]}^{i}-\omega_{[a}^{i0}e_{b]}^{0}+\delta_{jk}\omega_{[a}^{ij}e_{b]}^{k}.
\]
However, after imposing the time gauge $e_{b}^{0}=0$ the middle term
vanishes:
\[
\hf T_{ab}^{i}=\partial_{[a}e_{b]}^{i}+\delta_{jk}\omega_{[a}^{ij}e_{b]}^{k}.
\]
Let us now plug in
\[
\omega_{a}^{ij}=-\epsilon_{l}^{ij}\Gamma_{a}^{l},
\]
to get
\begin{align*}
\hf T_{ab}^{i} & =\partial_{[a}e_{b]}^{i}-\delta_{jk}\epsilon_{l}^{ij}\Gamma_{[a}^{l}e_{b]}^{k}\\
 & =\partial_{[a}e_{b]}^{i}+\epsilon_{kl}^{i}\Gamma_{[a}^{k}e_{b]}^{l}\\
 & \equiv D_{[a}e_{b]}^{i},
\end{align*}
where we have defined the \emph{covariant derivative} $D_{a}$, which
acts on $\mfg$-valued 1-forms $e_{b}^{i}$ as
\[
D_{a}e_{b}^{i}\equiv\partial_{a}e_{b}^{i}+\epsilon_{kl}^{i}\Gamma_{a}^{k}e_{b}^{l}.
\]
The equation $D_{a}e_{b}^{i}=0$ can be seen as the definition of
$\Gamma_{a}^{i}$ in terms of $e_{a}^{i}$, just as $\d_{\omega}e^{I}=0$
defines $\omega^{IJ}$ in terms of $e^{I}$.

In index-free notation, the spatial torsion equation of motion is
simply
\[
\T=\d_{\Ga}\ee=\d\ee+\left[\Ga,\ee\right]=0,
\]
where 
\[
\T\equiv\hf T_{ab}^{i}\ta_{i}\d x^{a}\wedge\d x^{b}
\]
is a $\mfg$-valued 2-form.

\subsubsection{The ``Electric Field''}

We now define the \emph{electric field} 2-form $\E$ as (half) the
commutator of two frame fields:
\[
\E\equiv\hf\left[\ee,\ee\right].
\]
This is analogous to the electric field in electromagnetism and Yang-Mills
theory, and indeed, the main reason for defining the Ashtekar variables
is to make gravity look like Yang-Mills theory.

In terms of components, we have
\[
\E\equiv\hf E_{ab}^{i}\ta_{i}\d x^{a}\wedge\d x^{b}\sp E_{ab}^{i}=\hf\left[\ee,\ee\right]_{ab}^{i}=\epsilon_{jk}^{i}e_{a}^{j}e_{b}^{k}.
\]
Alternatively, starting from the definition $\Et_{i}^{c}\equiv\hf\ept^{abc}\epsilon_{ijk}e_{a}^{j}e_{b}^{k}$
of the densitized triad, we multiply both sides by $\ept_{cde}$ and
get:
\[
\ept_{cde}\Et_{i}^{c}=\hf\left(\ept_{cde}\ept^{abc}\right)\epsilon_{ijk}e_{a}^{j}e_{b}^{k}=\hf\left(\delta_{d}^{a}\delta_{e}^{b}-\delta_{e}^{a}\delta_{d}^{b}\right)\epsilon_{ijk}e_{a}^{j}e_{b}^{k}=\epsilon_{ijk}e_{d}^{j}e_{e}^{k},
\]
which gives us the electric field in terms of the densitized triad:
\[
E_{ab}^{i}=\ept_{abc}\delta^{ij}\Et_{j}^{c}.
\]
Note that in the definition we ``undensitize'' the densitized triad,
which is a tensor density of weight $-1$, by contracting it with
the Levi-Civita tensor density, which has weight $1$. The 2-form
$\E$ is thus a proper tensor.

Now, since $\E=\left[\ee,\ee\right]/2$, we have
\begin{equation}
\d_{\Ga}\E=\hf\d_{\Ga}\left[\ee,\ee\right]=\left[\d_{\Ga}\ee,\ee\right]=\left[\T,\ee\right]=0.\label{eq:dGammaE}
\end{equation}
Therefore, just like the frame field $\ee$, the electric field $\E$
is also torsionless with respect to the connection $\Ga$.

\subsection{\label{subsec:The-Action-in}The Action in Terms of the Ashtekar
Variables}

\subsubsection{The Curvature}

The spacetime components $F_{\mu\nu}^{IJ}$ of the curvature 2-form,
related to the partially-index-free quantity $F^{IJ}$ by
\[
F^{IJ}\equiv\hf F_{\mu\nu}^{IJ}\d x^{\mu}\wedge\d x^{\nu},
\]
are
\[
\hf F_{\mu\nu}^{IJ}=\partial_{[\mu}\omega_{\nu]}^{IJ}+\eta_{KL}\omega_{[\mu}^{IK}\omega_{\nu]}^{LJ}.
\]
Let us write the 3+1 decomposition in spacetime:
\[
\hf F_{0c}^{IJ}=\partial_{[0}\omega_{c]}^{IJ}+\eta_{KL}\omega_{[0}^{IK}\omega_{c]}^{LJ},
\]
\[
\hf F_{bc}^{IJ}=\partial_{[b}\omega_{c]}^{IJ}+\eta_{KL}\omega_{[b}^{IK}\omega_{c]}^{LJ}.
\]
We can further decompose it in the internal space, remembering that
$\eta_{00}=-1$, $\eta_{ij}=\delta_{ij}$ and $\omega^{00}=0$:
\[
\hf F_{0c}^{0k}=\partial_{[0}\omega_{c]}^{0k}+\delta_{mn}\omega_{[0}^{0m}\omega_{c]}^{nk},
\]
\[
\hf F_{0c}^{kl}=\partial_{[0}\omega_{c]}^{kl}-\omega_{[0}^{k0}\omega_{c]}^{0l}+\delta_{mn}\omega_{[0}^{km}\omega_{c]}^{nl},
\]
\[
\hf F_{bc}^{0k}=\partial_{[b}\omega_{c]}^{0k}+\delta_{mn}\omega_{[b}^{0m}\omega_{c]}^{nk},
\]
\[
\hf F_{bc}^{kl}=\partial_{[b}\omega_{c]}^{kl}-\omega_{[b}^{k0}\omega_{c]}^{0l}+\delta_{mn}\omega_{[b}^{km}\omega_{c]}^{nl}.
\]
Plugging the definitions of $\Gamma_{a}^{i}$ and $K_{a}^{i}$ into
these expressions, we obtain:
\[
-\hf F_{0c}^{0k}=\partial_{[0}K_{c]}^{k}+\epsilon_{pq}^{k}K_{[0}^{p}\Gamma_{c]}^{q},
\]
\[
-\hf F_{0c}^{kl}=\epsilon_{p}^{kl}\partial_{[0}\Gamma_{c]}^{p}-K_{[0}^{k}K_{c]}^{l}+\Gamma_{[0}^{k}\Gamma_{c]}^{l},
\]
\[
-\hf F_{bc}^{0k}=\partial_{[b}K_{c]}^{k}+\epsilon_{pq}^{k}K_{[b}^{p}\Gamma_{c]}^{q},
\]
\[
-\hf F_{bc}^{kl}=\epsilon_{p}^{kl}\partial_{[b}\Gamma_{c]}^{p}-K_{[b}^{k}K_{c]}^{l}+\Gamma_{[b}^{k}\Gamma_{c]}^{l}.
\]
Note that, in arriving at these expressions, we obtained terms proportional
to $\delta^{kl}$, but they must vanish, since $F^{kl}$ must be anti-symmetric
in $k,l$.

Now we are finally ready to plug the curvature into the action. For
clarity, we define
\[
S=\frac{1}{\gamma}\int\d t\int\d^{3}x\left(L_{1}+L_{2}+L_{3}\right),
\]
where
\[
L_{1}\equiv-\hf\ept^{abc}e_{a}^{i}e_{b}^{j}\left(\delta_{ik}\delta_{jl}F_{0c}^{kl}+\gamma\epsilon_{ijk}F_{0c}^{0k}\right),
\]
\[
L_{2}\equiv-\hf N^{d}\ept^{abc}e_{d}^{i}e_{a}^{j}\left(\delta_{ik}\delta_{jl}F_{bc}^{kl}+\gamma\epsilon_{ijk}F_{bc}^{0k}\right),
\]
\[
L_{3}\equiv\hf N\ept^{abc}e_{a}^{i}\left(\delta_{ik}F_{bc}^{0k}-\hf\gamma\epsilon_{ikl}F_{bc}^{kl}\right).
\]
Let us calculate these terms one by one.

\subsubsection{$L_{1}$: The Kinetic Term and the Gauss Constraint}

Plugging the curvature into $L_{1}$, we find:
\begin{align*}
L_{1} & =\ept^{abc}e_{a}^{i}e_{b}^{j}\left(\delta_{ik}\delta_{jl}\left(-\hf F_{0c}^{kl}\right)+\gamma\epsilon_{ijk}\left(-\hf F_{0c}^{0k}\right)\right)\\
 & =\ept^{abc}e_{a}^{i}e_{b}^{j}\left(\delta_{ik}\delta_{jl}\left(\epsilon_{p}^{kl}\partial_{[0}\Gamma_{c]}^{p}-K_{[0}^{k}K_{c]}^{l}+\Gamma_{[0}^{k}\Gamma_{c]}^{l}\right)+\gamma\epsilon_{ijk}\left(\partial_{[0}K_{c]}^{k}+\epsilon_{pq}^{k}K_{[0}^{p}\Gamma_{c]}^{q}\right)\right)\\
 & =\hf\ept^{abc}e_{a}^{i}e_{b}^{j}\delta_{ik}\delta_{jl}\left(\epsilon_{p}^{kl}\left(\partial_{0}\Gamma_{c}^{p}-\partial_{c}\Gamma_{0}^{p}\right)-K_{0}^{k}K_{c}^{l}+K_{c}^{k}K_{0}^{l}+\Gamma_{0}^{k}\Gamma_{c}^{l}-\Gamma_{c}^{k}\Gamma_{0}^{l}\right)+\\
 & \qquad+\hf\gamma\ept^{abc}\epsilon_{ijk}e_{a}^{i}e_{b}^{j}\left(\partial_{0}K_{c}^{k}-\partial_{c}K_{0}^{k}+\epsilon_{pq}^{k}\left(K_{0}^{p}\Gamma_{c}^{q}-K_{c}^{p}\Gamma_{0}^{q}\right)\right).
\end{align*}
The densitized triad appears in both lines of $L_{1}$:
\begin{align*}
L_{1} & =\hf\epsilon^{ijm}\Et_{m}^{c}\delta_{ik}\delta_{jl}\left(\epsilon_{p}^{kl}\left(\partial_{0}\Gamma_{c}^{p}-\partial_{c}\Gamma_{0}^{p}\right)-K_{0}^{k}K_{c}^{l}+K_{c}^{k}K_{0}^{l}+\Gamma_{0}^{k}\Gamma_{c}^{l}-\Gamma_{c}^{k}\Gamma_{0}^{l}\right)+\\
 & \qquad+\gamma\Et_{k}^{c}\left(\partial_{0}K_{c}^{k}-\partial_{c}K_{0}^{k}+\epsilon_{pq}^{k}\left(K_{0}^{p}\Gamma_{c}^{q}-K_{c}^{p}\Gamma_{0}^{q}\right)\right)\\
 & =\hf\Et_{m}^{c}\left(2\delta_{p}^{m}\left(\partial_{0}\Gamma_{c}^{p}-\partial_{c}\Gamma_{0}^{p}\right)+\epsilon_{kl}^{m}\left(\Gamma_{0}^{k}\Gamma_{c}^{l}-\Gamma_{c}^{k}\Gamma_{0}^{l}-K_{0}^{k}K_{c}^{l}+K_{c}^{k}K_{0}^{l}\right)\right)+\\
 & \qquad+\gamma\Et_{k}^{c}\left(\partial_{0}K_{c}^{k}-\partial_{c}K_{0}^{k}+\epsilon_{pq}^{k}\left(K_{0}^{p}\Gamma_{c}^{q}-K_{c}^{p}\Gamma_{0}^{q}\right)\right)\\
 & =\Et_{p}^{c}\partial_{0}\Gamma_{c}^{p}+\Gamma_{0}^{p}\partial_{c}\Et_{p}^{c}+\Et_{m}^{c}\epsilon_{kl}^{m}\left(\Gamma_{0}^{k}\Gamma_{c}^{l}-K_{0}^{k}K_{c}^{l}\right)+\\
 & \qquad+\gamma\Et_{k}^{c}\partial_{0}K_{c}^{k}+\gamma K_{0}^{k}\partial_{c}\Et_{k}^{c}+\gamma\Et_{k}^{c}\epsilon_{pq}^{k}\left(K_{0}^{p}\Gamma_{c}^{q}-K_{c}^{p}\Gamma_{0}^{q}\right)\\
 & =\Et_{k}^{c}\partial_{0}\Gamma_{c}^{k}+\Gamma_{0}^{k}\partial_{c}\Et_{k}^{c}+\gamma\Et_{k}^{c}\partial_{0}K_{c}^{k}+\gamma K_{0}^{k}\partial_{c}\Et_{k}^{c}+\Et_{k}^{c}\epsilon_{ij}^{k}\left(\Gamma_{0}^{i}\Gamma_{c}^{j}-K_{0}^{i}K_{c}^{j}\right)+\gamma\Et_{k}^{c}\epsilon_{ij}^{k}\left(K_{0}^{i}\Gamma_{c}^{j}-K_{c}^{i}\Gamma_{0}^{j}\right)\\
 & =\Et_{k}^{c}\partial_{0}\left(\Gamma_{c}^{k}+\gamma K_{c}^{k}\right)+\left(\Gamma_{0}^{k}+\gamma K_{0}^{k}\right)\partial_{c}\Et_{k}^{c}+\epsilon_{ij}^{k}\Et_{k}^{c}\left(\Gamma_{0}^{i}\Gamma_{c}^{j}-K_{0}^{i}K_{c}^{j}+\gamma\Gamma_{0}^{i}K_{c}^{j}+\gamma K_{0}^{i}\Gamma_{c}^{j}\right)\\
 & =\Et_{k}^{c}\partial_{0}\left(\Gamma_{c}^{k}+\gamma K_{c}^{k}\right)+\left(\Gamma_{0}^{i}+\gamma K_{0}^{i}\right)\partial_{c}\Et_{i}^{c}+\Gamma_{0}^{i}\epsilon_{ij}^{k}\Et_{k}^{c}\left(\Gamma_{c}^{j}+\gamma K_{c}^{j}\right)-K_{0}^{i}\epsilon_{ij}^{k}\Et_{k}^{c}\left(K_{c}^{j}-\gamma\Gamma_{c}^{j}\right)\\
 & =\Et_{k}^{c}\partial_{0}\left(\Gamma_{c}^{k}+\gamma K_{c}^{k}\right)+\Gamma_{0}^{i}\left(\partial_{c}\Et_{i}^{c}+\epsilon_{ij}^{k}\left(\Gamma_{c}^{j}+\gamma K_{c}^{j}\right)\Et_{k}^{c}\right)+\gamma K_{0}^{i}\left(\partial_{c}\Et_{i}^{c}-\epsilon_{ij}^{k}\Et_{k}^{c}\left(\frac{1}{\gamma}K_{c}^{j}-\Gamma_{c}^{j}\right)\right)\\
 & =\Et_{k}^{c}\partial_{0}\left(\Gamma_{c}^{k}+\gamma K_{c}^{k}\right)+\\
 & \qquad+\left(\Gamma_{0}^{i}-\frac{1}{\gamma}K_{0}^{i}\right)\left(\partial_{c}\Et_{i}^{c}+\epsilon_{ij}^{k}\left(\Gamma_{c}^{j}+\gamma K_{c}^{j}\right)\Et_{k}^{c}\right)+\left(\frac{1}{\gamma}+\gamma\right)K_{0}^{i}\left(\partial_{c}\Et_{i}^{c}+\epsilon_{ij}^{k}\Gamma_{c}^{j}\Et_{k}^{c}\right),
\end{align*}
where we used the identity $\epsilon_{kl}^{m}\epsilon_{p}^{kl}=2\delta_{p}^{m}$,
integrated by parts the expressions $\Et_{p}^{c}\partial_{c}\Gamma_{0}^{p}$
and $\gamma\Et_{k}^{c}\partial_{c}K_{0}^{k}$, and then relabeled
indices and rearranged terms. Finally, we plug in the Ashtekar-Barbero
connection:
\[
A_{c}^{k}\equiv\Gamma_{c}^{k}+\gamma K_{c}^{k},
\]
define two Lagrange multipliers:
\[
\lambda^{i}\equiv\Gamma_{0}^{i}-\frac{1}{\gamma}K_{0}^{i}\sp\alpha^{i}\equiv\left(\frac{1}{\gamma}+\gamma\right)K_{0}^{i},
\]
and the \emph{Gauss constraint}:
\begin{equation}
G_{i}\equiv\partial_{c}\Et_{i}^{c}+\epsilon_{ij}^{k}A_{c}^{j}\Et_{k}^{c}.\label{eq:Gauss-index}
\end{equation}
The complete expression can now be written simply as:
\begin{equation}
L_{1}=\Et_{k}^{c}\partial_{0}A_{c}^{k}+\lambda^{i}G_{i}+\left(\frac{1}{\gamma}+\gamma\right)K_{0}^{i}\left(\partial_{c}\Et_{i}^{c}+\epsilon_{ij}^{k}\Gamma_{c}^{j}\Et_{k}^{c}\right).\label{eq:L_1}
\end{equation}
The first term is clearly a \emph{kinetic term}, indicating that $A_{c}^{k}$
and $\Et_{k}^{c}$ are \emph{conjugate variables}. The second term
imposes the Gauss constraint, which, as we will see in Section \ref{subsec:The-Constraints-as},
generates $\SUT$ gauge transformations. As for the third term, we
will show in the next subsection that it vanishes by the definition
of $\Gamma_{c}^{j}$.

\subsubsection{The Gauss Constraint in Index-Free Notation}

We can write the Gauss constraint in index-free notation. The covariant
differential of $\E$ in terms of the connection $\A$ is given by
\[
\d_{\A}\E\equiv\d\E+\left[\A,\E\right].
\]
The components of this 3-form, defined as usual by
\[
\d_{\A}\E=\frac{1}{6}\left(\d_{\A}\E\right)_{abc}^{i}\ta_{i}\d x^{a}\wedge\d x^{b}\wedge\d x^{c},
\]
are given by
\begin{align*}
\left(\d_{\A}\E\right)_{abc}^{i} & =\left(\d\E\right)_{abc}^{i}+\left[\A,\E\right]_{abc}^{i}\\
 & =3\left(\partial_{[a}E_{bc]}^{i}+\epsilon_{jk}^{i}A_{[a}^{j}E_{bc]}^{k}\right)\\
 & =3\left(\partial_{[a}\left(\ept_{bc]d}\delta^{il}\Et_{l}^{d}\right)+\epsilon_{jk}^{i}A_{[a}^{j}\left(\ept_{bc]d}\delta^{kl}\Et_{l}^{d}\right)\right)\\
 & =3\ept_{d[bc}\left(\delta^{il}\partial_{a]}\Et_{l}^{d}+\epsilon_{jk}^{i}A_{a]}^{j}\delta^{kl}\Et_{l}^{d}\right).
\end{align*}
Plugging in, we see that
\[
\d_{\A}\E=\hf\ept_{d[bc}\left(\delta^{il}\partial_{a]}\Et_{l}^{d}+\epsilon_{jk}^{i}A_{a]}^{j}\delta^{kl}\Et_{l}^{d}\right)\ta_{i}\d x^{a}\wedge\d x^{b}\wedge\d x^{c}.
\]
Next, we use the relation
\[
\d x^{a}\wedge\d x^{b}\wedge\d x^{c}=\ept^{abc}\d x^{1}\wedge\d x^{2}\wedge\d x^{3}\equiv\ept^{abc}\d^{3}x,
\]
along with the identity
\[
\ept_{dbc}\ept^{abc}=2\delta_{d}^{a},
\]
to find that
\begin{align*}
\d_{\A}\E & =\hf\ept_{dbc}\left(\delta^{il}\partial_{a}\Et_{l}^{d}+\epsilon_{jk}^{i}A_{a}^{j}\delta^{kl}\Et_{l}^{d}\right)\ta_{i}\ept^{abc}\d^{3}x\\
 & =\delta_{d}^{a}\left(\delta^{il}\partial_{a}\Et_{l}^{d}+\epsilon_{jk}^{i}A_{a}^{j}\delta^{kl}\Et_{l}^{d}\right)\ta_{i}\d^{3}x\\
 & =\left(\partial_{a}\Et_{i}^{a}+\epsilon_{ij}^{k}A_{a}^{j}\Et_{k}^{a}\right)\ta^{i}\d^{3}x.
\end{align*}
Finally, we \emph{smear} this 3-form inside a 3-dimensional integral,
with a Lagrange multiplier $\la\equiv\lambda^{i}\ta_{i}$:
\[
\int\la\cdot\d_{\A}\E=\int\lambda^{i}\left(\partial_{a}\Et_{i}^{a}+\epsilon_{ij}^{k}A_{a}^{j}\Et_{k}^{a}\right)\d^{3}x.
\]
We thus see that demanding $\d_{\A}\E=0$ is equivalent to demanding
that (\ref{eq:Gauss-index}) vanishes:
\[
\G=\d_{\A}\E=0\soossp G_{i}\equiv\partial_{a}\Et_{i}^{a}+\epsilon_{ij}^{k}A_{a}^{j}\Et_{k}^{a}=0.
\]
Let us also write (\ref{eq:dGammaE}) with indices in the same way,
replacing $A_{a}^{j}$ with $\Gamma_{a}^{j}$:
\begin{equation}
\d_{\Ga}\E=0\soossp\partial_{a}\Et_{i}^{a}+\epsilon_{ij}^{k}\Gamma_{a}^{j}\Et_{k}^{a}=0.\label{eq:dGammaE2}
\end{equation}
Taking the difference of the two constraints, we get
\begin{align*}
\d_{\A}\E-\d_{\Ga}\E & =\left(\d\E+\left[\Ga+\gamma\K,\E\right]\right)-\left(\d\E+\left[\Ga,\E\right]\right)\\
 & =\gamma\left[\K,\E\right]\\
 & =0,
\end{align*}
or with indices,
\[
\left[\K,\E\right]=0\soosp\epsilon_{ki}^{j}K_{a}^{i}\Et_{j}^{a}=0.
\]
Now, the extrinsic curvature with two spatial indices is symmetric:
\[
K_{ab}=K_{\left(ab\right)}\sp K_{\left[ab\right]}=0.
\]
It is related to $K_{a}^{i}$ by
\[
K_{ab}=K_{a}^{i}e_{b}^{j}\delta_{ij}.
\]
Thus, the condition that its anti-symmetric part vanishes is
\[
K_{[ab]}=K_{[a}^{i}e_{b]}^{j}\delta_{ij}=0.
\]
Contracting with $\det\left(e\right)\epsilon^{klm}e_{k}^{a}e_{l}^{b}$,
we get
\begin{align*}
0 & =\det\left(e\right)\epsilon^{klm}e_{k}^{a}e_{l}^{b}K_{[a}^{i}e_{b]}^{j}\delta_{ij}\\
 & =\hf\det\left(e\right)\epsilon^{klm}e_{k}^{a}e_{l}^{b}\left(K_{a}^{i}e_{b}^{j}-K_{b}^{i}e_{a}^{j}\right)\delta_{ij}\\
 & =\hf\det\left(e\right)\epsilon^{klm}\left(\delta_{l}^{j}e_{k}^{a}K_{a}^{i}-\delta_{k}^{j}e_{l}^{b}K_{b}^{i}\right)\delta_{ij}\\
 & =\hf\det\left(e\right)\epsilon^{klm}\left(\delta_{il}e_{k}^{a}K_{a}^{i}-\delta_{ik}e_{l}^{a}K_{a}^{i}\right)\\
 & =\hf\det\left(e\right)\epsilon^{klm}\left(\delta_{il}e_{k}^{a}K_{a}^{i}+\delta_{il}e_{k}^{a}K_{a}^{i}\right)\\
 & =\det\left(e\right)\epsilon^{klm}\delta_{il}e_{k}^{a}K_{a}^{i}\\
 & =\det\left(e\right)\epsilon_{i}^{mk}e_{k}^{a}K_{a}^{i}\\
 & =\epsilon_{i}^{mk}K_{a}^{i}\Et_{k}^{a}.
\end{align*}
Therefore, $G_{k}=0$ is also equivalent to $K_{\left[ab\right]}=0$.
Yet another way to write this constraint, in index-free notation,
is to define a new quantity \cite{Freidel2019}
\begin{equation}
\P\equiv\d_{\A}\ee,\label{eq:new-quantity}
\end{equation}
such that
\[
\d_{\A}\E=\hf\d_{\A}\left[\ee,\ee\right]=\left[\d_{\A}\ee,\ee\right]=\left[\P,\ee\right].
\]

Finally, given (\ref{eq:dGammaE2}) we can simplify (\ref{eq:L_1})
to
\[
L_{1}=\Et_{k}^{c}\partial_{0}A_{c}^{k}+\lambda^{i}G_{i}.
\]

\subsubsection{$L_{2}$: The Vector (Spatial Diffeomorphism) Constraint}

Plugging the curvature into $L_{2}$, we find:
\begin{align*}
L_{2} & =N^{d}\ept^{abc}e_{d}^{i}e_{a}^{j}\left(\delta_{ik}\delta_{jl}\left(-\hf F_{bc}^{kl}\right)+\gamma\epsilon_{ijk}\left(-\hf F_{bc}^{0k}\right)\right)\\
 & =N^{d}\ept^{abc}e_{d}^{i}e_{a}^{j}\left(\delta_{ik}\delta_{jl}\left(\epsilon_{p}^{kl}\partial_{b}\Gamma_{c}^{p}+\Gamma_{b}^{k}\Gamma_{c}^{l}-K_{b}^{k}K_{c}^{l}\right)+\gamma\epsilon_{ijk}\left(\partial_{b}K_{c}^{k}+\epsilon_{pq}^{k}K_{b}^{p}\Gamma_{c}^{q}\right)\right)\\
 & =N^{d}\ept^{abc}e_{d}^{i}e_{a}^{j}\left(\epsilon_{ijk}\partial_{b}A_{c}^{k}+\delta_{ik}\delta_{jl}\left(\Gamma_{b}^{k}\Gamma_{c}^{l}-K_{b}^{k}K_{c}^{l}\right)+\gamma\left(\delta_{ip}\delta_{jq}-\delta_{iq}\delta_{jp}\right)K_{b}^{p}\Gamma_{c}^{q}\right)\\
 & =N^{d}\ept^{abc}e_{d}^{i}e_{a}^{j}\left(\epsilon_{ijk}\partial_{b}A_{c}^{k}+\delta_{il}\delta_{jm}\left(\left(\Gamma_{b}^{l}\Gamma_{c}^{m}-K_{b}^{l}K_{c}^{m}\right)+\gamma\left(K_{b}^{l}\Gamma_{c}^{m}+\Gamma_{b}^{l}K_{c}^{m}\right)\right)\right).
\end{align*}
The curvature 2-form of the Ashtekar-Barbero connection, for which
we will also use the letter $F$ but with only one internal index,
is defined as:
\[
\hf F_{bc}^{k}\equiv\partial_{[b}A_{c]}^{k}+\hf\epsilon_{lm}^{k}A_{b}^{l}A_{c}^{m}.
\]
Expanding $A_{c}^{k}\equiv\Gamma_{c}^{k}+\gamma K_{c}^{k}$ and contracting
with $\ept^{abc}\epsilon_{ijk}$, we get
\begin{align*}
\hf\ept^{abc}\epsilon_{ijk}F_{bc}^{k} & =\ept^{abc}\epsilon_{ijk}\left(\partial_{b}A_{c}^{k}+\hf\epsilon_{lm}^{k}\left(\Gamma_{b}^{l}+\gamma K_{b}^{l}\right)\left(\Gamma_{c}^{m}+\gamma K_{c}^{m}\right)\right)\\
 & =\ept^{abc}\left(\epsilon_{ijk}\partial_{b}A_{c}^{k}+\hf\left(\delta_{il}\delta_{jm}-\delta_{im}\delta_{jl}\right)\left(\Gamma_{b}^{l}\Gamma_{c}^{m}+\gamma\Gamma_{b}^{l}K_{c}^{m}+\gamma K_{b}^{l}\Gamma_{c}^{m}+\gamma^{2}K_{b}^{l}K_{c}^{m}\right)\right)\\
 & =\ept^{abc}\multibrl{\epsilon_{ijk}\partial_{b}A_{c}^{k}+\hf\delta_{il}\delta_{jm}\left(\Gamma_{b}^{l}\Gamma_{c}^{m}-\Gamma_{b}^{m}\Gamma_{c}^{l}\right)+}\\
 & \qquad\multibrr{+\hf\delta_{il}\delta_{jm}\left(\gamma\left(\Gamma_{b}^{l}K_{c}^{m}+K_{b}^{l}\Gamma_{c}^{m}-\Gamma_{b}^{m}K_{c}^{l}-K_{b}^{m}\Gamma_{c}^{l}\right)+\gamma^{2}\left(K_{b}^{l}K_{c}^{m}-K_{b}^{m}K_{c}^{l}\right)\right)}\\
 & =\ept^{abc}\left(\epsilon_{ijk}\partial_{b}A_{c}^{k}+\delta_{il}\delta_{jm}\left(\Gamma_{b}^{l}\Gamma_{c}^{m}+\gamma\left(K_{b}^{l}\Gamma_{c}^{m}+\Gamma_{b}^{l}K_{c}^{m}\right)+\gamma^{2}K_{b}^{l}K_{c}^{m}\right)\right).
\end{align*}
Therefore
\begin{align*}
 & \ept^{abc}\left(\hf\epsilon_{ijk}F_{bc}^{k}-\delta_{il}\delta_{jm}\left(1+\gamma^{2}\right)K_{b}^{l}K_{c}^{m}\right)=\\
 & =\ept^{abc}\left(\epsilon_{ijk}\partial_{b}A_{c}^{k}+\delta_{il}\delta_{jm}\left(\left(\Gamma_{b}^{l}\Gamma_{c}^{m}-K_{b}^{l}K_{c}^{m}\right)+\gamma\left(K_{b}^{l}\Gamma_{c}^{m}+\Gamma_{b}^{l}K_{c}^{m}\right)\right)\right).
\end{align*}
Plugging into $L_{2}$, we get
\[
L_{2}=N^{d}\ept^{abc}e_{d}^{i}e_{a}^{j}\left(\hf\epsilon_{ijk}F_{bc}^{k}-\delta_{il}\delta_{jm}\left(1+\gamma^{2}\right)K_{b}^{l}K_{c}^{m}\right).
\]
For the next step, we use the identity
\[
\ept^{abc}e_{a}^{j}=e_{p}^{b}e_{q}^{c}\epsilon^{jpq}\det\left(e\right).
\]
Plugging in, we obtain
\begin{align*}
L_{2} & =N^{d}e_{d}^{i}e_{p}^{b}e_{q}^{c}\epsilon^{jpq}\det\left(e\right)\left(\hf\epsilon_{ijk}F_{bc}^{k}-\delta_{il}\delta_{jm}\left(1+\gamma^{2}\right)K_{b}^{l}K_{c}^{m}\right)\\
 & =N^{d}e_{d}^{i}e_{p}^{b}e_{q}^{c}\det\left(e\right)\left(\hf\left(\delta_{k}^{p}\delta_{i}^{q}-\delta_{i}^{p}\delta_{k}^{q}\right)F_{bc}^{k}-\epsilon_{m}^{pq}\delta_{il}\left(1+\gamma^{2}\right)K_{b}^{l}K_{c}^{m}\right)\\
 & =N^{d}\det\left(e\right)\left(\hf e_{d}^{i}\left(e_{p}^{b}e_{i}^{c}F_{bc}^{p}-e_{i}^{b}e_{q}^{c}F_{bc}^{q}\right)-e_{d}^{i}e_{p}^{b}e_{q}^{c}\epsilon_{m}^{pq}\delta_{il}\left(1+\gamma^{2}\right)K_{b}^{l}K_{c}^{m}\right)\\
 & =N^{d}\det\left(e\right)\left(\hf\left(e_{p}^{b}F_{bd}^{p}-e_{q}^{c}F_{dc}^{q}\right)-e_{d}^{i}e_{p}^{b}e_{q}^{c}\epsilon_{m}^{pq}\delta_{il}\left(1+\gamma^{2}\right)K_{b}^{l}K_{c}^{m}\right)\\
 & =N^{d}\det\left(e\right)\left(e_{p}^{b}F_{bd}^{p}-e_{d}^{i}e_{p}^{b}e_{q}^{c}\epsilon_{m}^{pq}\delta_{il}\left(1+\gamma^{2}\right)K_{b}^{l}K_{c}^{m}\right)\\
 & =-N^{a}\det\left(e\right)\left(e_{p}^{b}F_{ab}^{p}+e_{a}^{i}e_{p}^{b}e_{q}^{c}\epsilon_{m}^{pq}\delta_{il}\left(1+\gamma^{2}\right)K_{b}^{l}K_{c}^{m}\right).
\end{align*}
Next, we use the definition of the densitized triad $\Et_{i}^{a}\equiv\det\left(e\right)e_{i}^{a}$:
\[
L_{2}=-N^{a}\left(\Et_{p}^{b}F_{ab}^{p}+\left(1+\gamma^{2}\right)e_{a}^{i}e_{p}^{b}\delta_{il}K_{b}^{l}\epsilon_{m}^{pq}K_{c}^{m}\Et_{q}^{c}\right).
\]
Recall that the Gauss constraint is equivalent to
\[
G_{i}=\gamma\epsilon_{ij}^{k}K_{c}^{j}\Et_{k}^{c},
\]
or, relabeling indices and rearranging,
\[
\epsilon_{m}^{pq}K_{c}^{m}\Et_{q}^{c}=-\frac{1}{\gamma}G^{p}.
\]
Plugging into $L_{2}$, we get
\begin{align*}
L_{2} & =-N^{a}\left(\Et_{p}^{b}F_{ab}^{p}-\left(\frac{1}{\gamma}+\gamma\right)e_{a}^{i}e_{p}^{b}\delta_{il}K_{b}^{l}G^{p}\right)\\
 & =-N^{a}\Et_{p}^{b}F_{ab}^{p}+\left(\frac{1}{\gamma}+\gamma\right)N_{a}K_{p}^{a}G^{p}.
\end{align*}
The part with $G^{p}$ is redundant -- the Gauss constraint is already
enforced by $L_{1}$, and we can combine the second term of $L_{2}$
with $L_{1}$ by redefining some fields. Thus we get 
\[
L_{2}=-N^{a}\Et_{p}^{b}F_{ab}^{p}.
\]
We can now define the \emph{vector (or momentum) constraint}:
\[
V_{a}\equiv\Et_{p}^{b}F_{ab}^{p}.
\]
Then $L_{2}$ simply enforces this constraint with the Lagrange multiplier
$N^{a}$:
\[
L_{2}=-N^{a}V_{a}.
\]
In Section \ref{subsec:The-Constraints-as} we will discuss how this
constraint is related to spatial diffeomorphisms.

\subsubsection{The Vector Constraint in Index-Free Notation}

We have
\begin{align*}
N^{i}\left[\ee,\F\right]_{i} & =N^{i}\epsilon_{ijk}e^{j}\wedge F^{k}\\
 & =N^{i}\epsilon_{ijk}e_{a}^{j}\d x^{a}\wedge\hf F_{bc}^{k}\d x^{b}\wedge\d x^{c}\\
 & =\hf N^{i}\epsilon_{ijk}e_{a}^{j}F_{bc}^{k}\d x^{a}\wedge\d x^{b}\wedge\d x^{c}\\
 & =\hf N^{d}e_{d}^{i}\epsilon_{ijk}e_{a}^{j}F_{bc}^{k}\d x^{a}\wedge\d x^{b}\wedge\d x^{c}\\
 & =\fr N^{d}E_{dak}F_{bc}^{k}\d x^{a}\wedge\d x^{b}\wedge\d x^{c}\\
 & =\fr N^{d}\left(2\ept_{dae}\Et_{k}^{e}\right)F_{bc}^{k}\d x^{a}\wedge\d x^{b}\wedge\d x^{c}\\
 & =\hf N^{d}\ept_{dae}\Et_{k}^{e}F_{bc}^{k}\d x^{a}\wedge\d x^{b}\wedge\d x^{c}\\
 & =\hf N^{d}\ept_{dae}\Et_{k}^{e}F_{bc}^{k}\ept^{abc}\d^{3}x\\
 & =-\hf N^{d}\ept^{abc}\ept_{ade}\Et_{k}^{e}F_{bc}^{k}\d^{3}x\\
 & =-\hf N^{d}\left(\delta_{d}^{b}\delta_{e}^{c}-\delta_{e}^{b}\delta_{d}^{c}\right)\Et_{k}^{e}F_{bc}^{k}\d^{3}x\\
 & =-N^{d}\delta_{d}^{[b}\delta_{e}^{c]}\Et_{k}^{e}F_{bc}^{k}\d^{3}x\\
 & =-N^{b}\Et_{k}^{c}F_{bc}^{k}\d^{3}x.
\end{align*}
Thus, in terms of differential forms, we can write the vector constraint
as
\[
\N\cdot\left[\ee,\F\right]=0.
\]

\subsubsection{$L_{3}$: The Scalar (Hamiltonian) Constraint}

Finally, we plug the curvature into the last term in the action:
\begin{align*}
L_{3} & =-N\ept^{abc}e_{a}^{i}\left(\delta_{ik}\left(-\hf F_{bc}^{0k}\right)-\hf\gamma\epsilon_{ikl}\left(-\hf F_{bc}^{kl}\right)\right)\\
 & =-N\ept^{abc}e_{a}^{i}\left(\delta_{ik}\left(\partial_{b}K_{c}^{k}+\epsilon_{pq}^{k}K_{b}^{p}\Gamma_{c}^{q}\right)-\hf\gamma\epsilon_{ikl}\left(\epsilon_{p}^{kl}\partial_{b}\Gamma_{c}^{p}-K_{b}^{k}K_{c}^{l}+\Gamma_{b}^{k}\Gamma_{c}^{l}\right)\right).
\end{align*}
Using the identity for $\ept^{abc}e_{a}^{i}$ in (\ref{eq:epsilon-e-id}),
we get
\begin{align*}
L_{3} & =-N\frac{\epsilon^{imn}\Et_{m}^{b}\Et_{n}^{c}}{\sqrt{\det\left(E\right)}}\left(\delta_{ik}\left(\partial_{b}K_{c}^{k}+\epsilon_{pq}^{k}K_{b}^{p}\Gamma_{c}^{q}\right)-\hf\gamma\epsilon_{ikl}\left(\epsilon_{p}^{kl}\partial_{b}\Gamma_{c}^{p}-K_{b}^{k}K_{c}^{l}+\Gamma_{b}^{k}\Gamma_{c}^{l}\right)\right)\\
 & =-N\frac{\epsilon^{imn}\Et_{m}^{b}\Et_{n}^{c}}{\sqrt{\det\left(E\right)}}\left(\delta_{ij}\partial_{b}K_{c}^{j}+\epsilon_{ijk}K_{b}^{j}\Gamma_{c}^{k}-\hf\gamma\left(2\delta_{ij}\partial_{b}\Gamma_{c}^{j}+\epsilon_{ijk}\left(\Gamma_{b}^{j}\Gamma_{c}^{k}-K_{b}^{j}K_{c}^{k}\right)\right)\right)\\
 & =N\frac{\epsilon^{imn}\Et_{m}^{b}\Et_{n}^{c}}{\sqrt{\det\left(E\right)}}\left(\gamma\delta_{ij}\partial_{b}\left(\Gamma_{c}^{j}-\frac{1}{\gamma}K_{c}^{j}\right)-\epsilon_{ijk}\left(K_{b}^{j}\Gamma_{c}^{k}+\hf\gamma\left(K_{b}^{j}K_{c}^{k}-\Gamma_{b}^{j}\Gamma_{c}^{k}\right)\right)\right).
\end{align*}
Substituting $K_{c}^{j}=\frac{1}{\gamma}\left(A_{c}^{j}-\Gamma_{c}^{j}\right)$,
we obtain
\begin{align*}
L_{3} & =N\frac{\epsilon^{imn}\Et_{m}^{b}\Et_{n}^{c}}{\sqrt{\det\left(E\right)}}\multibrl{\gamma\delta_{ij}\partial_{b}\left(\Gamma_{c}^{j}-\frac{1}{\gamma^{2}}\left(A_{c}^{j}-\Gamma_{c}^{j}\right)\right)+}\\
 & \qquad\multibrr{-\epsilon_{ijk}\left(\frac{1}{\gamma}\left(A_{b}^{j}-\Gamma_{b}^{j}\right)\Gamma_{c}^{k}+\hf\gamma\left(\frac{1}{\gamma}\left(A_{b}^{j}-\Gamma_{b}^{j}\right)\frac{1}{\gamma}\left(A_{c}^{k}-\Gamma_{c}^{k}\right)-\Gamma_{b}^{j}\Gamma_{c}^{k}\right)\right)}\\
 & =-N\frac{\epsilon_{i}^{mn}\Et_{m}^{b}\Et_{n}^{c}}{\gamma\sqrt{\det\left(E\right)}}\left(\left(\partial_{b}A_{c}^{i}+\frac{1}{2}\epsilon_{jk}^{i}A_{b}^{j}A_{c}^{k}\right)-\left(1+\gamma^{2}\right)\left(\partial_{b}\Gamma_{c}^{i}+\hf\epsilon_{jk}^{i}\Gamma_{b}^{j}\Gamma_{c}^{k}\right)\right)\\
 & =-N\frac{\epsilon_{i}^{mn}\Et_{m}^{b}\Et_{n}^{c}}{2\gamma\sqrt{\det\left(E\right)}}\left(F_{bc}^{i}-\left(1+\gamma^{2}\right)R_{bc}^{i}\right),
\end{align*}
where we identified the curvature of the Ashtekar-Barbero connection:
\[
\hf F_{bc}^{i}\equiv\partial_{[b}A_{c]}^{i}+\hf\epsilon_{jk}^{i}A_{b}^{j}A_{c}^{k},
\]
as well as the curvature of the spin connection:
\[
\hf R_{bc}^{i}\equiv\partial_{[b}\Gamma_{c]}^{i}+\hf\epsilon_{jk}^{i}\Gamma_{b}^{j}\Gamma_{c}^{k}.
\]
If we define the \emph{scalar} (or \emph{Hamiltonian}) \emph{constraint}:
\begin{equation}
C\equiv-\frac{\epsilon_{i}^{mn}\Et_{m}^{b}\Et_{n}^{c}}{2\gamma\sqrt{\det\left(E\right)}}\left(F_{bc}^{i}-\left(1+\gamma^{2}\right)R_{bc}^{i}\right),\label{eq:scalar-R}
\end{equation}
then $L_{3}$ is simply
\[
L_{3}=NC,
\]
and it imposes the scalar constraint via the Lagrange multiplier $N$.
The scalar constraint generates the time evolution of the theory,
that is, from one spatial slice to another.

\subsubsection{The Scalar Constraint in Terms of the Extrinsic Curvature}

We can rewrite the scalar constraint in another way which is more
commonly encountered. First we plug $A_{b}^{j}=\Gamma_{b}^{j}+\gamma K_{b}^{j}$
into the curvature of the Ashtekar-Barbero connection:
\begin{align*}
F_{bc}^{i} & =2\partial_{[b}\left(\Gamma_{c]}^{i}+\gamma K_{c]}^{i}\right)+\epsilon_{jk}^{i}\left(\Gamma_{b}^{j}+\gamma K_{b}^{j}\right)\left(\Gamma_{c}^{k}+\gamma K_{c}^{k}\right)\\
 & =2\left(\partial_{[b}\Gamma_{c]}^{i}+\hf\epsilon_{jk}^{i}\Gamma_{b}^{j}\Gamma_{c}^{k}\right)+\gamma^{2}\epsilon_{jk}^{i}K_{b}^{j}K_{c}^{k}+2\gamma\left(\partial_{[b}K_{c]}^{i}+\epsilon_{jk}^{i}\Gamma_{b}^{j}K_{c}^{k}\right)\\
 & =R_{bc}^{i}+\gamma^{2}\epsilon_{jk}^{i}K_{b}^{j}K_{c}^{k}+2\gamma D_{[b}\left(\Gamma\right)K_{c]}^{i},
\end{align*}
where we defined the covariant derivative of $K_{c}^{i}$ with respect
to the spin connection:
\[
D_{[b}\left(\Gamma\right)K_{c]}^{i}\equiv\partial_{[b}K_{c]}^{i}+\epsilon_{jk}^{i}\Gamma_{b}^{j}K_{c}^{k}.
\]
Now, since the spin connection $\Gamma_{a}^{k}$ is compatible with
the triad, we have:
\[
\hf T_{ab}^{i}\equiv D_{a}\left(\Gamma\right)e_{b}^{i}\equiv\partial_{a}e_{b}^{i}+\epsilon_{kl}^{i}\Gamma_{a}^{k}e_{b}^{l}=0.
\]
However,
\[
D_{a}\left(\Gamma\right)\left(e_{b}^{i}e_{i}^{b}\right)=D_{a}\left(\Gamma\right)\left(3\right)=0\soosp e_{i}^{b}D_{a}\left(\Gamma\right)e_{b}^{i}=-e_{b}^{i}D_{a}\left(\Gamma\right)e_{i}^{b}.
\]
Therefore
\begin{align*}
0 & =e_{b}^{i}D_{a}\left(\Gamma\right)e_{i}^{b}\\
 & =-e_{i}^{b}D_{a}\left(\Gamma\right)e_{b}^{i}\\
 & =-e_{i}^{b}\left(\partial_{a}e_{b}^{i}+\epsilon_{kl}^{i}\Gamma_{a}^{k}e_{b}^{l}\right)\\
 & =e_{b}^{i}\left(\partial_{a}e_{i}^{b}+\epsilon_{ik}^{l}\Gamma_{a}^{k}e_{l}^{b}\right),
\end{align*}
and we obtain that
\[
D_{a}\left(\Gamma\right)e_{i}^{b}\equiv\partial_{a}e_{i}^{b}+\epsilon_{ik}^{l}\Gamma_{a}^{k}e_{l}^{b}=0.
\]
Multiplying by $\det\left(e\right)$ and using the definition of $\Et_{i}^{b}$,
we get
\[
D_{a}\left(\Gamma\right)\Et_{i}^{b}\equiv\partial_{a}\Et_{i}^{b}+\epsilon_{ik}^{l}\Gamma_{a}^{k}\Et_{l}^{b}=0.
\]
Thus, when we contract $F_{bc}^{i}$ with $\epsilon_{i}^{mn}\Et_{m}^{b}\Et_{n}^{c}$,
we can insert $\gamma\epsilon_{i}^{mn}\Et_{n}^{c}$ into the covariant
derivative: 
\begin{align*}
\epsilon_{i}^{mn}\Et_{m}^{b}\Et_{n}^{c}F_{bc}^{i} & =\epsilon_{i}^{mn}\Et_{m}^{b}\Et_{n}^{c}\left(R_{bc}^{i}+\gamma^{2}\epsilon_{jk}^{i}K_{b}^{j}K_{c}^{k}\right)+2\Et_{m}^{b}D_{b}\left(\Gamma\right)\left(\gamma\epsilon_{i}^{mn}K_{c}^{i}\Et_{n}^{c}\right)\\
 & =\epsilon_{i}^{mn}\Et_{m}^{b}\Et_{n}^{c}\left(R_{bc}^{i}+\gamma^{2}\epsilon_{jk}^{i}K_{b}^{j}K_{c}^{k}\right)-2\Et_{m}^{b}D_{b}\left(\Gamma\right)G^{m},
\end{align*}
where we used
\[
G^{m}=-\gamma\epsilon_{i}^{mn}K_{c}^{i}\Et_{n}^{c}.
\]
Rearranging terms, we get
\[
\epsilon_{i}^{mn}\Et_{m}^{b}\Et_{n}^{c}R_{bc}^{i}=\epsilon_{i}^{mn}\Et_{m}^{b}\Et_{n}^{c}\left(F_{bc}^{i}-\gamma^{2}\epsilon_{jk}^{i}K_{b}^{j}K_{c}^{k}\right)+2\Et_{m}^{b}D_{b}\left(\Gamma\right)G^{m}.
\]
Plugging into $C$, we obtain
\begin{align*}
C & =-\frac{1}{2\gamma\sqrt{\det\left(E\right)}}\left(\epsilon_{i}^{mn}\Et_{m}^{b}\Et_{n}^{c}F_{bc}^{i}-\left(1+\gamma^{2}\right)\epsilon_{i}^{mn}\Et_{m}^{b}\Et_{n}^{c}R_{bc}^{i}\right)\\
 & =-\frac{1}{2\gamma\sqrt{\det\left(E\right)}}\left(\epsilon_{i}^{mn}\Et_{m}^{b}\Et_{n}^{c}F_{bc}^{i}-\left(1+\gamma^{2}\right)\left(\epsilon_{i}^{mn}\Et_{m}^{b}\Et_{n}^{c}\left(F_{bc}^{i}-\gamma^{2}\epsilon_{jk}^{i}K_{b}^{j}K_{c}^{k}\right)+2\Et_{m}^{b}D_{b}\left(\Gamma\right)G^{m}\right)\right)\\
 & =\frac{\gamma\epsilon_{i}^{mn}\Et_{m}^{b}\Et_{n}^{c}}{2\sqrt{\det\left(E\right)}}\left(F_{bc}^{i}-\left(1+\gamma^{2}\right)\epsilon_{jk}^{i}K_{b}^{j}K_{c}^{k}\right)+\frac{\left(1+\gamma^{2}\right)\Et_{m}^{b}}{\gamma\sqrt{\det\left(E\right)}}D_{b}\left(\Gamma\right)G^{m}.
\end{align*}
Now, if the Gauss constraint is satisfied, then the second term is
redundant, and we can get rid of it. We obtain the familiar expression
for the scalar or Hamiltonian constraint:
\[
C=\frac{\gamma\epsilon_{i}^{mn}\Et_{m}^{b}\Et_{n}^{c}}{2\sqrt{\det\left(E\right)}}\left(F_{bc}^{i}-\left(1+\gamma^{2}\right)\epsilon_{jk}^{i}K_{b}^{j}K_{c}^{k}\right).
\]

\subsubsection{The Scalar Constraint in Index-Free Notation}

Finally, let us write the scalar constraint in index-free notation.
We first use the identity (\ref{eq:epsilon-e-id}):
\[
\frac{\epsilon^{imn}\Et_{m}^{b}\Et_{n}^{c}}{\sqrt{\det\left(E\right)}}=\ept^{abc}e_{a}^{i}.
\]
Plugging in, and ignoring the overall factor of $\gamma$, we get
\[
C=\hf\ept^{abc}\delta_{il}e_{a}^{l}\left(F_{bc}^{i}-\left(1+\gamma^{2}\right)\epsilon_{jk}^{i}K_{b}^{j}K_{c}^{k}\right).
\]
Now, from our definition of the graded dot product (see Section \ref{subsec:The-Graded-Dot})
we have:
\begin{align*}
\ee\cdot\F & =\delta_{il}e^{l}\wedge F^{i}\\
 & =\delta_{il}\left(e_{a}^{l}\d x^{a}\right)\left(\hf F_{bc}^{i}\wedge\d x^{b}\wedge\d x^{c}\right)\\
 & =\hf\delta_{il}e_{a}^{l}F_{bc}^{i}\d x^{a}\wedge\d x^{b}\wedge\d x^{c}\\
 & =\hf\delta_{il}e_{a}^{l}F_{bc}^{i}\ept^{abc}\d^{3}x.
\end{align*}
Furthermore, from our definition (\ref{eq:triple-product}) of the
triple product we have
\begin{align*}
\ee\cdot\left[\K,\K\right] & =\delta_{il}\epsilon_{jk}^{i}e^{l}\wedge K^{j}\wedge K^{k}\\
 & =\delta_{il}\epsilon_{jk}^{i}\left(e_{a}^{l}\d x^{a}\right)\wedge\left(K_{b}^{j}\d x^{b}\right)\wedge\left(K_{c}^{k}\d x^{c}\right)\\
 & =\delta_{il}\epsilon_{jk}^{i}e_{a}^{l}K_{b}^{j}K_{c}^{k}\d x^{a}\wedge\d x^{b}\wedge\d x^{c}\\
 & =\delta_{il}\epsilon_{jk}^{i}e_{a}^{l}K_{b}^{j}K_{c}^{k}\ept^{abc}\d^{3}x.
\end{align*}
Thus we can write
\[
C\equiv\ee\cdot\left(\F-\frac{1+\gamma^{2}}{2}\left[\K,\K\right]\right).
\]
If we smear this 3-form inside a 3-dimensional integral, with a Lagrange
multiplier $N$, we get the appropriate expression for the scalar
constraint.

Furthermore, let us consider again the new quantity defined in (\ref{eq:new-quantity}):
\begin{align*}
\P & \equiv\d_{\A}\ee=\d\ee+\left[\A,\ee\right]\\
 & =\d\ee+\left[\Ga+\gamma\K,\ee\right]\\
 & =\d_{\Ga}\ee+\gamma\left[\K,\ee\right]\\
 & =\gamma\left[\K,\ee\right],
\end{align*}
since $\d_{\Ga}\ee=0$. Thus we can write
\[
\ee\cdot\left[\K,\K\right]=\K\cdot\left[\K,\ee\right]=\frac{1}{\gamma}\K\cdot\P,
\]
and the scalar constraint becomes
\[
C\equiv\ee\cdot\F-\hf\left(\frac{1}{\gamma}+\gamma\right)\K\cdot\P.
\]
In this form of the constraint, it is clear that it is automatically
satisfied if $\F=\P=0$.

Similarly, for (\ref{eq:scalar-R}),
\[
C=-\frac{\epsilon_{i}^{mn}\Et_{m}^{b}\Et_{n}^{c}}{2\gamma\sqrt{\det\left(E\right)}}\left(F_{bc}^{i}-\left(1+\gamma^{2}\right)R_{bc}^{i}\right),
\]
we again use (\ref{eq:epsilon-e-id}) to get, ignoring the overall
factor of $-1/\gamma$,
\[
C=\hf\ept^{abc}\delta_{il}e_{a}^{l}\left(F_{bc}^{i}-\left(1+\gamma^{2}\right)R_{bc}^{i}\right).
\]
Then, we have as before
\[
\ee\cdot\F=\hf\delta_{il}e_{a}^{l}F_{bc}^{i}\ept^{abc}\d^{3}x,
\]
and similarly
\[
\ee\cdot\RR=\hf\delta_{il}e_{a}^{l}R_{bc}^{i}\ept^{abc}\d^{3}x,
\]
where
\[
\RR\equiv\d_{\Ga}\Ga=\d\Ga+\hf\left[\Ga,\Ga\right].
\]
The scalar constraint can thus be written simply as
\[
C=\ee\cdot\left(\F-\left(1+\gamma^{2}\right)\RR\right).
\]

\subsection{\label{subsec:The-Symplectic-Potential-Ashtekar}The Symplectic Potential}

Above, we found the symplectic potential \ref{eq:symplectic-Holst}
of the Holst action:
\[
\Theta=\fr\int_{\Sigma}\left(\star+\frac{1}{\gamma}\right)e_{I}\wedge e_{J}\wedge\delta\omega^{IJ}.
\]
Let us rewrite it in terms of the 3-dimensional internal indices,
using the 3-dimesional internal-space Levi-Civita symbol $\epsilon_{ijk}\equiv\epsilon_{0ijk}$:
\begin{align*}
\Theta & =\fr\int_{\Sigma}\left(\hf\epsilon_{IJKL}e^{I}\wedge e^{J}\wedge\delta\omega^{KL}+\frac{1}{\gamma}e_{I}\wedge e_{J}\wedge\delta\omega^{IJ}\right)\\
 & =\fr\int_{\Sigma}\left(\epsilon_{ijk}e^{0}\wedge e^{i}\wedge\delta\omega^{jk}+\epsilon_{ijk}e^{i}\wedge e^{j}\wedge\delta\omega^{0k}+\frac{2}{\gamma}e_{0}\wedge e_{i}\wedge\delta\omega^{0i}+\frac{1}{\gamma}e_{i}\wedge e_{j}\wedge\delta\omega^{ij}\right).
\end{align*}
Since $e^{0}=0$ on $\Sigma$ due to the time gauge, the two terms
with $e^{0}$ vanish and we are left with:
\[
\Theta=\fr\int_{\Sigma}\left(\epsilon_{ijk}e^{i}\wedge e^{j}\wedge\delta\omega^{0k}+\frac{1}{\gamma}e_{i}\wedge e_{j}\wedge\delta\omega^{ij}\right).
\]
Recall that we defined the electric field as
\[
\E\equiv\hf\left[\ee,\ee\right],
\]
or with indices
\[
E_{i}=\epsilon_{ijk}e^{j}\wedge e^{k}\soossp e_{i}\wedge e_{j}=\hf\epsilon_{ijk}E^{k}.
\]
Thus our symplectic potential becomes
\[
\Theta=\frac{1}{4\gamma}\int_{\Sigma}E_{i}\wedge\delta\left(\hf\epsilon_{jk}^{i}\omega^{jk}+\gamma\omega^{0i}\right).
\]
We identify here the dual spin connection and extrinsic curvature
defined in Section \ref{subsec:The-Ashtekar-Barbero-Connection}:
\[
\Gamma_{a}^{i}\equiv-\hf\epsilon_{jk}^{i}\omega_{a}^{jk}\sp K_{a}^{i}\equiv-\omega_{a}^{0i},
\]
so the expression in parentheses is none other than the (minus) Ashtekar-Barbero
connection:
\[
\A\equiv\Ga+\gamma\K\soosp A_{a}^{i}\equiv\Gamma_{a}^{i}+\gamma K_{a}^{i}=-\hf\epsilon_{jk}^{i}\omega_{a}^{jk}-\gamma\omega_{a}^{0i}.
\]
Ignoring the irrelevant overall factor, the symplectic potential now
reaches its final form
\begin{equation}
\Theta=\int_{\Sigma}\E\cdot\delta\A.\label{eq:symplectic-potential}
\end{equation}

\subsection{\label{subsec:The-Constraints-as}The Constraints as Generators of
Symmetries}

Let $\CC$ be the space of smooth connections on $\Sigma$. The kinematical
(unconstrained) phase space of 3+1-dimensional gravity is given by
the cotangent bundle $\PP\equiv T^{*}\CC$. To get the physical (that
is, gauge-invariant) phase space, we must perform symplectic reductions
with respect to the constraints. These constraints are best understood
in their smeared form as generators of gauge transformations. The
smeared \emph{Gauss constraint} can be written as
\[
\GG\left(\al\right)\equiv\hf\int_{\Sigma}\la\cdot\d_{\A}\E,
\]
where $\la$ is a $\mfg$-valued 0-form. This constraint generates
the infinitesimal $G$ gauge transformations:
\[
\left\{ \A,\GG\left(\la\right)\right\} \propto\d_{\A}\la\sp\left\{ \E,\GG\left(\la\right)\right\} \propto\left[\E,\la\right].
\]
The smeared \emph{vector constraint} is given by
\[
\VV\left(\xi\right)\equiv\int_{\Sigma}\N\cdot\left[\ee,\F\right],
\]
where $\xi^{a}$ is a spatial vector and the Lagrange multiplier $N^{i}\equiv\xi^{a}e_{a}^{i}$
is a $\mfg$-valued 0-form. From the Gauss and vector constraints
we may construct the \emph{diffeomorphism constraint}:
\[
\DD\left(\xi\right)\equiv\VV\left(\xi\right)-\GG\left(\xi\lrc\A\right),
\]
where $\xi\lrc\A\equiv\xi^{a}A_{a}^{i}\ta_{i}$ is an interior product
(see Footnote \ref{fn:The-interior-product}). This constraint generates
the infinitesimal spatial diffeomorphism transformations
\[
\left\{ \A,\DD\left(\xi\right)\right\} \propto\LLL_{\xi}\A\sp\left\{ \E,\DD\left(\xi\right)\right\} \propto\LLL_{\xi}\E,
\]
where $\LLL_{\xi}$ is the \emph{Lie derivative}.

\subsection{\label{subsec:Summary-Ashtekar}Summary}

\subsubsection{The Ashtekar Action with Indices}

We have found that the \emph{Ashtekar action }of classical loop gravity
is:
\[
S=\frac{1}{\gamma}\int\d t\int_{\Sigma}\d^{3}x\left(\Et_{i}^{a}\partial_{t}A_{a}^{i}+\lambda^{i}G_{i}+N^{a}V_{a}+NC\right),
\]
where:
\begin{itemize}
\item $\gamma$ is the \emph{Barbero-Immirzi parameter},
\item $\Sigma$ is a 3-dimensional spatial slice,
\item $a,b,c,\ldots\in\left\{ 1,2,3\right\} $ are spatial indices on $\Sigma$,
\item $i,j,k,\ldots\in\left\{ 1,2,3\right\} $ are indices in the Lie algebra
$\mfg$,
\item $\Et_{i}^{a}\equiv\det\left(e\right)e_{i}^{a}$ is the \emph{densitized
triad}, a rank $\left(1,0\right)$ tensor of density weight $-1$,
where $e_{i}^{a}$ is the inverse \emph{frame field} (or \emph{triad}),
related to the inverse spatial metric $g^{ab}$ via $g^{ab}=e_{i}^{a}e_{j}^{b}\delta^{ij}$,
\item $A_{a}^{i}$ is the \emph{Ashtekar-Barbero connection},
\item $\partial_{t}$ is the derivative with respect to the time coordinate
$t$, such that each spatial slice is at a constant value of $t$,
\item $\lambda^{i}$, $N^{a}$ and $N$ are Lagrange multipliers,
\item $G_{i}\equiv\partial_{a}\Et_{i}^{a}+\epsilon_{ij}^{k}A_{a}^{j}\Et_{k}^{a}$
is the \emph{Gauss constraint},
\item $V_{a}\equiv\Et_{i}^{b}F_{ab}^{i}$ is the \emph{vector (or momentum
or diffeomorphism) constraint}, where $F_{ab}^{i}$ is the \emph{curvature
}of the Ashtekar-Barbero connection:
\[
F_{ab}^{i}\equiv\partial_{a}A_{b}^{i}-\partial_{b}A_{a}^{i}+\epsilon_{jk}^{i}A_{a}^{j}A_{b}^{k}.
\]
\item $C$ is the\emph{ scalar (or Hamiltonian) constraint}, defined as
\[
C\equiv\frac{\epsilon_{i}^{mn}\Et_{m}^{a}\Et_{n}^{b}}{2\sqrt{\det\left(\Et\right)}}\left(F_{ab}^{i}-\left(1+\gamma^{2}\right)\epsilon_{jk}^{i}K_{a}^{j}K_{b}^{k}\right),
\]
where $K_{a}^{i}$ is the \emph{extrinsic curvature}.
\end{itemize}
From the first term in the action, we see that the connection and
densitized triad are conjugate variables, and they form the Poisson
algebra
\[
\left\{ A_{a}^{i}\left(x\right),A_{b}^{j}\left(y\right)\right\} =\left\{ \Et_{i}^{a}\left(x\right),\Et_{j}^{b}\left(y\right)\right\} =0,
\]
\[
\left\{ A_{a}^{i}\left(x\right),\Et_{j}^{b}\left(y\right)\right\} =\gamma\delta_{j}^{i}\delta_{a}^{b}\delta\left(x,y\right).
\]

\subsubsection{The Ashtekar Action in Index-Free Notation}

In index-free notation, the action takes the form
\[
S=\frac{1}{\gamma}\int\d t\int_{\Sigma}\left(\E\cdot\partial_{t}\A+\la\cdot\left[\ee,\P\right]+\N\cdot\left[\ee,\F\right]+N\left(\ee\cdot\F-\hf\left(\frac{1}{\gamma}+\gamma\right)\K\cdot\P\right)\right),
\]
where now:
\begin{itemize}
\item $\E\equiv\hf\left[\ee,\ee\right]$ is the electric field 2-form, defined
in terms of the densitized triad and the frame field as 
\[
\E\equiv\hf E_{ab}^{i}\ta_{i}\d x^{a}\wedge\d x^{b}\soosp E_{ab}^{i}=\ept_{abc}\delta^{ij}\Et_{j}^{c}=\epsilon_{jk}^{i}e_{a}^{j}e_{b}^{k}.
\]
\item $\A\equiv A_{a}^{i}\ta_{i}\d x^{a}$ is the $\mfg$-valued Ashtekar-Barbero
connection 1-form.
\item $\ee\equiv e_{a}^{i}\ta_{i}\d x^{a}$ is the $\mfg$-valued frame
field 1-form.
\item $\P\equiv\d_{\A}\ee$ is a $\mfg$-valued 2-form.
\item The Gauss constraint is $\la\cdot\left[\ee,\P\right]=\la\cdot\d_{\A}\E$
where the Lagrange multiplier $\la$ is a $\mfg$-valued 0-form.
\item The vector (or momentum or diffeomorphism) constraint is $\N\cdot\left[\ee,\F\right]$
where the Lagrange multiplier $\N$ is a $\mfg$-valued 0-form and
$\F$ is the $\mfg$-valued curvature 2-form
\[
\F\equiv\d_{\A}\A\equiv\d\A+\hf\left[\A,\A\right].
\]
\item The scalar (or Hamiltonian) constraint is $N\left(\ee\cdot\F-\hf\left(\frac{1}{\gamma}+\gamma\right)\K\cdot\P\right)$
where the Lagrange multiplier $N$ is a 0-form, \textbf{not }valued
in $\mfg$, and $\K\equiv K_{a}^{i}\ta_{i}\d x^{a}$ is the $\mfg$-valued
extrinsic curvature 1-form.
\end{itemize}
The symplectic potential, in index-free notation, is
\[
\Theta=\int_{\Sigma}\E\cdot\delta\A.
\]

\section{\label{sec:Cosmic-Strings-in}Cosmic Strings in 3+1 Dimensions}

We now generalize the discussion of point particles in 2+1 dimensions,
which we derived in Chapter \ref{sec:Point-Particles-in}, to one
more dimension, obtaining (cosmic) strings in 3+1 dimensions.

\subsection{\label{sec:Delta-function-proof}Proof that $\protect\d^{2}\phi=2\pi\delta^{\left(2\right)}\left(r\right)$}

We begin by proving a relation between $\d^{2}\phi$ and the Dirac
delta 2-form in cylindrical coordinates, similar to the one we derived
in Section \ref{sec:Delta-Functions-Solid-Angles}.

Let us define a cylinder $\Sigma$ with coordinates $\left(r,\phi,z\right)$
such that 
\[
r\in\left[0,R\right]\sp\phi\in\left[0,2\pi\right)\sp z\in\left[-\frac{L}{2},+\frac{L}{2}\right].
\]
Furthermore, let 
\[
f\equiv f_{r}\thinspace\d r+f_{\phi}\thinspace\d\phi+f_{z}\thinspace\d z
\]
be a test 1-form such that
\begin{equation}
\partial_{\phi}f_{z}\left(r=0\right)=0.\label{eq:delta-con}
\end{equation}
The condition (\ref{eq:delta-con}) means that that value of the 1-form
on the string itself, $f\left(r=0\right)$, is the same for each value
of $\phi$. This certainly makes sense, as different values of $\phi$
at $r=0$ (for a particular choice of $z$) correspond to the same
point.

We define a 2-form distribution $\delta^{\left(2\right)}\left(r\right)$
such that 
\begin{equation}
\int_{\Sigma}f\wedge\delta^{\left(2\right)}\left(r\right)=\int_{\left\{ r=0\right\} }f,\label{eq:delta-def}
\end{equation}
where $\left\{ r=0\right\} $ is the line along the $z$ axis. Let
us now show that the 2-form $\d^{2}\phi$ satisfies this definition.

Using the graded Leibniz rule we have, since $f$ is a 1-form, 
\[
f\wedge\d^{2}\phi=\d f\wedge\d\phi-\d\left(f\wedge\d\phi\right).
\]

Integrating this on $\Sigma$, we get 
\begin{equation}
\int_{\Sigma}f\wedge\d^{2}\phi=\int_{\Sigma}\d f\wedge\d\phi-\int_{\Sigma}\d\left(f\wedge\d\phi\right).\label{eq:int}
\end{equation}
The second integral in (\ref{eq:int}) can easily be integrated using
Stokes' theorem: 
\[
\int_{\Sigma}\d\left(f\wedge\d\phi\right)=\int_{\partial\Sigma}f\wedge\d\phi=\int_{\partial\Sigma}\left(f_{r}\thinspace\d r+f_{z}\thinspace\d z\right)\wedge\d\phi.
\]
The boundary of the cylinder consists of three parts: 
\[
\partial\Sigma=\left\{ r=R\right\} \cup\left\{ z=-\frac{L}{2}\right\} \cup\left\{ z=+\frac{L}{2}\right\} .
\]
Note that $\d r=0$ for the first part and $\d z=0$ for the second
and third; thus 
\begin{equation}
\int_{\Sigma}\d\left(f\wedge\d\phi\right)=\int_{\left\{ r=R\right\} }f_{z}\thinspace\d z\wedge\d\phi+\int_{\left\{ z=\pm L/2\right\} }f_{r}\thinspace\d r\wedge\d\phi.\label{eq:dfdphi}
\end{equation}
As for the first integral in (\ref{eq:int}), we have 
\begin{align*}
\int_{\Sigma}\d f\wedge\d\phi & =\int_{\Sigma}\d\left(f_{r}\thinspace\d r+f_{\phi}\thinspace\d\phi+f_{z}\thinspace\d z\right)\wedge\d\phi\\
 & =\int_{\Sigma}\left(\partial_{z}f_{r}\thinspace\d z\wedge\d r+\partial_{r}f_{z}\d r\wedge\d z\right)\wedge\d\phi.
\end{align*}
For the first term we find 
\begin{align*}
\int_{\Sigma}\partial_{z}f_{r}\thinspace\d z\wedge\d r\wedge\d\phi & =\int_{\phi=0}^{2\pi}\int_{r=0}^{R}\left(\int_{z=-L/2}^{+L/2}\partial_{z}f_{r}\thinspace\d z\right)\d r\wedge\d\phi\\
 & =\int_{\phi=0}^{2\pi}\int_{r=0}^{R}\left(f_{r}\left(z=+\frac{L}{2}\right)-f_{r}\left(z=-\frac{L}{2}\right)\right)\d r\wedge\d\phi\\
 & =\int_{\left\{ z=\pm L/2\right\} }f_{r}\thinspace\d r\wedge\d\phi,
\end{align*}
where the orientation of the boundary at $z=-L/2$ is chosen to be
opposite to that at $z=+L/2$, and for the second term we find 
\begin{align*}
\int_{\Sigma}\partial_{r}f_{z}\d r\wedge\d z\wedge\d\phi & =\int_{\phi=0}^{2\pi}\int_{z=-L/2}^{+L/2}\left(\int_{r=0}^{R}\partial_{r}f_{z}\d r\right)\d z\wedge\d\phi\\
 & =\int_{\phi=0}^{2\pi}\int_{z=-L/2}^{+L/2}\left(f_{z}\left(r=R\right)-f_{z}\left(r=0\right)\right)\d z\wedge\d\phi\\
 & =\int_{\left\{ r=R\right\} }f_{z}\thinspace\d z\wedge\d\phi-\int_{\phi=0}^{2\pi}\int_{z=-L/2}^{+L/2}f_{z}\left(r=0\right)\d z\wedge\d\phi.
\end{align*}
In conclusion, given (\ref{eq:dfdphi}) we see that 
\[
\int_{\Sigma}\d f\wedge\d\phi=\int_{\Sigma}\d\left(f\wedge\d\phi\right)-\int_{\phi=0}^{2\pi}\int_{z=-L/2}^{+L/2}f_{z}\left(r=0\right)\d z\wedge\d\phi,
\]
and therefore (\ref{eq:int}) becomes
\[
\int_{\Sigma}f\wedge\d^{2}\phi=-\int_{\phi=0}^{2\pi}\int_{z=-L/2}^{+L/2}f_{z}\left(r=0\right)\d z\wedge\d\phi.
\]
Finally, due to the condition (\ref{eq:delta-con}), we can rewrite
this as: 
\begin{align*}
\int_{\Sigma}f\wedge\d^{2}\phi & =\int_{z=-L/2}^{+L/2}\int_{\phi=0}^{2\pi}\left(f_{z}\left(r=0\right)\d\phi\right)\d z\\
 & =2\pi\int_{z=-L/2}^{+L/2}f_{z}\left(r=0\right)\d z.
\end{align*}
Noting that $\d\phi=\d r=0$ along the line $\left\{ r=0\right\} $,
we see that 
\[
\int_{\left\{ r=0\right\} }f=\int_{\left\{ r=0\right\} }f_{z}\thinspace\d z,
\]
and thus we find that 
\begin{equation}
\int_{\Sigma}f\wedge\d^{2}\phi=2\pi\int_{\left\{ r=0\right\} }f.\label{eq:d2phi-delta}
\end{equation}
Given (\ref{eq:delta-def}), we see that indeed $\d^{2}\phi=2\pi\delta^{\left(2\right)}\left(r\right)$,
as we wanted to prove.

Note that the delta 2-form distribution may be written as
\[
\delta^{\left(2\right)}\left(r\right)=\delta\left(r\right)\d x\wedge\d y=\delta\left(r\right)r\thinspace\d r\wedge\d\phi,
\]
where $\delta\left(r\right)$ is the usual 1-dimensional delta function.
Therefore we have
\[
\d^{2}\phi=2\pi\delta\left(r\right)r\thinspace\d r\wedge\d\phi.
\]

\subsection{The Frame Field and Spin Connection}

To describe a cosmic string in 3+1 dimensions, we use cylindrical
coordinates $\left(t,r,\phi,z\right)$ with the (infinite) string
lying along the $z$ axis. The metric will be similar to (\ref{eq:2p1-metric}),
except that now the parameter $S$ will cause periodicity in the $z$
direction instead of the $t$ direction:
\[
\d s^{2}=-\d t^{2}+\frac{\d r^{2}}{\left(1-M\right)^{2}}+r^{2}\thinspace\d\phi^{2}+\left(\d z+S\thinspace\d\phi\right)^{2}.
\]
Similar to the 2+1D case, we can define
\begin{equation}
T\equiv t\sp R\equiv\frac{r}{1-M}\sp\Phi\equiv\left(1-M\right)\phi\sp Z\equiv z+S\phi,\label{eq:flat-coords-3p1}
\end{equation}
and the metric becomes flat:
\[
\d s^{2}=-\d T^{2}+\d R^{2}+R^{2}\thinspace\d\Phi^{2}+\d Z^{2},
\]
with the periodicity conditions
\[
\Phi\sim\Phi+2\pi\left(1-M\right)\sp Z\sim Z+2\pi S.
\]
Since we do not have a time shift, we will not create closed timelike
curves, which may potentially violate causality. Instead, the periodicity
is in the $Z$ direction. When we foliate spacetime into 3-dimensional
spatial slices in order to go to the Hamiltonian formulation, $Z$
will play the same role that $T$ played in the 2+1D case (which did
\textbf{not} involve a foliation).

We proceed as in the 2+1D case. First we define the frame fields:
\begin{equation}
e^{0}=\d T\sp e^{1}=\d R\sp e^{2}=R\thinspace\d\Phi\sp e^{3}=\d Z.\label{eq:frame-field}
\end{equation}
Unfortunately, the trick we used in 2+1D, which allowed us to have
only one internal index in the spin connection, does not work in 3+1D,
since the Hodge dual will simply turn 2 indices into $4-2=2$ indices.
Thus, we must use the following definition of the torsion 2-form:
\[
T^{I}\equiv\d_{\omega}e^{I}=\d e^{I}+\udi{\omega}IJ\wedge e^{J}.
\]
The four components of the torsion are:
\[
T^{0}=\udi{\omega}01\wedge\d R+\udi{\omega}02\wedge R\thinspace\d\Phi+\udi{\omega}03\wedge\d Z,
\]
\[
T^{1}=\udi{\omega}10\wedge\d T+\udi{\omega}12\wedge R\thinspace\d\Phi+\udi{\omega}13\wedge\d Z,
\]
\[
T^{2}=\d R\wedge\d\Phi+\udi{\omega}20\wedge\d T+\udi{\omega}21\wedge\d R+\udi{\omega}23\wedge\d Z,
\]
\[
T^{3}=\udi{\omega}30\wedge\d T+\udi{\omega}31\wedge\d R+\udi{\omega}32\wedge R\thinspace\d\Phi.
\]
In order for the torsion to vanish, all of the components of $\udi{\omega}IJ$
must be set to zero except
\[
\udi{\omega}21=\d\Phi,
\]
which is needed in order to cancel the $\d R\wedge\d\Phi$ term in
$T^{2}$. Note that, since the metric on the internal space is flat,
we have that $\udi{\omega}21=\omega^{21}=-\omega^{12}=\d\Phi$. Finally,
we go back to the original coordinates using (\ref{eq:flat-coords-3p1}):
\begin{equation}
\udi{\omega}21=\left(1-M\right)\d\phi=-\udi{\omega}12\sp e^{0}=\d t\sp e^{1}=\frac{\d r}{1-M}\sp e^{2}=r\thinspace\d\phi\sp e^{3}=\d z+S\thinspace\d\phi.\label{eq:spin-frame}
\end{equation}

Then the torsion becomes
\[
T^{0}=T^{1}=T^{2}=0\sp T^{3}=S\thinspace\d^{2}\phi=2\pi S\delta\left(r\right)\d r\wedge\d\phi.
\]
We may also calculate the curvature of the spin connection, which
is defined as
\[
\udi RIJ\equiv\d_{\omega}\udi{\omega}IJ=\d\udi{\omega}IJ+\udi{\omega}IK\wedge\udi{\omega}KJ.
\]
Its components will all be zero, except
\[
\udi R12=-\udi R21=-\left(1-M\right)\d^{2}\phi=-2\pi\left(1-M\right)\delta\left(r\right)\d r\wedge\d\phi.
\]

\subsection{The Foliation of Spacetime and the Ashtekar Variables}

To go to the Hamiltonian formulation, we perform the 3+1 split and
impose the time gauge, as detailed in Section \ref{subsec:The-3+1-Split}:
\[
e^{0}=N\thinspace\d t\sp e^{i}=N^{i}\thinspace\d t+e_{a}^{i}\thinspace\d x^{a}.
\]
From (\ref{eq:spin-frame}) we see that we are already in the time
gauge, and the lapse and shift are trivial, $N=1$ and $N^{i}=0$,
as one would indeed expect from a flat spacetime.

Since the spatial slices are 3-dimensional, we can now use the trick
of turning two internal-space indices into one by defining the dual
spin connection as in the 2+1-dimensional case:
\[
\Gamma^{i}\equiv-\hf\udi{\epsilon}i{jk}\omega^{jk}.
\]
Since the only non-zero components of $\omega^{ij}$ are $\omega^{21}=-\omega^{12}=\left(1-M\right)\d\phi$,
we get:
\[
\Gamma^{1}=\Gamma^{2}=0\sp\Gamma^{3}=\left(1-M\right)\d\phi.
\]
The frame field on each spatial slice is simply
\[
e^{1}=\frac{\d r}{1-M}\sp e^{2}=r\thinspace\d\phi\sp e^{3}=\d z+S\thinspace\d\phi.
\]
We can now use index-free notation again:
\[
\Ga=\left(1-M\right)\ta_{3}\thinspace\d\phi\sp\ee=\frac{\ta_{1}\thinspace\d r}{1-M}+\ta_{3}\thinspace\d z+\left(S\ta_{3}+r\ta_{2}\right)\d\phi.
\]
The torsion will be
\[
\T\equiv\d_{\Ga}\ee=\d\ee+\left[\Ga,\ee\right]=S\ta_{3}\d^{2}\phi=2\pi S\thinspace\delta\left(r\right)\ta_{3}\thinspace\d r\wedge\d\phi.
\]
As we derived above, the first Ashtekar variable is the electric field
$\E$, defined as
\[
\E\equiv\hf\left[\ee,\ee\right]\soosp E^{i}=\hf\udi{\epsilon}i{jk}e^{j}\wedge e^{k}.
\]
Calculating it, we get
\[
\E=\frac{\left(r\ta_{3}-S\ta_{2}\right)\d r\wedge\d\phi+\left(\ta_{1}+\ta_{2}\right)\d z\wedge\d r}{1-M}+r\ta_{1}\d\phi\wedge\d z.
\]
The second Ashtekar variables is the Ashtekar-Barbero connection $\A$,
defined as
\[
\A\equiv\Ga+\gamma\K,
\]
where $\gamma$ is the \emph{Barbero-Immirzi parameter} and $\K$
is the extrinsic curvature, defined as
\[
K_{a}^{i}\equiv\omega_{a}^{i0}.
\]
In our case, it is clear that the extrinsic curvature vanishes; this
makes sense, as we are on equal-time slices in an essentially flat
spacetime. Therefore the Ashtekar connection is in fact identical
to the dual spin connection:
\[
\A=\Ga=\left(1-M\right)\ta_{3}\thinspace\d\phi.
\]
The curvatures of these connections are:
\[
\RR\equiv\d_{\Ga}\Ga=\d\Ga+\hf\left[\Ga,\Ga\right],
\]
\[
\F\equiv\d_{\A}\A=\d\A+\hf\left[\A,\A\right],
\]
and they are both equal to:
\[
\RR=\F=\left(1-M\right)\ta_{3}\thinspace\d^{2}\phi=2\pi\left(1-M\right)\delta\left(r\right)\ta_{3}\thinspace\d r\wedge\d\phi.
\]
We may define $\m\equiv\left(1-M\right)\ta_{3}$ and $\s\equiv S\ta_{3}$,
where $\ta_{3}$ takes the role that $\ta_{0}$ had in the 2+1D case.
Then we may write
\[
\Ga=\A=\m\thinspace\d\phi\sp\ee=\frac{\ta_{1}\thinspace\d r}{1-M}+\ta_{3}\thinspace\d z+\left(\s+r\ta_{2}\right)\d\phi,
\]
\[
\T=\P=2\pi\s\thinspace\delta\left(r\right)\thinspace\d r\wedge\d\phi\sp\RR=\F=2\pi\m\thinspace\delta\left(r\right)\d r\wedge\d\phi.
\]

\subsection{The Dressed Quantities}

As in the 2+1-dimensional case, in order to make the connection and
frame field invariant under a gauge transformation, we must dress
them. Then we get:
\[
\Ga=\A=h^{-1}\m h\thinspace\d\phi+h^{-1}\d h,
\]
\[
\ee=h^{-1}\left(\d\x+\left(\s+\left[\m,\x\right]\right)\d\phi\right)h,
\]
\[
\E=h^{-1}\left(\hf\left[\d\x,\d\x\right]+\left[\d\x,\left(\s+\left[\m,\x\right]\right)\d\phi\right]\right)h,
\]
\[
\RR=\F=2\pi h^{-1}\m h\thinspace\delta\left(r\right)\d r\wedge\d\phi,
\]
\[
\T=\P=2\pi h^{-1}\left(\s+\left[\m,\x\right]\right)h\thinspace\delta\left(r\right)\d r\wedge\d\phi.
\]
We see that we have essentially obtained exactly the same expressions
as for the point particle in 2+1 dimensions -- unsurprisingly, since
we explicitly started with the same metric as in 2+1 dimensions, only
with one more dimension. A string is simply a delta-function source
of curvature and torsion in an otherwise flat and torsionless spacetime.
Just as we did in Section \ref{subsec:The-Dressed-Quantities}, we
can write the curvature and torsion in terms of ``momentum'' $\p$
and ``angular momentum'' $\j$:

\[
\RR=\F=2\pi\p\thinspace\delta\left(r\right)\d r\wedge\d\phi\sp\T=\P=2\pi\j\thinspace\delta\left(r\right)\d r\wedge\d\phi.
\]

\clearpage{}

\part{\label{part:3P1-discrete}3+1 Dimensions: The Discrete Theory}

\section{\label{sec:The-Discrete-Geometry-3P1}The Discrete Geometry}

\subsection{The Cellular Decomposition and Its Dual}

The cellular decomposition in the 3+1-dimensional case is similar
to the one we had in the 2+1-dimensional case, as described in Section
\ref{subsec:The-Cellular-Decomposition}, except that there is, of
course, one more dimension. Each element of the cellular decomposition
$\Delta$ is uniquely dual to an element of the dual cellular decomposition
$\Delta^{*}$. \emph{Cells }$c$ are now 3-dimensional but still dual
to \emph{nodes }$c^{*}$, \emph{sides} $s$ form the 2-dimensional
boundaries of the cells and are dual to \emph{links }$s^{*}$ which
connect the nodes, \emph{edges }$e$ form the 1-dimensional boundaries
of the sides and are dual to \emph{faces }$e^{*}$, and \emph{vertices
}$v$ are dual to \emph{volumes }$v^{*}$. This is summarized in the
following table:
\begin{center}
\begin{tabular}{|c|c|c|}
\hline 
$\Delta$ &  & $\Delta^{*}$\tabularnewline
\hline 
\hline 
0-cells (\emph{vertices}) $v$ & dual to & 3-cells (\emph{volumes}) $v^{*}$\tabularnewline
\hline 
1-cells (\emph{edges}) $e$ & dual to & 2-cells (\emph{faces}) $e^{*}$\tabularnewline
\hline 
2-cells \emph{(sides)} $s$ & dual to & 1-cells (\emph{links}) $s^{*}$\tabularnewline
\hline 
3-cells \emph{(cells)} $c$ & dual to & 0-cells (\emph{nodes}) $c^{*}$\tabularnewline
\hline 
\end{tabular}
\par\end{center}

\begin{figure}[H]
\begin{centering}
\includegraphics[width=1\textwidth]{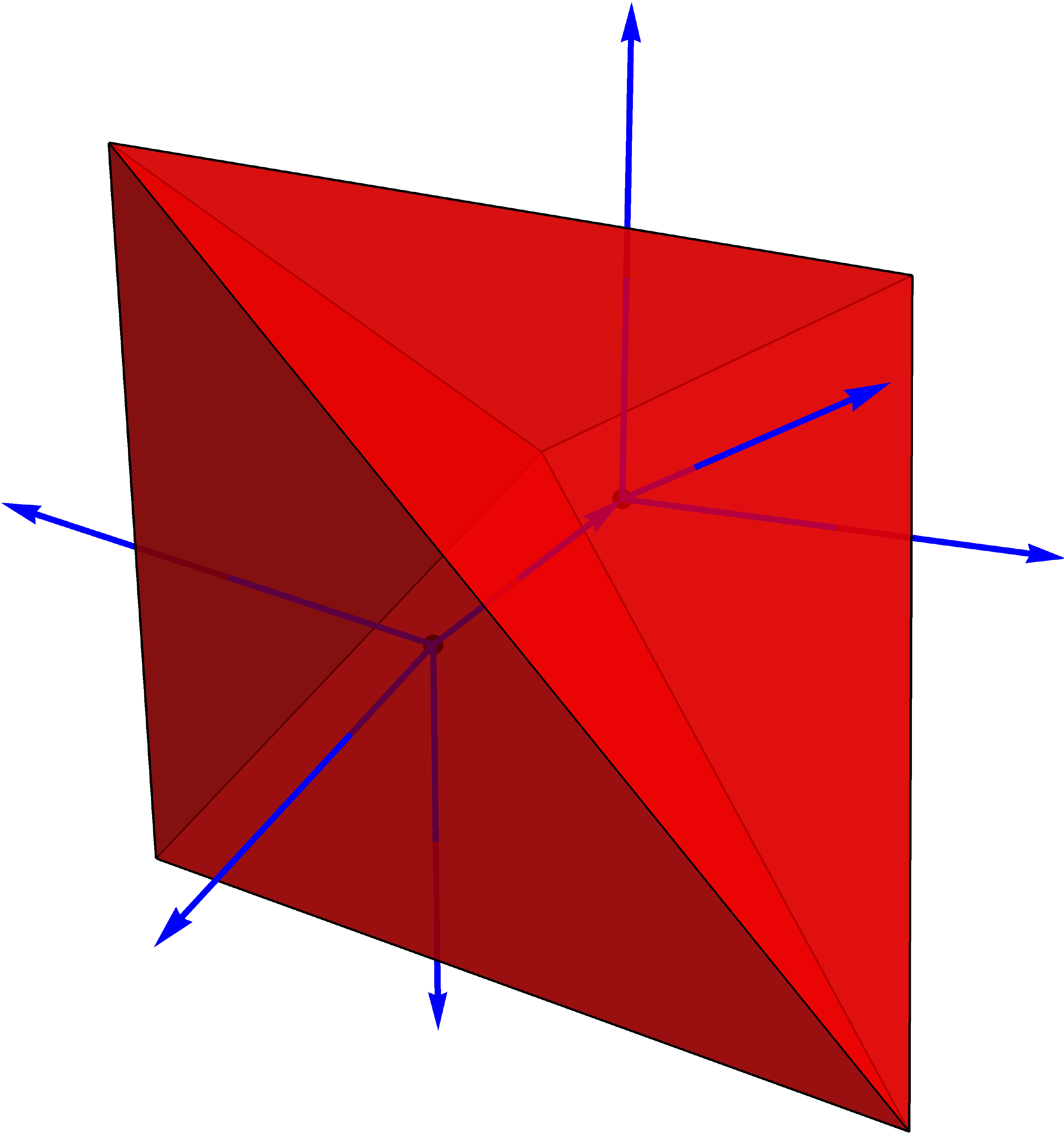}
\par\end{centering}
\caption{\label{fig:3P1TwoTetras}A simple discretization with two cells (in
red) and a dual spin network (in blue). Here, the discretization is
the simplest possible one: a \emph{simplicial complex}, where the
cells are 3-simplices (tetrahedrons) and their faces are 2-simplices
(triangles). The edges are 1-simplices (line segments), and the vertices
are, of course, 0-simplices (points). However, our formalism also
allows the cells to be arbitrary convex polyhedra and the faces to
be arbitrary convex polygons. From the figure, it should be clear
that each of the two cells is dual to a node (located inside it),
and each of the faces is dual to a link, with the face shared by the
two cells dual to the link connecting the two nodes. This is further
illustrated by Figure \ref{fig:3P1MiddleLink}.}
\end{figure}

We will write:
\begin{itemize}
\item $c=\left(s_{1},\ldots,s_{n}\right)$ to indicate that the boundary
of the cell $c$ is composed of the $n$ sides $s_{1},\ldots,s_{n}$.
\item $s=\left(e_{1},\ldots,e_{n}\right)$ to indicate that the boundary
of the side $s$ is composed of the $n$ edges $e_{1},\ldots,e_{n}$.
\item $s=\left(cc'\right)$ to indicate that the side $s$ is shared by
the two cells $c,c'$.
\item $s^{*}=\left(cc'\right)^{*}$ to indicate that the link $s^{*}$ (dual
to the side $s$) connects the two nodes $c^{*}$ and $c^{\prime*}$
(dual to the cells $c,c'$).
\item $e=\left(c_{1},\ldots,c_{n}\right)$ to indicate the the edge $e$
is shared by the $n$ cells $c_{1},\ldots,c_{n}$.
\item $e=\left(s_{1},\ldots,s_{n}\right)$ to indicate the the edge $e$
is shared by the $n$ sides $s_{1},\ldots,s_{n}$.
\item $e=\left(vv'\right)$ to indicate that the edge $e$ connects the
two vertices $v,v'$.
\end{itemize}
This construction is illustrated in Figure \ref{fig:3P1TwoTetras}.

\begin{figure}[H]
\begin{centering}
\includegraphics[width=1\textwidth]{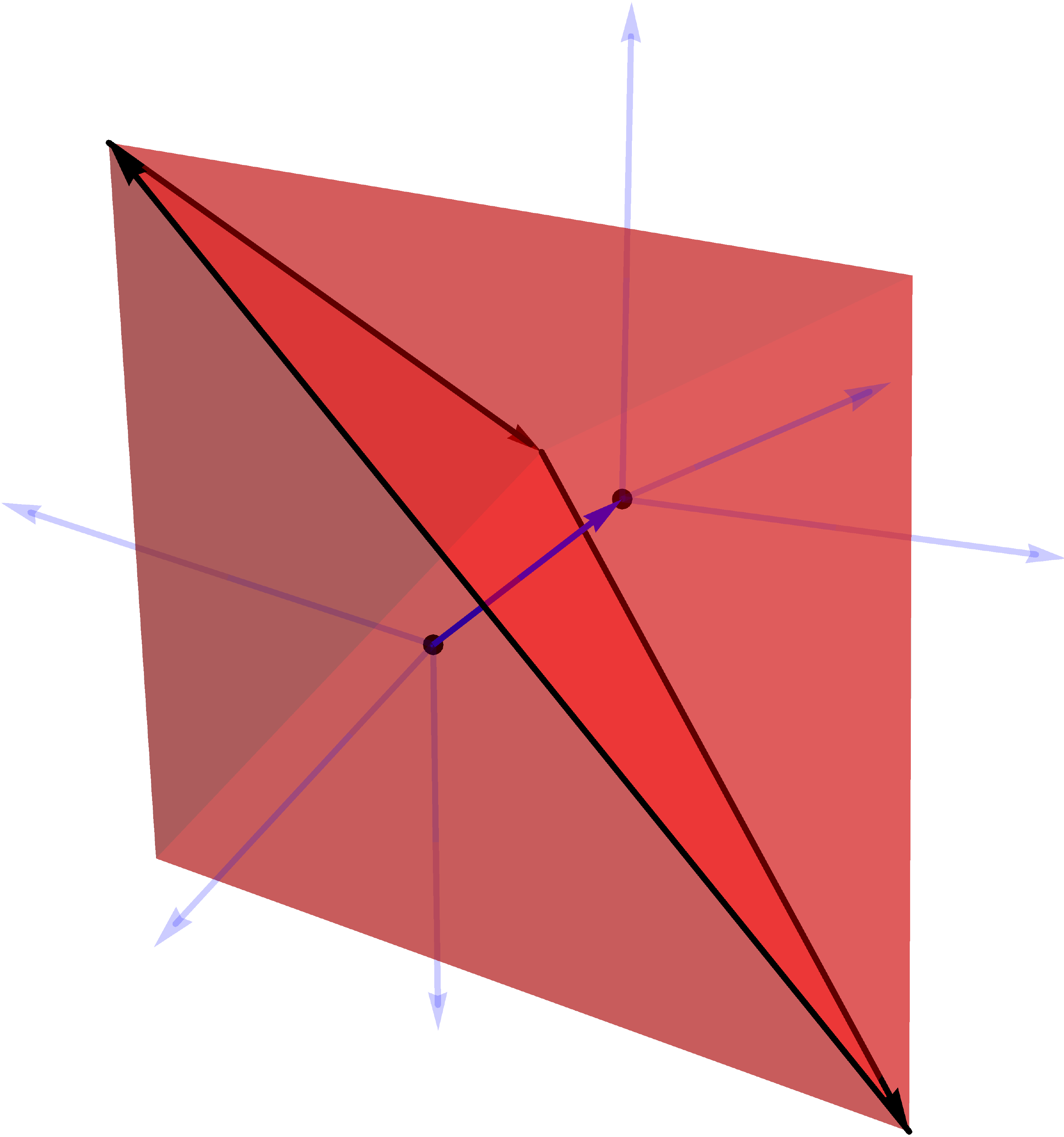}
\par\end{centering}
\caption{\label{fig:3P1MiddleLink}For clarity, we have highlighted the middle
link, the two nodes it connects, the (triangular) face shared by the
two cells, and the edges on the boundary on that face (which are,
as illustrated, oriented) -- and dimmed everything else.}
\end{figure}

In the 3+1-dimensional case, it would make sense to generalize the
disks we defined in the 2+1-dimensional case to cylinders which surround
the edges. This construction is left for future work; in this thesis,
we will not worry about regularizing the singularities, and instead
just use holonomies to probe the curvature and torsion in an indirect
manner, as will be shown below.

\subsection{Truncating the Geometry to the Edges}

The connection and frame field inside each cell $c$ -- that is,
on the \emph{interior }of the cell, but \textbf{not }on the edges
and vertices -- are taken to be
\[
\A\bl_{c}=h_{c}^{-1}\d h_{c}\sp\ee\bl_{c}=h_{c}^{-1}\d\x_{c}h_{c}\soosp\E\bl_{c}=\hf\left[\ee,\ee\right]\bl_{c}=h_{c}^{-1}\left[\d\x_{c},\d\x_{c}\right]h_{c},
\]
in analogy with the 2+1D case. Since they correspond to
\[
\F\equiv\d_{\A}\A=0\sp\P\equiv\d_{\A}\ee=0,
\]
they trivially solve all of the constraints:
\[
\left[\ee,\P\right]=\left[\ee,\F\right]=\ee\cdot\F-\hf\left(\frac{1}{\gamma}+\gamma\right)\K\cdot\P=0.
\]
We also impose that $\F$ and $\P$ are non-zero only on the edges:
\begin{equation}
\F=\sum_{e}\p_{e}\delta\left(e\right)\sp\T=\sum_{e}\j_{e}\delta\left(e\right),\label{eq:F,T}
\end{equation}
where $\delta\left(e\right)$ is a 2-form delta function such that
for any 1-form $f$
\[
\int_{\Sigma}f\wedge\delta\left(e\right)=\int_{e}f,
\]
as discussed in Chapter \ref{sec:Cosmic-Strings-in}, and $\p_{e}$
and $\j_{e}$ are constant algebra elements encoding the curvature
and torsion, respectively, on each edge\footnote{We absorbed the factor of $2\pi$ into $\p_{e}$ and $\j_{e}$ for
brevity.}. These distributional curvature and torsion describe a network of
\emph{cosmic strings}: 1-dimensional topological defects carrying
curvature and torsion in an otherwise flat spacetime. The expressions
(\ref{eq:F,T}) are derived from first principles in Chapter \ref{sec:Cosmic-Strings-in}.

This construction describes a \emph{piecewise-flat-and-torsionless
geometry}; the cells are flat and torsionless, and the curvature and
torsion are located only on the edges of the cells. We may interpret
the 1-skeleton $\Gamma$, the set of all edges in the cellular decomposition
$\Delta$, as a network of cosmic strings.

The reason for considering this particular geometry comes from the
assumption that the geometry can only be probed by taking loops of
holonomies along the spin network. Imagine a 3-dimensional slice $\Sigma$
with arbitrary geometry. We first embed a spin network $\Gamma^{*}$,
which can be any graph, in $\Sigma$. Then we draw a dual graph, $\Gamma$,
such that each edge of $\Gamma$ passes through exactly one loop of
$\Gamma^{*}$. We take a holonomy along each of the loops of $\Gamma^{*}$,
and encode the result on the edges of $\Gamma$. The resulting discrete
geometry is exactly the one we described above, and it is completely
equivalent to the continuous geometry with which we started, since
the holonomies along the spin network cannot tell the difference between
the continuous geometry and the discrete one.

In short, given a choice of a particular spin network, an arbitrary
continuous geometry may be reduced to an equivalent discrete geometry,
given by a network of cosmic strings, one for each loop of the spin
network.

\section{\label{sec:Discretizing-the-Symplectic}Discretizing the Symplectic
Potential}

\subsection{First Step: From Continuous to Discrete Variables}

We start with the symplectic potential obtained in (\ref{eq:symplectic-potential}),
\[
\Theta=-\int_{\Sigma}\E\cdot\delta\A.
\]
Using the identity
\[
\delta\A\bl_{c}=h_{c}^{-1}\left(\d\De h_{c}\right)h_{c},
\]
the potential becomes
\[
\Theta=-\sum_{c}\int_{c}\left[\d\x_{c},\d\x_{c}\right]\cdot\d\De h_{c}.
\]
To use Stokes' theorem in the first integral, we note that
\[
\left[\d\x_{c},\d\x_{c}\right]\cdot\d\De h_{c}=\d\left(\left[\d\x_{c},\d\x_{c}\right]\cdot\De h_{c}\right)=\d\left(\left[\x_{c},\d\x_{c}\right]\cdot\d\De h_{c}\right),
\]
hence we can write
\[
\left[\d\x_{c},\d\x_{c}\right]\cdot\d\De h_{c}=\left(1-\lambda\right)\d\left(\left[\d\x_{c},\d\x_{c}\right]\cdot\De h_{c}\right)+\lambda\d\left(\left[\x_{c},\d\x_{c}\right]\cdot\d\De h_{c}\right),
\]
so
\[
\Theta=-\sum_{c}\int_{\partial c}\left(\left(1-\lambda\right)\left[\d\x_{c},\d\x_{c}\right]\cdot\De h_{c}+\lambda\left[\x_{c},\d\x_{c}\right]\cdot\d\De h_{c}\right).
\]
As in the 2+1D case, we have a family of polarizations corresponding
to different values of $\lambda\in\left[0,1\right]$.

\subsection{Second Step: From Cells to Sides}

Next we decompose the boundary $\partial c$ of each cell $c=\left(s_{1},\ldots,s_{n}\right)$
into sides $s_{1},\ldots,s_{n}$. Each side $s=\left(cc'\right)$
will have exactly two contributions, one from the cell $c$ and another,
with opposite sign, from the cell $c'$. We thus rewrite $\Theta$
as 
\[
\Theta=-\sum_{\left(cc'\right)}\int_{\left(cc'\right)}\left(I_{c'}-I_{c}\right),
\]
where
\[
I_{c}\equiv\left(1-\lambda\right)\left[\d\x_{c},\d\x_{c}\right]\cdot\De h_{c}+\lambda\left[\x_{c},\d\x_{c}\right]\cdot\d\De h_{c}.
\]
Now, in the 3+1D case we have the continuity conditions, derived in
the same way as in Section \ref{subsec:The-Continuity-Conditions}:
\[
h_{c'}=h_{c'c}h_{c}\sp\x_{c'}=h_{c'c}(\x_{c}-\x_{c}^{c'})h_{cc'},
\]
\[
\De h_{c'}=\De\left(h_{c'c}h_{c}\right)=h_{c'c}\left(\De h_{c}-\De h_{c}^{c'}\right)h_{cc'},
\]
\[
\d\De h_{c'}=h_{c'c}\d\De h_{c}h_{cc'},
\]
\[
\d\x_{c'}=\d\left(h_{c'c}(\x_{c}-\x_{c}^{c'})h_{cc'}\right)=h_{c'c}\d\x_{c}h_{cc'},
\]
where all of the conditions are valid only on the side $s=\left(cc'\right)$.
Using these conditions, we find that
\begin{align*}
I_{c'} & =\left(1-\lambda\right)\left[\d\x_{c'},\d\x_{c'}\right]\cdot\De h_{c'}+\lambda\left[\x_{c'},\d\x_{c'}\right]\cdot\d\De h_{c'}\\
 & =\left(1-\lambda\right)\left[\d\x_{c},\d\x_{c}\right]\cdot\left(\De h_{c}-\De h_{c}^{c'}\right)+\lambda\left[\x_{c}-\x_{c}^{c'},\d\x_{c}\right]\cdot\d\De h_{c}.
\end{align*}
Comparing with $I_{c}$, we see that many terms cancel, and we are
left with
\[
\Theta=\sum_{\left(cc'\right)}\int_{\left(cc'\right)}\left(\left(1-\lambda\right)\left[\d\x_{c},\d\x_{c}\right]\cdot\De h_{c}^{c'}+\lambda\left[\x_{c}^{c'},\d\x_{c}\right]\cdot\d\De h_{c}\right).
\]
Since $\De h_{c}^{c'}$ and $\x_{c}^{c'}$ are constant, we may rewrite
this as
\[
\Theta=\sum_{\left(cc'\right)}\left(\left(1-\lambda\right)\De h_{c}^{c'}\cdot\int_{\left(cc'\right)}\left[\d\x_{c},\d\x_{c}\right]+\lambda\x_{c}^{c'}\cdot\int_{\left(cc'\right)}\left[\d\x_{c},\d\De h_{c}\right]\right).
\]
Now, in order to use Stokes' theorem again, we can write
\[
\left[\d\x_{c},\d\x_{c}\right]=\d\left[\x_{c},\d\x_{c}\right],
\]
and
\[
\left[\d\x_{c},\d\De h_{c}\right]=\d\left[\x_{c},\d\De h_{c}\right]=-\d\left[\d\x_{c},\De h_{c}\right],
\]
which we write, defining an additional polarization parameter $\mu\in\left[0,1\right]$,
as
\[
\left[\d\x_{c},\d\De h_{c}\right]=\left(1-\mu\right)\d\left[\x_{c},\d\De h_{c}\right]-\mu\d\left[\d\x_{c},\De h_{c}\right].
\]
The symplectic potential now becomes
\[
\Theta=\sum_{\left(cc'\right)}\left(\left(1-\lambda\right)\De h_{c}^{c'}\cdot\int_{\partial\left(cc'\right)}\left[\x_{c},\d\x_{c}\right]+\lambda\x_{c}^{c'}\cdot\int_{\partial\left(cc'\right)}\left(\left(1-\mu\right)\left[\x_{c},\d\De h_{c}\right]-\mu\left[\d\x_{c},\De h_{c}\right]\right)\right),
\]
and it describes a two-parameter family of potentials for each value
of $\lambda\in\left[0,1\right]$ and $\mu\in\left[0,1\right]$.

\subsection{\label{subsec:Third-Step:-From}Third Step: From Sides to Edges}

The boundary $\partial s$ of each side $s=\left(e_{1},\ldots,e_{n}\right)$
is composed of edges $e$. Conversely, each edge $e=\left(s_{1},\ldots,s_{N_{e}}\right)$
is part of the boundary of $N_{e}$ different sides, which we label
in sequential order $s_{i}\equiv\left(c_{i}c_{i+1}\right)$ for $i\in\left\{ 1,\ldots,N_{e}\right\} $,
with the convention that $c_{N_{e}+1}$ is the same as $c_{1}$ after
encircling the edge $e$ once. Note that this sequence of sides is
dual to a loop of links $s_{i}^{*}=\left(c_{i}c_{i+1}\right)^{*}$
around the edge $e$. Then we can rearrange the integrals as follows:
\[
\sum_{\left(cc'\right)}\int_{\partial\left(cc'\right)}=\sum_{e}\int_{e}\sum_{i=1}^{N_{e}}.
\]
The potential becomes
\[
\Theta=\sum_{e}\int_{e}\sum_{i=1}^{N_{e}}I_{c_{i}c_{i+1}},
\]
where
\[
I_{c_{i}c_{i+1}}\equiv\left(1-\lambda\right)\De h_{c_{i}}^{c_{i+1}}\cdot\left[\x_{c_{i}},\d\x_{c_{i}}\right]+\lambda\x_{c_{i}}^{c_{i+1}}\cdot\left(\left(1-\mu\right)\left[\x_{c_{i}},\d\De h_{c_{i}}\right]-\mu\left[\d\x_{c_{i}},\De h_{c_{i}}\right]\right).
\]
We would like to perform a final integration using Stokes' theorem.
For this we again need to somehow cancel some elements, as we did
before. However, since there are now $N_{e}$ different contributions,
we cannot use the continuity conditions between each pair of adjacent
cells, since in order to get cancellations, all terms must have the
same base point (subscript).

One option is to choose a particular cell and trace everything back
to that cell. However, this forces us to choose a specific cell for
each edge. A more symmetric solution involves splitting each holonomy
$h_{c_{i}c_{i+1}}$, which goes from from $c_{i}^{*}$ to $c_{i+1}^{*}$,
into two holonomies -- first going from $c_{i}^{*}$ to (some arbitrary
point $e_{0}$ on) $e$ and then back to $c_{i+1}^{*}$, using the
recipe given in Section \ref{subsec:Holonomies-and}: 
\[
h_{c_{i}c_{i+1}}=h_{c_{i}e}h_{ec_{i+1}}\sp\x_{c_{i}}^{c_{i+1}}=\x_{c_{i}}^{e}\oplus\x_{e}^{c_{i+1}}=\x_{c_{i}}^{e}+h_{c_{i}e}\x_{e}^{c_{i+1}}h_{ec_{i}}=h_{c_{i}e}\left(\x_{e}^{c_{i+1}}-\x_{e}^{c_{i}}\right)h_{ec_{i}}.
\]
From this we find that
\[
\De h_{c_{i}}^{c_{i+1}}=h_{c_{i}e}\left(\De h_{e}^{c_{i+1}}-\De h_{e}^{c_{i}}\right)h_{ec_{i}}.
\]
Therefore
\begin{align*}
I_{c_{i}c_{i+1}} & =\left(1-\lambda\right)h_{c_{i}e}\left(\De h_{e}^{c_{i+1}}-\De h_{e}^{c_{i}}\right)h_{ec_{i}}\cdot\left[\x_{c_{i}},\d\x_{c_{i}}\right]+\\
 & \qquad+\lambda h_{c_{i}e}\left(\x_{e}^{c_{i+1}}-\x_{e}^{c_{i}}\right)h_{ec_{i}}\cdot\left(\left(1-\mu\right)\left[\x_{c_{i}},\d\De h_{c_{i}}\right]-\mu\left[\d\x_{c_{i}},\De h_{c_{i}}\right]\right).
\end{align*}
Furthermore, we have the usual continuity conditions between a cell
$c_{i}$ and the edge $e$:
\[
\x_{c_{i}}=h_{c_{i}e}\left(\x_{e}-\x_{e}^{c_{i}}\right)h_{ec_{i}}\sp\d\x_{c_{i}}=h_{c_{i}e}\d\x_{e}h_{ec_{i}},
\]
\[
h_{c_{i}}=h_{c_{i}e}h_{e}\sp\De h_{c_{i}}=h_{c_{i}e}\left(\De h_{e}-\De h_{e}^{c_{i}}\right)h_{ec_{i}}\sp\d\De h_{c_{i}}=h_{c_{i}e}\d\De h_{e}h_{ec_{i}}.
\]
If we had a cylinder around the edge $e$, then these conditions would
have been valid on the boundary between the cylinder and the cell.
However, in the case we are considering here, the cylinder has zero
radius, so these conditions are instead valid on the edge $e$ itself.

Plugging in, we get
\begin{align*}
I_{c_{i}c_{i+1}} & =\left(1-\lambda\right)\left(\De h_{e}^{c_{i+1}}-\De h_{e}^{c_{i}}\right)\cdot\left[\x_{e}-\x_{e}^{c_{i}},\d\x_{e}\right]+\\
 & \qquad+\lambda\left(\x_{e}^{c_{i+1}}-\x_{e}^{c_{i}}\right)\cdot\left(\left(1-\mu\right)\left[\x_{e}-\x_{e}^{c_{i}},\d\De h_{e}\right]-\mu\left[\d\x_{e},\De h_{e}-\De h_{e}^{c_{i}}\right]\right).
\end{align*}
Now we sum over all the terms, and take anything that does not depend
on $i$ out of the sum and anything that is constant out of the integral.
We get
\[
\Theta=\sum_{e}\left(\Theta_{e}+\sum_{i=1}^{N_{e}}\Theta_{e}^{c_{i}c_{i+1}}\right),
\]
where
\begin{align*}
\Theta_{e} & \equiv\left(1-\lambda\right)\int_{e}\left[\x_{e},\d\x_{e}\right]\cdot\sum_{i=1}^{N_{e}}\left(\De h_{e}^{c_{i+1}}-\De h_{e}^{c_{i}}\right)+\\
 & \qquad+\lambda\left(\left(1-\mu\right)\int_{e}\left[\x_{e},\d\De h_{e}\right]-\mu\int_{e}\left[\d\x_{e},\De h_{e}\right]\right)\cdot\sum_{i=1}^{N_{e}}\left(\x_{e}^{c_{i+1}}-\x_{e}^{c_{i}}\right),
\end{align*}
\begin{align*}
\Theta_{e}^{c_{i}c_{i+1}} & \equiv-\left(1-\lambda\right)\left[\x_{e}^{c_{i}},\int_{e}\d\x_{e}\right]\cdot\left(\De h_{e}^{c_{i+1}}-\De h_{e}^{c_{i}}\right)+\\
 & \qquad-\lambda\left(\x_{e}^{c_{i+1}}-\x_{e}^{c_{i}}\right)\cdot\left(\left(1-\mu\right)\left[\x_{e}^{c_{i}},\int_{e}\d\De h_{e}\right]-\mu\left[\int_{e}\d\x_{e},\De h_{e}^{c_{i}}\right]\right).
\end{align*}
Note that $\Theta_{e}$ exists uniquely for each edge, while $\Theta_{e}^{c_{i}c_{i+1}}$
exists uniquely for each combination of edge $e$ and side $\left(c_{i}c_{i+1}\right)$.

\subsection{The Edge Potential}

\begin{figure}[H]
\begin{centering}
\includegraphics[width=1\textwidth]{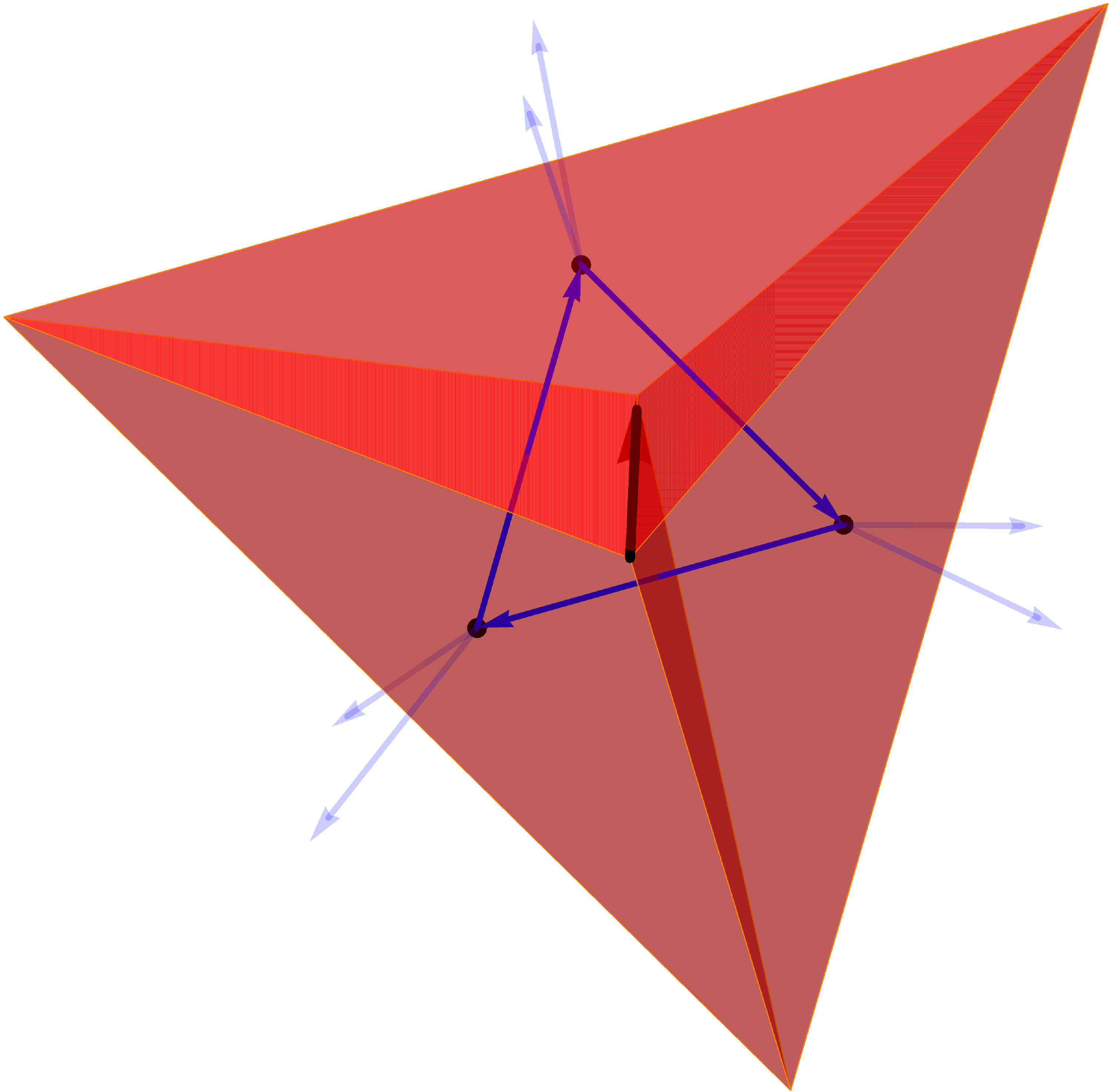}
\par\end{centering}
\caption{\label{fig:3P1Loop}Three cells (tetrahedrons, in red) dual to three
nodes (in black). The cells share three faces (highlighted) which
are dual to three oriented links (blue arrows) connecting the nodes.
The three links form a loop, which goes around the single edge shared
by all three cells (the thick black arrow in the middle). By taking
a holonomy around the loop $\left(c_{1}c_{2}c_{3}c_{1}\right)$, we
can detect the curvature and torsion encoded in the middle edge.}
\end{figure}

In $\Theta_{e}$, we notice that both sums are telescoping -- each
term cancels out one other term, and we are left with only the first
and last term:
\begin{align*}
\sum_{i=1}^{N_{e}}\left(\De h_{e}^{c_{i+1}}-\De h_{e}^{c_{i}}\right) & =\left(\De h_{e}^{c_{2}}-\De h_{e}^{c_{1}}\right)+\left(\De h_{e}^{c_{3}}-\De h_{e}^{c_{2}}\right)+\cdots+\left(\De h_{e}^{c_{N_{e}+1}}-\De h_{e}^{c_{N_{e}}}\right)\\
 & =\De h_{e}^{c_{N_{e}+1}}-\De h_{e}^{c_{1}},
\end{align*}
\begin{align*}
\sum_{i=1}^{N_{e}}\left(\x_{e}^{c_{i+1}}-\x_{e}^{c_{i}}\right) & =\left(\x_{e}^{c_{2}}-\x_{e}^{c_{1}}\right)+\left(\x_{e}^{c_{3}}-\x_{e}^{c_{2}}\right)+\cdots+\left(\x_{e}^{c_{N_{e}+1}}-\x_{e}^{c_{N_{e}}}\right)\\
 & =\x_{e}^{c_{N_{e}+1}}-\x_{e}^{c_{1}}.
\end{align*}
Now, $c_{N_{e}+1}$ is the same as $c_{1}$ after encircling $e$
once. So, if the geometry is completely flat and torsionless, we can
just say that $\Theta_{e}$ vanishes. However, if the edge carries
curvature and/or torsion, then after winding around the edge once,
the rotational and translational holonomies should detect them. This
is illustrated in Figure \ref{fig:3P1Loop}. We choose to label this
as follows:
\begin{equation}
\De h_{e}^{c_{N_{e}+1}}-\De h_{e}^{c_{1}}\equiv\delta\M_{e}\sp\x_{e}^{c_{N_{e}+1}}-\x_{e}^{c_{1}}\equiv\SS_{e}.\label{eq:holo-MS}
\end{equation}
The values of $\M_{e}$ and $\SS_{e}$ in (\ref{eq:holo-MS}) are
directly related\footnote{To find the exact relation, we should regularize the edges using cylinders,
just as we regularized the vertices using disks in the 2+1D case,
which then allowed us to find a relation between the holonomies and
the mass and spin of the particles. We leave this calculation for
future work.} to the values of $\p_{e}$ and $\j_{e}$ in (\ref{eq:F,T}), which
determine the momentum and angular momentum of the string that lies
on the edge $e$. We may interpret (\ref{eq:holo-MS}) in two ways.
Either we first find $\M_{e}$ and $\SS_{e}$ by calculating the difference
of holonomies, as defined in (\ref{eq:holo-MS}), and then define
$\p_{e}$ and $\j_{e}$ in (\ref{eq:F,T}) as functions of these quantities
-- or, conversely, we start with strings that have well-defined momentum
and angular momentum $\p_{e}$ and $\j_{e}$, and then define $\M_{e}$
and $\SS_{e}$ as appropriate functions of $\p_{e}$ and $\j_{e}$.

Unfortunately, aside from this simplification, it does not seem possible
to simplify $\Theta_{e}$ any further, since there is no obvious way
to write the integrands as exact 1-forms. The only thing left for
us to do, therefore, is to call the integrals by names\footnote{Our definition of $\X_{e}$ here alludes to the definition of ``angular
momentum'' in \cite{Freidel:2013bfa}, and is analogous to the ``vertex
flux'' $\X_{v}$ we defined in the 2+1D case in Section \ref{subsec:The-Particle-Potential}.
Similarly, the definition of $\De H_{e}$ (below) is analogous to
the ``vertex holonomy'' $H_{v}$ we defined in the 2+1D case. Also
note that the definitions of $\De H_{e}$, $\De H_{e}^{1}$, and $\De H_{e}^{2}$,
which are 1-forms on field space, define the holonomies $H_{e}$,
$H_{e}^{1}$ and $H_{e}^{2}$ themselves only implicitly; see Footnote
\ref{fn:H_v-foot}.}:
\[
\X_{e}\equiv\int_{e}\left[\x_{e},\d\x_{e}\right]\sp\De H_{e}^{1}\equiv\int_{e}\left[\x_{e},\d\De h_{e}\right]\sp\De H_{e}^{2}\equiv\int_{e}\left[\d\x_{e},\De h_{e}\right],
\]
and write:
\[
\Theta_{e}=\left(1-\lambda\right)\X_{e}\cdot\delta\M_{e}+\lambda\SS_{e}\cdot\left(\left(1-\mu\right)\De H_{e}^{1}-\mu\De H_{e}^{2}\right).
\]
In fact, since both $H_{e}^{1}$ and $H_{e}^{2}$ are conjugate to
the same variable $\SS_{e}$, we might as well collect them into a
single variable:
\[
\De H_{e}\equiv\left(1-\mu\right)\De H_{e}^{1}-\mu\De H_{e}^{2},
\]
so that the choice of parameter $\mu\in\left[0,1\right]$ simply chooses
how much of $H_{e}^{1}$ compared to $H_{e}^{2}$ is used this variable.
We obtain:
\[
\Theta_{e}=\left(1-\lambda\right)\X_{e}\cdot\delta\M_{e}+\lambda\SS_{e}\cdot\De H_{e}.
\]
This term is remarkably similar to the vertex potential we found in
the second line of (\ref{eq:Theta-v-corner}) in the analysis of corner
modes in 2+1D. Indeed, the derivation here is very much analogous
to the derivation of Chapter \ref{subsec:Corner-Modes}. This term
encodes the dynamics of the curvature and torsion on each edge $e$.
In fact, if we perform a change of variables:
\[
\left(1-\lambda\right)\X_{e}\mt\X_{e}\sp\lambda\SS_{e}\mt-\left(\SS_{e}+\left[\M_{e},\X_{e}\right]\right),
\]
we obtain precisely the same term that we obtained in (\ref{eq:Theta_v-simplified}):
\[
\Theta_{e}=\X_{e}\cdot\delta\M_{e}-\left(\SS_{e}+\left[\M_{e},\X_{e}\right]\right)\cdot\De H_{e}.
\]

\subsection{The Link Potential}

The term $\Theta_{e}^{c_{i}c_{i+1}}$, defined at the end of Section
\ref{subsec:Third-Step:-From}, is easily integrable. Since we don't
need the telescoping sum anymore, we can simplify this term by returning
to the original variables:
\[
\x_{e}^{c_{i+1}}-\x_{e}^{c_{i}}=h_{ec_{i}}\x_{c_{i}}^{c_{i+1}}h_{c_{i}e}\sp\De h_{e}^{c_{i+1}}-\De h_{e}^{c_{i}}=h_{ec_{i}}\De h_{c_{i}}^{c_{i+1}}h_{c_{i}e},
\]
so it becomes
\begin{align*}
\Theta_{e}^{c_{i}c_{i+1}} & =-\left(1-\lambda\right)\left[\x_{e}^{c_{i}},\int_{e}\d\x_{e}\right]\cdot h_{ec_{i}}\De h_{c_{i}}^{c_{i+1}}h_{c_{i}e}+\\
 & \qquad-\lambda h_{ec_{i}}\x_{c_{i}}^{c_{i+1}}h_{c_{i}e}\cdot\left(\left(1-\mu\right)\left[\x_{e}^{c_{i}},\int_{e}\d\De h_{e}\right]-\mu\left[\int_{e}\d\x_{e},\De h_{e}^{c_{i}}\right]\right).
\end{align*}
We also have the usual inversion relations (see Section \ref{subsec:The-Cartan-Decomposition})
\[
h_{c_{i}e}\x_{e}^{c_{i}}h_{ec_{i}}=-\x_{c_{i}}^{e}\sp,h_{c_{i}e}\De h_{e}^{c_{i}}h_{ec_{i}}=-\De h_{c_{i}}^{e},
\]
so we can further simplify to:
\begin{align*}
\Theta_{e}^{c_{i}c_{i+1}} & =\left(1-\lambda\right)\left[\x_{c_{i}}^{e},h_{c_{i}e}\left(\int_{e}\d\x_{e}\right)h_{ec_{i}}\right]\cdot\De h_{c_{i}}^{c_{i+1}}+\\
 & \qquad+\lambda\x_{c_{i}}^{c_{i+1}}\cdot\left(\left(1-\mu\right)\left[\x_{c_{i}}^{e},h_{c_{i}e}\left(\int_{e}\d\De h_{e}\right)h_{ec_{i}}\right]+\mu\left[h_{c_{i}e}\left(\int_{e}\d\x_{e}\right)h_{ec_{i}},\De h_{c_{i}}^{e}\right]\right).
\end{align*}
Next, we assume that the edge $e$ starts at the vertex $v$ and ends
at the vertex $v'$, i.e. $e=\left(vv'\right)$. Then we can evaluate
the integrals explicitly:
\[
\int_{e}\d\De h_{e}=\De h_{e}\left(v'\right)-\De h_{e}\left(v\right)\sp\int_{e}\d\x_{e}=\x_{e}\left(v'\right)-\x_{e}\left(v\right).
\]
Now, let $h_{vv'}$ and $\x_{v}^{v'}$ be the rotational and translational
holonomies along the edge $e$, that is, from $v$ to $v'$. Then
we can split them so that they also pass through a point $e_{0}$
on the edge $e$, as follows:
\[
h_{vv'}=h_{ve}h_{ev'}\soosp\De h_{v}^{v'}=h_{ve}\left(\De h_{e}^{v'}-\De h_{e}^{v}\right)h_{ev},
\]
\[
\x_{v}^{v'}=\x_{v}^{e}\oplus\x_{e}^{v'}=h_{ve}\left(\x_{e}^{v'}-\x_{e}^{v}\right)h_{ev},
\]
in analogy with (\ref{eq:vvp-split-h}) and (\ref{eq:vvp-split-x}).
Given that\footnote{Remember that here we are \textbf{not }dealing with dressed holonomies
as we did in the 2+1D case, so $h_{e}\left(e_{0}\right)=1$ and $\x_{e}\left(e_{0}\right)=0$!} $h_{e}\left(v\right)=h_{ev}$ and $\x_{e}\left(v\right)=\x_{e}^{v}$,
the integrals may now be written as
\[
\int_{e}\d\De h_{e}=h_{ev}\De h_{v}^{v'}h_{ve}\sp\int_{e}\d\x_{e}=h_{ev}\x_{v}^{v'}h_{ve}.
\]
Moreover, since $h_{c_{i}e}h_{ev}=h_{c_{i}v}$, we have
\[
h_{c_{i}e}\left(\int_{e}\d\De h_{e}\right)h_{ec_{i}}=h_{c_{i}e}\left(h_{ev}\De h_{v}^{v'}h_{ve}\right)h_{ec_{i}}=h_{c_{i}v}\De h_{v}^{v'}h_{vc_{i}},
\]
\[
h_{c_{i}e}\left(\int_{e}\d\x_{e}\right)h_{ec_{i}}=h_{c_{i}e}\left(h_{ev}\x_{v}^{v'}h_{ve}\right)h_{ec_{i}}=h_{c_{i}v}\x_{v}^{v'}h_{vc_{i}}.
\]
With this, we may simplify $\Theta_{e}^{c_{i}c_{i+1}}$ to
\begin{align*}
\Theta_{e}^{c_{i}c_{i+1}} & =\left(1-\lambda\right)\left[\x_{c_{i}}^{e},h_{c_{i}v}\x_{v}^{v'}h_{vc_{i}}\right]\cdot\De h_{c_{i}}^{c_{i+1}}+\\
 & \qquad+\lambda\x_{c_{i}}^{c_{i+1}}\cdot\left(\left(1-\mu\right)\left[\x_{c_{i}}^{e},h_{c_{i}v}\De h_{v}^{v'}h_{vc_{i}}\right]+\mu\left[h_{c_{i}v}\x_{v}^{v'}h_{vc_{i}},\De h_{c_{i}}^{e}\right]\right).
\end{align*}

\subsection{Holonomies and Fluxes}

Finally, in order to relate this to the spin network phase space,
as we did in the 2+1D case, we need to identify holonomies and fluxes.
From the 2+1D case, we know that the fluxes are in fact also holonomies
-- but they are translational, not rotational, holonomies. $h_{c_{i}c_{i+1}}$
is by definition the rotational holonomy on the link $\left(c_{i}c_{i+1}\right)^{*}$,
and $\x_{c_{i}}^{c_{i+1}}$ is is by definition the translational
holonomy on the link $\left(c_{i}c_{i+1}\right)^{*}$, so it's natural
to simply define
\[
H_{c_{i}c_{i+1}}\equiv h_{c_{i}c_{i+1}}\sp\X_{c_{i}}^{c_{i+1}}\equiv\x_{c_{i}}^{c_{i+1}}.
\]
We should also define holonomies and fluxes on the sides dual to the
links. By inspection, the flux on the side $\left(c_{i}c_{i+1}\right)$
must be
\[
\XXt_{c_{i}}^{c_{i+1}}\equiv\left[\x_{c_{i}}^{v},h_{c_{i}v}\x_{v}^{v'}h_{vc_{i}}\right].
\]
Note that this expression depends only on the source cell $c_{i}$
and not on the target cell $c_{i+1}$, just as the analogous flux
in the 2+1D case only depended on the source cell. This is an artifact
of using the continuity conditions to write everything in terms of
the source cell in order to make the expression integrable (and this
is why we symmetrized the potential in Chapter (\ref{subsec:Corner-Modes})
-- the same can be done here, of course). The first term in the commutator
is $\x_{c_{i}}^{v}$, the translational holonomy from the node $c_{i}^{*}$
to the vertex $v$, the starting point of $e$. The second term contains
$\x_{v}^{v'}$, the translational holonomy along the edge $e$.

As for holonomies on the sides -- again, since we initially had two
ways to integrate, we also have two different ways to define holonomies.
However, as above, since both holonomies are conjugate to the same
flux, $\X_{c_{i}}^{c_{i+1}}$, there is really no reason to differentiate
them. Therefore we just define implicitly:
\[
\De\Ht_{c_{i}}^{c_{i+1}}\equiv\left(1-\mu\right)\left[\x_{c_{i}}^{v},h_{c_{i}v}\De h_{v}^{v'}h_{vc_{i}}\right]+\mu\left[h_{c_{i}v}\x_{v}^{v'}h_{vc_{i}},\De h_{c_{i}}^{v}\right],
\]
and the choice of parameter $\mu\in\left[0,1\right]$ simply determines
how much of this holonomy comes from each polarization. We finally
get:
\[
\Theta_{e}^{c_{i}c_{i+1}}=\left(1-\lambda\right)\XXt_{c_{i}}^{c_{i+1}}\cdot\De H_{c_{i}}^{c_{i+1}}+\lambda\X_{c_{i}}^{c_{i+1}}\cdot\De\Ht_{c_{i}}^{c_{i+1}}.
\]
This is exactly\footnote{Aside from the relative sign, which comes from the fact that in the
beginning we were writing a 3-form instead of a 2-form as an exact
form, and plays no role here since each term describes a separate
phase space.} the same term we obtained in the 2+1D case, (\ref{eq:Theta-corner})!
It represents a holonomy-flux phase space on each link. For $\lambda=0$
the holonomies are on links and the fluxes are on their dual sides,
while for the dual polarization $\lambda=1$ the fluxes are on the
links and the holonomies are on the sides, in analogy with the two
polarization we found in the 2+1D case.

\subsection{Summary}

We have obtained the following discrete symplectic potential:
\[
\Theta=\sum_{e}\left(\left(1-\lambda\right)\X_{e}\cdot\delta\M_{e}+\lambda\SS_{e}\cdot\De H_{e}+\sum_{i=1}^{N_{e}}\left(\left(1-\lambda\right)\XXt_{c_{i}}^{c_{i+1}}\cdot\De H_{c_{i}}^{c_{i+1}}+\lambda\X_{c_{i}}^{c_{i+1}}\cdot\De\Ht_{c_{i}}^{c_{i+1}}\right)\right),
\]
where for each edge $e$:
\begin{itemize}
\item $\left\{ c_{1},\ldots,c_{N_{e}}\right\} $ are the $N_{e}$ cells
around the edge,
\item $\X_{e}\equiv\int_{e}\left[\x_{e},\d\x_{e}\right]$ is the ``edge
flux'',
\item $\M_{e}$, defined implicitly by $\delta\M_{e}\equiv\De h_{e}^{c_{N_{e}+1}}-\De h_{e}^{c_{1}}$,
represents the curvature on the edge,
\item $\De H_{e}\equiv\int_{e}\left(\left(1-\mu\right)\left[\x_{e},\d\De h_{e}\right]-\mu\left[\d\x_{e},\De h_{e}\right]\right)$
is the ``edge holonomy'',
\item $\SS_{e}\equiv\x_{e}^{c_{N_{e}+1}}-\x_{e}^{c_{1}}$ represents the
torsion on the edge,
\item $\XXt_{c_{i}}^{c_{i+1}}\equiv\left[\x_{c_{i}}^{v},h_{c_{i}v}\x_{v}^{v'}h_{vc_{i}}\right]$
is the flux on the side $\left(c_{i}c_{i+1}\right)$ shared by the
cells $c_{i}$ and $c_{i+1}$,
\item $H_{c_{i}c_{i+1}}\equiv h_{c_{i}c_{i+1}}$ is the holonomy on the
link $\left(c_{i}c_{i+1}\right)^{*}$ dual to the side $\left(c_{i}c_{i+1}\right)$,
\item $\X_{c_{i}}^{c_{i+1}}\equiv\x_{c_{i}}^{c_{i+1}}$ is the flux on the
link $\left(c_{i}c_{i+1}\right)^{*}$,
\item $\Ht_{c_{i}}^{c_{i+1}}$, defined implicitly by $\De\Ht_{c_{i}}^{c_{i+1}}\equiv\left(1-\mu\right)\left[\x_{c_{i}}^{v},h_{c_{i}v}\De h_{v}^{v'}h_{vc_{i}}\right]+\mu\left[h_{c_{i}v}\x_{v}^{v'}h_{vc_{i}},\De h_{c_{i}}^{v}\right]$,
is the holonomy on the side $\left(c_{i}c_{i+1}\right)$.
\end{itemize}
We interpret this as the phase space of a spin network $\Gamma^{*}$
coupled to a network of cosmic strings $\Gamma$, with mass and spin
related to the curvature and torsion.

\section{\label{sec:Conclusions}Conclusions}

\subsection{Summary of Our Results}

In this thesis, we performed a rigorous piecewise-flat-and-torsionless
discretization of classical general relativity in the first-order
formulation, in both 2+1 and 3+1 dimensions, carefully keeping track
of curvature and torsion via holonomies. We showed that the resulting
phase space is precisely that of spin networks, the quantum states
of discrete spacetime in loop quantum gravity, with additional degrees
of freedom called edge modes, which result from the discretization
itself and possess their own unique symmetries.

The main contributions of this work are as follows.

\textbf{1. It establishes, for the first time, a rigorous proof of
the equivalence between spin networks and piecewise-flat geometries
with curvature and torsion degrees of freedom.}
\begin{quote}
In the 2+1-dimensional case, each node of the spin network is dual
to a 2-dimensional cell, and each link connecting two nodes is dual
to the edge shared by the two corresponding cells. A loop of links
(or a face) is dual to a vertex of the cellular decomposition. These
vertices are the locations where point particles reside, and by examining
the value of the holonomies along the loop dual to a vertex, we learn
about the curvature and torsion induced by the particle at the vertex
by virtue of the Einstein equation.

In the 3+1-dimensional case, the situation is quite similar. Each
node of the spin network is dual to a 3-dimensional cell, and each
link connecting two nodes is dual to the side shared by the two corresponding
cells. A loop of links (or a face) is dual to an edge of the cellular
decomposition. These edges are the locations where strings reside,
and by examining the value of the holonomies along the loop dual to
an edge, we learn about the curvature and torsion induced by the string
at the edge by virtue of the Einstein equation.

Equivalently, if we assume that the only way to detect curvature and
torsion is by looking at appropriate holonomies on the loops of the
spin networks, then we may interpret our result as taking some arbitrary
continuous geometry, not necessarily generated by particles or strings,
truncating it, and encoding it on the vertices in 2+1D or edges in
3+1D. The holonomies cannot tell the difference between a continuous
geometry and a singular geometry; they can only tell us about the
total curvature and torsion inside the loop.

As the spin networks are the quantum states of space and geometry
is loop quantum gravity, our results illustrate a precise way in which
these states can be assigned classical spatial geometries.
\end{quote}
\textbf{2. It demonstrates that careful consideration of edge modes
is crucial both for the purpose of this proof and for future work
in the field of loop quantum gravity.}
\begin{quote}
Indeed, in Chapter \ref{sec:The-Constraints-as} we analyzed the symmetries
generated by the constraints and found that the \textbf{entire }symplectic
form, not just the spin network terms but also the edge mode (or particle)
term -- which depends on the extra degrees of freedom $h_{v}\left(v\right)$
and $\x_{v}\left(v\right)$ and the curvature and torsion encoded
in the variables $\M_{v}$ and $\SS_{v}$ -- is invariant under this
transformation.

Furthermore, in Section \ref{subsec:The-Particle-Potential} we saw
that the edge mode term transforms in a well-defined way under both
right translations, corresponding to gauge transformations, and left
translations, corresponding to an additional symmetry which did not
exist in the continuous theory.

One could choose a particular gauge where the edge modes are ``frozen'',
that is, $h_{v}\left(v\right)=1$ and $\x_{v}\left(v\right)=0$. In
this case, the entire particle term vanishes (in the limit where the
disks are shrunk, $v_{0}\to v$). Freezing the edge modes makes exactly
as much sense as any other gauge-fixing: it is convenient, but completely
masks an important symmetry of our theory.
\end{quote}
\textbf{3. It set the stage for collaboration between the loop quantum
gravity community and theoretical physicists working on edge modes
from other perspectives, such as quantum electrodynamics, non-abelian
gauge theories, and classical gravity.}
\begin{quote}
This is especially important because loop quantum gravity has always
been considered somewhat of a ``fringe'' theory. The study of edge
modes has become increasingly popular in recent years among physicists
who work on more ``mainstream'' theories, and interaction between
the loop quantum gravity community and the physicists working on edge
modes would no doubt be mutually beneficial for both communities.
\end{quote}
\textbf{4. It further developed the idea, introduced in \cite{Dupuis:2017otn}
and later expanded in \cite{Teleparallel}, that spin networks have
a dual description related to teleparallel gravity, where gravity
is encoded in torsion instead of curvature degrees of freedom.}
\begin{quote}
In fact, we found a whole spectrum of theories for $\lambda\in\left[0,1\right]$,
with the cases $\lambda=0$ (usual loop gravity), $\lambda=1$ (dual/teleparallel
loop gravity), and possibly $\lambda=1/2$ (Chern-Simons theory) being
of particular importance. We also analyzed the discrete constraints
in detail in both polarizations. Importantly, we have discovered that
this duality exists in both 2+1 and 3+1 dimensions, although we did
not study it in detail in the latter case. The existence of this dual
formulation of loop gravity may open up new avenues of research that
have so far been unexplored, and new ways to tackle long-standing
open problems in loop quantum gravity.
\end{quote}

\subsection{Future Plans}

In this thesis we presented a very detailed analysis of the 2+1-dimensional
toy model, which is, of course, simpler than the realistic 3+1-dimensional
case. This analysis was performed with the philosophy that the 2+1D
toy model can provide deep insights about the 3+1D theory.

Indeed, as we have seen in Parts \ref{part:3P1-cont} and \ref{part:3P1-discrete},
many structures from the 2+1D case, such as the cellular decomposition
and its relation to the spin network, the rotational and translational
holonomies and their properties, and the singular matter sources,
can be readily generalized to the 3+1D with minimal modifications.
Thus, the results of Parts \ref{part:2P1-cont} and \ref{part:2P1-discrete}
should be readily generalizable as well. As we have seen, we indeed
obtain the same symplectic potential in both cases, which is not surprising
-- since we used the same structures in both.

However, the 3+1-dimensional case presents many challenges which would
require much more work, far beyond the scope of this thesis, to overcome.
Here we present some suggestions for possible research directions
in 3+1D. Note that there are also many things one could explore in
the 2+1D case, but we choose to focus on 3+1D since it is the physically
relevant case. Of course, in many cases it would be beneficial to
try introducing new structures (e.g. a cosmological constant) in the
2+1D case first, since the lessons learned from the toy theory may
then be employed in the realistic theory -- as we, indeed, did in
this thesis.

\textbf{1. Proper treatment of the singularities}
\begin{quote}
In the 2+1D case, we carefully treated the 0-dimensional singularities,
the point particles, by regularizing them with disks. This introduced
many complications, but also ensured that our results were completely
rigorous. In the 3+1D case, we skipped this crucial part, and instead
jumped right to the end by assuming the results we had in 2+1D apply
to the 3+1D case as well.

It would be instructive to repeat this in 3+1D and carefully treat
the 1-dimensional singularities, the cosmic strings, by regularizing
them with cylinders. Of course, this calculation will be much more
involved than the one we did in 2+1D, as we now have to worry not
only about the boundary of the disk but about the various boundaries
of the cylinder. In particular, we must also regularize the vertices
by spheres such that the top and bottom of each cylinder start on
the surface of a sphere; this is further necessary in order to understand
what happens at the points where several strings meet.

In attempts to perform this calculation, we encountered many mathematical
and conceptual difficulties, which proved to be impossible to overcome
within the scope of this thesis. Therefore, we leave it to future
work.
\end{quote}
\textbf{2. Proper treatment of the edge modes}
\begin{quote}
In the 2+1D case we analyzed the edge modes in detail, in particular
by studying their role in the symplectic potential in both the continuous
and discrete cases. However, in the 3+1D case we again skipped this
and instead assumed our results from 2+1D still hold. In future work,
we plan to perform a rigorous study of the edge modes in 3+1D, including
their role in the symplectic potential and the new symmetries they
generate.
\end{quote}
\textbf{3. Introducing a cosmological constant}
\begin{quote}
In this thesis, we greatly simplified the calculation in 3+1 dimensions
by imposing that the geometry inside the cells is flat, mimicking
the 2+1-dimensional case. A more complicated case, but still probably
doable within our framework, is incorporating a cosmological constant,
which will then impose that the cells are homogeneously curved rather
than flat. In this case, it would be instructive to perform the calculation
in the 2+1D toy model first, and then generalize it to 3+1D.
\end{quote}
\textbf{4. Including point particles}
\begin{quote}
Cosmic strings in 3+1D have a very similar mathematical structure
to point particles in 2+1D (compare Chapters \ref{sec:Point-Particles-in}
and \ref{sec:Cosmic-Strings-in}). For this reason, we used string-like
defects as our sources of curvature and torsion in 3+1D, which then
allowed us to generalize our results from 2+1D in a straightforward
way. An important, but extremely complicated, modification would be
to allow point particles in 3+1D as well.

More precisely, in 2+1D, we added sources for the curvature and torsion
constraints, which are 2-forms. This is equivalent to adding matter
sources on the right-hand side of the Einstein equation. Since these
distributional sources are 2-form delta functions on a 2-dimensional
spatial slice, they pick out 0-dimensional points, which we interpreted
as a particle-like defects.

In 3+1D, we again added sources for the curvature and torsion, which
in this case are \textbf{not }constraints, but rather imposed by hand
to vanish. Since these distributional sources are 2-form delta functions
on a 3-dimensional spatial slice, they pick out 1-dimensional strings.
What we should actually do is add sources for the three constraints
-- Gauss, vector, and scalar -- which are 3-forms. The 3-form delta
functions will then pick out 0-dimensional points.

However, doing this would introduce several difficulties, both mathematical
and conceptual. Perhaps the most serious problem would be that in
3 dimensions, once cannot place a vertex inside a loop. Indeed, in
2 dimensions, a loop encircling a vertex cannot be shrunk to a point,
as it would have to pass through the vertex. Similarly, in 3 dimensions,
a loop encircling an edge cannot be shrunk to a point without passing
through the edge. Therefore, in these cases it makes sense to say
that the vertex or edge is inside the loop.

However, in 3 dimensions there is no well-defined way in which a vertex
can be said to be inside a loop; any loop can always be shrunk to
a point without passing through any particular vertex. Hence, it is
unclear how holonomies on the loops of the spin network would be able
to detect the curvature induced by a point particle at a vertex. Solving
this problem might require generalizing the concept of spin networks
to allow for higher-dimensional versions of holonomies.
\end{quote}
\textbf{5. Taking Lorentz boosts into account}
\begin{quote}
In 2+1D, we split spacetime into 2-dimensional slices of equal time,
but we left the internal space 2+1-dimensional. The internal symmetry
group was then the full Lorentz group. However, in 3+1D, we not only
split spacetime into 3-dimensional slices of equal time, we did the
same to the internal space as well, and imposed the time gauge $e_{a}^{0}=0$.
The internal symmetry group thus reduced from the Lorentz group to
the rotation group.

Although this 3+1 split of the internal space is standard in 3+1D
canonical loop gravity, one may still wonder what happened to the
boosts, and whether we might be missing something important by assuming
that the variables on each cell are related to those on other cells
only by rotations, and not by a full Lorentz transformation. This
analysis might prove crucial for capturing the full theory of gravity
in 3+1D in our formalism, and in particular, for considering forms
of matter other than cosmic strings.
\end{quote}
\textbf{6. Motivating a relation to teleparallel gravity}
\begin{quote}
In both 2+1D and 3+1D, we found that the discrete phase space carries
two different polarizations. In 2+1D, we motivated an interpretation
where one polarization corresponds to usual general relativity and
the other to teleparallel gravity, an equivalent theory where gravity
is encoded in torsion instead of curvature degrees of freedom. In
the future we plan to motivate a similar relation between the two
polarizations in 3+1D.
\end{quote}
\textbf{7. Analyzing the discrete constraints}
\begin{quote}
In 2+1D, we provided a detailed analysis of the discrete Gauss and
curvature constraints, and the symmetries that they generate. We would
like to provide a similar analysis of the discrete Gauss, vector,
and scalar constraints in the 3+1D case. This will allow us to better
understand the discrete structure we have found, and in particular,
its relation to edge modes symmetries.
\end{quote}
\newpage{}

\bibliographystyle{Utphys}
\phantomsection\addcontentsline{toc}{section}{\refname}\bibliography{Barak-Shoshany-PhD-Thesis}

\end{document}